\documentclass{jfm}%
\usepackage{graphicx}
\usepackage{amssymb}
\usepackage{amsmath}
\usepackage{xcolor}
\usepackage{breqn}
\usepackage{bm}
\usepackage{tikz}
\usepackage[boxed]{algorithm2e}
\usepackage{array}
\usepackage[version=4]{mhchem}
\usepackage{siunitx}
\usepackage{multirow}
\usepackage{longtable}
\usepackage{pdflscape}
\usepackage[mathscr]{euscript}
\usepackage{stackrel}
\usepackage{longtable,tabularx}
\usepackage{hyperref}
\usepackage{titlesec}

\titleformat{\paragraph}
{\normalfont\normalsize\itshape}{\theparagraph}{1em}{}
\titlespacing*{\paragraph}
{0pt}{3.25ex plus 1ex minus .2ex}{1.5ex plus .2ex}

\ifCUPmtlplainloaded \else
\checkfont{eurm10}
\iffontfound
\IfFileExists{upmath.sty}
{\typeout{^^JFound AMS Euler Roman fonts on the system,
using the 'upmath' package.^^J}
}
{\typeout{^^JFound AMS Euler Roman fonts on the system, but you
dont seem to have the}
\typeout{'upmath' package installed. JFM.cls can take advantage
of these fonts,^^Jif you use 'upmath' package.^^J}

}
\else

\fi
\fi
\ifCUPmtlplainloaded \else
\checkfont{msam10}
\iffontfound
\IfFileExists{amssymb.sty}
{\typeout{^^JFound AMS Symbol fonts on the system, using the
'amssymb' package.^^J}
  \let\leq=\leqslant
  \let\geq=\geqslant
}{}
\fi
\fi
\ifCUPmtlplainloaded \else
\IfFileExists{amsbsy.sty}
{\typeout{^^JFound the 'amsbsy' package on the system, using it.^^J}
}
{\providecommand\boldsymbol[1]{\mbox{\boldmath $##1$}}}
\fi

\providecommand{\FMTeXButton}[1]{#1}
\FMTeXButton{\author[N. Sharan and J. Bellan]{Nek Sharan$^{1}$ and Josette Bellan$^{1,2}$}}
\FMTeXButton{\affiliation{$^{1}$California Institute of Technology, Pasadena, CA 91125, USA\\
$^{2}$Jet Propulsion Laboratory, California Institute of Technology, Pasadena,
CA 91109, USA}}

\begin{document}
\title{Investigation of high-pressure turbulent jets using direct numerical simulation}
\author{Nek Sharan$^{1}$\thanks{now at CCS-2, Los Alamos National Laboratory, Los Alamos, NM 87544, USA} and Josette Bellan$^{1,2}$}
\date{\today }
\maketitle

\begin{abstract}
Direct numerical simulations of free round jets at a Reynolds number ($Re_{D}$)
of $5000$, based on jet diameter ($D$) and jet-exit bulk velocity ($U_{e}$),
are performed to study jet turbulence characteristics at
supercritical pressures. The jet consists of Nitrogen ($\mathrm{N_{2}}$) that
is injected into $\mathrm{N_{2}}$ at same temperature. To understand
turbulent mixing, a passive scalar is transported with the flow at unity Schmidt number.
Two sets of inflow conditions that model jets issuing from either a smooth
contraction nozzle (laminar inflow) or a long pipe nozzle (turbulent
inflow) are considered. By changing one parameter at a time, the simulations
examine the jet-flow sensitivity to the thermodynamic condition (characterized in terms of the compressibility factor
($Z$) and the normalized isothermal compressibility), inflow condition, and ambient pressure ($p_{\infty}$) spanning perfect- to real-gas
conditions. The inflow affects flow statistics in the near-field
(containing the potential core closure and the transition region) as well as
further downstream (containing fully-developed flow with self-similar
statistics) at both atmospheric and supercritical $p_{\infty}$. The sensitivity to
inflow is larger in the transition region, where the laminar-inflow jets
exhibit dominant coherent structures that produce higher mean strain rates and
higher turbulent kinetic energy than in turbulent-inflow jets.
Decreasing $Z$ at a fixed supercritical 
$p_{\infty}$ enhances pressure and density fluctuations
(non-dimensionalized by local mean pressure and density, respectively), but the effect on velocity
fluctuations depends also on local flow dynamics. When $Z$ 
is reduced, large mean strain rates in the 
transition region of laminar-inflow jets significantly enhance velocity
fluctuations (non-dimensionalized by local mean velocity) and scalar mixing, whereas the effects 
are minimal in jets from turbulent inflow.

\end{abstract}\vspace{-0.15cm}
%


\begin{keywords} turbulent round jets; high-pressure conditions;
supercritical mixing; direct numerical simulation\vspace{-0.15cm}
\end{keywords}

\section{Introduction\label{sec:Introduction}}

Fuel injection and turbulent mixing in numerous applications, \textit{e.g.}
diesel, gas turbine, and liquid-rocket engines, occur at pressures and
temperatures that may exceed the critical values of injected fuel and
oxidizer. At such high pressure (high $p$), species properties are
significantly different from those at atmospheric $p$. Flow development,
mixed-fluid composition and thermal field evolution under supercritical $p$ is
characterized by strong non-linear coupling among dynamics, transport
properties, and thermodynamics \cite[e.g.][]{okong2002direct,okong2002directAIAA,masi2013multi} 
that influences power generation, soot formation, and thermal efficiency of the engines.

The current state-of-the-art in modeling such flows is considerably more
advanced than the experimental diagnostics that may produce reliable data for model
evaluation under such conditions. Indeed, high-order turbulence statistics at engine-relevant
high-$p$ conditions are difficult to measure and, as of now, remain
unavailable. Table \ref{tab:exp_data} lists a sample of supercritical round-jet experimental studies
and the flow conditions considered in those experiments. All studies provide
only a qualitative assessment of the jet-flow turbulence, highlighting the
challenge of obtaining high-fidelity measurements under these conditions.
Additionally, several input parameters necessary to perform
corresponding numerical simulations are not always reported. Accurate
simulations not only require a careful choice of equation of state,
multi-species mass and thermal diffusion models, and, at high-Reynolds numbers, 
subgrid-scale models, but also a matching inflow and boundary conditions to the 
experiment that are not always available. A large Reynolds number ($Re_{D}$) 
multi-species simulation involves several models, a fact which complicates isolation 
of individual model errors and a reliable study of jet turbulence characteristics. 
Moreover, jet turbulence and its sensitivity to flow parameters at
supercritical conditions is not well understood even in a simple
single-species setting. Indeed, previous high-$p$ studies mostly examined 
temporal shear layer configurations \cite[e.g.][]{okong2002direct,okong2002directAIAA,masi2013multi,sciacovelli2019influence}. 
A few studies of spatially evolving turbulent jets \cite[e.g.][]{gnanaskandan2017numerical,gnanaskandan2018side} 
have focused on large-eddy simulation (LES) modeling and on direct numerical simulation (DNS) of binary-species diffusion, 
but did not address the influence of ambient pressure and
thermodynamic departure from perfect gas on jet turbulence.
The present study fills that void by performing direct
numerical simulations of single-species round jets at various ambient (chamber) pressure
($p_{\infty}$), compressibility factor ($Z$) and inflow conditions.

Effects of (dynamics-based) compressibility, defined in terms of various (convective, turbulence, 
gradient, deformation) Mach numbers, on free-shear flows
have been investigated at perfect-gas conditions in numerous
studies, e.g. \cite{papamoschou1988compressible}, \cite{lele1994compressibility}, \cite{vreman1996compressible}, \cite{freund2000compressibility} and \cite{pantano2002study}.
In general, an increase in this compressibility, referred to here as dynamic 
compressibility, is associated with reduced
turbulence kinetic energy (t.k.e.) and reduced momentum-thickness
growth rate in shear layers. The reduction is attributed to decrease
in t.k.e production resulting from reduced pressure fluctuations in
the pressure-strain term \cite[]{vreman1996compressible}. For homogeneous
shear flow, the rapid-distortion-theory results of \cite{simone1997effect}
showed that the t.k.e. change with dynamic compressibility depends
on a non-dimensional time based on the mean strain rate. These studies
also found that dynamic compressibility influences t.k.e. largely by altering
the `structure' of turbulence and less so by the dilatational terms
in the t.k.e. equation. Real-gas effects at high pressure introduce
a different type of compressibility, a thermodynamics-based compressibility
characterized by
\begin{equation}
Z\equiv\frac{p}{\left(\rho R_{\mathrm{u}}T/m\right)},\label{eq:comp_factor}
\end{equation}
where $\rho$ is the density, $T$ denotes the temperature, $R_{u}$
is the universal gas constant and $m$ is the species molar mass. Unlike
non-dimensional parameters in fluid dynamics, such as the Reynolds, Prandtl and
Schmidt numbers which measure the relative importance of two different physical
phenomena, $Z$ measures the physical effects of intermolecular forces
and finite volume of gas molecules. In this study, using $Z$ as one
of the important non-dimensional thermodynamic parameters, the effects of thermodynamic 
compressibility on jet spread rate and t.k.e. production are examined to determine the
physical mechanism by which changes in $Z$ influence jet-flow turbulence.

\addtocounter{table}{-1}
\begin{landscape}

\begin{center}
\begin{table*}
\begin{centering}
\begin{longtable}{>{\centering}p{5.75cm}>{\centering}p{4cm}ccccc>{\centering}p{2.5cm}}
\hline 
Reference & Species 

(injected + chamber) & $U_{e}$ (m/s) & $T_{\mathrm{r,ch}}$ & $T_{\mathrm{r,inj}}$ & $P_{\mathrm{r,ch}}$ & $P_{\mathrm{r,inj}}$ & $Re_{D}$ $\times10^{3}$ $\left(\rho_{e}U_{e}D/\mu\right)$\tabularnewline
\hline 
\hline 
\cite{newman1971behavior} & $\mathrm{L}\mathrm{CO_{2}}+\mathrm{CO_{2}}/\mathrm{N}_{2}$ & 2.0 - 4.0 & 0.97 - 1.09 & 0.97 & 0.86 - 1.23 & NA & $\sim20$ - $30$\tabularnewline
\hline 
\cite{woodward1996raman} & $\mathrm{LN}_{2}+\mathrm{N}_{2}/\mathrm{He}$ & $\sim1.8$ - $2.2$ & 2.21 - 2.46 & 0.70 - 0.91 & 0.83 - 2.03 & NA & $3.4$ - $4.1$\tabularnewline
\hline 
\multirow{3}{5.75cm}{\quad{}\quad{}\quad{} \cite{h1998atomization}} & \multirow{2}{4cm}{\quad{}\quad{}\quad{}\enskip{}$\mathrm{LN}_{2}+\mathrm{N}_{2}$} & 1 & 2.38 & 0.83 & 0.59 - 1.18 & NA & $\sim18$ - $19$\tabularnewline
\cline{3-8} 
 &  & 1.3 & 1.98 & 0.71 & 0.83 - 2.03 & NA & $\sim21$ - $23$\tabularnewline
\cline{2-8} 
 & $\mathrm{LN}_{2}+\mathrm{He}$ & 1.7 & \textcolor{blue}{2.31} & 0.66 & \textcolor{blue}{1.62 - 2.44} & NA & $\sim23$ - $24$\tabularnewline
\hline 
\cite{oschwald1999supercritical} & $\mathrm{LN}_{2}+\mathrm{N}_{2}$ & 5.0 - 20.0 & 2.36 & 0.79 - 1.11 & 1.17 - 1.76 & NA & $115$ - $340$ \tabularnewline
\hline 
\cite{chehroudi2002visual} & $\mathrm{LN}_{2}+\mathrm{N}_{2}$ & 10.0 - 15.0 & 2.38 & 0.71 - 0.87 & 0.23 - 2.74 & NA & $25$ - $75$\tabularnewline
\hline 
\cite{mayer2003raman} & $\mathrm{LN}_{2}+\mathrm{N}_{2}$ & 1.8 - 5.4 & 2.36 & 1.0 - 1.11 & 3.95 - 5.98 & NA & $\sim47$ - $157$\tabularnewline
\hline 
\cite{segal2008subcritical} & Fluoroketone + $\mathrm{N}_{2}$ & 7.0 - 25.0 & \textcolor{blue}{0.66 - 1.07}  & 0.68 - 1.28 & \textcolor{blue}{0.05 - 1.86} & 0.2 - 2.2 & $11$ - $42$\tabularnewline
\hline 
\cite{roy2013disintegrating} & Fluoroketone + $\mathrm{N}_{2}$ & 7.07 - 30.0 & \textcolor{blue}{0.69 - 1.09}  & 1.0 - 1.31 & \textcolor{blue}{1.26 - 1.88} & 1.34 - 1.98 & NA\tabularnewline
\hline 
\multirow{1}{5.75cm}{\quad{}\quad{}\quad{} \cite{falgout2015evidence}} & \multirow{1}{4cm}{\quad{}\quad{}\enskip{}Dodecane + Air} & NA & \textcolor{blue}{0.7 \& 1.4} & 0.55 & \textcolor{blue}{1.6 \& 3.2} & 82.55 & NA\tabularnewline
\hline 
\multirow{2}{5.75cm}{\cite{muthukumaran2016initial,muthukumaran2016mixing}} & Fluoroketone + $\mathrm{N}_{2}$ & 0.86 - 7.5 & \textcolor{blue}{0.82 - 1.03} & 0.99 - 1.07 & \textcolor{blue}{0.81 - 1.34} & NA & NA\tabularnewline
\cline{2-8} 
 & Fluoroketone + $\mathrm{He}$ & 0.82 - 19.0 & \textcolor{blue}{0.82 - 1.05} & 0.98 - 1.07 & \textcolor{blue}{0.72 - 1.34} & NA & NA\tabularnewline
\hline 
\multirow{3}{5.75cm}{\quad{}\quad{}\cite{baab2016speed,baab2018quantitative}} & n-hexane + $\mathrm{N}_{2}$ & \textasciitilde{} 91 & \textcolor{blue}{0.58} & 1.24 & \textcolor{blue}{1.65} & 1.81 & $120$\tabularnewline
\cline{2-8} 
 & n-pentane + $\mathrm{N}_{2}$ & 76 \& 96 & \textcolor{blue}{0.63} & 1.28 \& 1.13 & \textcolor{blue}{1.48} & 1.62 \& 1.61 & $121$ - $139$\tabularnewline
\cline{2-8} 
 & Fluoroketone + $\mathrm{N}_{2}$ & 41 \& 72 & \textcolor{blue}{0.67} & 1.13 & \textcolor{blue}{1.34 \& 2.11} & 2.11 & $172$ - $272$\tabularnewline
\hline 
\multirow{2}{5.75cm}{\quad{}\quad{} \cite{poursadegh2017fuel}} & Propane + $\mathrm{N}_{2}$ & NA & \textcolor{blue}{0.9 - 1.35} & 0.9 - 0.93 & \textcolor{blue}{0.7 - 1.18} & 4.7 & NA\tabularnewline
\cline{2-8} 
 & Propane + $\mathrm{N}_{2}$ & NA & \textcolor{blue}{1.35} & 1.06 & \textcolor{blue}{1.3} & 4.7 & NA\tabularnewline
\hline 
\multirow{2}{5.75cm}{\quad{}\quad{}\quad{} \cite{gao2019injection}} & RP-3 kerosene + Air & NA & \textcolor{blue}{0.45} & 0.96 - 1.17 & \textcolor{blue}{0.042} & 0.84 - 1.88 & NA\tabularnewline
\cline{2-8} 
 & $\mathrm{N}_{2}$ + Air & $\sim254.8$ - $2374.2$ & \textcolor{blue}{2.28} & 4.91 - 6.02 & \textcolor{blue}{0.029} & 0.59 - 1.32 & $\sim87.7$ - $341$\tabularnewline
\hline 
\end{longtable}
\par\end{centering}

\caption{High-pressure round jet experimental studies. $U_{e}=$ jet-exit bulk
velocity, $T_{\mathrm{r,ch}}=$ chamber reduced temperature, $T_{\mathrm{r,inj}}=$
injectant reduced temperature, $P_{\mathrm{r,ch}}=$ chamber reduced
pressure, $P_{\mathrm{r,inj}}=$ injectant reduced pressure, $\rho_{e}=$
jet-exit (or injectant) fluid density, NA = not available. $\sim$ denotes values
not provided in the reference but deduced from other parameters. Numbers
in \textcolor{blue}{blue} denote reduced chamber conditions based
on injectant critical temperature and pressure. \label{tab:exp_data}}
\end{table*}

\par\end{center}

\end{landscape}

Turbulent free-shear flow computations are sensitive to the choices
of initial/inflow conditions, domain size and numerical discretization
\cite[]{balaras2001self,mattner2011large,sharan2018mixing}. In particular,
several experimental \cite[e.g.][]{wygnanski1986large,slessor1998turbulent,mi2001influence}
and computational \cite[e.g.][]{ghosal1997numerical,boersma1998numerical,grinstein2001vortex}
studies have observed near- as well as far-field flow sensitivity
to inflow conditions, supporting the theoretical arguments of \cite{george1989self}
on existence of various self-similar states determined by the initial/inflow
condition. Experimental jet-flow studies typically use a smooth contraction
nozzle or a long straight pipe to initialize jet flows \cite[]{mi2001influence}.
The smooth contraction nozzle produces a laminar inflow with `top-hat'
velocity profile, whereas the long straight pipe produces a fully-developed
turbulent inflow. Both inflow cases are studied here, first,
to examine the sensitivity of presumably existing self-similar states to thermodynamic 
conditions and, second, to determine how the effects
of $p_{\infty}$ and $Z$ are influenced by inflow change.
While it is
well-known that perfect-gas jets attain a self-similar state, the
equivalent information for compressible real-gas jets is unclear. Additionally,
conclusions from the studies of inflow effects on incompressible jets
\cite[e.g.][]{boersma1998numerical} need not necessarily extend to compressible jets, and
therefore, the inflow effects on compressible real-gas jets is explored in this study.

The present study addresses both perfect-gas jet flows, for which theoretical
\cite[e.g.][]{morris1983viscous,michalke1984survey} and experimental \cite[e.g.][]{wygnanski1969some,panchapakesan1993turbulence,hussein1994velocity}
results exist, and high-$p$ supercritical jets, for which
detailed turbulence statistics similar to those of perfect-gas jets do
not exist, as discussed above. Accurate high-$p$ numerical simulations that correctly
account for the non-linear coupling of thermodynamic variables with
mass and thermal diffusion are challenging. \cite{masi2013multi}
used a multi-species model \cite[previously proposed by][]{okong2002direct} to account for these non-linear effects
and used the model for DNS of temporal
mixing layers. The present study applies that model to single-species
spatially-developing jet flows, as a precursor to multi-species jet
simulations. The results from this study provide a database to compare
and contrast turbulence statistics from anticipated high-$p$ multi-species
jet calculations and to initiate studies to validate LES models
for supercritical flows \cite[e.g.][]{TB2010,schmitt2010large,selle2010large,TB2011}.
A recent single-species round jet DNS study \cite[]{ries2017numerical}
examined turbulence statistics and heat transport in a supercritical
cold jet using the low-Mach-number equations that decouple pressure
and density calculation to neglect the acoustic and compressibility effects.
In contrast, the present study solves the fully compressible equations for
jets at a variety of thermodynamic and inflow conditions.

The paper is organized as follows. The governing equations for single-species flow at atmospheric 
and supercritical $p_{\infty}$ are discussed in \S \ref{sec:governing_eqn}. The numerical discretization and computational
setup are described in \S \ref{sub:Numerical-details}.
Details of the boundary conditions and the two inflow conditions considered in this study are
provided in \S \ref{sec:inflow_boundary_conditions}. The results are presented and
discussed in \S \ref{sec:Results}: \S \ref{sub:Case1to4_comparison} provides  an assessment of 
the effects of $p_{\infty}$ and $Z$ at a fixed supercritical $p_{\infty}$ and jet-exit (inflow) 
bulk velocity $U_e$; the influence of $p_{\infty}$ at a fixed $Z$ is examined in 
\S \ref{sub:Case3n5_comparison}; the effects of $p_{\infty}$ and $Z$ at a fixed jet-exit (inflow) 
Mach number $Ma_e$, to distinguish them from the cases with a fixed $U_e$, is investigated in \S \ref{sub:Mach0p6_results}; 
\S \ref{sub:Inflow-effects} evaluates the effects of inflow change at atmospheric and supercritical $p_{\infty}$.
A discussion of the observed results and conclusions are provided
in 
\S \ref{sec:Conclusions}. In addition, a validation of the equation of state and the transport coefficient
models used in this study at high pressures is presented in Appendix \ref{sec:validation_EoS_transport}, a grid convergence study is described in Appendix \ref{sec:Grid-resolution} and a validation of the perfect-gas simulation results against experimental data is discussed in Appendix \ref{sub:Case-1-results}. 

\section{Flow conditions and governing equations \label{sec:governing_eqn}}

\begin{figure}
\begin{centering}
\includegraphics[width=6.9cm]{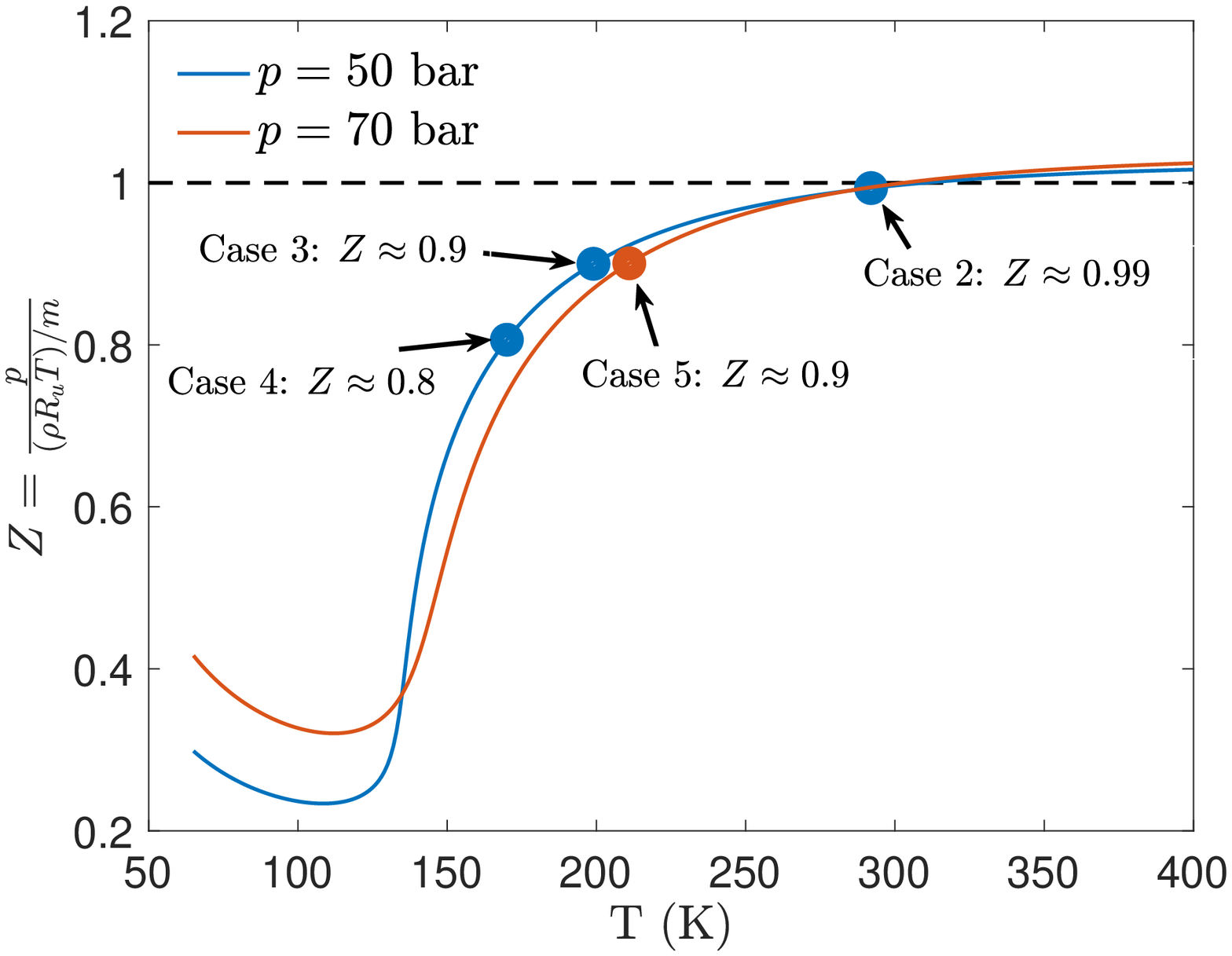}\includegraphics[width=6.9cm]{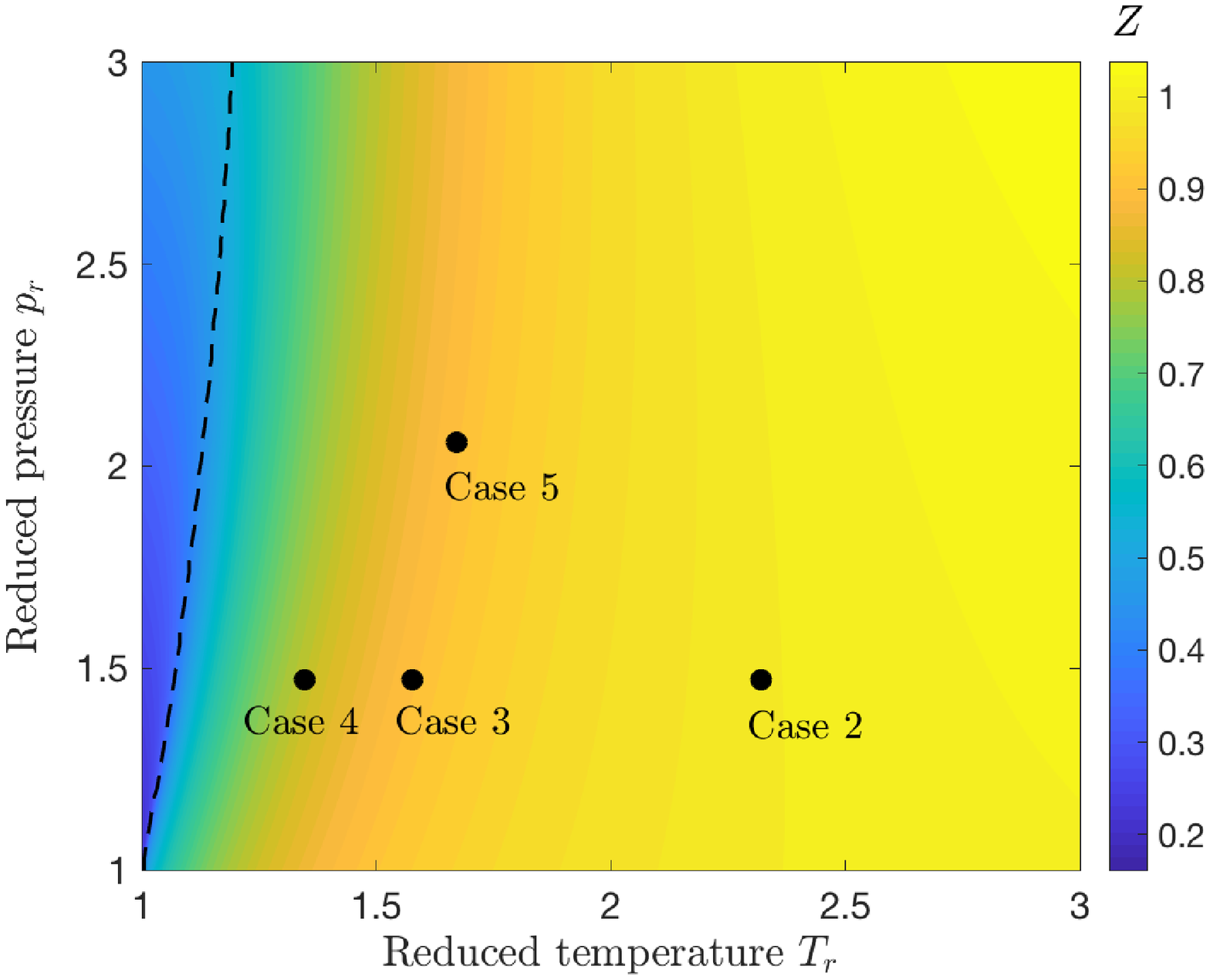}
\par\end{centering}

\begin{centering}
\qquad{}(a)\qquad{}\qquad{}\qquad{}\qquad{}\qquad{}\qquad{}\qquad{}\qquad{}\qquad{}\qquad{}\qquad{}(b)
\par\end{centering}

\caption{(a) Compressibility factor of $\mathrm{N_{2}}$ at $50$ and $70$ bar
pressure. Blue and red markers denote the ambient conditions for various
high-$p$ cases of table \ref{tab:Summary_of_cases}. (b) $p$-$T$ diagram 
of $\mathrm{N_{2}}$ at supercritical conditions. The reduced temperature is $T_r=T/T_c$ 
and the reduced pressure is $p_r=p/p_c$, where the critical temperature $T_c=126.2~$K and 
critical pressure $p_c=33.98~$bar for $\mathrm{N_{2}}$. The contour illustrates the 
distribution of $Z$ and the black dashed line is the Widom line defined as the locus
of the maximum isobaric heat capacity. Black markers denote the location of various
high-$p$ cases of table \ref{tab:Summary_of_cases} on the thermodynamic state plane. \label{fig:comp_factor}}
\end{figure}

Table \ref{tab:Summary_of_cases} summarizes the thermodynamic conditions for
the present numerical simulations. Various flow conditions are considered
to examine influences of high-$p$ thermodynamics
and inflow conditions on round-jet flow statistics. All conditions, simply called \textquotedblleft
cases\textquotedblright, simulate
single-species $\mathrm{N_{2}}$ jets issuing into a quiescent chamber at 
a $Re_{D}$ of 5000. 
In each case, the injected and ambient (chamber) fluid temperature and pressure
have the same value, i.e., the jet injects into a chamber fluid that is as dense
as the injected fluid. Figure \ref{fig:comp_factor}(a) shows $Z$ of pure N$_{2}$ 
for a temperature range at $p=50$ bar and $p=70$ bar with the ambient thermodynamic 
state of various high-$p$ cases denoted by markers. Figure \ref{fig:comp_factor}(b) 
shows the locations of those cases on the supercritical $p$-$T$ diagram of N$_{2}$ 
and their proximity to the Widom line (depicted as dashed black line) which emanates 
from the critical point and divides the supercritical regime into a liquid-like fluid 
at smaller $T_r$ and a gas-like fluid at larger $T_r$ \cite[]{simeoni2010widom,banuti2017similarity}. 
For Case 2, at the ($p_{\infty},T_{\infty}$) conditions, $Z\approx0.994;$ for Case 3, $Z\approx0.9;$
while for Case 4, $Z\approx0.8$, thus, representing significant departure from 
perfect-gas behavior. Cases 2 to 4 investigate the effect of $Z$; Case 2 is furthest from the Widom line, whereas Case 4 is the closest. Case 5 
compared against Case 3 examines the influence of $p_{\infty}$ at constant $Z$. To further characterize the real-gas effects in various cases, the values of isothermal compressibility, $\beta_{T}$, and isentropic (or adiabatic) compressibility, $\beta_{S}$, for ambient condition in various cases are listed in table \ref{tab:Isothermal-compressibility}. $\beta_{T}$ and $\beta_{S}$ can be obtained as a function of $Z$ using
\begin{equation}
\beta_{T}=-\frac{1}{V}\left(  \frac{\partial V}{\partial p}\right)  _{T}%
=\frac{1}{p}-\frac{1}{Z}\left(  \frac{\partial Z}{\partial p}\right)  _{T},
\end{equation}%
\begin{equation}
\beta_{S}=-\frac{1}{V}\left(  \frac{\partial V}{\partial p}\right)  _{S}%
=\frac{\beta_{T}}{\gamma}%
\end{equation}
where $V$ is the volume, $\gamma$ denotes the ratio of
the heat capacity at constant pressure to the heat capacity at constant volume
and $S$ denotes the entropy. $\beta_{T}$ and $\beta_{S}$ are dimensional quantities with units of inverse of pressure, and a direct
comparison of their values across various cases turns out not to be very informative. However, for a perfect gas, $\beta_{T}=1/p$, and the real-gas effect at ambient conditions may be isolated from $\beta_{T}$ by
examining $\beta_{T}-1/p_{\infty}$ non-dimensionalized using $p_{\infty}$.
The non-dimensional quantity $p_{\infty}\left(\beta_{T}-1/p_{\infty}\right)$ is listed in table \ref{tab:Isothermal-compressibility} and will be used to explain the pressure/density fluctuations observed among various cases in \S \ref{sec:Results}.

The jet-exit Mach number listed in table \ref{tab:Summary_of_cases} is $Ma_e=U_{e}/c_{\infty}$, where $U_{e}$ is the jet-exit (inflow) bulk velocity and 
$c_{\infty}$ denotes the ambient sound speed. The bulk velocity is formally defined in \S \ref{sec:Inflow-condition}. 
To simulate jets with identical inflow mean velocity for a perturbation type (laminar/turbulent), the same value of $U_{e}$ is used
in Cases 1--5, 1T, 2T and 4T. Thus, the differences in $Ma_e$ across those cases arise from the variation in 
$c_{\infty}$ due to different ambient thermodynamic conditions. To examine the influence of $p_{\infty}$ and $Z$ at a fixed $Ma_e$, Cases 2M and 4M are considered with same (laminar) inflow and ambient thermodynamic conditions as Cases 2 and 4, respectively, but with $U_{e}$ varied to obtain a $Ma_e$ of $0.6$, which is the value used in Case 1.

Cases 1T, 2T and 4T examine the influence of inflow perturbations through
comparisons against Cases 1, 2 and 4, respectively. Numerical results from 
increasingly finer grid resolutions, denoted by $N_{x}\times N_{y}\times N_{z}$, 
are used to ensure grid convergence, as discussed in Appendix \ref{sec:Grid-resolution}.
Results from the finest grid simulation of each case are discussed in \S \ref{sec:Results}.
The significance of factor $\mathcal{F}$ in table \ref{tab:Summary_of_cases} is explained in \S \ref{sec:transport_properties}.

The governing equations are the set of conservation equations and the equation
of state; this equation set is complemented by the transport properties.

\noindent 
\begin{table*}
\begin{centering}
\begin{tabular}{cccccccc}
\hline 
\multirow{2}{*}{Case (description)} & \multirow{2}{*}{$N_{x}\times N_{y}\times N_{z}$} & $p_{\infty}$ & $T_{\mathrm{ch}}$$\left(=T_{\mathrm{inj}}\right)$ & \multirow{2}{*}{$Z$} & \multirow{2}{*}{$\mathcal{F}$} & \multirow{2}{*}{$Ma_e$} & Inflow\tabularnewline
 &  & (bar) & (K) &  &  &  & perturbation\tabularnewline
\hline 
\multirow{3}{*}{1 (atm-$p$) } & $240\times216\times216$ & \multirow{3}{*}{1} & \multirow{3}{*}{293} & \multirow{3}{*}{1.0} & \multirow{3}{*}{6.5} & \multirow{3}{*}{0.6} & \multirow{3}{*}{$0.004U_{e}$ (lam)}\tabularnewline
\cline{2-2} 
 & $320\times288\times288$ &  &  &  &  &  & \tabularnewline
\cline{2-2} 
 & $400\times320\times320$ &  &  &  &  &  & \tabularnewline
\hline 
\multirow{4}{*}{2 (high-$p\left(50\right)$; $Z\approx1$) } & $240\times216\times216$ & \multirow{4}{*}{50} & \multirow{4}{*}{293} & \multirow{4}{*}{0.99} & \multirow{4}{*}{309.4} & \multirow{4}{*}{0.58} & \multirow{4}{*}{$0.004U_{e}$ (lam)}\tabularnewline
\cline{2-2} 
 & $320\times288\times288$ &  &  &  &  &  & \tabularnewline
\cline{2-2} 
 & $400\times320\times320$ &  &  &  &  &  & \tabularnewline
\cline{2-2} 
 & $480\times360\times360$ &  &  &  &  &  & \tabularnewline
\hline 
\multirow{2}{*}{3 (high-$p\left(50\right)$; $Z\approx0.9$) } & $400\times320\times320$ & \multirow{2}{*}{50} & \multirow{2}{*}{199} & \multirow{2}{*}{0.9} & \multirow{2}{*}{641.4} & \multirow{2}{*}{0.73} & \multirow{2}{*}{$0.004U_{e}$ (lam)}\tabularnewline
\cline{2-2} 
 & $480\times360\times360$ &  &  &  &  &  & \tabularnewline
\hline 
\multirow{3}{*}{4 (high-$p\left(50\right)$; $Z\approx0.8$) } & $400\times320\times320$ & \multirow{3}{*}{50} & \multirow{3}{*}{170} & \multirow{3}{*}{0.8} & \multirow{3}{*}{895.7} & \multirow{3}{*}{0.82} & \multirow{3}{*}{$0.004U_{e}$ (lam)}\tabularnewline
\cline{2-2} 
 & $480\times360\times360$ &  &  &  &  &  & \tabularnewline
\cline{2-2} 
 & $560\times408\times408$ &  &  &  &  &  & \tabularnewline
\hline 
\multirow{2}{*}{5 (high-$p\left(70\right)$; $Z\approx0.9$) } & $400\times320\times320$ & \multirow{2}{*}{70} & \multirow{2}{*}{211} & \multirow{2}{*}{0.9} & \multirow{2}{*}{774.1} & \multirow{2}{*}{0.69} & \multirow{2}{*}{$0.004U_{e}$ (lam)}\tabularnewline
\cline{2-2} 
 & $480\times360\times360$ &  &  &  &  &  & \tabularnewline
\hline 
2M (high-$p\left(50\right)$; $Z\approx1$)  & $400\times320\times320$ & 50 & 293 & 0.99 & 309.4 & 0.6 & $0.004U_{e}$ (lam)\tabularnewline
\hline 
4M (high-$p\left(50\right)$; $Z\approx0.8$)  & $480\times360\times360$ & 50 & 170 & 0.8 & 895.7 & 0.6 & $0.004U_{e}$ (lam)\tabularnewline
\hline 
1T (atm-$p$) & $400\times320\times320$ & 1 & 293 & 1.0 & 6.5 & 0.6 & pipe-flow turb\tabularnewline
\hline 
2T (high-$p\left(50\right)$; $Z\approx1$)  & $400\times320\times320$ & 50 & 293 & 0.99 & 309.4 & 0.58 & pipe-flow turb\tabularnewline
\hline 
4T (high-$p\left(50\right)$; $Z\approx0.8$)  & $480\times360\times360$ & 50 & 170 & 0.8 & 895.7 & 0.82 & pipe-flow turb\tabularnewline
\hline 
\end{tabular}
\par\end{centering}

\caption{Summary of the parameters for numerical simulations. The subscripts ``$\mathrm{inj}$'' and ``$\mathrm{ch}$'' denote the injection and chamber conditions,
respectively. $p_{ch}\equiv p_{\infty}$ and $T_{ch}\equiv T_{\infty}$. ``lam" associated with a inflow perturbation denotes laminar conditions. Suffix ``M" in the name
of a case, e.g. 2M and 4M, denotes high-$p$ cases with same Mach number as Case 1. Suffix ``T" in the name of a case, e.g. 1T, 2T and 4T, denotes turbulent inflow cases. \label{tab:Summary_of_cases}}
\end{table*}

\noindent 

\begin{table*}
\begin{centering}
\begin{tabular}{c>{\centering}m{1cm}cc>{\centering}m{1cm}c}
\hline 
\multirow{2}{*}{Case (description)} & \multirow{2}{1cm}{\centering{}$Z$} & $\beta_{T}\times10^{5}$ & \multirow{2}{*}{$p_{\infty}\left(\beta_{T}-\frac{1}{p_{\infty}}\right)$} & \multirow{2}{1cm}{\centering{}$\gamma$} & $\beta_{S}\times10^{5}$\tabularnewline
 &  & $\left(\mathrm{Pa}^{-1}\right)$ &  & & $\left(\mathrm{Pa}^{-1}\right)$ \tabularnewline
\hline 
1 (atm-$p$)  & 1 & 1.00000 & 0.000 & 1.4 & 0.71429\tabularnewline
\hline 
2 (high-$p\left(50\right)$; $Z\approx1$)  & 0.99 & 0.01995 & -0.002 & 1.49 & 0.01339\tabularnewline
\hline 
3 (high-$p\left(50\right)$; $Z\approx0.9$) & 0.9 & 0.02190 & 0.095 & 1.69 & 0.01296\tabularnewline
\hline 
4 (high-$p\left(50\right)$; $Z\approx0.8$)  & 0.8 & 0.02488 & 0.244 & 1.96 & 0.01269\tabularnewline
\hline 
5 (high-$p\left(70\right)$; $Z\approx0.9$)  & 0.9 & 0.01524 & 0.067 & 1.74 & 0.00876\tabularnewline
\hline 
\end{tabular}
\par\end{centering}

\caption{Values of various thermodynamic quantities for ambient conditions in Cases 1 to 5. Cases 2M and 4M have same values as Cases 2 and 4, respectively, and Cases 1T, 2T and 4T have same values as Cases 1, 2 and 4, respectively. \label{tab:Isothermal-compressibility}}
\end{table*}

\vspace{-0.5cm}
\subsection{Conservation equations\label{sec:conservation}}

The compressible flow equations for conservation of mass, momentum, energy, and a passive scalar, solved in this study, are
\begin{equation} 
\frac{\partial\rho}{\partial t}+\frac{\partial}{\partial x_{j}}\left[  \rho u_{j}\right]  =0, \label{eq:eq1}%
\end{equation}%
\begin{equation} 
\frac{\partial}{\partial t}\left(  \rho u_{i}\right)  +\frac{\partial }{\partial x_{j}}\left[  \rho u_{i}u_{j}+p\delta_{ij}-\sigma_{ij}\right]  =0, \label{eq:eq2}%
\end{equation}%
\begin{equation} \frac{\partial}{\partial t}\left(  \rho e_{t}\right)  +\frac{\partial }{\partial x_{j}}\left[  \left(  \rho e_{t}+p\right)  u_{j}-u_{i}\sigma _{ij}+q_{j}\right]  =0, \label{eq:eq3}%
\end{equation}%
\begin{equation} 
\frac{\partial}{\partial t}\left(  \rho \xi\right)  +\frac{\partial }{\partial x_{j}}\left[  \rho \xi\ u_{j}+J_{j}\right]  =0, \label{eq:eq4}%
\end{equation} 
where $t$ denotes the time, $\left(x_{1},x_{2},x_{3}\right)\equiv\left(x,y,z\right)$ are the Cartesian directions, subscripts $i$ and $j$ refer to the spatial coordinates, $u_{i}$ is the velocity, $p$ is the pressure, $\delta_{ij}$ is the Kronecker delta,  $e_{t}=e+u_{i}u_{i}/2$ is the total energy (\textit{i.e.}, internal energy, $e$, plus kinetic energy), $\xi\in\left[0,1\right]$ is a passive scalar transported with the flow, $\sigma_{ij}$ is the Newtonian viscous stress tensor%
\begin{equation} 
\sigma_{ij}=\mu\left(  2S_{ij}-\frac{2}{3}S_{kk}\delta_{ij}\right), \qquad S_{ij}=\frac{1}{2}\left(  \frac{\partial u_{i}}{\partial x_{j}} +\frac{\partial u_{j}}{\partial x_{i}} \right), \label{eq:stress}%
\end{equation}
where $\mu$ is the viscosity, $S_{ij}$ is the strain-rate tensor, and $q_{j}=-\lambda\thinspace\partial T/\partial x_{j}$ and $J_{j}=-\mathscr{D}\thinspace\partial\xi/\partial x_{j}$  are the heat flux and scalar diffusion flux in $j$-direction, respectively. $\lambda$  is the thermal conductivity and $\mathscr{D}=\mu/Sc$ is the scalar diffusivity, where $Sc$ denotes the Schmidt number. The injected fluid is assigned a  scalar value, $\xi$, of 1, whereas the chamber fluid a value of 0. The passive scalar is not a physical species, and is only used as a surrogate quantity to track the injected fluid in this simple single-species flow.

\subsection{Equation of state\label{sec:EOS}}

For the near-atmospheric-$p$ simulations (Cases 1 and 1T), the perfect
gas equation of state (EOS) is applicable, given by 
\begin{equation}
p=\frac{\rho R_{\mathrm{u}}T}{m}.\label{eq:PerfectGas_EOS}%
\end{equation}

For the high-$p$ simulations (Cases 2--5, 2M, 4M, 2T and 4T), the conservation equations 
are coupled with a Peng-Robinson (PR) EOS
\begin{equation}
p=\frac{R_{\mathrm{u}}T}{\left(v_{\mathrm{PR}}-b_{\mathrm{mix}}\right)}-\frac{a_{\mathrm{mix}}}{\left(v_{\mathrm{PR}}^{2}+2b_{\mathrm{mix}}v_{\mathrm{PR}}-b_{\mathrm{mix}}^{2}\right)}.\label{eq:RealGas_EOS}%
\end{equation}
The molar PR volume $v_{\mathrm{PR}}=V-v_{\mathrm{s}}$, where the
molar volume $V=m/\rho$. $v_{\mathrm{s}}$ denotes the volume shift
introduced to improve the accuracy of the PR EOS at high pressures
\cite[]{okong2002directAIAA,harstad1997efficient}. $a_{mix}$ and
$b_{mix}$ are functions of $T$ and the molar fraction $X_{\alpha}$ -- here
$X_{\alpha}=1$ -- and are obtained from expressions previously 
published \cite[Appendix B]{sciacovelli2019influence}.

\subsection{Transport properties\label{sec:transport_properties}}

For the near-atmospheric-$p$ simulations (Cases 1 and 1T), the viscosity is modeled as a power law
\begin{equation}
\mu=\mu_{R}\left(\frac{T}{T_{R}}\right)^{n} \label{eq:mu_1bar}%
\end{equation}
with $n=2/3$ and the reference viscosity being $\mu_{R}=\rho_{e}U_{e}D/Re_{D}$,
where $\rho_{e}$ and $U_{e}$ are the jet-exit fluid density and
jet-exit bulk velocity, respectively, and the reference temperature is $T_{R}=293\thinspace\mathrm{K}$.
The thermal conductivity is $\lambda=\mu C_{p}/\mathrm{Pr}$, where Prandtl
number $\mathrm{Pr}=0.7$ (as typical of 1 bar flows), the ratio of
specific heats $\gamma=1.4$, and the isobaric heat capacity $C_{p}=\gamma R_{\mathrm{u}}/\left(\gamma-1\right)$
is assumed.

For real gases in high-$p$ simulations (Cases 2--5, 2M, 4M, 2T and 4T), the physical viscosity, $\mu_{\mathrm{ph}}$, and thermal conductivity,
$\lambda_{\mathrm{ph}}$, are calculated using the Lucas method \cite[Chapter 9]{poling2001properties}
and the Stiel-Thodos method \cite[Chapter 10]{poling2001properties},
respectively, as a function of the local thermodynamic conditions. 
The computational viscosity, $\mu$, and thermal conductivity,
$\lambda$, are obtained by scaling $\mu_{\mathrm{ph}}$ and
$\lambda_{\mathrm{ph}}$ with a factor $\mathcal{F}=\mu_{R}/\mu_{\mathrm{ph},\infty}$,
\textit{i.e.} $\mu=\mathcal{F}\mu_{\mathrm{ph}}$ and $\lambda=\mathcal{F}\lambda_{\mathrm{ph}}$,
to allow simulations at the specified $Re_{D}$ of 5000.
The ambient physical viscosity ($\mu_{\mathrm{ph},\infty}$) is 
$\mu_{\mathrm{ph}}$ at the pressure $p_{\infty}$ and the temperature
$T_{\mathrm{ch}}$ of respective cases. This procedure ensures that $\mathrm{Pr}$, which 
is computed as a function of the local thermodynamic variables, has the 
physically correct value. The scalar diffusivity is obtained from $\mathscr{D}=\mu/Sc$, 
where unity Schmidt number is assumed in all cases. A validation of the transport and thermodynamic properties 
calculated from the above methods is presented in Appendix \ref{sec:validation_EoS_transport}.

The $\mathcal{F}$ values for all cases
are listed in table \ref{tab:Summary_of_cases}. As an example, for Case 1,
$\mathcal{F}=\mu_{R}/\mu_{\mathrm{ph},\infty}\approx6.5$ ($\mu_{R}=\rho_{e}U_{e}D/\operatorname{Re}%
_{D}=1.136\times10^{-4}\,\mathrm{Pa.s}$ and $\mu_{\mathrm{ph},\infty}%
=1.757\times10^{-5}\,\mathrm{Pa.s}$ at $p_{\infty}=1\,$bar and $T_{\mathrm{ch}}=293\,$K), and for Case 2,
$\mathcal{F}\approx309.4$ ($\mu_{R}=5.715\times10^{-3}\,\mathrm{Pa.s}$ and
$\mu_{\mathrm{ph},\infty}=1.847\times10^{-5}\,\mathrm{Pa.s}$ at $p_{\infty}=50\,$bar and $T_{\mathrm{ch}}=293\,$K).
$\mathcal{F}$ is larger in Case 2 compared to Case 1 because of the larger density
$\rho_{e}$ at $50$ bar that requires a larger $\mu_{R}$ for a fixed
$\operatorname{Re}_{D}$, while the physical viscosity $\mu
_{\mathrm{ph},\infty}$ remains relatively unchanged with increase in $p$.

\section{Numerical aspects\label{sec:computational}}

\subsection{Computational domain and numerical method\label{sub:Numerical-details}}

\begin{figure}
\begin{centering}
\includegraphics[width=15cm]{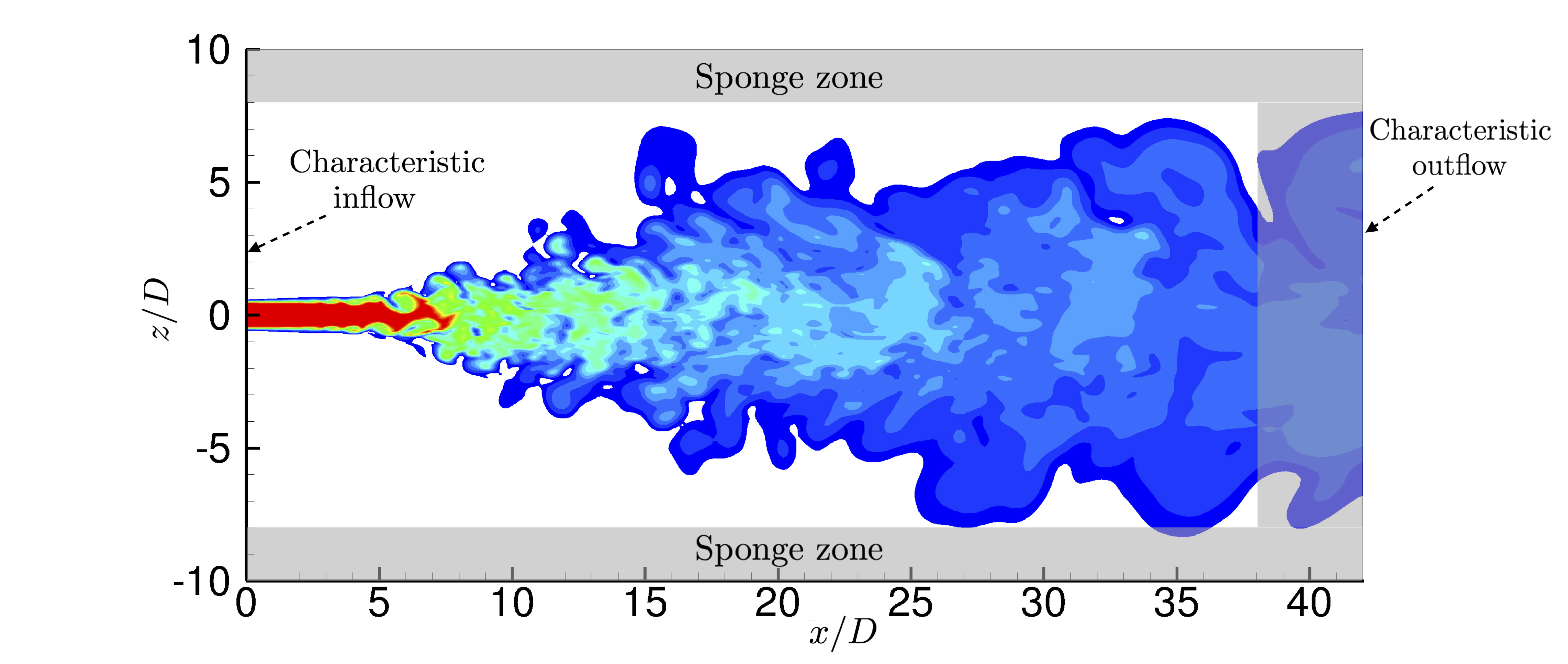}
\par\end{centering}

\caption{A 2D schematic showing the extent of computational domain in axial
and radial direction, and the boundary conditions applied at various
boundaries. \label{fig:jet_2d_schematic_bc}}
\end{figure}

For notation simplicity, $\left(  x_{1},x_{2},x_{3}\right)  \equiv\left(
x,y,z\right)  $ is adopted for axis labels. $\left(u_{1},u_{2},u_{3}\right)$ 
denote the Cartesian velocity components, whereas $\left(u,v,w\right)$ 
denote the axial, radial and azimuthal velocity.
The computational domain extends to $42D$ in the axial ($x$-)direction
and $20D$ in the $y$- and $z$-direction including the sponge zones,
as shown schematically in a $x$-$z$ plane of figure \ref{fig:jet_2d_schematic_bc}. 
The boundary conditions are discussed in \S \ref{sub:boundary_cond}.

Spatial derivatives are approximated using the sixth-order compact
finite-difference scheme and time integration uses the classical explicit 
fourth-order Runge-Kutta method. To avoid
unphysical accumulation of energy at high wavenumbers, resulting
from the use of non-dissipative spatial discretization, the conservative
variables are filtered every five time steps using an explicit eighth-order
filter \cite[]{kennedy1994several}. The derivative approximations and filter operations
over non-uniform stretched grids and polar grids (for post-processing
and inflow generation) uses the generalized-coordinate formulation
\cite[e.g.][]{sharan2016time,sharan2018time}.

To obtain the numerical solution, the conservation
equations are first solved at each time step. With $\rho$ and $e=e_{t}-u_{i}u_{i}/2$ obtained
from the conservation equations and $T$ computed iteratively from $e,$ the EOS
is used to calculate $p$ \cite[]{okong2002directAIAA}.

\noindent 

\subsection{Boundary and inflow
conditions\label{sec:inflow_boundary_conditions}}

\subsubsection{Boundary conditions\label{sub:boundary_cond}}

The outflow boundary in the axial direction
and all lateral boundaries have sponge zones \cite[]{bodony2006analysis}
with non-reflecting outflow Navier-Stokes characteristic boundary
conditions (NSCBC) \cite[]{poinsot1992boundary} at the boundary faces.
Sponge zones at each outflow boundary have a width of $10\%$ of the
domain length normal to the boundary face. The sponge strength at
each boundary decreases quadratically with distance normal to the
boundary. The performance of one-dimensional NSCBC \cite[]{poinsot1992boundary,okong2002consistent}
as well as its three-dimensional extension \cite[]{lodato2008three}
by inclusion of transverse terms were also evaluated without the sponge
zones; they permit occasional spurious reflections into the domain and,
therefore, the use of sponge zones was deemed necessary.

\subsubsection{Inflow conditions\label{sec:Inflow-condition}}

The role of initial/inflow conditions on free-shear flow development
as well as the asymptotic (self-similar) state attained by the flow
at atmospheric conditions is well recognized \cite[]{george2004role,boersma1998numerical,sharan2019turbulent}.
To examine the high-$p$ jet-flow sensitivity to initial conditions,
two types of inflows are considered, portraying either a jet exiting
a smooth contracting nozzle or a jet exiting a long pipe. The former
produces laminar inflow conditions with top-hat jet-exit mean velocity
profile whereas the latter produces turbulent inflow conditions of
fully-developed pipe flow \cite[]{mi2001influence}.

Cases 1--5, 2M and 4M employ laminar inflow conditions with velocity profile
at the inflow plane given by \cite[e.g.][]{michalke1984survey}
\begin{equation}
u(r)=\frac{U_{e}}{2}\left(1-\tanh\left[\frac{r-r_{0}}{2\theta_{0}}\right]\right),\label{eq:lam_inflow_mean_vel}
\end{equation}
where $r=\sqrt{y^{2}+z^{2}}$, the jet exit radius is $r_{0}=D/2$ and
the momentum thickness is $\theta_{0}=0.04r_{0}$. Small random perturbations
with maximum amplitude of $0.004U_{e}$, as listed in table \ref{tab:Summary_of_cases},
are superimposed on the inflow velocity profile to trigger jet flow
transition to turbulence. Perturbations are only added to the velocity field.

Cases 1T, 2T and 4T utilize turbulent inflow conditions, typical of
jets exiting a long pipe. The inflow is generated using the approach of 
\cite{klein2003digital} modified to accommodate circular-pipe inflow geometry. This approach
generates inflow statistics matching a prescribed mean velocity and
Reynolds stress tensor, using the method of \cite{lund1998generation},
with fine-scale perturbations possessing a prescribed spatial correlation
length scale. The mean velocity and Reynolds stress
profiles are here specified from the fully-developed pipe flow DNS results
of \cite{eggels1994fully}, where the Reynolds number, based on pipe
diameter and bulk velocity, of $5300$ is close to the jet Reynolds
number of present study. The bulk velocity is defined as
\begin{equation}
U_{b}=\frac{1}{\pi\left(D/2\right)^{2}}\stackrel[0]{D/2}{\int}2\pi ru\thinspace\mathrm{d}r.\label{eq:bulk_vel}
\end{equation}
For small values of $\theta_0$ in (\ref{eq:lam_inflow_mean_vel}), $U_b$ for laminar inflow cases is approximately equal to $U_e$. 
$U_{b}$ in Cases 1T, 2T and 4T is chosen to be equal to $U_e$ of Cases 1, 2 and 4, respectively, to allow fair one-to-one comparisons between them.
Since $U_e$ has the same value for Cases 1--5, the bulk inflow velocity is approximately the same for Cases 1--5, 1T, 2T and 4T. The choice of the correlation
length scale determines the energy distribution among various spatial
scales. Increasing the length scale leads to more dominant large-scale
structures. Since the turbulent inflow simulations are aimed at examining
the influence of fully-developed fine-scale inflow turbulence on jet
statistics, a small isotropic value of $L/D=0.1$ is assumed for the
correlation length scale, this value being marginally larger than
the finest scale in the velocity spectra of figures 7(a-c) in \cite{eggels1994fully}.

\begin{figure}
\begin{centering}
\includegraphics[width=8cm]{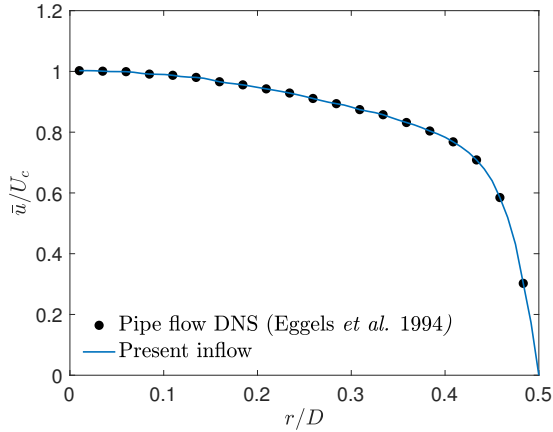}
\par\end{centering}

\caption{Inflow mean velocity normalized by the centerline velocity for the
(pseudo-)turbulent inflow compared against the pipe flow DNS results
of \cite{eggels1994fully}. \label{fig:inflow_mean_vel_comp}}
\end{figure}

\begin{figure}
\begin{centering}
\includegraphics[width=6.9cm]{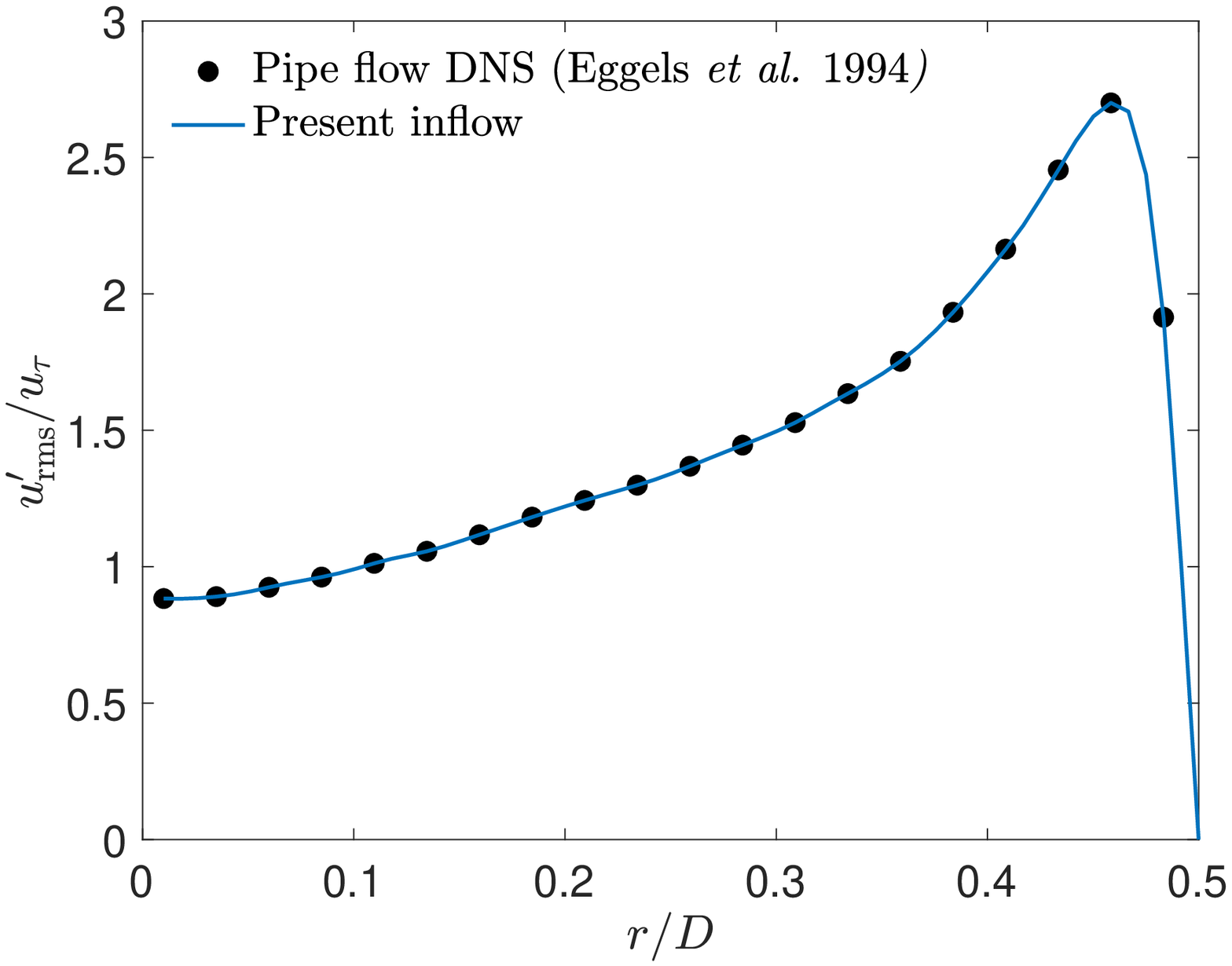}\includegraphics[width=6.9cm]{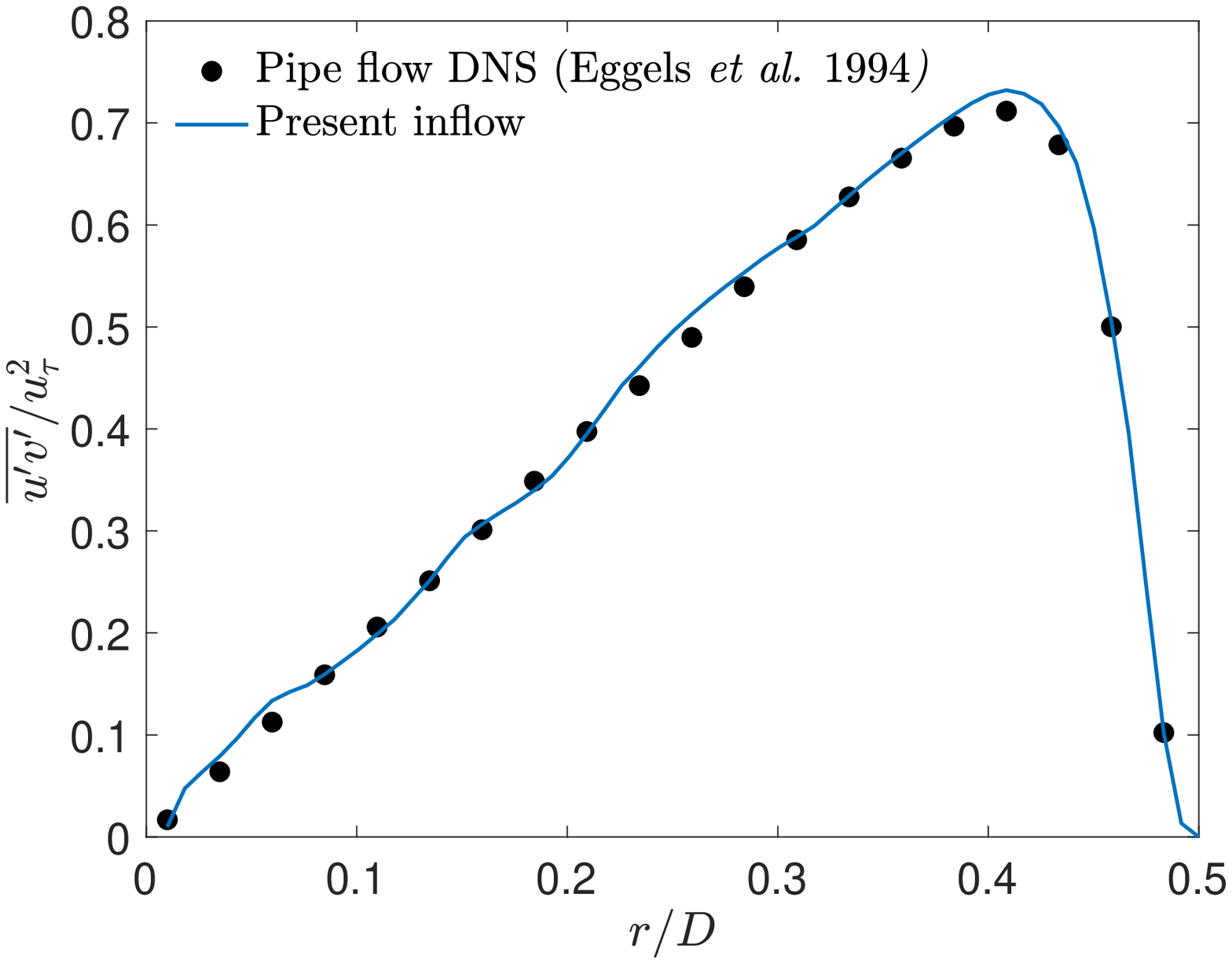}
\par\end{centering}

\begin{centering}
(a)\qquad{}\qquad{}\qquad{}\qquad{}\qquad{}\qquad{}\qquad{}\qquad{}\qquad{}\qquad{}\qquad{}\qquad{}(b)
\par\end{centering}



\caption{(a) R.m.s. axial velocity fluctuation, $u'_{\textrm{rms}}$, and (b) the Reynolds stress, $\overline{u^{'}v^{'}}$, normalized by the wall friction velocity, $u_{\tau}$, from the (pseudo-)turbulent inflow compared against the pipe flow DNS results of \cite{eggels1994fully}.\label{fig:inflow_Rey_stress_comp}}
\end{figure}

Figures \ref{fig:inflow_mean_vel_comp} and \ref{fig:inflow_Rey_stress_comp}
validate the turbulent inflow implementation. In figure \ref{fig:inflow_mean_vel_comp},
the mean axial velocity from the present turbulent inflow is compared against the pipe
flow DNS results \cite[case DNS(E) of][]{eggels1994fully}. Figure
\ref{fig:inflow_Rey_stress_comp} illustrates a similar comparison of the
r.m.s. axial-velocity fluctuation, $u'_{\textrm{rms}}$, and the 
Reynolds stress, $\overline{u^{'}v^{'}}$. The radial- and the azimuthal-velocity fluctuations, $v'_{\textrm{rms}}$ and $w'_{\textrm{rms}}$, compare similarly well with the respective DNS profiles, and have been omitted for brevity.
The overbar $\left(\bar{\bullet}\right)$
denotes mean quantities, calculated by an average over time and azimuthal
($\theta$) coordinate, given by a discrete approximation of
\begin{equation}
\bar{u}(x,r)=\frac{1}{2\pi}\intop_{0}^{2\pi}\left(\frac{1}{\left(t_{2}-t_{1}\right)}\intop_{t_{1}}^{t_{2}}u(x,r,\theta,t)\mathrm{d}t\right)\mathrm{d}\theta.\label{eq:average_defn}
\end{equation}
For all results in this study, the time average is performed over time steps in the interval $t_{1}=1000\leq tU_{e}/D\leq4000=t_{2}$. The r.m.s.
fluctuations are calculated from
\begin{equation}
u'_{\mathrm{rms}}=\sqrt{\left\langle \left\langle \left(u-\left\langle u\right\rangle _{t}\right)^{2}\right\rangle _{t}\right\rangle _{\theta}}=\sqrt{\left\langle \left\langle u^{2}\right\rangle _{t}-\left\langle u\right\rangle _{t}^{2}\right\rangle _{\theta}},\label{eq:rms_fluc}
\end{equation}
where $\left\langle \cdot\right\rangle _{t}$ and $\left\langle \cdot\right\rangle _{\theta}$
denotes the time and azimuthal averages, respectively. 
Using the notation $\left\langle \cdot\right\rangle _{t}$ and $\left\langle \cdot\right\rangle _{\theta}$,
(\ref{eq:average_defn}) can be written as $\bar{u}=\left\langle \left\langle u\right\rangle _{t}\right\rangle _{\theta}$.

The method described in \cite{klein2003digital}
assumes a Cartesian grid with uniform spacing, where the periodic
directions, along which averages are computed to determine mean quantities,
are aligned with the Cartesian directions. The round-jet inflow considered
here has circular orifice, where the azimuthal direction is periodic,
which is not aligned with a Cartesian direction. Therefore, the fluctuations
are computed on a polar grid and, then, interpolated to the Cartesian
inflow grid.

\noindent 

\noindent 

\vspace{-1.0cm}
\section{Results\label{sec:Results}}

The influence of $p_{\infty}$ and $Z$ on the laminar-inflow jet behavior is examined
first in \S \ref{sub:Case1to4_comparison}. Then, the effects of $p_{\infty}$ at a fixed $Z$ of 0.9
are investigated in \S \ref{sub:Case3n5_comparison}. To
differentiate between the effects of dynamic and thermodynamic
compressibility, the influence of $p_{\infty}$ and $Z$ at a fixed $Ma_e$ of 0.6 is studied 
in \S \ref{sub:Mach0p6_results}. Finally, the effect of the inflow condition -- laminar versus turbulent -- is
addressed in \S \ref{sub:Inflow-effects}, first as a baseline for the fully
compressible atmospheric-$p$ conditions in \S \ref{sub:Case1n1T_comparison}
and then at high-$p$ conditions in \S \ref{sub:Inflow-effects-at-highP}.

To provide confidence in the numerical formulation and discretization, a
validation of Case 1, which obeys the perfect-gas EOS, against experimental results 
is presented in Appendix \ref{sub:Case-1-results}; additionally, those results permit
comparisons with high-$p$ flow results where relevant.

\subsection{Effects of high pressure and compressibility factor\label{sub:Case1to4_comparison}}

The influence of $p_{\infty}$ (from atmospheric
to supercritical) on jet-flow dynamics and mixing is here examined by comparing
results from Cases 1 and 2. Further, the effects of $Z$ at supercritical $p_{\infty}$ are examined by comparing results
from Cases 2 to 4. As indicated in table \ref{tab:Summary_of_cases},
in each case the fluid in the injected jet is as dense as the ambient (or chamber)
fluid. The inflow bulk velocity, defined by (\ref{eq:bulk_vel}),
is the same for all cases. As a result, the inflow bulk momentum
varies with change in inflow density.

\subsubsection{Mean axial velocity and spreading rate\label{subsub:meanVel_hfRad_Case1to4}}

\begin{figure}
\begin{centering}
\includegraphics[width=6.9cm]{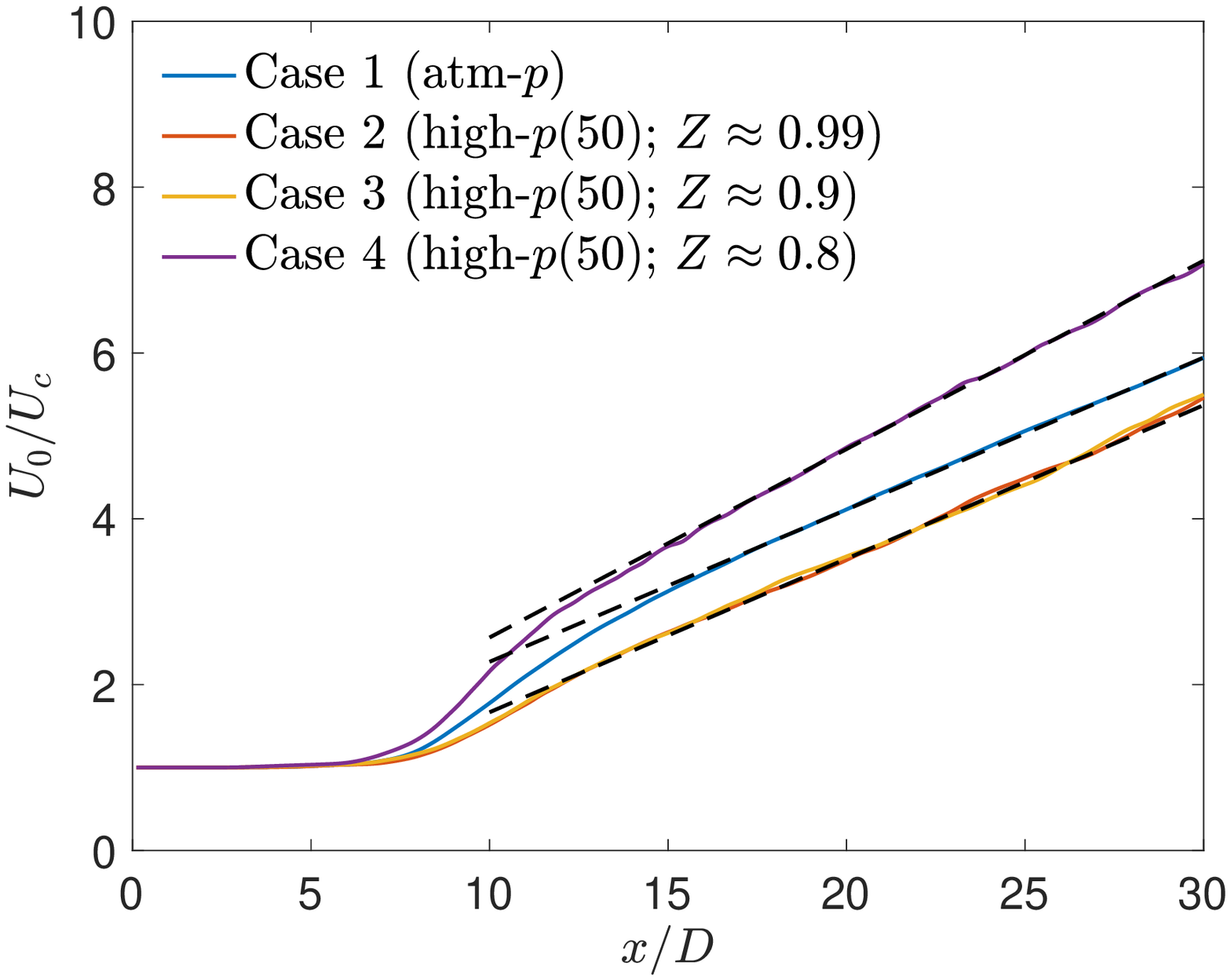}\includegraphics[width=6.9cm]{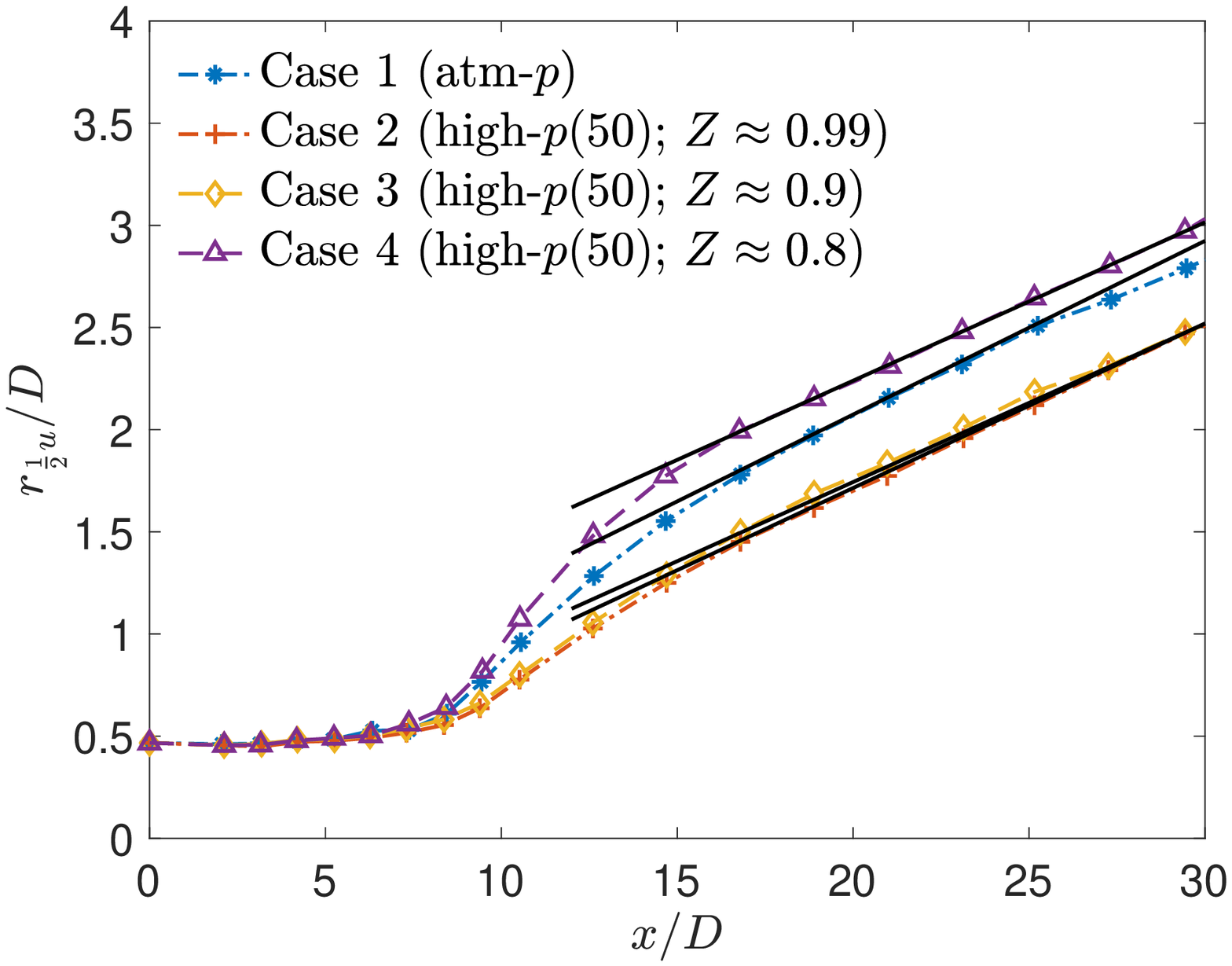}
\par\end{centering}

\begin{centering}
(a)\qquad{}\qquad{}\qquad{}\qquad{}\qquad{}\qquad{}\qquad{}\qquad{}\qquad{}\qquad{}\qquad{}\qquad{}(b)
\par\end{centering}

\caption{Case 1--4 comparisons: Streamwise variation of the (a) inverse of
centerline mean axial velocity ($U_{c}$) normalized by the jet-exit
centerline velocity ($U_{0}$) and (b) velocity half radius ($r_{\frac{1}{2}u}$).
The black dashed lines in (a) are given by equation (\ref{eq:centerline_vel})
using $B_{u}=5.5$, $x_{0u}=-2.4D$ for Case 1, $B_{u}=5.4$, $x_{0u}=D$
for Cases 2 \& 3, and $B_{u}=4.4$, $x_{0u}=-1.3D$ for Case 4. The
black solid lines in (b) are given by: 
$r_{\frac{1}{2}u}/D=0.085\left(x/D+4.4\right)$ for Case 1, 
$r_{\frac{1}{2}u}/D=0.0805\left(x/D+1.3\right)$ for Case 2, 
$r_{\frac{1}{2}u}/D=0.0775\left(x/D+2.5\right)$ for Case 3, and 
$r_{\frac{1}{2}u}/D=0.077\left(x/D+8.9\right)$ for Case 4. 
\label{fig:invUc_hfRad_Cases1to4}}
\end{figure}

The inverse of the centerline mean axial velocity, $U_{c}(x)$ $\left(=\bar{u}(x,0)\right)$, normalized by the  
the jet-exit centerline velocity, $U_0$  $\left(=U_{c}(0)\right)$,
for Cases 1 to 4 is presented in figure \ref{fig:invUc_hfRad_Cases1to4}(a).
For the laminar inflow cases, which have a top-hat jet-exit mean velocity profile, 
$U_{0}=U_{e}$, and since Cases 1--5 use same $U_{e}$, $U_{0}$ is the same for all 
cases in figure \ref{fig:invUc_hfRad_Cases1to4}(a).
To our knowledge, figure \ref{fig:invUc_hfRad_Cases1to4}(a) demonstrates for the first time that
supercritical jets in the Mach number range $\left[ 0.58,\thinspace 0.82 \right]$, see table \ref{tab:Summary_of_cases}, attain self-similarity. This finding 
differs from the self-similarity observed in the low-Mach-number results of \cite{ries2017numerical}, 
where the compressibility effects were ignored and the conservation equations did not use the pressure calculated from the EOS. 
In contrast, the fully compressible equations solved in the present study use 
the strongly non-linear EOS which contributes to the
thermodynamic-variable fluctuations, and self-similarity
is not an obvious outcome.

In figure \ref{fig:invUc_hfRad_Cases1to4}(a), the potential core length 
is approximately the same in all cases,
but the velocity decay rates differ among cases in both the transition
and the fully-developed self-similar regions. In the transition region
($7\lesssim x/D\lesssim15$), the mean axial-velocity decay, assessed
by the slope of the lines in figure \ref{fig:invUc_hfRad_Cases1to4}(a),
decreases with increasing $p_{\infty}$ from $1$ bar (Case 1) to $50$
bar (Case 2), remains approximately the same with decrease in $Z$
from $0.99$ (Case 2) to $0.9$ (Case 3), and increases significantly
with further decrease in $Z$ to $0.8$ (Case 4). In the self-similar
region, the decay rates are quantified by the inverse of $B_{u}$, defined through equation (\ref{eq:centerline_vel}).
$1/B_{u}$ increases from $1/5.5$ for Case 1 to $1/5.4$ for Case
2 \& 3 and to $1/4.4$ for Case 4. Lines with slope $1/B_{u}$ are
shown as black dashed lines in figure \ref{fig:invUc_hfRad_Cases1to4}(a). 

Figure \ref{fig:invUc_hfRad_Cases1to4}(b) compares the velocity half-radius
($r_{\frac{1}{2}u}$) among Cases 1--4. In the transition region
($7\lesssim x/D\lesssim15$), the jet spread defined by the half-radius
is larger for Case 1 than Case 2. The profiles are nearly identical
for Cases 2 and 3, and Case 4 shows a significantly larger jet spread
than the other cases. In the self-similar region, the linear spread can be
described by the black solid lines of figure \ref{fig:invUc_hfRad_Cases1to4}(b); 
the equations describing the solid lines are included in the figure caption.
The self-similar spread rate decreases from Case 1 to Case 4. The
decrease is relatively small from Case 2 to Case 3, and negligible
from Case 3 to Case 4. Variation of $\xi_{c}/\xi_{0}$ and $r_{\frac{1}{2}\xi}/D$
(not shown here for brevity) are similar to
those of the velocity field in figure \ref{fig:invUc_hfRad_Cases1to4}.

The decay of $U_{c}$, observed in figure \ref{fig:invUc_hfRad_Cases1to4}(a),
is a result of the concurrent processes of: (a) transfer of kinetic energy from the mean field
to fluctuations, (b) transport of mean kinetic energy away from the
centerline as more ambient fluid is entrained, and (c) mean viscous dissipation.
These processes interact as follows. The
entrainment of ambient fluid (initially at rest) into the jet enhances the momentum
and kinetic energy of the ambient fluid. Transport of
momentum/energy from the jet core facilitates the ambient-fluid entrainment and
jet spread. As a result, a wider jet spread is associated with a larger decay in
$U_{c}$. Therefore, the profiles for various cases look similar in figures
\ref{fig:invUc_hfRad_Cases1to4}(a) and (b). Production term of the t.k.e. equation quantifies
the loss of mean kinetic energy to turbulent fluctuations and mean strain
rate magnitude is proportional to the mean viscous dissipation. The variation of $U_{c}$ across
various cases in figure \ref{fig:invUc_hfRad_Cases1to4}(a) follows the variation
of t.k.e. production and mean strain rate magnitude, as discussed in \S \ref{sub:summary_inflowEffects}. 

The considerably larger decay of $U_{c}$ in Case 4 compared to other cases is at this point 
conjectured to be a coupled effect of its proximity to the Widom line, i.e. the thermodynamic 
state (see figure \ref{fig:comp_factor}), and the mean strain rates generated in the flow, i.e. the dynamic 
state, that depends on the thermodynamic state, the inflow condition (laminar versus turbulent), and the jet-exit (inflow) Mach number. The proximity to the Widom line
determines the departure from perfect-gas behavior and the relative magnitude of pressure fluctuations across various cases, as further discussed in \S \ref{subsub:Case1to4_prsFluc}. Prior to examining the role of pressure flucutations in the unique behavior of Case 4, an evaluation of the consistency of $U_{c}$ decay with the kinetic energy transfer from the mean field
to fluctuations is performed by next examining the velocity fluctuations and their self-similarity.
\vspace{-0.45cm}

\noindent 

\noindent 
\begin{figure}
\begin{centering}
\includegraphics[width=9cm]{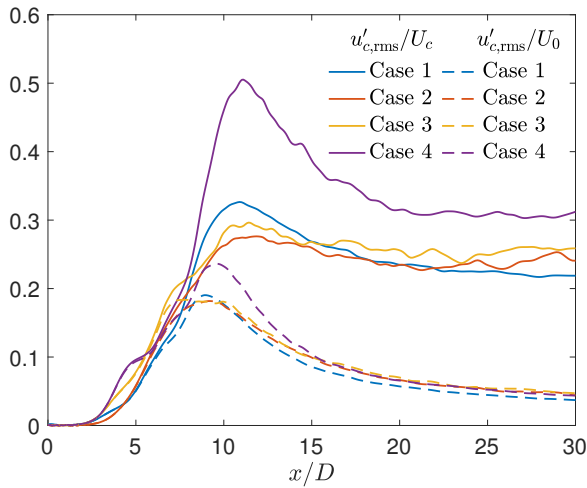}
\par\end{centering}

\caption{Case 1--4 comparisons: Streamwise variation of the centerline r.m.s.
axial-velocity fluctuation ($u'_{c,\textrm{rms}}$) normalized by
the centerline mean value, $U_{c}$, and jet-exit mean value, $U_{0}$.
\label{fig:rms_fluc_Cases1to4}}
\end{figure}

\noindent 
\begin{figure}
\begin{centering}
\includegraphics[width=6.9cm]{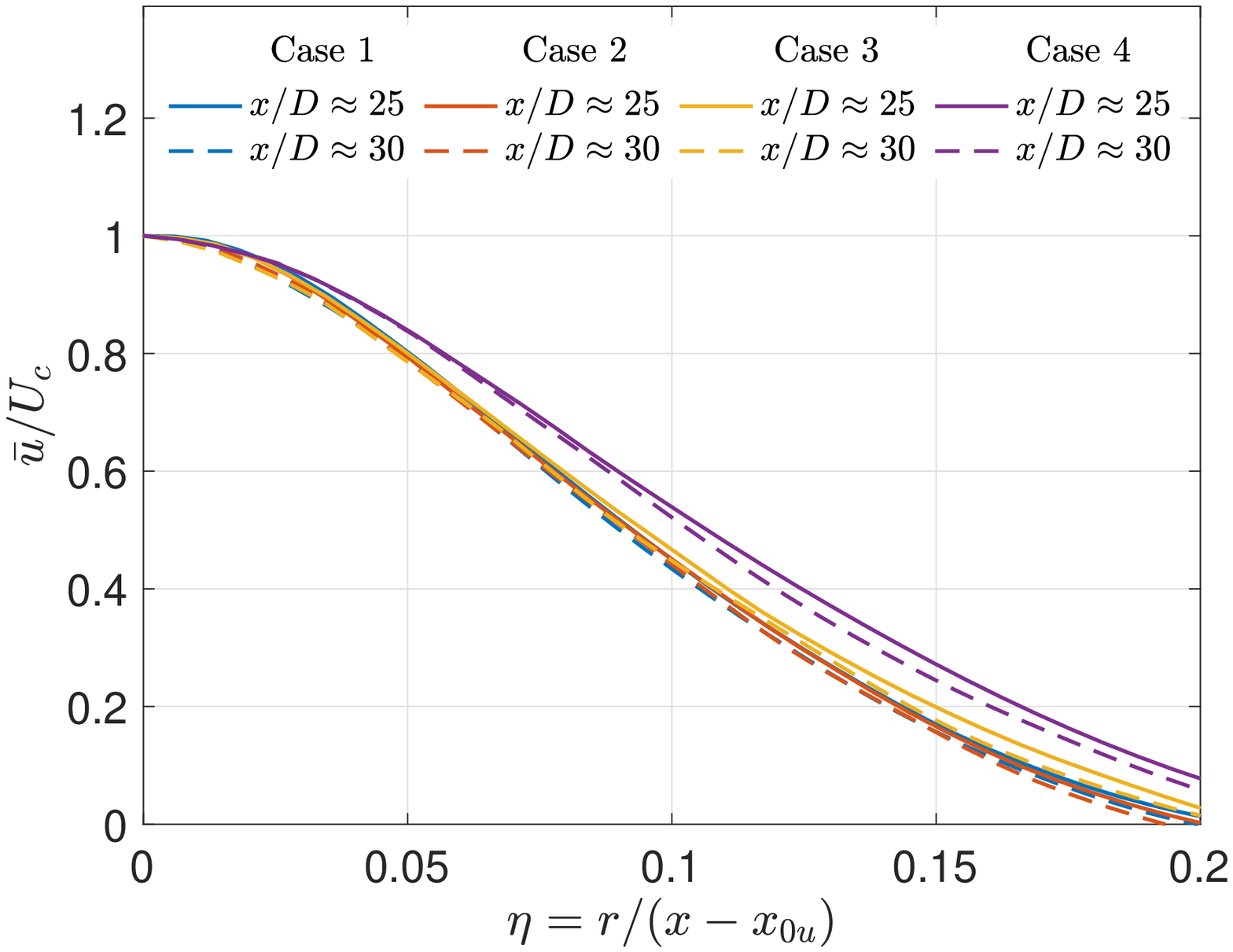}\includegraphics[width=6.9cm]{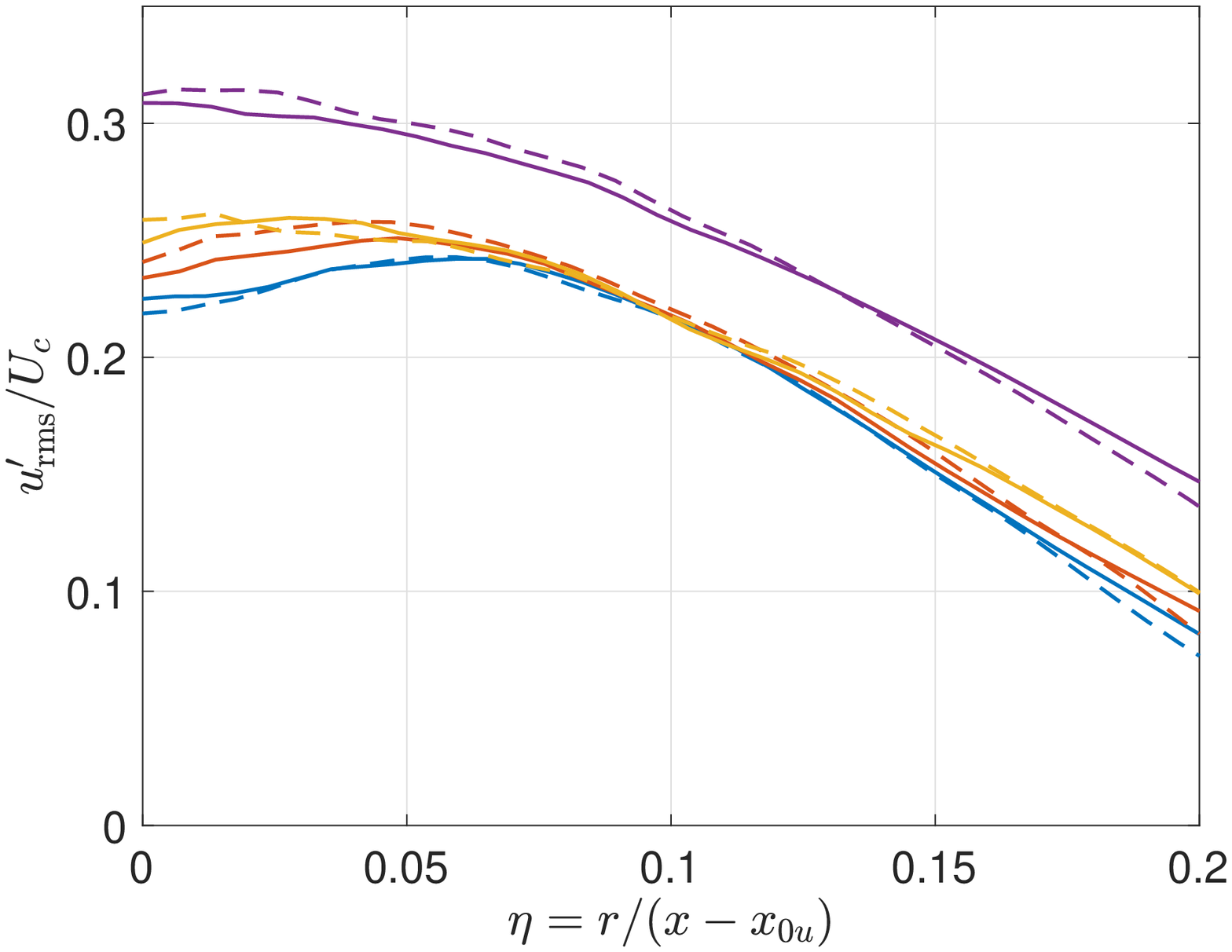}
\par\end{centering}

\begin{centering}
(a)\qquad{}\qquad{}\qquad{}\qquad{}\qquad{}\qquad{}\qquad{}\qquad{}\qquad{}\qquad{}\qquad{}\qquad{}(b)
\par\end{centering}

\begin{centering}
\includegraphics[width=6.9cm]{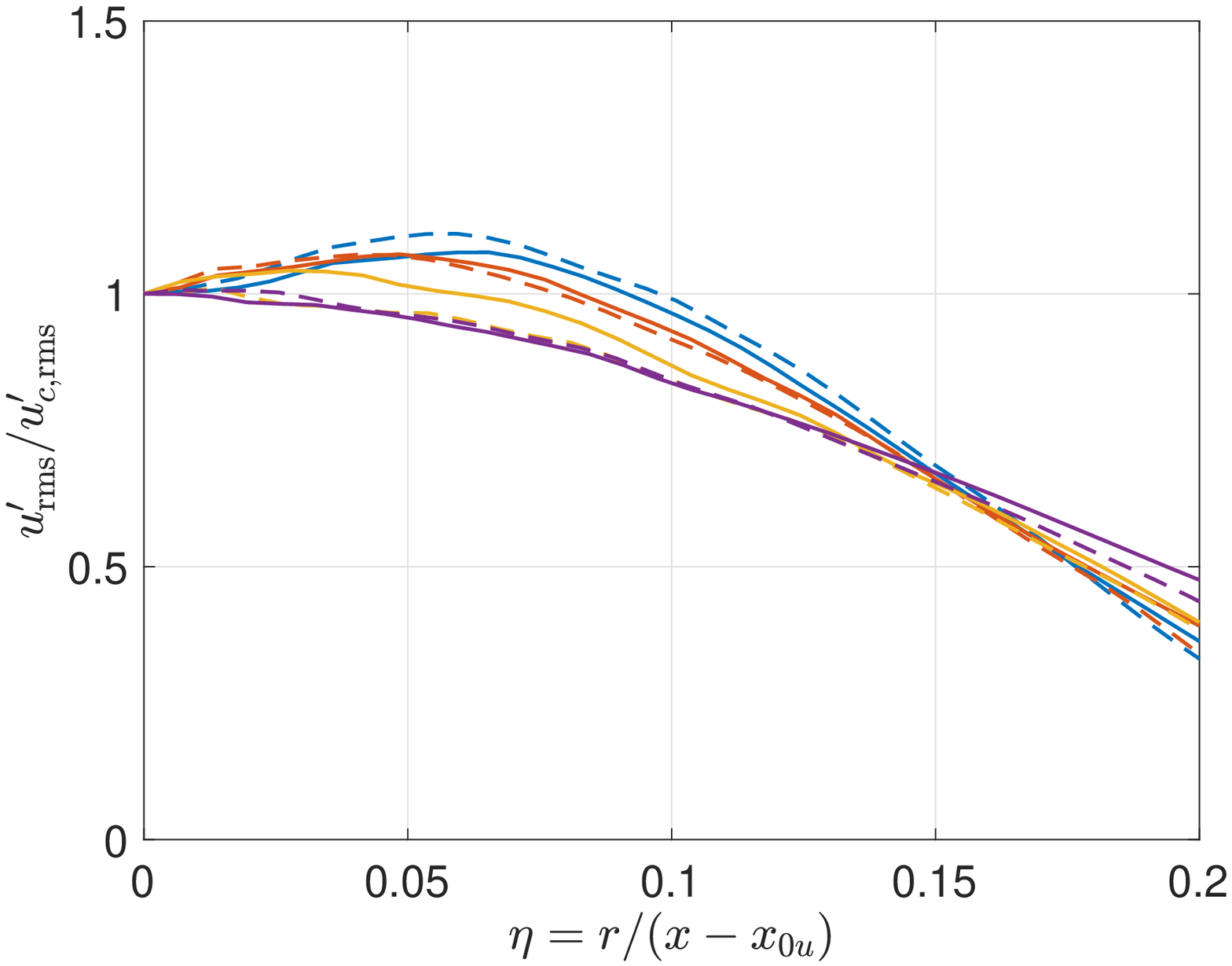}\includegraphics[width=6.9cm]{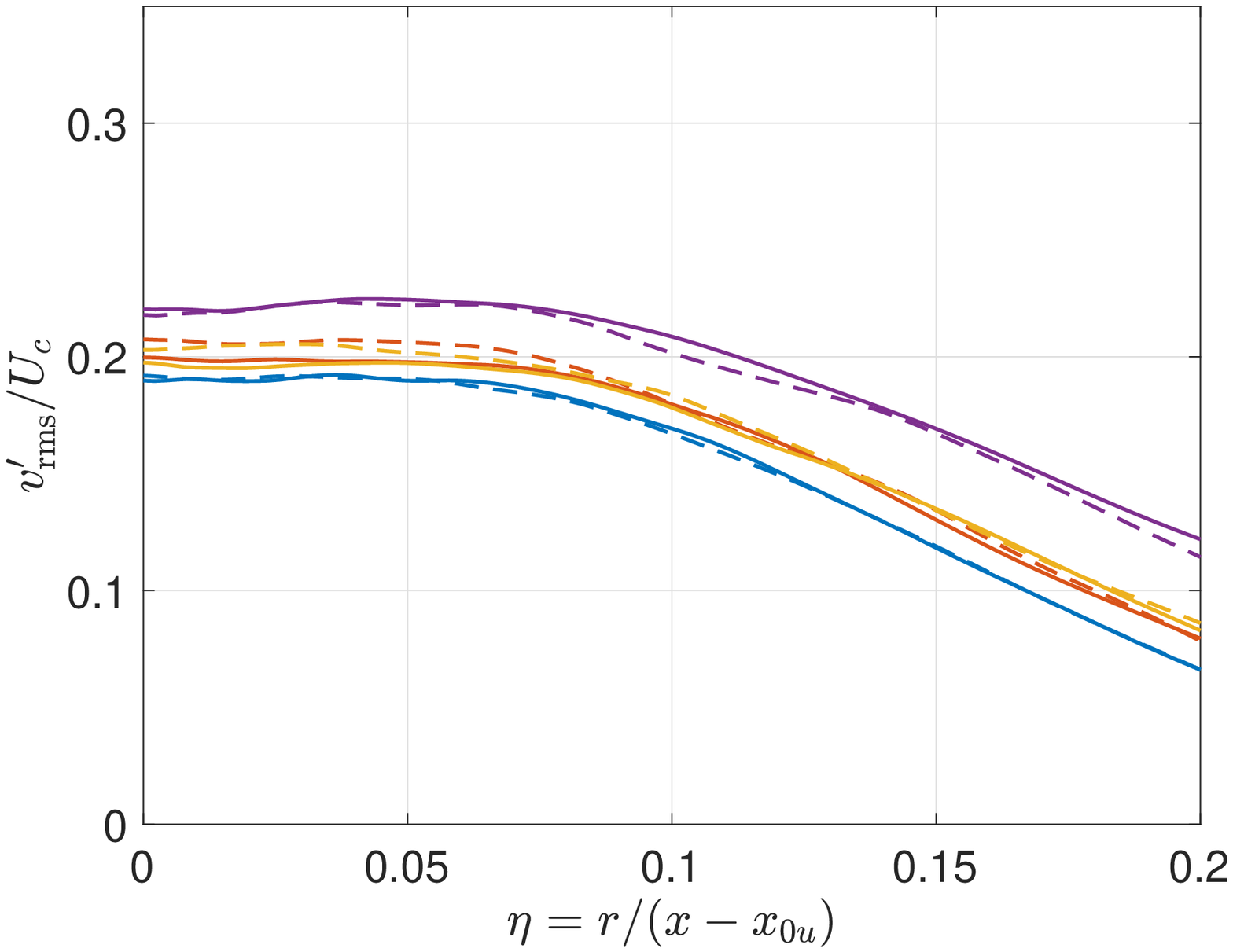}
\par\end{centering}

\begin{centering}
(c)\qquad{}\qquad{}\qquad{}\qquad{}\qquad{}\qquad{}\qquad{}\qquad{}\qquad{}\qquad{}\qquad{}\qquad{}(d)
\par\end{centering}

\begin{centering}
\includegraphics[width=6.9cm]{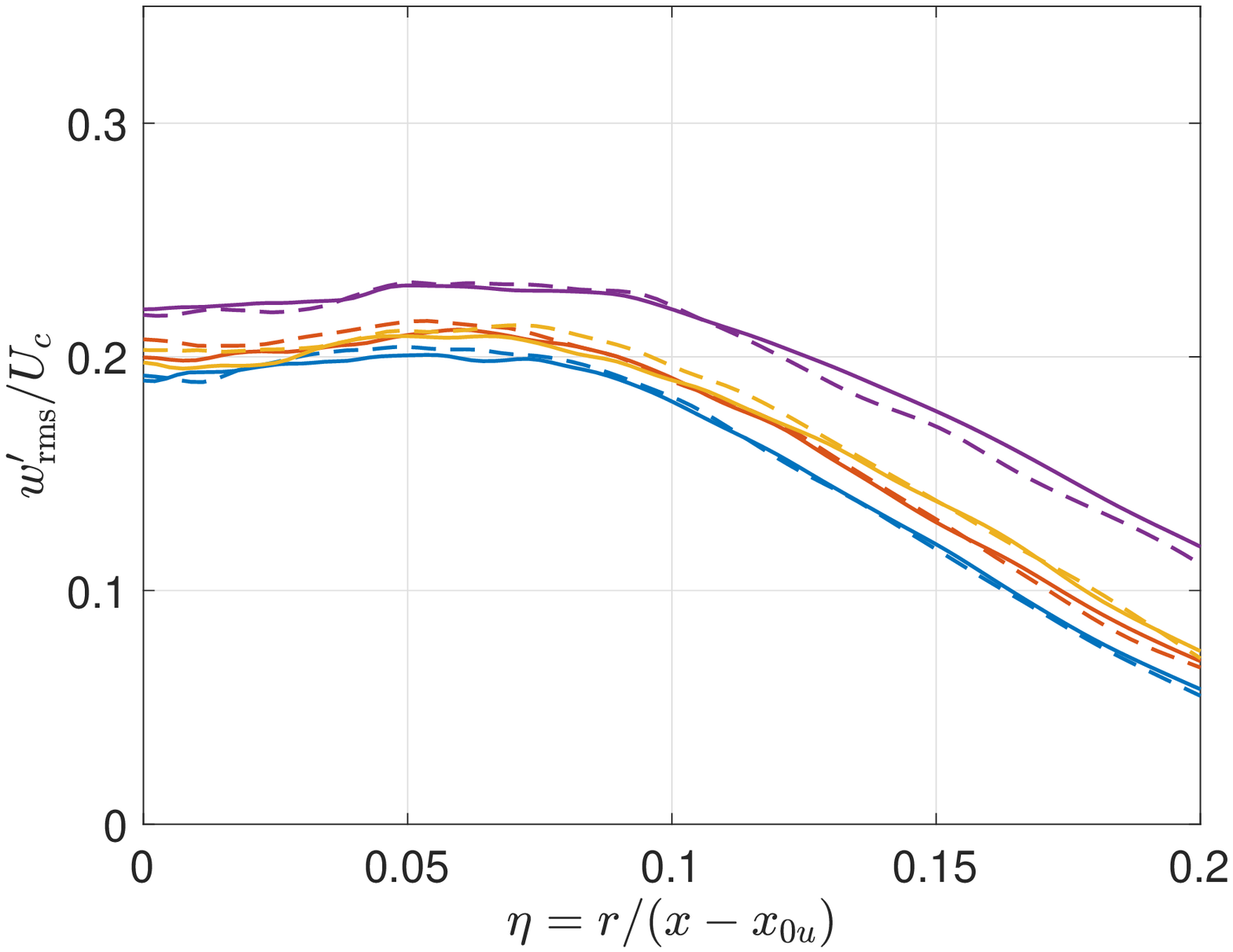}\includegraphics[width=6.9cm]{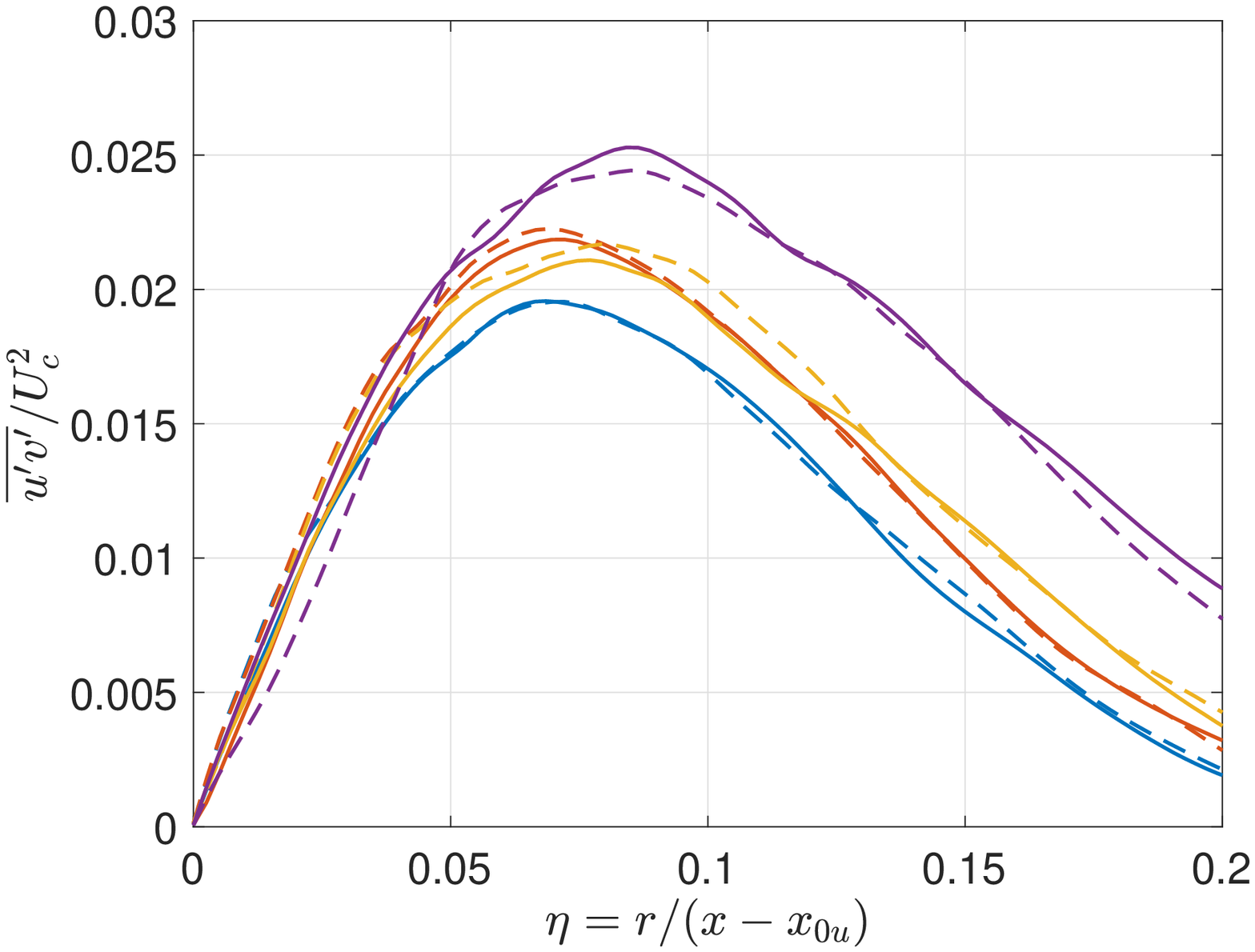}
\par\end{centering}

\begin{centering}
(e)\qquad{}\qquad{}\qquad{}\qquad{}\qquad{}\qquad{}\qquad{}\qquad{}\qquad{}\qquad{}\qquad{}\qquad{}(f)
\par\end{centering}

\caption{Case 1--4 comparisons: Radial profiles of (a) mean axial velocity
($\bar{u}$) normalized by the centerline mean axial velocity ($U_{c}$),
(b) r.m.s. axial velocity fluctuations ($u_{\mathrm{rms}}^{'}$) normalized
by the centerline mean axial velocity, (c) r.m.s. axial velocity fluctuations
($u_{\mathrm{rms}}^{'}$) normalized by the centerline r.m.s. axial
velocity fluctuations ($u_{c,\mathrm{rms}}^{'}$), (d) normalized
r.m.s. radial velocity fluctuations ($v_{\mathrm{rms}}^{'}$), (e)
normalized r.m.s. azimuthal velocity fluctuations ($w_{\mathrm{rms}}^{'}$),
and (f) normalized Reynolds stress ($\overline{u^{'}v^{'}}$) at various
axial locations. The legend is the same for all plots. \label{fig:Case_1to4_vel_stat}}
\end{figure}

\subsubsection{Velocity fluctuations and self-similarity\label{subsub:velFluc_Case1to4}}

The centerline r.m.s. axial-velocity fluctuation is 
depicted in figure \ref{fig:rms_fluc_Cases1to4} for Cases 1--4 with two different normalizations.
Since $U_{0}$ has the same value for Cases 1--4, the normalization with $U_{0}$ compares the absolute fluctuation
magnitude among various cases. On the other hand, the normalization with $U_{c}$ shows the fluctuation magnitude
with respect to the local mean value. Larger $u'_{c,\textrm{rms}}/U_{c}$ values
are expected to indicate greater local transfer of mean kinetic energy to fluctuations.
Accordingly, larger $u'_{c,\textrm{rms}}/U_{c}$ in figure \ref{fig:rms_fluc_Cases1to4}
should imply a higher slope ($U_{c}$ decay rate) in
the corresponding region in figure \ref{fig:invUc_hfRad_Cases1to4}(a).
Case 4, which has the largest $u'_{c,\textrm{rms}}/U_{c}$ among all cases in
both the transition and the self-similar region, also exhibits largest
slopes (decay rates) in figure \ref{fig:invUc_hfRad_Cases1to4}(a).
Case 1 has larger $u'_{c,\textrm{rms}}/U_{c}$ than Cases 2 and 3
in the transition region, and, accordingly, higher decay rates in
that region in figure \ref{fig:invUc_hfRad_Cases1to4}(a). In the
self-similar region, $u'_{c,\textrm{rms}}/U_{c}$ in Cases 2 and 3
are marginally larger than in Case 1, and, accordingly, the self-similar $U_c$
decay rates of Cases 2 and 3 are marginally higher. This confirms that the decay in $U_c$
is consistently reflected in the magnitude of $u'_{c,\textrm{rms}}/U_{c}$.

The linear mean axial-velocity decay and the linear jet-spread rate,
downstream of the transition region, in figure \ref{fig:invUc_hfRad_Cases1to4}
indicates the self-similarity of the mean axial velocity. The self-similarity
of mean axial velocity and Reynolds stresses is further examined from their radial variation
in figure \ref{fig:Case_1to4_vel_stat}. Figure \ref{fig:Case_1to4_vel_stat}(a)
shows the radial profiles of $\bar{u}/U_{c}$
from Cases 1--4 at $x/D\approx25$ (solid lines) and $30$ (dashed
lines). In all cases, profiles at the two axial locations show minimal
differences, suggesting that $\bar{u}/U_{c}$ has attained
self-similarity. The self-similar mean velocity/scalar profile is commonly 
expressed as \cite[e.g.][]{mi2001influence,xu2002effect}
\begin{equation}
\bar{u}(x,r)=U_{c}\left(x\right)f\left(\eta\right),\qquad\bar{\xi}(x,r)=\xi_{c}\left(x\right)g\left(\eta\right),\label{eq:self_similar_profiles}
\end{equation}
where $f\left(\eta\right)$ and $g\left(\eta\right)$ are similarity
functions, often described by Gaussian distributions,
\begin{equation}
f\left(\eta\right)=\exp\left(-A_{u}\eta^{2}\right),\qquad g\left(\eta\right)=\exp\left(-A_{\xi}\eta^{2}\right),\label{eq:self_similar_Gauss_dist}
\end{equation}
where $A_{u}$ and $A_{\xi}$ are constants, here determined from a least-squares fit 
of the simulation data. The least-squares procedure applied to $x/D\approx30$
profiles of figure \ref{fig:Case_1to4_vel_stat}(a) yields $A_{u}=79.5$ for Cases 1 and 2, $A_{u}=77.2$ for
Case 3, and $A_{u}=64.4$ for Case 4. Thus, increasing $p$ from $1$
bar (Case 1) to $50$ bar (Case 2) has minimal influence on the radial
variation of the self-similar axial-velocity profile. A decrease in $Z$
from $0.99$ (Case 2) to $0.9$ (Case 3) and then to $0.8$ (Case
4) at $p_{\infty}=50$ bar increases $\bar{u}/U_{c}$ at a fixed $\eta$.

The radial variation of normalized r.m.s. velocity fluctuations at $x/D\approx25$
and $30$ are compared for Cases 1--4 in figures \ref{fig:Case_1to4_vel_stat}(b)--(e).
In all figures, the profiles at $x/D\approx25$ (solid lines)
and $30$ (dashed lines) show minimal difference, and hence the r.m.s.
velocity fluctuations can be considered self-similar around $x/D\approx25$.
$u'_{\textrm{rms}}/U_{c}$, shown in figure \ref{fig:Case_1to4_vel_stat}(b),
increases in the vicinity of centerline with increase in $p_{\infty}$
from $1$ bar (Case 1) to $50$ bar (Case 2), but the differences diminish
with increase in $\eta$. A decrease in $Z$ from $0.99$ (Case 2) to
$0.9$ (Case 3) marginally increases $u'_{\textrm{rms}}/U_{c}$ at
both small and large $\eta$. Further decrease in $Z$ from $0.9$
(Case 3) to $0.8$ (Case 4) shows significant increase in $u'_{\textrm{rms}}/U_{c}$
at all $\eta$-locations. $u'_{\textrm{rms}}/u'_{c,\textrm{rms}}$,
plotted in figure \ref{fig:Case_1to4_vel_stat}(c) shows that the  
fluctuations increase with radial distance near
the centerline in Case 1, with maximum at $\eta\approx0.07$. The
location of the maximum (in terms of $\eta$) recedes towards the
centerline progressively in Cases 2 and 3. Case 4 does not exhibit
an off-axis maximum and $u'_{\textrm{rms}}/u'_{c,\textrm{rms}}$ decreases
monotonically with $\eta$, highlighting the peculiarity with respect to Cases 2 and 3.

Additionally, $v_{\textrm{rms}}^{\prime}/U_{c}$ and $w_{\text{rms}}^{\prime}/U_{c}$, shown
in figures \ref{fig:Case_1to4_vel_stat}(d) and (e), respectively,
increase from Case 1 to 4. The increase is marginal from Case 1 to
3, but significant in Case 4. Axisymmetry of a round-jet flow requires
that $v'_{\textrm{rms}}$ and $w'_{\textrm{rms}}$ be equal at the
centerline, which is nearly true for all cases in figures \ref{fig:Case_1to4_vel_stat}(d)
and (e). Comparable profiles of $\overline{u'v'}/U_{c}^{2}$
in figure \ref{fig:Case_1to4_vel_stat}(f)
at $x/D\approx25$ and $30$ suggest that $\overline{u'v'}/U_{c}^{2}$
attains self-similarity around $x/D\approx25$ in Cases 1--4. $\overline{u'v'}/U_{c}^{2}$
is similar for Cases 1--4 in the vicinity of the centerline but the
profiles differ at larger $\eta$, where Case 4 values are considerably
larger than the other cases.

\noindent 
\begin{figure}
\begin{centering}
\includegraphics[width=6.9cm]{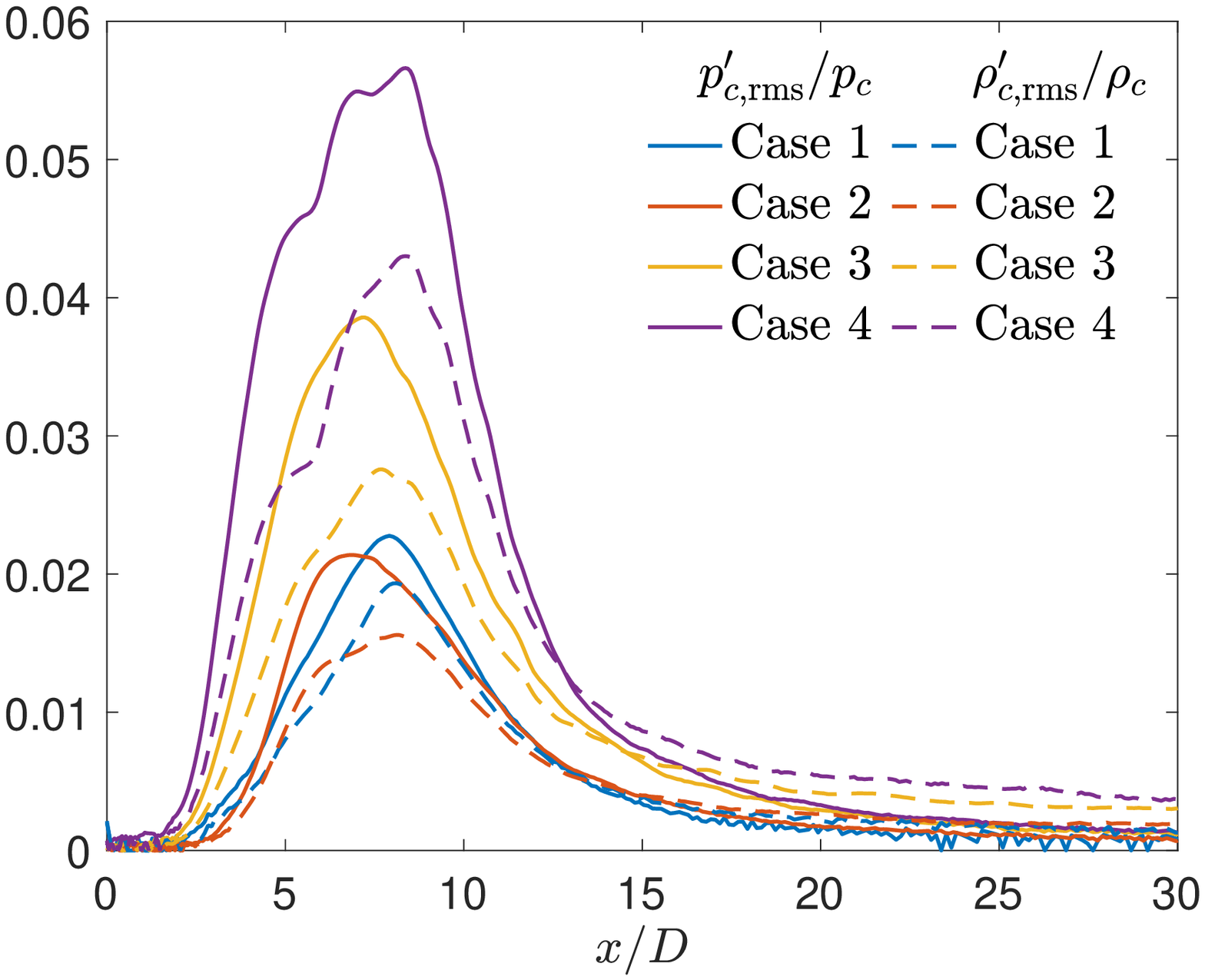}\includegraphics[width=6.9cm]{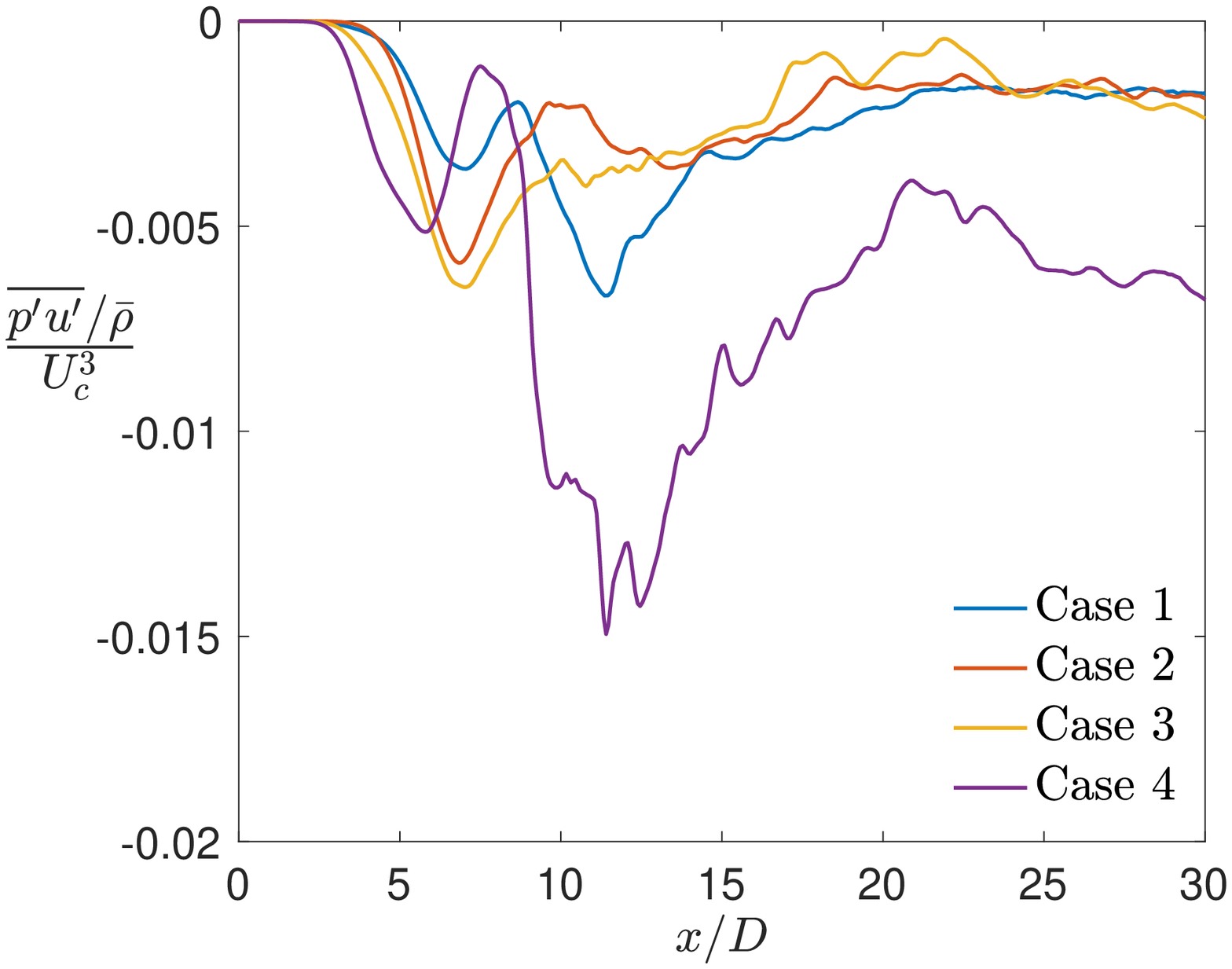}
\par\end{centering}

\begin{centering}
(a)\qquad{}\qquad{}\qquad{}\qquad{}\qquad{}\qquad{}\qquad{}\qquad{}\qquad{}\qquad{}\qquad{}\qquad{}(b)
\par\end{centering}

\begin{centering}
\includegraphics[width=6.9cm]{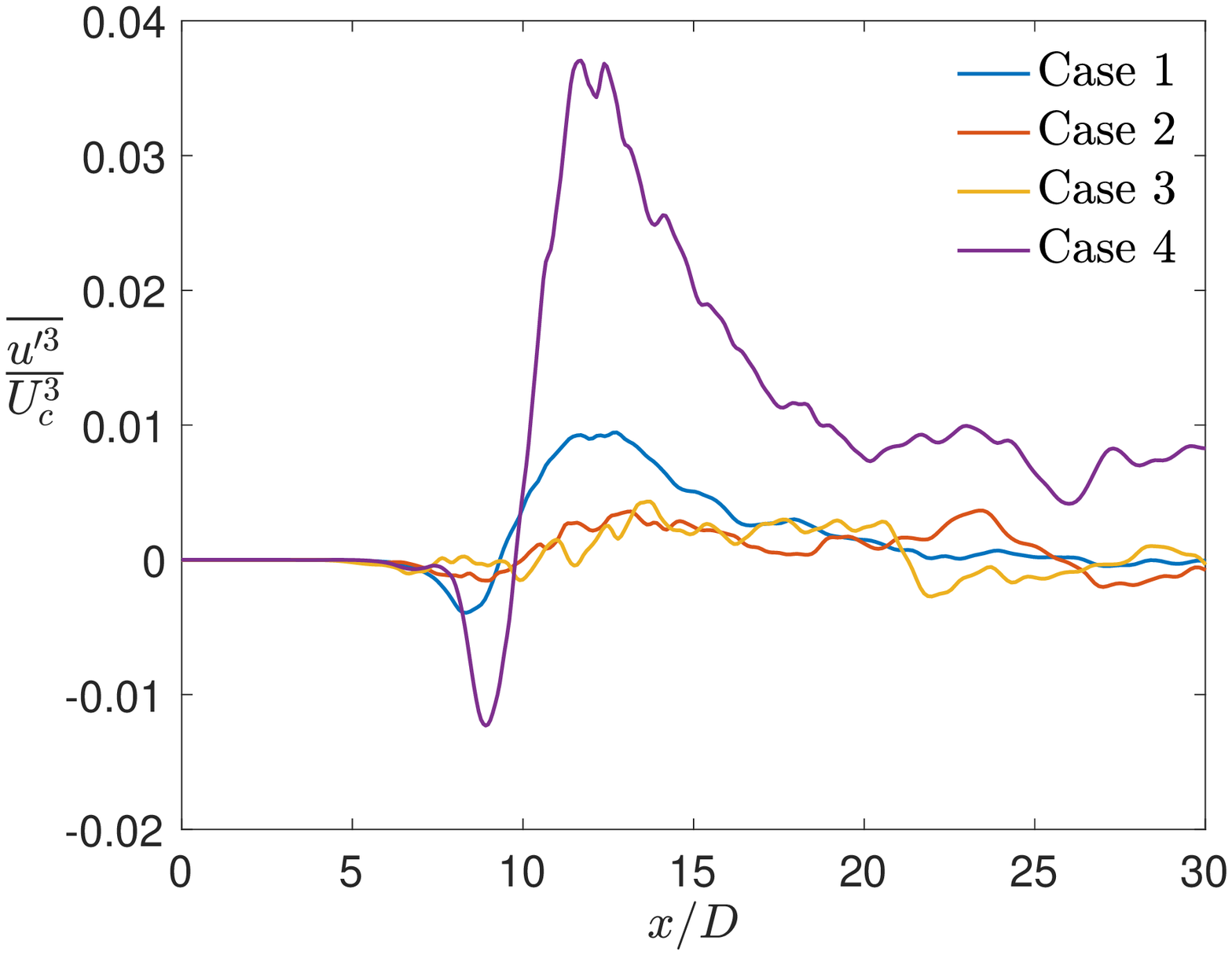}
\par\end{centering}

\begin{centering}
(c)
\par\end{centering}

\caption{Case 1--4 comparisons: Streamwise variation of the (a) centerline
r.m.s. pressure and density fluctuations, denoted by $p'_{c,\mathrm{rms}}$
and $\rho'_{c,\mathrm{rms}}$, respectively, normalized by the centerline
mean pressure ($p_{c}$) and density ($\rho_{c}$), respectively,
(b) normalized fluctuating pressure-axial velocity correlation $\left(\overline{p'u'}/\bar{\rho}U_{c}^{3}\right)$,
and (c) normalized third-order velocity moment $\left(\overline{u^{\prime3}}/U_{c}^{3}\right)$.
\label{fig:centerline_pfluc_pvelCorl}}
\end{figure}

\noindent 
\begin{figure}
\begin{centering}
\includegraphics[width=6.9cm]{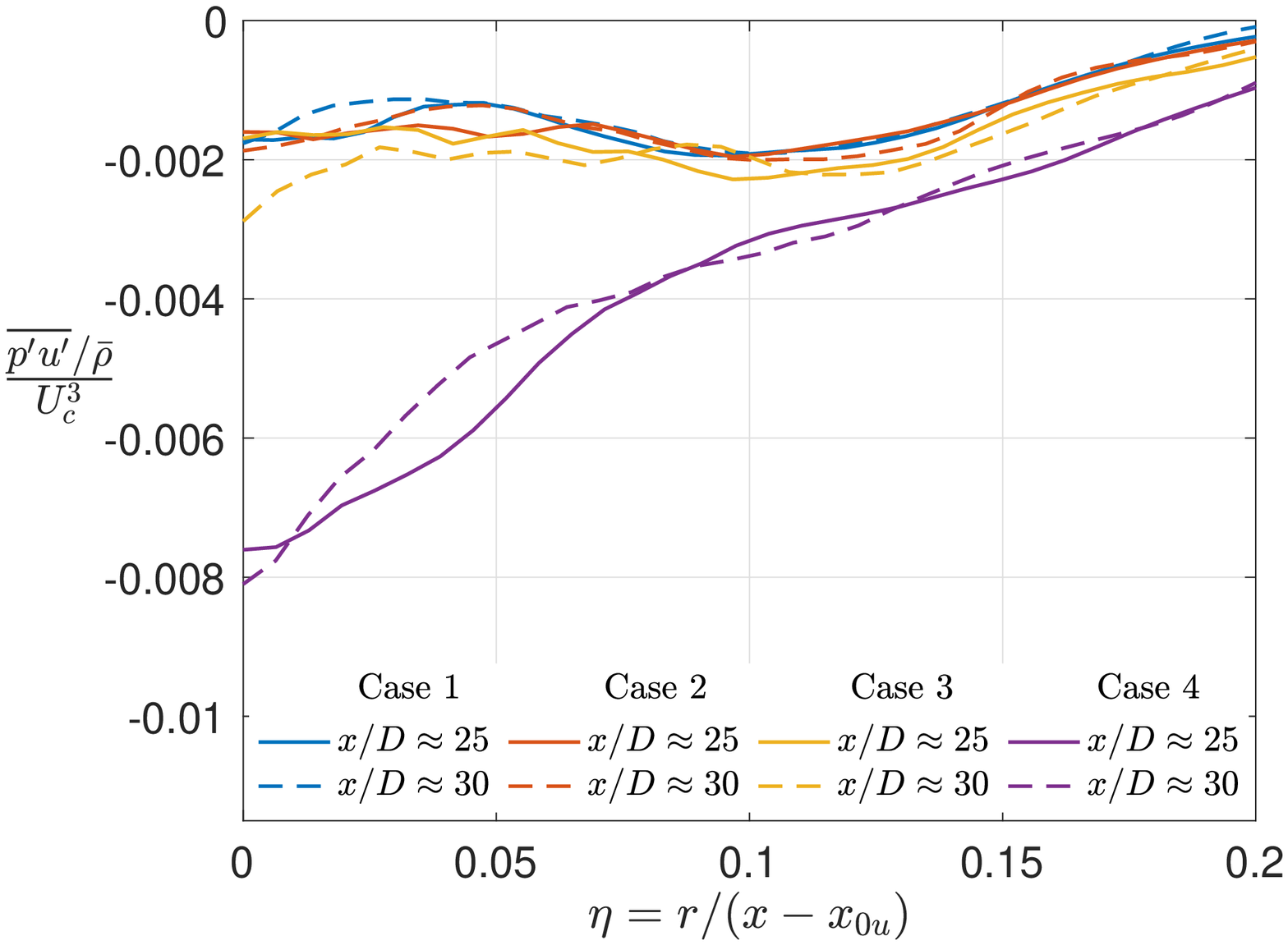}\includegraphics[width=6.9cm]{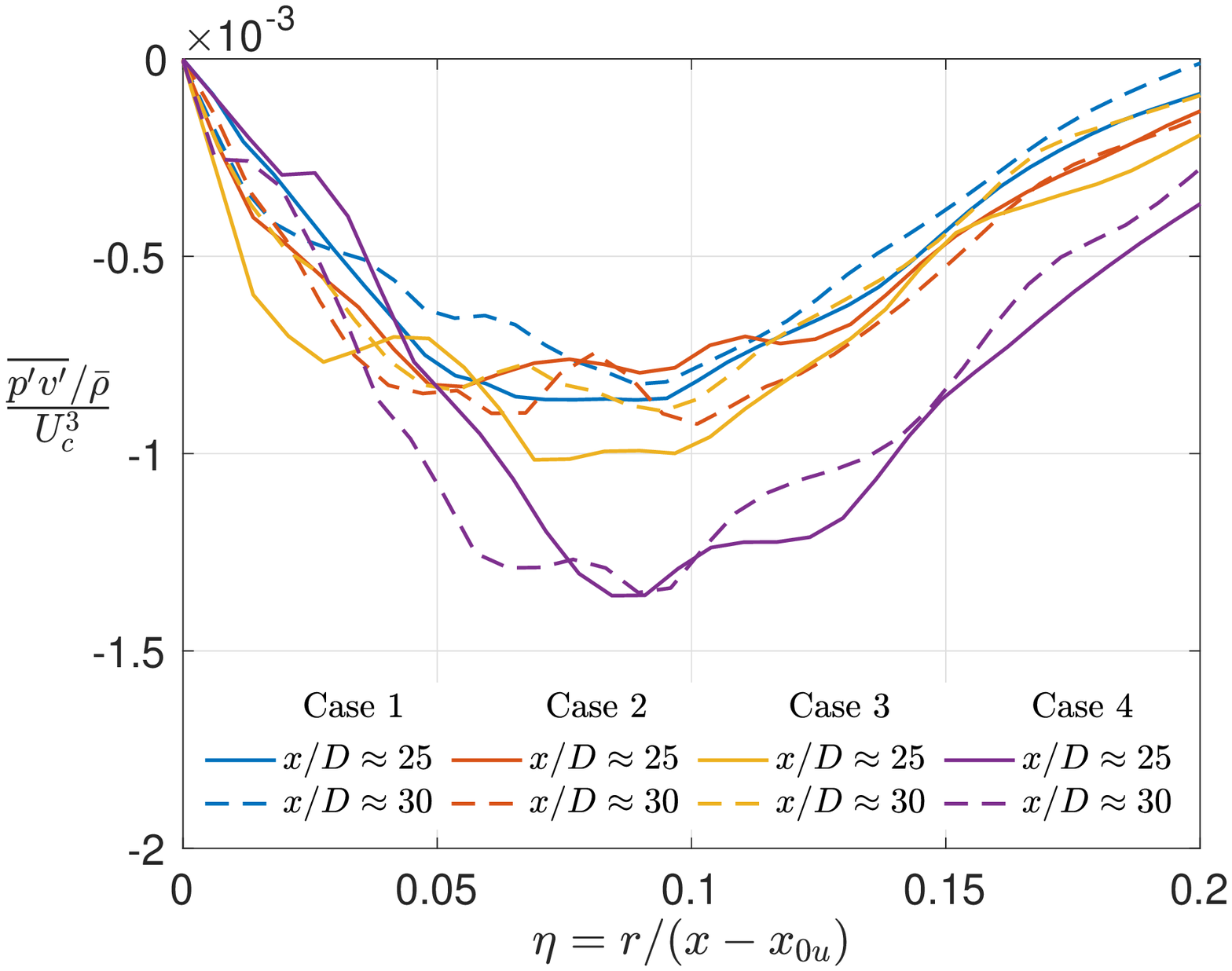}
\par\end{centering}

\begin{centering}
(a)\qquad{}\qquad{}\qquad{}\qquad{}\qquad{}\qquad{}\qquad{}\qquad{}\qquad{}\qquad{}\qquad{}\qquad{}(b)
\par\end{centering}

\caption{Case 1--4 comparisons: Radial profiles of the (a) normalized pressure-axial
velocity correlation and (b) normalized pressure-radial velocity correlation
at $x/D\approx25$ and $30$. \label{fig:prs_vel_correlation_Cases1to4}}
\end{figure}

\subsubsection{Pressure and density fluctuations, pressure-velocity correlation, and third-order velocity moments\label{subsub:Case1to4_prsFluc}}

The differences in mean axial-velocity for various cases, observed
in figure \ref{fig:invUc_hfRad_Cases1to4}, is consistent with
the differences in velocity fluctuations, examined in the previous section. Larger velocity
fluctuations imply greater transfer of energy from the mean field
to fluctuations, resulting in greater decay of mean velocity. The 
differences in velocity fluctuations with $p_{\infty}$ and $Z$, 
however, remain to be explained, and this topic is addressed next.

\begin{figure}
\begin{centering}
\includegraphics[width=6.9cm]{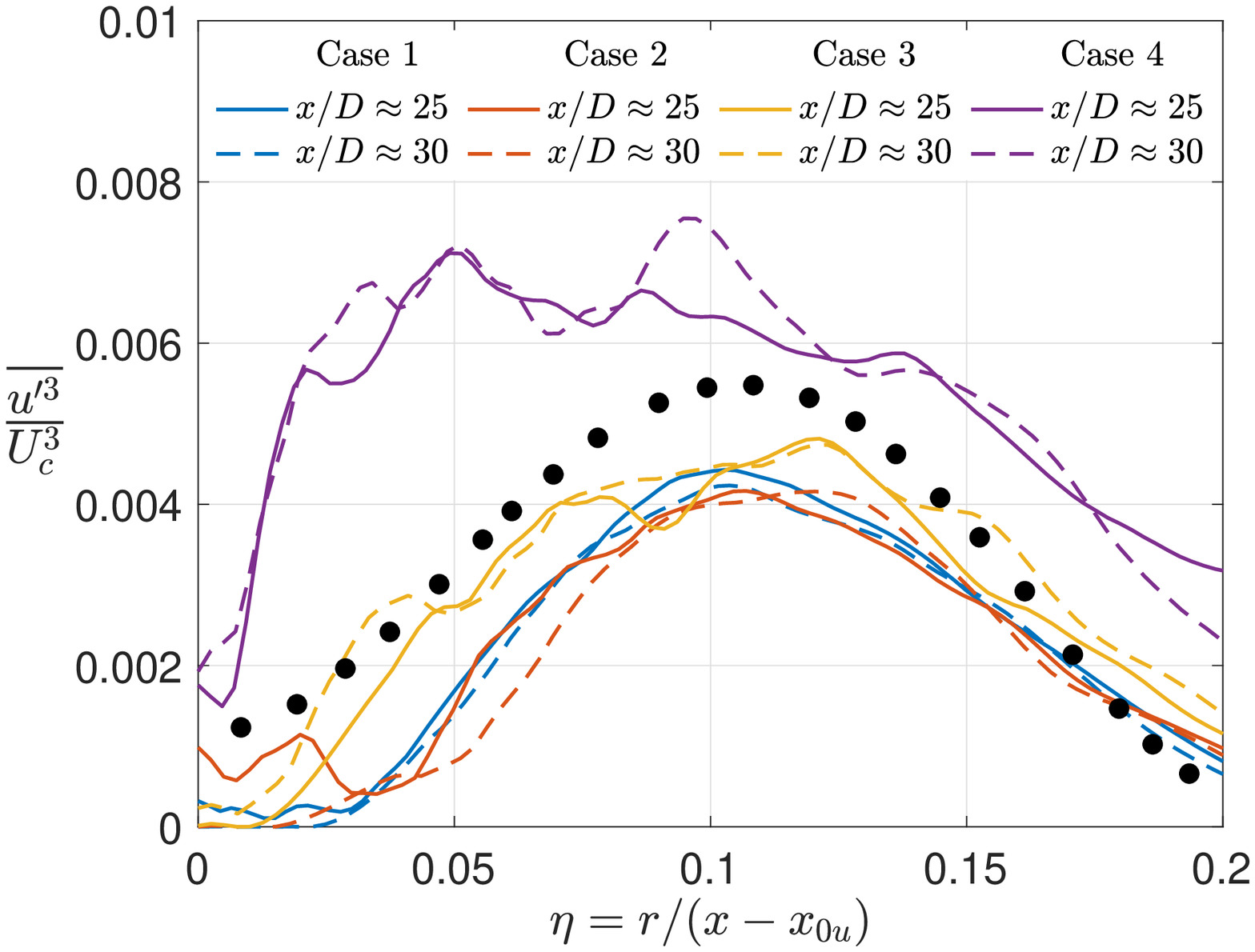}\includegraphics[width=6.9cm]{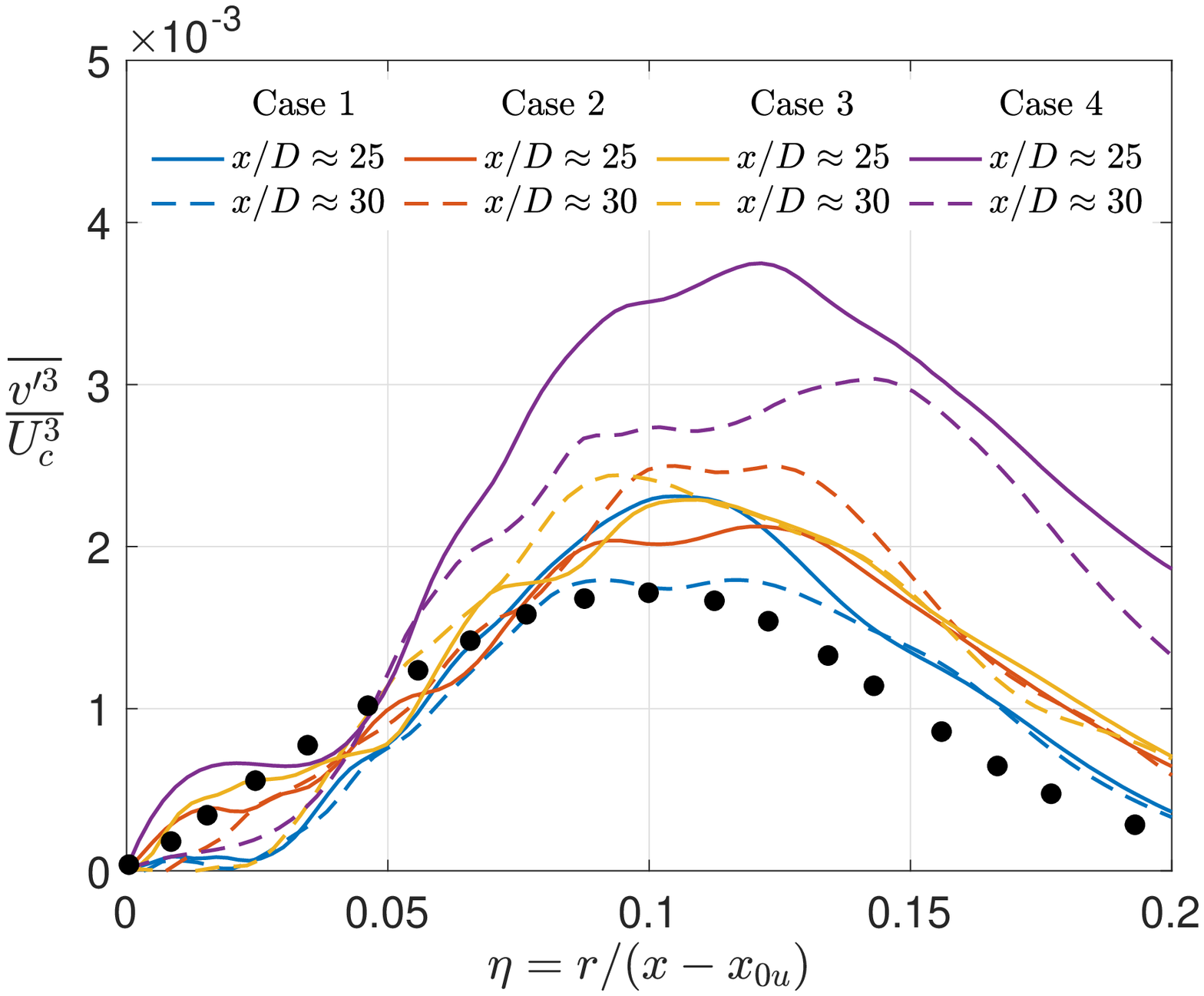}
\par\end{centering}

\begin{centering}
(a)\qquad{}\qquad{}\qquad{}\qquad{}\qquad{}\qquad{}\qquad{}\qquad{}\qquad{}\qquad{}\qquad{}\qquad{}(b)
\par\end{centering}

\begin{centering}
\includegraphics[width=6.9cm]{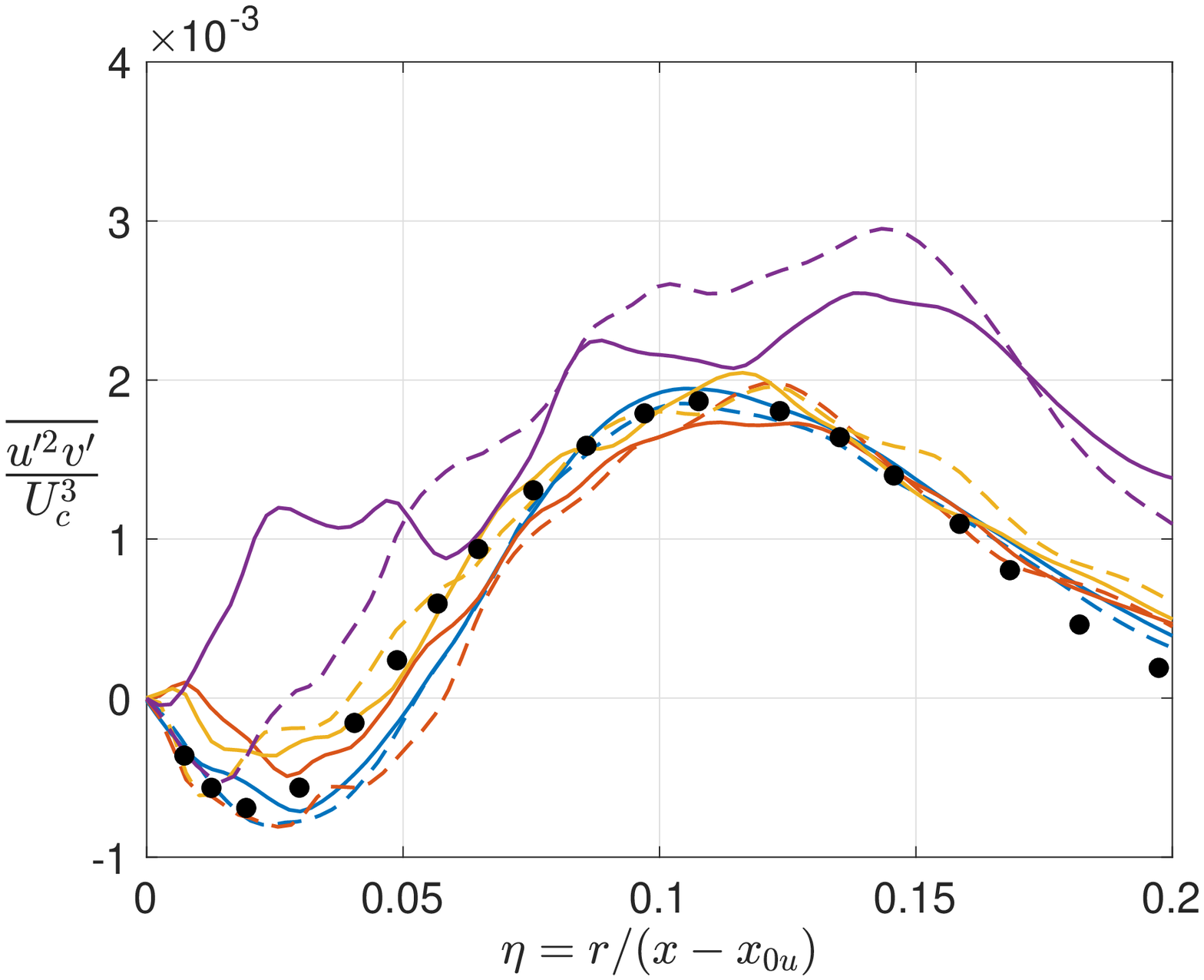}\includegraphics[width=6.9cm]{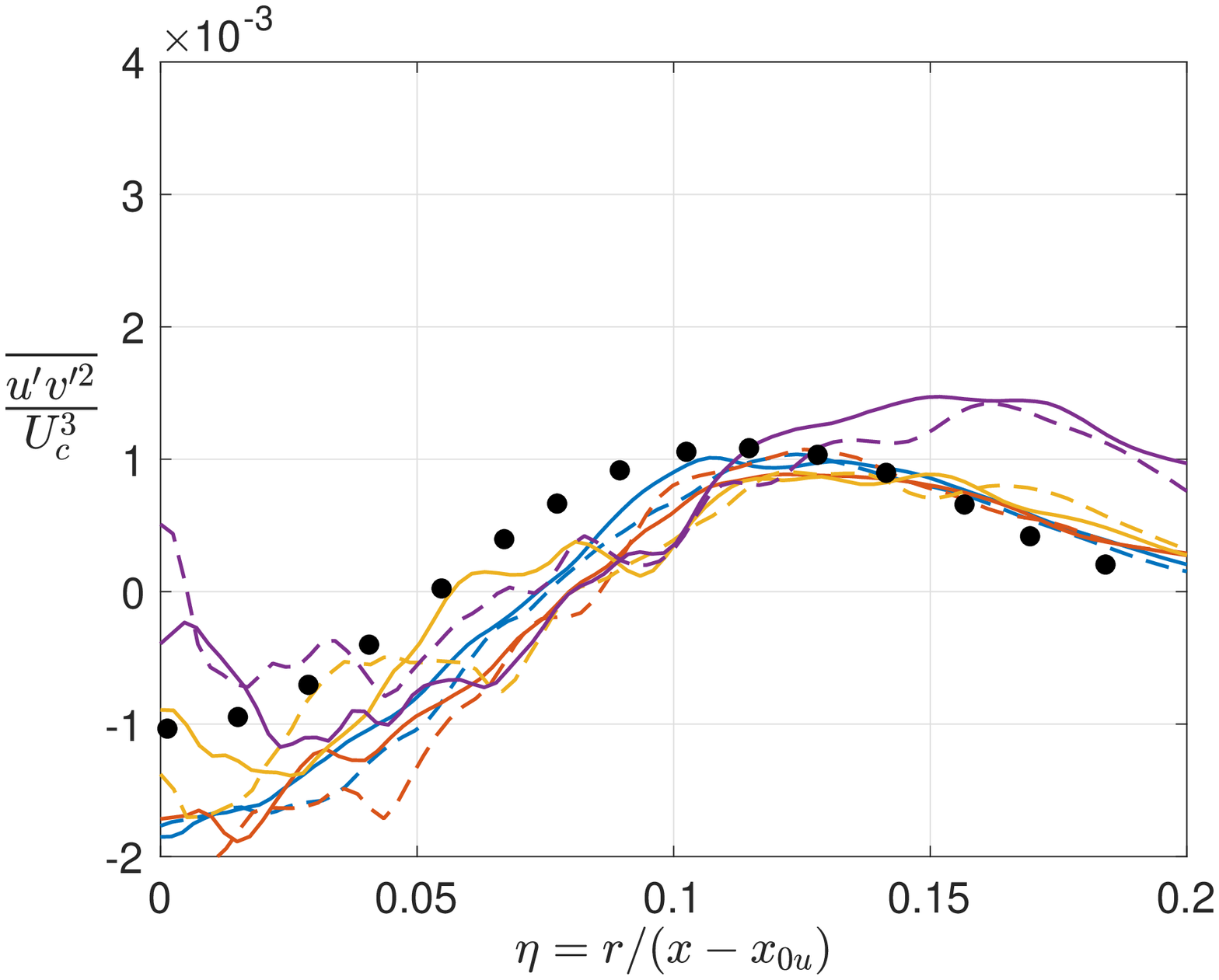}
\par\end{centering}

\begin{centering}
(c)\qquad{}\qquad{}\qquad{}\qquad{}\qquad{}\qquad{}\qquad{}\qquad{}\qquad{}\qquad{}\qquad{}\qquad{}(d)
\par\end{centering}

\begin{centering}
\includegraphics[width=6.9cm]{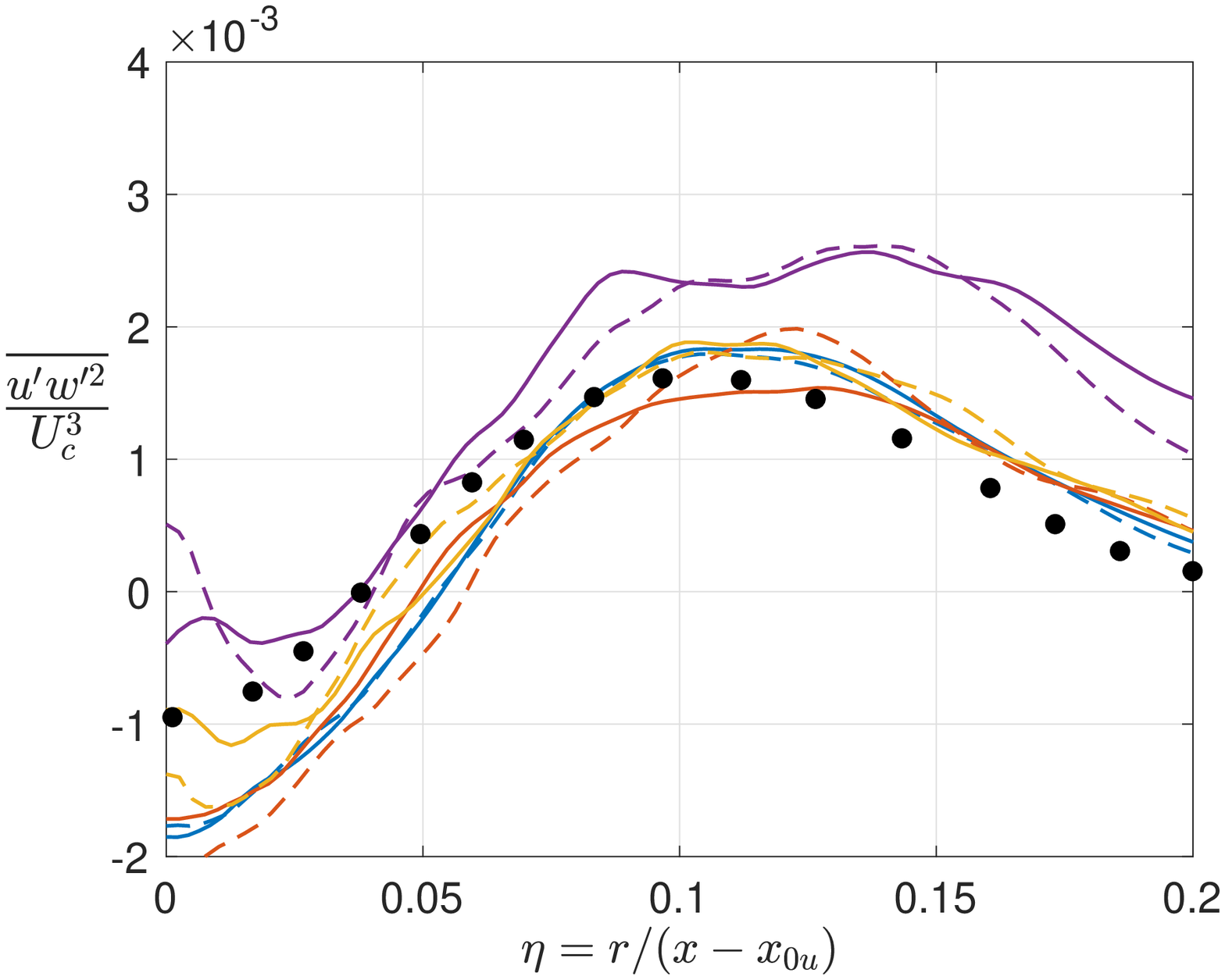}\includegraphics[width=6.9cm]{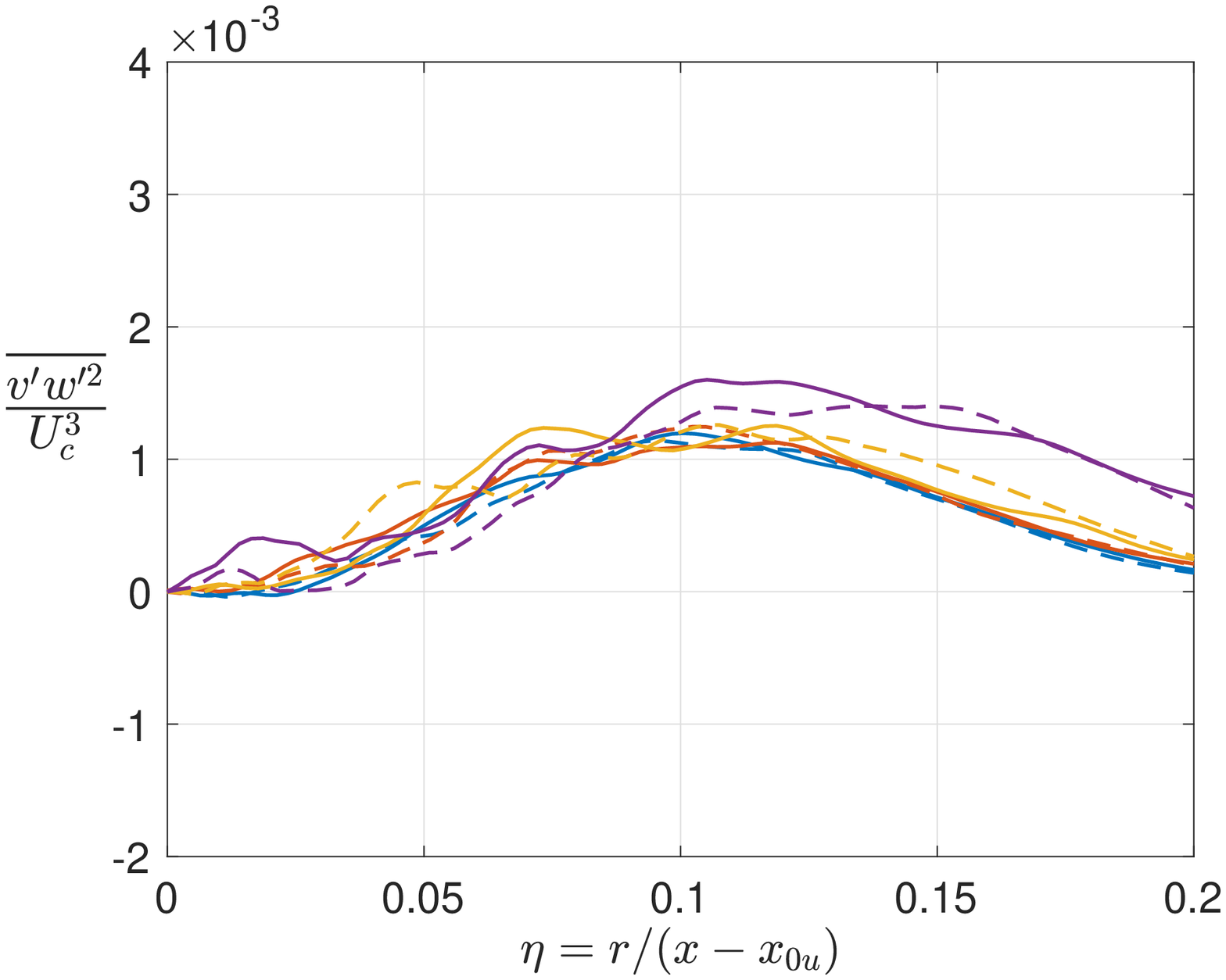}
\par\end{centering}

\begin{centering}
(e)\qquad{}\qquad{}\qquad{}\qquad{}\qquad{}\qquad{}\qquad{}\qquad{}\qquad{}\qquad{}\qquad{}\qquad{}(f)
\par\end{centering}

\caption{Case 1--4 comparisons: Radial profiles of the normalized third-order
velocity moments. (a) $\overline{u^{\prime3}}/U_{c}^{3}$, (b) $\overline{v^{\prime3}}/U_{c}^{3}$,
(c) $\overline{u^{\prime2}v'}/U_{c}^{3}$, (d) $\overline{u'v^{\prime2}}/U_{c}^{3}$,
(e) $\overline{u'w^{\prime2}}/U_{c}^{3}$ and (f) $\overline{v'w^{\prime2}}/U_{c}^{3}$.
The black markers $\bullet$ show profiles from the experiments
of \cite{panchapakesan1993turbulence}. \label{fig:TripleVel_corr_Cases1to4}}
\end{figure}

Gradients of the pressure/density fluctuations, pressure-velocity correlations and third-order velocity moments
determine the transport terms in the Reynolds stress and t.k.e. equations 
\cite[e.g.][]{panchapakesan1993turbulence,hussein1994velocity}, and hence their role in 
causing the differences observed in velocity fluctuations 
of Cases 1--4 
(figures \ref{fig:rms_fluc_Cases1to4} and \ref{fig:Case_1to4_vel_stat})  is
examined in figures \ref{fig:centerline_pfluc_pvelCorl}, \ref{fig:prs_vel_correlation_Cases1to4},
and \ref{fig:TripleVel_corr_Cases1to4}.

The axial variation of the centerline r.m.s. pressure and density fluctuations normalized using 
centerline mean values, $p_{c}$ and $\rho_{c}$, respectively, are 
compared in figure \ref{fig:centerline_pfluc_pvelCorl}(a) for Cases 1--4.
The normalization provides information on the fluctuation magnitude with respect to local pressure and
density (thermodynamic state). In all cases, $p'_{c,\textrm{rms}}/p_{c}$
and $\rho'_{c,\textrm{rms}}/\rho_{c}$ have a maximum in the transition
region and asymptote to a constant value in the self-similar region.
$p'_{c,\textrm{rms}}/p_{c}$ exceed $\rho'_{c,\textrm{rms}}/\rho_{c}$ in 
the transition region and vice versa in the self-similar region. Increasing 
$p_{\infty}$ from $1$ bar (Case 1) to $50$ bar (Case 2) slightly reduces
$p'_{c,\textrm{rms}}/p_{c}$ (shown as solid lines) and $\rho'_{c,\textrm{rms}}/\rho_{c}$
(shown as dashed lines) at all centerline locations, whereas decreasing $Z$
from $0.99$ (Case 2) to $0.9$ (Case 3) and then to $0.8$ (Case
4) increases $p'_{c,\textrm{rms}}/p_{c}$ and $\rho'_{c,\textrm{rms}}/\rho_{c}$
significantly. The variation of $p'_{c,\textrm{rms}}/p_{c}$ and $\rho'_{c,\textrm{rms}}/\rho_{c}$
follows the variation of $p_{\infty}\left(\beta_{T}-1/p_{\infty}\right)$, listed in table \ref{tab:Isothermal-compressibility},
that measures the real-gas effects at ambient thermodynamic condition. The large 
value of $p_{\infty}\left(\beta_{T}-1/p_{\infty}\right)$ for Case 4 concurs with 
the large $p'_{c,\textrm{rms}}/p_{c}$ and $u'_{c,\textrm{rms}}/U_{c}$ observed in Case 4,
a fact which indicates that the large $U_c$ decay and jet spread in Case 4 is a result of
the real-gas effects due to its proximity to the Widom line. 

While $p_{\infty}\left(\beta_{T}-1/p_{\infty}\right)$ explains the behavior of Case 4 with respect to other cases, it does not explain the behavior of Case 3 with respect to Case 1 in the transition region of the flow. Larger $p_{\infty}\left(\beta_{T}-1/p_{\infty}\right)$ in Case 3 may suggest larger $p'_{c,\textrm{rms}}/p_{c}$ and $u'_{c,\textrm{rms}}/U_{c}$ in Case 3 compared to Case 1. But, while $p'_{c,\textrm{rms}}/p_{c}$ is larger in Case 3 than in Case 1 at all axial locations, $u'_{c,\textrm{rms}}/U_{c}$ in Case 1 exceeds Case 3 in the transition region resulting in larger $U_c$ decay and jet spread in Case 1 than in Case 3. To understand this discrepancy, the centerline variation of fluctuating pressure-axial velocity correlation, 
$\overline{p'u'}$, whose axial gradient determines t.k.e. diffusion due to 
pressure fluctuation transport in the t.k.e. equation, is illustrated in 
figure \ref{fig:centerline_pfluc_pvelCorl}(b). 
Large local changes in $\overline{p'u'}$ increase
the turbulent transport term magnitude in the t.k.e. equation. The
$\overline{p'u'}$ values are non-positive at all centerline locations
for all cases, implying that a positive pressure fluctuation (higher
than the mean) is correlated with negative velocity fluctuation (lower
than the mean) and vice versa. $\frac{\overline{p'u'}/\bar{\rho}}{U_{c}^{3}}$
profiles in figure \ref{fig:centerline_pfluc_pvelCorl}(b) for all
cases have a local minimum in the near field $5\lesssim x/D\lesssim9$ and
downstream of that minimum, the variations in $\frac{\overline{p'u'}/\bar{\rho}}{U_{c}^{3}}$
are much larger in Case 4 compared to Case 1, which itself exhibits larger variations than in Cases 2 and 3. In the
region $9\lesssim x/D\lesssim15$, $\frac{\overline{p'u'}/\bar{\rho}}{U_{c}^{3}}$
profiles exhibit distinct minima (with large negative values) in Cases
1 and 4, whereas the profiles of Cases 2 and 3 smoothly approach a
near-constant value. The larger variation of $\frac{\overline{p'u'}/\bar{\rho}}{U_{c}^{3}}$
in this region in Cases 1 and 4 coincides with the larger $U_c$ decay and jet spread in figure \ref{fig:invUc_hfRad_Cases1to4}
and larger $u'_{c,\textrm{rms}}/U_{c}$ in figure \ref{fig:rms_fluc_Cases1to4}
for those cases. This indicates that higher $p'_{c,\textrm{rms}}/p_{c}$ does not
guarantee higher $u'_{c,\textrm{rms}}/U_{c}$, instead $u'_{c,\textrm{rms}}/U_{c}$ follows the behavior of $\frac{\overline{p'u'}/\bar{\rho}}{U_{c}^{3}}$, whose axial gradient determines the turbulent transport term in the Reynolds stress equation governing $u'_{c,\textrm{rms}}$ and the t.k.e. equation. In \S \ref{sub:summary_inflowEffects}, it is further observed that the variations in t.k.e. turbulent transport agrees with the variations in the t.k.e. production resulting from the structural change in turbulence due to thermodynamic conditions. The differences in $u'_{c,\textrm{rms}}/U_{c}$ behavior of Cases 3 and 4 with respect to Case 1 suggests that a relatively large change in thermodynamic condition from a perfect gas, as in Case 4, is required to effect a large change in $u'_{c,\textrm{rms}}/U_{c}$.
On the centerline, the fluctuating pressure-radial
velocity correlation, $\overline{p'v'}$, is null.

The centerline variation of $\overline{u^{\prime3}}/U_{c}^{3}$
is examined in figure \ref{fig:centerline_pfluc_pvelCorl}(c) for Cases 1--4.
The increase in $p_{\infty}$ from 1 bar (Case 1) to 50 bar (Case 2) reduces
the overall axial variations of $\overline{u^{\prime3}}/U_{c}^{3}$, whereas the decrease
in $Z$ from $0.99$ (Case 2) to $0.9$ (Case 3) at 50 bar pressure
has minimal influence on $\overline{u^{\prime3}}/U_{c}^{3}$ behavior.
Further decrease of $Z$ to $0.8$ (Case 4) significantly enhances
variations in $\overline{u^{\prime3}}/U_{c}^{3}$, indicating the 
significance of its gradient in the t.k.e. equation with proximity
to the Widom line. 
Similar to figure \ref{fig:centerline_pfluc_pvelCorl}(b),
regions of large variations in $\overline{u^{\prime3}}/U_{c}^{3}$ concur
with the regions of large changes in mean axial-velocity and large
velocity fluctuations seen in figures \ref{fig:invUc_hfRad_Cases1to4}
and \ref{fig:rms_fluc_Cases1to4}, respectively.

To complete the physical picture, the radial variation of fluctuating pressure-velocity correlations and
third-order velocity moments at $x/D\approx25$ and $30$ from Cases
1--4 is compared in figures \ref{fig:prs_vel_correlation_Cases1to4}
and \ref{fig:TripleVel_corr_Cases1to4}, respectively. Both $\frac{\overline{p'u'}/\bar{\rho}}{U_{c}^{3}}$
and $\frac{\overline{p'v'}/\bar{\rho}}{U_{c}^{3}}$ 
exhibit negative values at all radial locations. $\frac{\overline{p'u'}/\bar{\rho}}{U_{c}^{3}}$
peaks in absolute magnitude at the centerline, whereas the peak of
$\frac{\overline{p'v'}/\bar{\rho}}{U_{c}^{3}}$ lies off-axis. The
radial variation of $\frac{\overline{p'u'}/\bar{\rho}}{U_{c}^{3}}$
in Case 4 is significantly larger than in other cases at all radial locations,
implying greater normalized t.k.e. diffusion flux due to pressure
fluctuation transport by axial-velocity fluctuations. The normalized
radial t.k.e. diffusion flux from pressure fluctuation transport,
$\frac{\overline{p'v'}/\bar{\rho}}{U_{c}^{3}}$, is similar near the
centerline for all cases but larger in magnitude in Case 4 for $\eta\gtrsim0.075$, indicating greater radial t.k.e. transport  in Case 4 that enhances entrainment and mixing at the edges of the jet shear layer. 
Similarly, the third-order velocity moments, representing the t.k.e. diffusion 
fluxes due to the transport of Reynolds stresses
by the fluctuating velocity field, depicted in figure \ref{fig:TripleVel_corr_Cases1to4},
highlight the difference in behavior in Case 4 from other cases. All non-zero third-order moments
from Cases 1--4 together with the experimental profiles from \cite{panchapakesan1993turbulence}
are shown in the figure; the \cite{panchapakesan1993turbulence} experiments
measured all third-order moments except $\overline{v'w^{\prime2}}$. Aside
from the large flux values in Case 4, a noticeable feature in correlations
$\overline{u^{\prime2}v'}$, $\overline{u'v^{\prime2}}$ and $\overline{u'w^{\prime2}}$
is their negative values near the centerline. The negative $\overline{u^{\prime2}v'}$
indicates a radial flux of the axial component of t.k.e., $\overline{u^{\prime2}}$,
towards the centerline. The smaller region of negative values with decrease
in $Z$ from Case 2 to Case 4 indicates a smaller
radial flux towards the centerline and a dominant radially outward
transport of $\overline{u^{\prime2}}$.\vspace{-0.75cm}

\noindent 

\subsubsection{Passive scalar mixing}

\begin{figure}
\begin{centering}
(a) \includegraphics[width=11cm]{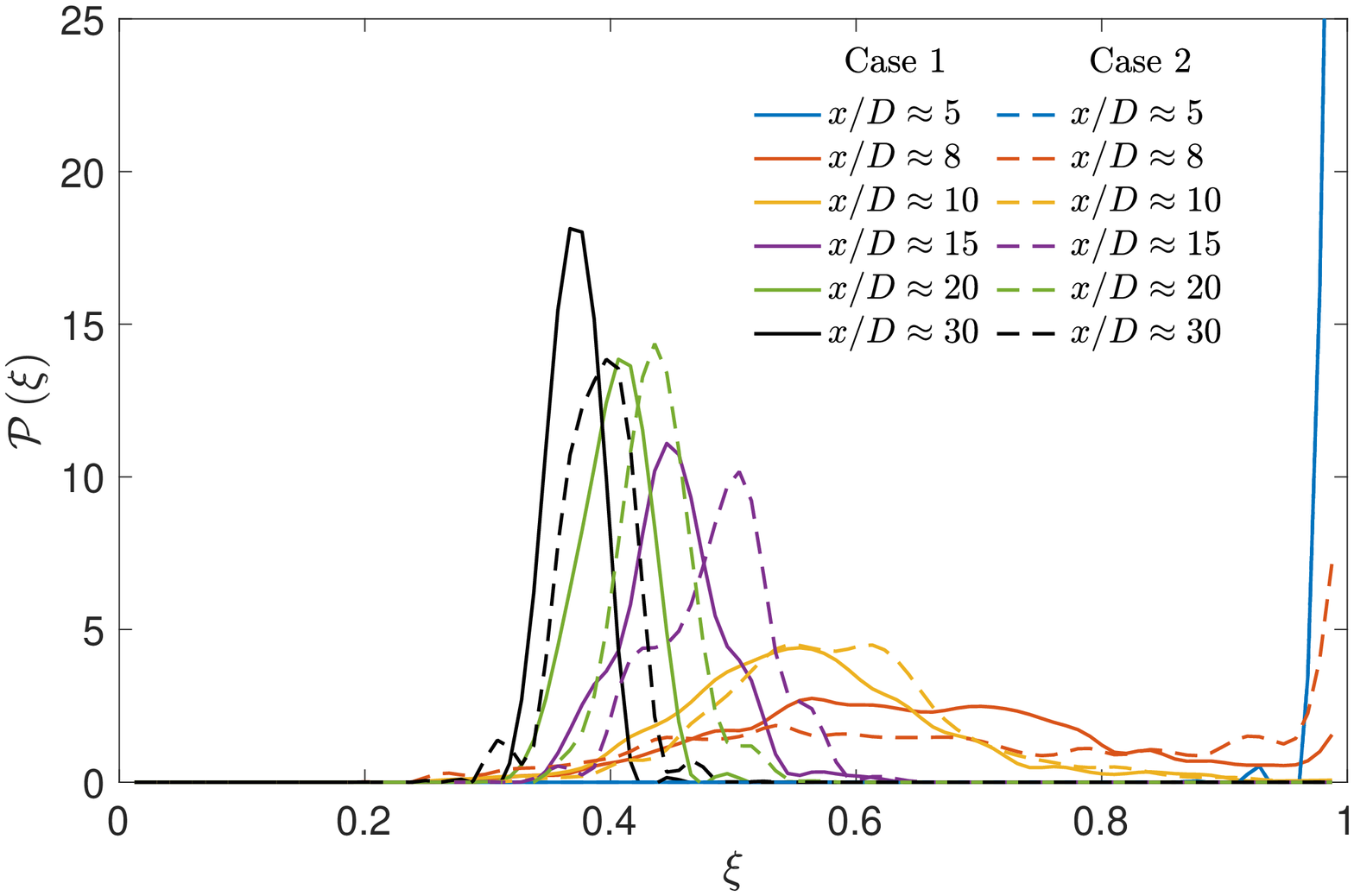}
\par\end{centering}

\begin{centering}
(b) \includegraphics[width=11cm]{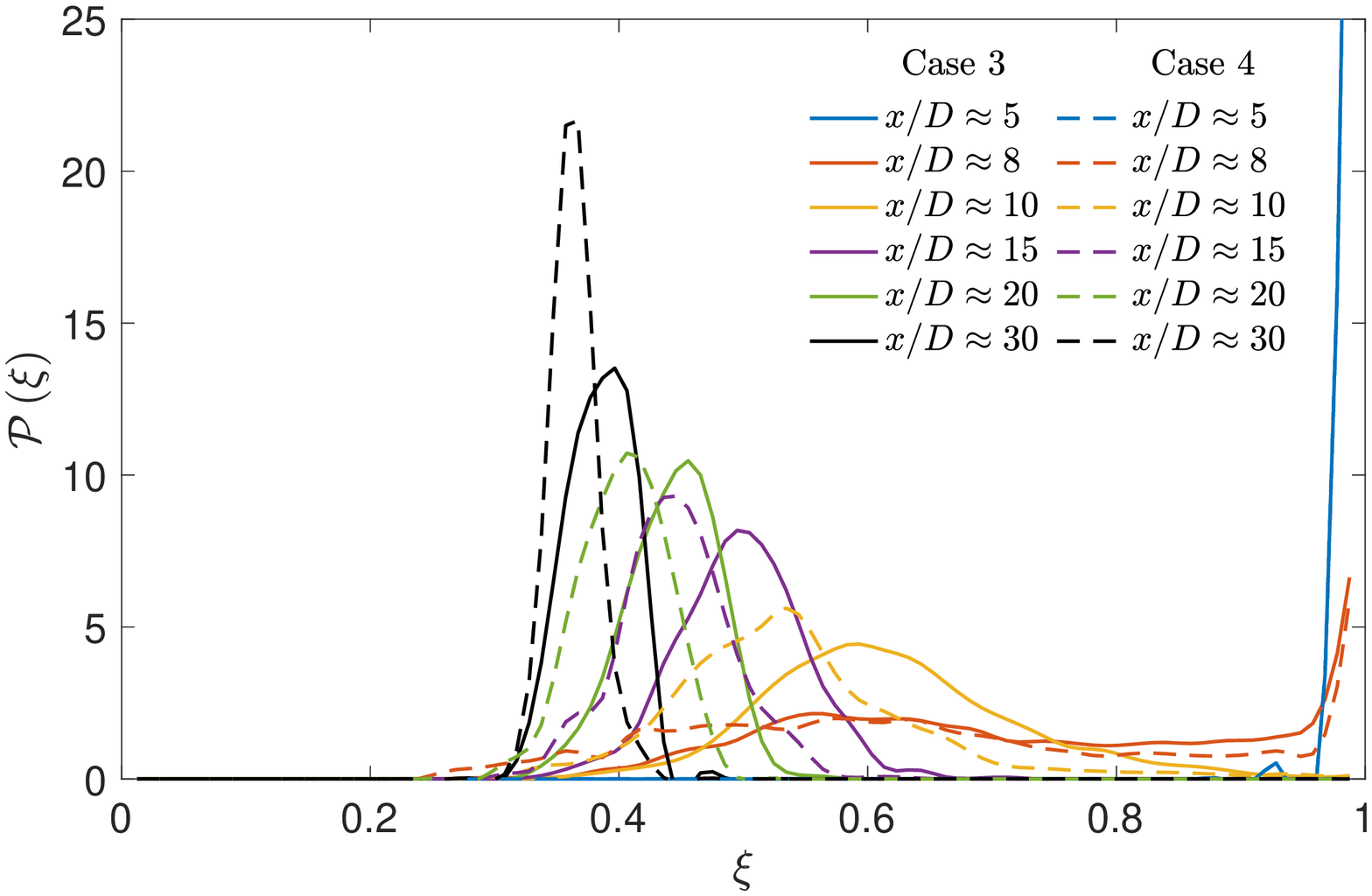}
\par\end{centering}

\caption{Scalar probability density function, $\mathcal{P}\left(\xi\right)$,
at various centerline axial locations for (a) Cases 1 and 2, and (b)
Cases 3 and 4. \label{fig:SclPDF_cases1to4}}
\end{figure}

\begin{figure}
\begin{centering}
\includegraphics[width=6.9cm]{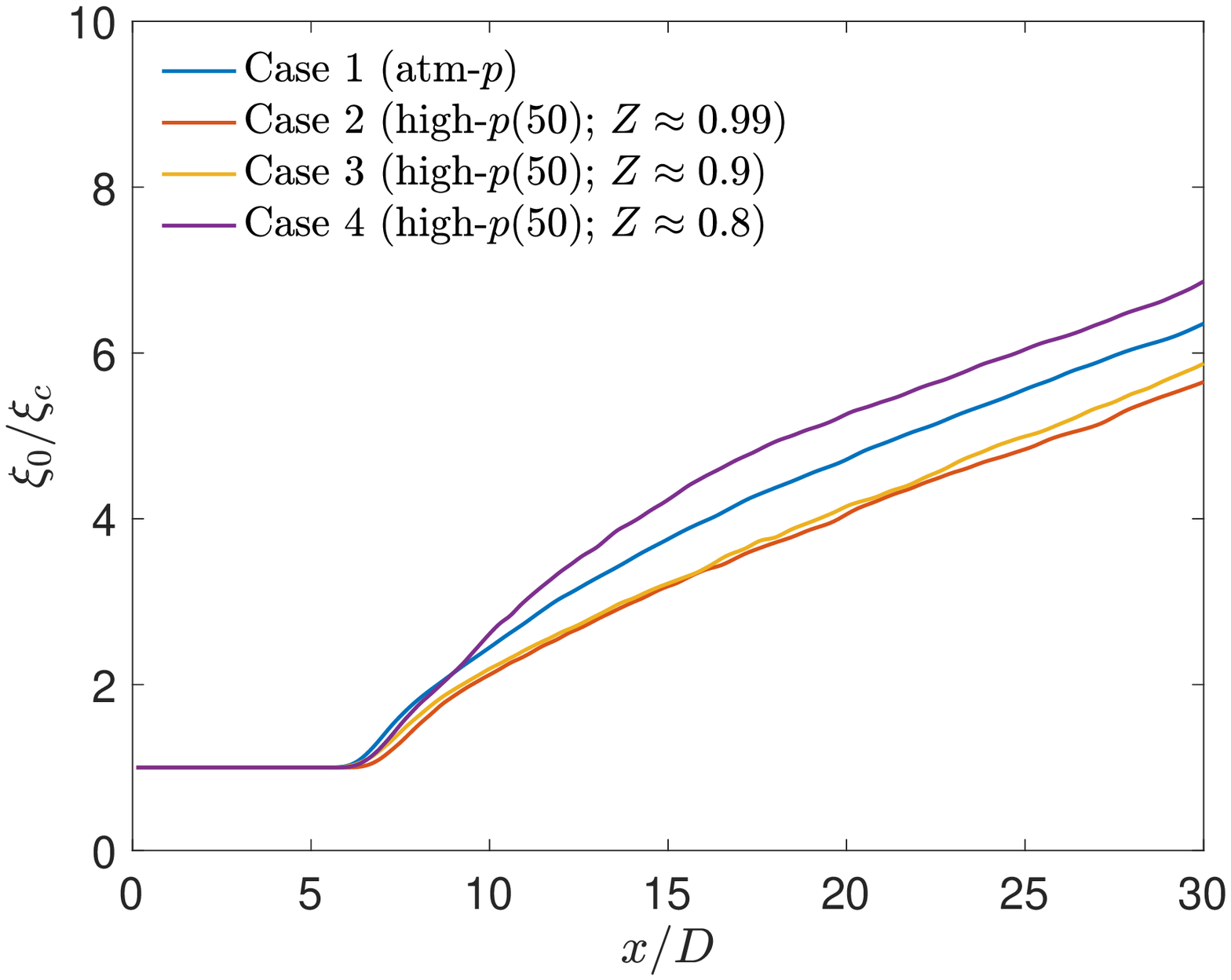}\includegraphics[width=6.9cm]{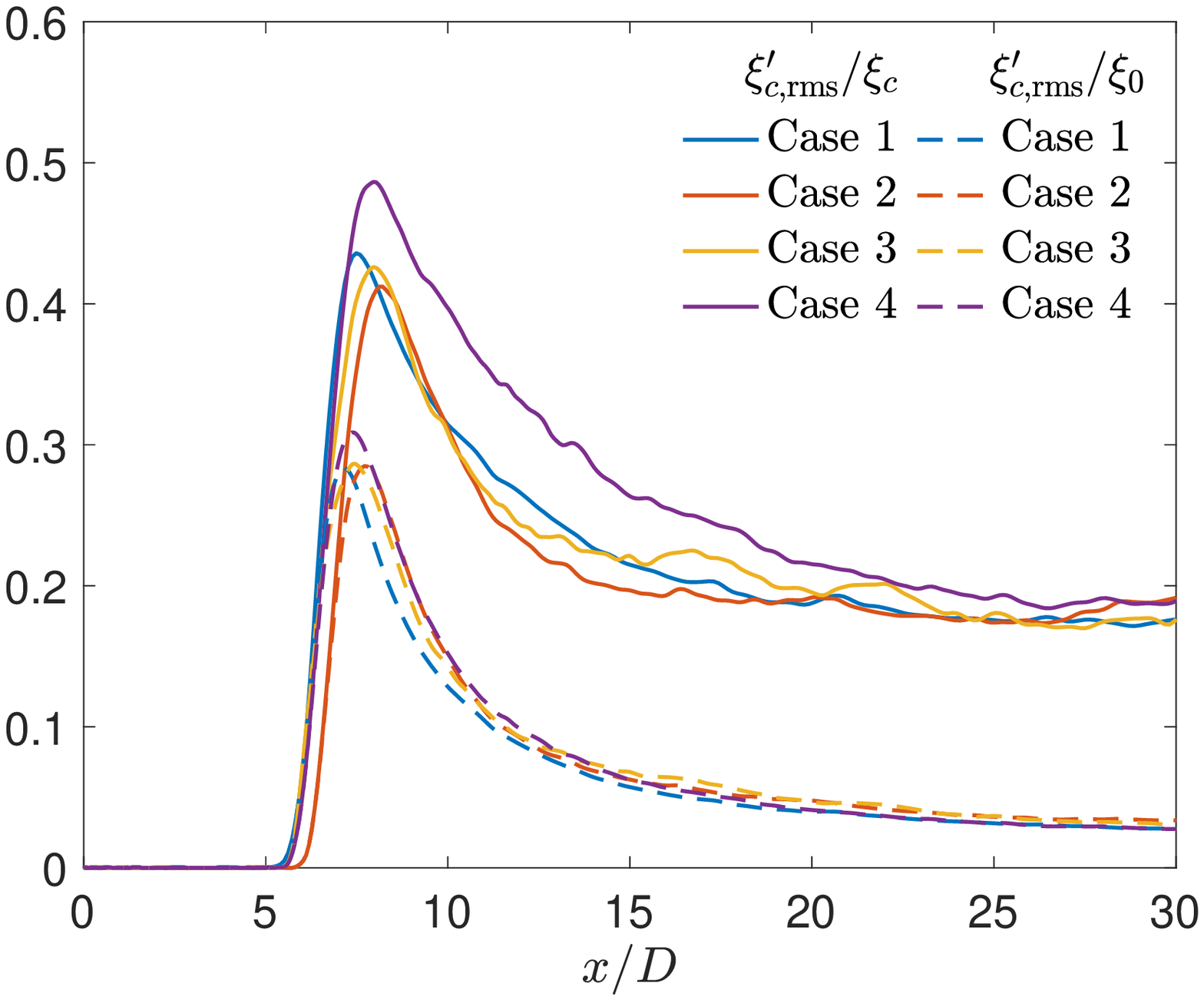}
\par\end{centering}

\begin{centering}
(a)\qquad{}\qquad{}\qquad{}\qquad{}\qquad{}\qquad{}\qquad{}\qquad{}\qquad{}\qquad{}\qquad{}\qquad{}(b)
\par\end{centering}

\caption{Case 1--4 comparisons: Streamwise variation of the (a) inverse of
centerline scalar concentration ($\xi_{c}$) normalized by the jet-exit
centerline value ($\xi_{0}$) and (b) centerline r.m.s. scalar fluctuation
($\xi'_{c,\textrm{rms}}$) normalized by the centerline mean value,
$\xi_{c}$, and jet-exit mean value, $\xi_{0}$. \label{fig:invScl_rmsFluc_Cases1to4}}
\end{figure}

To assess whether there are mixing differences among Cases 1--4 emulating those of the velocity field,
the one-point scalar probability density function (p.d.f.) is examined in figure \ref{fig:SclPDF_cases1to4} at various centerline locations. The p.d.f., $\mathcal{P}\left(  \xi\right) $, is defined such that
\begin{equation}
\intop_{0}^{1}\mathcal{P}\left(\tilde{\xi}\right)\mathrm{d}\tilde{\xi}=1\qquad\textrm{and}\qquad\bar{\xi}=\intop_{0}^{1}\tilde{\xi}\mathcal{P}\left(\tilde{\xi}\right)\mathrm{d}\tilde{\xi}.\label{eq:SclPDF_defn}
\end{equation} 
Comparisons between
Cases 1 and 2, shown in figure \ref{fig:SclPDF_cases1to4}(a), evaluate
the effect of $p_{\infty}$ increase at approximately same $Z$ value. At $x/D\approx5$,
the centerline contains pure jet fluid $\left(\xi=1\right)$ in both
cases. The potential core closes downstream of $x/D\approx5$, and
the p.d.f. at $x/D\approx8$ in both cases shows a wide spread with
mixed-fluid concentration in the range $0.3\lesssim\xi\lesssim1$.
Velocity/pressure statistics in the transition region of Cases 1 and
2 differ significantly, as observed in figures \ref{fig:invUc_hfRad_Cases1to4},
\ref{fig:rms_fluc_Cases1to4} and \ref{fig:centerline_pfluc_pvelCorl}(a).
Similarly, $\overline{\xi}$ and $\xi_{\textrm{rms}}^{\prime}$ vary in the
transition region yielding differences in mixed-fluid composition
and $\mathcal{P}\left(  \xi\right)$. In the transition region, the mean scalar concentration
decays at a faster rate in Case 1 than in Case 2, as shown in figure
\ref{fig:invScl_rmsFluc_Cases1to4}(a). As a result, downstream of
$x/D\approx10$, the p.d.f. peaks are closer to the jet pure fluid
concentration $\left(\xi=1\right)$ in Case 2 than in Case 1, indicating lesser 
mixing in Case 2 compared to Case 1. For
$x/D\geq10$, the absolute scalar fluctuations $\xi'_{c,\textrm{rms}}/\xi_{0}$
are slightly larger in Case 2 than in Case 1, despite smaller normalized local
fluctuation $\xi'_{c,\textrm{rms}}/\xi_{c}$ in Case 2 between $10 \geq x/D \geq 20$, as shown in
figure \ref{fig:invScl_rmsFluc_Cases1to4}(b). Larger $\xi_{c,\textrm{rms}%
}^{\prime}/\xi_{0}$ implies wider p.d.f. profiles with smaller peaks, indicating larger fluctuations in the mixed-fluid composition and, hence, greater mixing in Case 2 compared to Case 1. However, due to the large jet width downstream of $x/D\approx15$, the scalar fluctuations at the centerline mix the already (partially) mixed fluid in the vicinity of the centerline and not the pure ambient fluid with the jet fluid. As a result, the p.d.f. peaks showing the mean mixed-fluid concentration continue to be closer to the jet pure fluid concentration $\left(\xi=1\right)$ in Case 2 than in Case 1.

Figure \ref{fig:SclPDF_cases1to4}(b) compares the scalar p.d.f. from
Cases 3 and 4 to examine the effect of $Z$ on mixing behavior at supercritical
pressure. Analogous to figure \ref{fig:SclPDF_cases1to4}(a),
the p.d.f. profiles at $x/D\approx5$ and $8$ are largely similar
between the two cases. Differences in peak scalar value, representing
the mean concentration, arise in the transition region, consistent
with the observations in figure \ref{fig:invScl_rmsFluc_Cases1to4}. Large
scalar fluctuations around $x/D\approx10$ in Case 4 compared to Case 3 leads 
to greater mixing in Case 4 in the transition region, resulting in the p.d.f. peaks 
that are closer to the jet pure fluid concentration $\left(\xi=1\right)$ in Case 3 than in Case 4.
At locations downstream of $x/D\approx15$, the slightly larger $\xi_{c,\textrm{rms}%
}^{\prime}/\xi_{0}$ in Case 3 compared to Case 4 leads to wider p.d.f.
profiles with smaller peaks in Case 3. However, the large jet width downstream of $x/D\approx15$ causes mixing of the already (partially) mixed fluid in the vicinity of the centerline and, hence, the p.d.f. peaks continue to be closer to the jet pure fluid concentration in Case 3 than in Case 4. \vspace{-0.5cm}

\noindent 

\noindent 

\subsubsection{Summary}
The examination of the influence of $p_{\infty}$ and $Z$ on flow statistics in
laminar-inflow jets at fixed $Re_{D}$ yields several conclusions. The velocity statistics (mean and fluctuations) attain self-similarity in high-$p$ compressible jets.
The flow exhibits sensitivity to $p_{\infty}$ 
and $Z$ in the transition as well as the self-similar region, with larger differences 
observed in the transition region. The normalized pressure and density flucutations in the flow follow the behavior of the non-dimensional quantity $p_{\infty}\left(\beta_{T}-1/p_{\infty}\right)$, listed in table \ref{tab:Isothermal-compressibility}, that provides a measure other than $Z$ to estimate departure from perfect gas. Proximity of the ambient flow conditions to the Widom line increases $p_{\infty}\left(\beta_{T}-1/p_{\infty}\right)$ and enhances the normalized pressure and density flucutations in the flow, especially in the transition region. The velocity behavior (mean and fluctuations) can be explained in terms of the spatial variation of the normalized pressure-velocity correlations and third-order velocity moments that determine the transport terms in the t.k.e. equation. Increase in velocity fluctuations also enhances the scalar fluctuations, leading to greater centerline mixing in Case 4 compared to the other cases.
%

\subsection{Effects of supercritical pressure at a fixed compressibility factor\label{sub:Case3n5_comparison}}

The above analysis examined the effects of $Z$ at a 
fixed supercritical $p_\infty$. 
To examine its counterpart, the influence of $p_\infty$
at a fixed $Z$ of $0.9$ is studied here by comparing results between Cases 3 and
5 (see table \ref{tab:Summary_of_cases}). The value of $p_{\infty}\left(\beta_{T}-1/p_{\infty}\right)$
is slightly larger in Case 3 compared to Case 5, a fact which according to the results of \S \ref{sub:Case1to4_comparison}
should lead to larger local pressure/density fluctuations in Case 3.

\begin{figure}
\begin{centering}
\includegraphics[width=6.9cm]{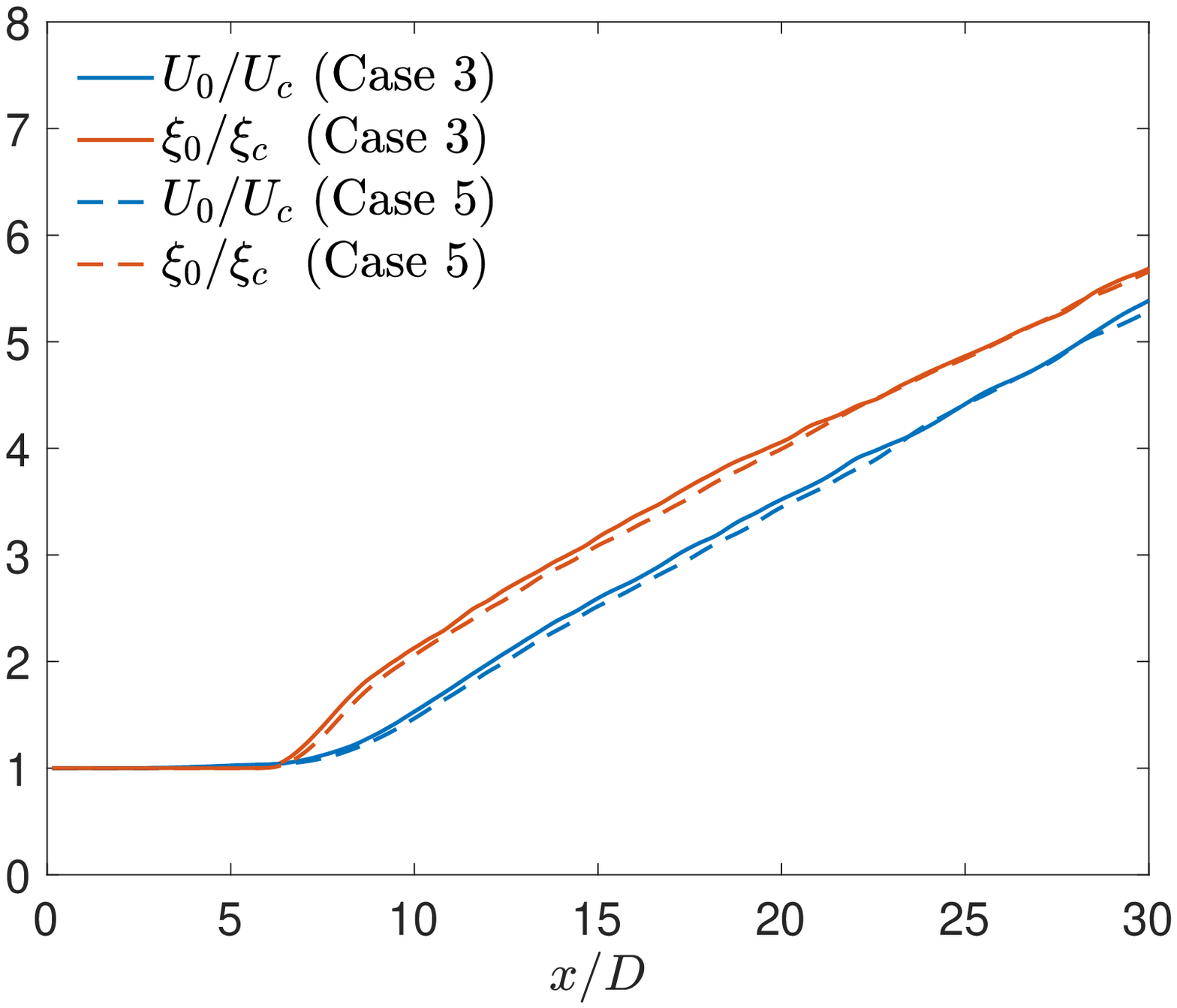}\includegraphics[width=6.9cm]{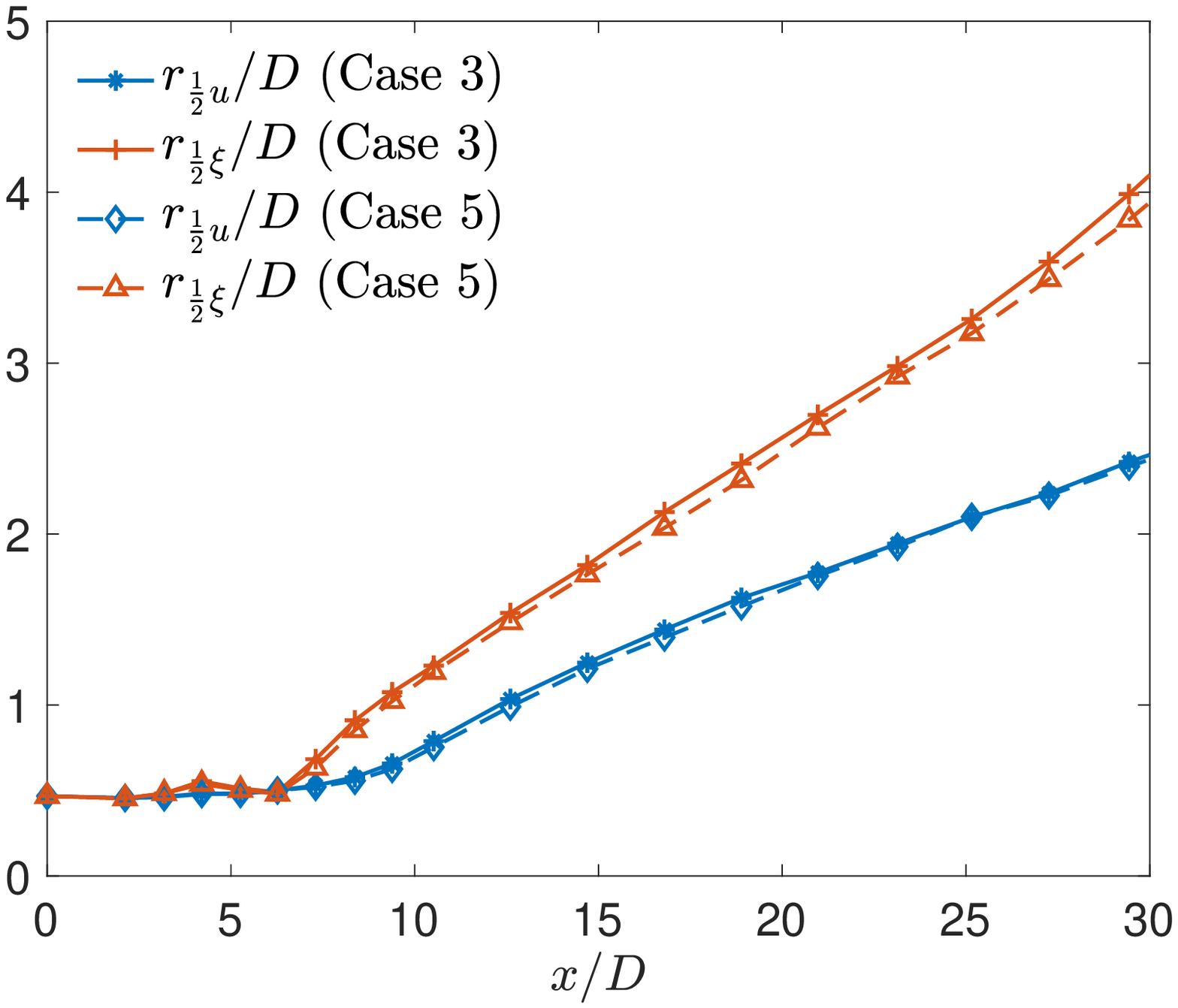}
\par\end{centering}

\begin{centering}
(a)\qquad{}\qquad{}\qquad{}\qquad{}\qquad{}\qquad{}\qquad{}\qquad{}\qquad{}\qquad{}\qquad{}\qquad{}(b)
\par\end{centering}

\caption{Case 3 and 5 comparisons: Streamwise variation of the (a) inverse
of centerline mean axial velocity ($U_{c}$) and scalar ($\xi_{c}$) normalized
by the jet-exit centerline values $U_{0}$ and $\xi_{0}$, respectively,
and (b) velocity and scalar half radius ($r_{\frac{1}{2}u}$ and $r_{\frac{1}{2}\xi}$).
\label{fig:invUc_hfRad_Cases3n5}}
\end{figure}

The differences in the centerline profiles of the mean axial velocity and mean scalar concentration between Cases 3 and 5, presented in figure \ref{fig:invUc_hfRad_Cases3n5}(a), are minimal in the
transition region and they diminish in the self-similar region. 
Similar to figure \ref{fig:invUc_hfRad_Cases3n5}(a),
minor differences are observed in the velocity 
half-radius ($r_{\frac{1}{2}u}$), shown in figure \ref{fig:invUc_hfRad_Cases3n5}(b), between Cases 3 and 5. 
In comparison, small but noticeable differences are observed in scalar half-radius ($r_{\frac{1}{2}\xi}$), 
where the jet spread in Case 3 is slightly larger than that
in Case 5.

\begin{figure}
\begin{centering}
\includegraphics[width=6.9cm]{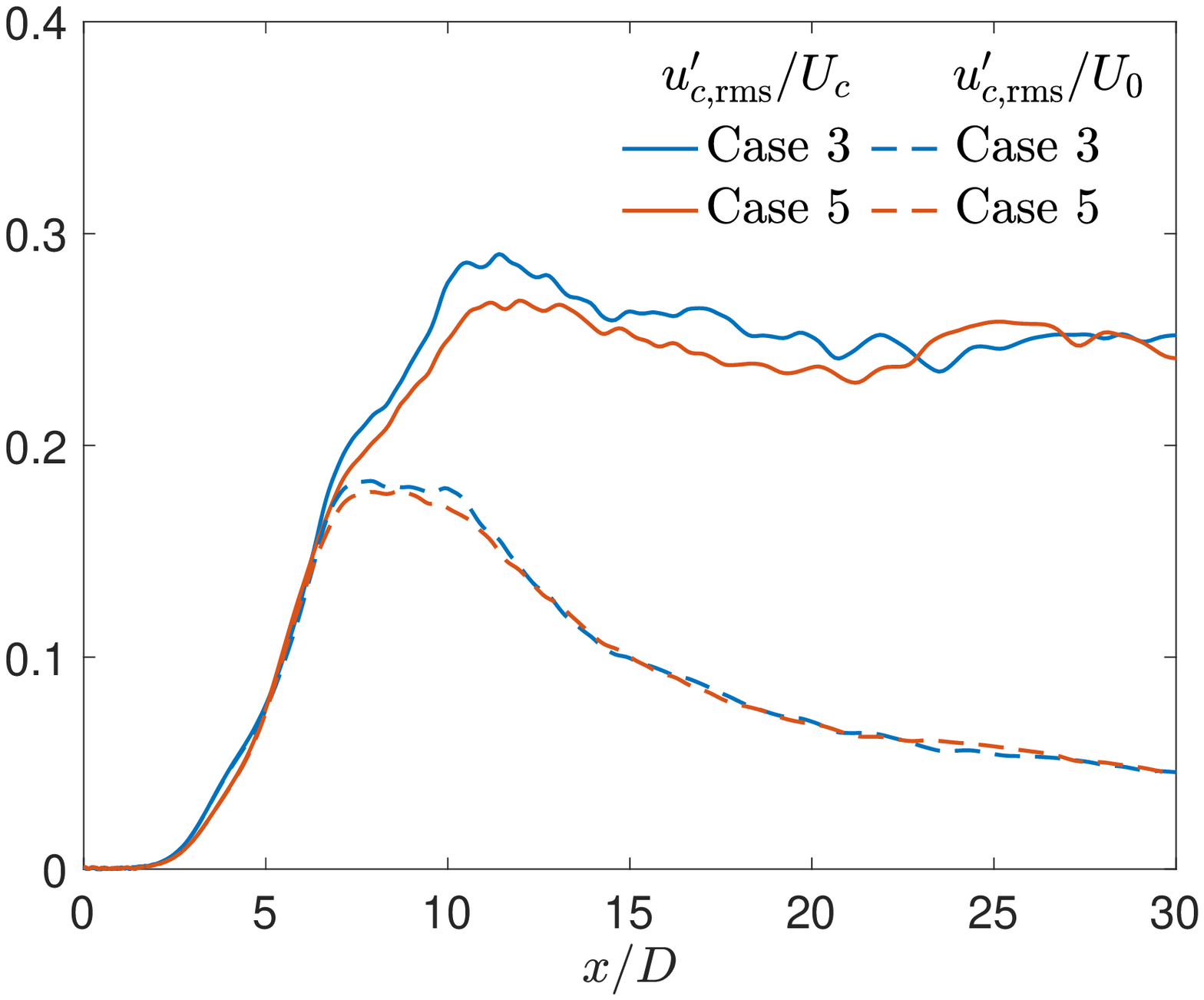}\includegraphics[width=6.9cm]{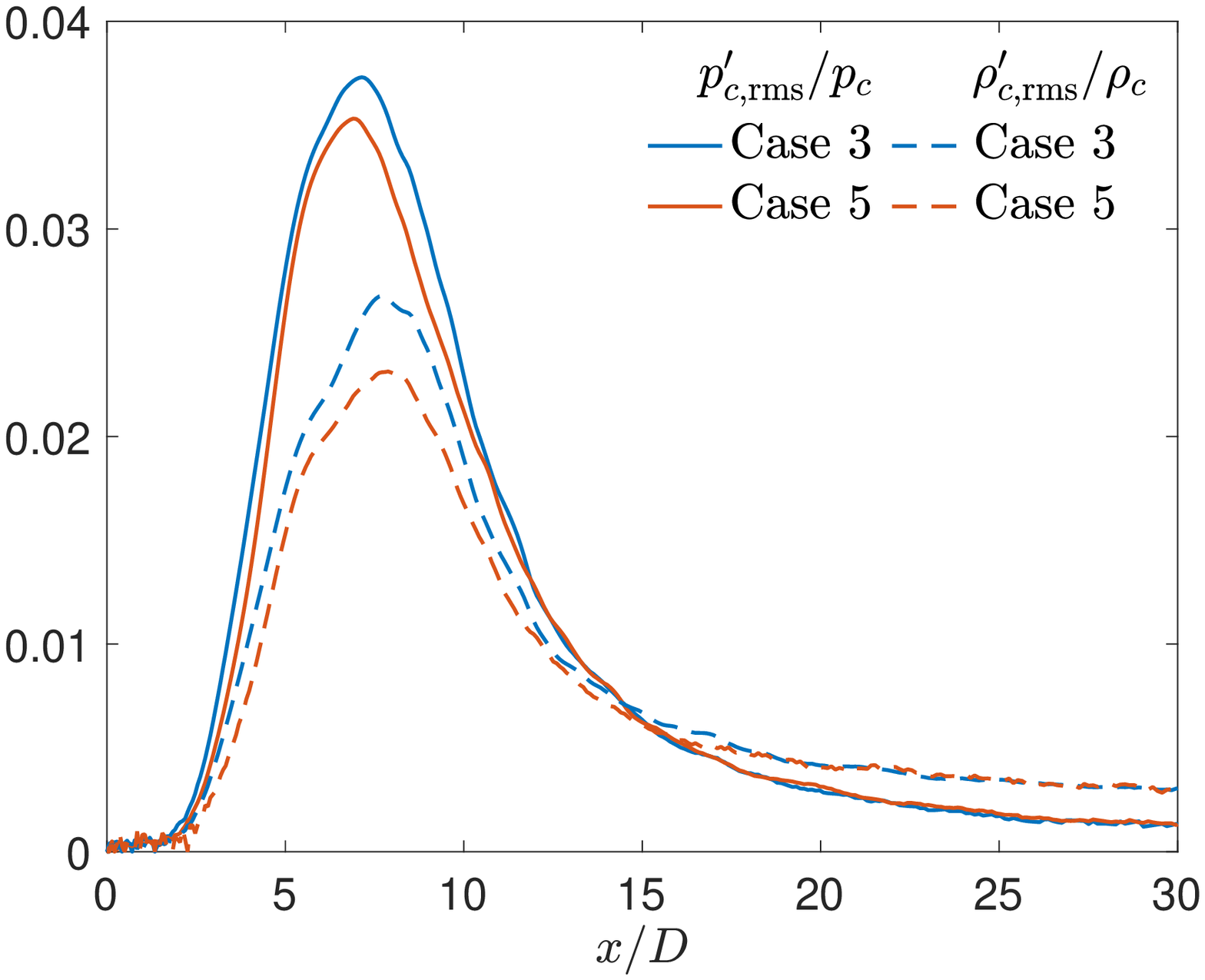}
\par\end{centering}

\begin{centering}
(a)\qquad{}\qquad{}\qquad{}\qquad{}\qquad{}\qquad{}\qquad{}\qquad{}\qquad{}\qquad{}\qquad{}\qquad{}(b)
\par\end{centering}

\caption{Case 3 and 5 comparisons: Streamwise variation of (a) the centerline
r.m.s. axial velocity fluctuations ($u_{c,\mathrm{rms}}^{'}$) normalized
by the centerline mean axial velocity ($U_{c}$) and centerline jet-exit
axial velocity ($U_{0}$), and (b) the centerline r.m.s. pressure
and density fluctuations, denoted by $p'_{c,\mathrm{rms}}$ and $\rho'_{c,\mathrm{rms}}$,
respectively, normalized by the centerline mean pressure ($p_{c}$)
and density ($\rho_{c}$), respectively. \label{fig:centerline_urms_prms_cases3n5}}
\end{figure}

To further examine the differences between Case 3 and Case 5, a comparison
of the normalized centerline velocity, pressure and density fluctuations
is shown in figure \ref{fig:centerline_urms_prms_cases3n5}. Centerline
r.m.s. axial-velocity fluctuations with two different normalizations
are compared in figure \ref{fig:centerline_urms_prms_cases3n5}(a).
$U_{0}$ has the same value for Cases 3 and 5, therefore, $u'_{c,\textrm{rms}}/U_{0}$
compares the absolute fluctuation magnitude. In contrast, $u'_{c,\textrm{rms}}/U_{c}$
depicts the fluctuation magnitude with respect to local mean value.
Case 3 exhibits slightly larger $u'_{c,\textrm{rms}}/U_{c}$ and $u'_{c,\textrm{rms}}/U_{0}$
in the transition region than Case 5, as expected from the slightly greater $U_c$ decay in Case 3 than Case 5 in figure \ref{fig:invUc_hfRad_Cases3n5}. The normalized pressure and density
fluctuations, illustrated in figure \ref{fig:centerline_urms_prms_cases3n5}(b),
show similarly that $p'_{c,\textrm{rms}}/p_{c}$ and $\rho'_{c,\textrm{rms}}/\rho_{c}$
are larger in Case 3 than Case 5 in the transition region of the flow. \vspace{-1.25cm}

\noindent 

\noindent 

\noindent 
\begin{figure}
\begin{centering}
\includegraphics[width=11cm]{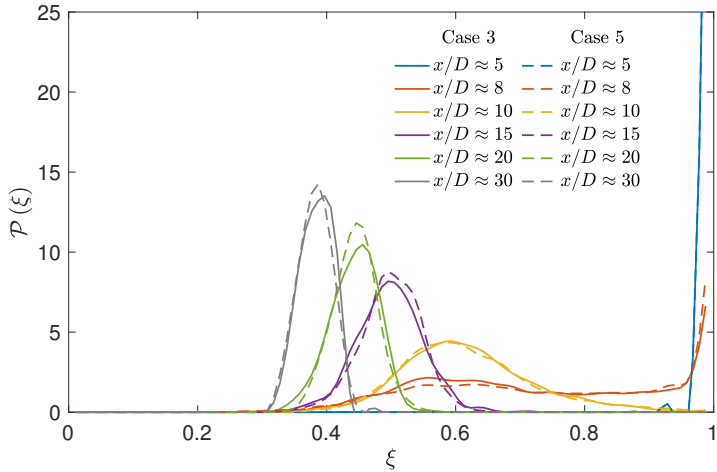}
\par\end{centering}

\caption{Case 3 and 5 comparisons: Scalar probability density function, $\mathcal{P}\left(\xi\right)$,
at various centerline axial locations. \label{fig:Scl_PDF_cases3n5}}
\end{figure}

Whether these slightly larger fluctuations in Case 3 lead to greater mixing is assessed using 
$\mathcal{P}\left(  \xi\right)$ in figure \ref{fig:Scl_PDF_cases3n5}. 
While the p.d.f.s
at various axial locations look nearly identical, at locations in
the transition region and downstream, i.e. $x/D\gtrsim10$, the p.d.f.
peaks are smaller and the profiles somewhat wider in Case 3, a fact which
indicates slightly larger scalar fluctuations and greater mixing than
in Case 5. This indicates that at same $Z$, larger velocity and thermodynamic
fluctuations lead to enhanced mixing.

These results demonstrate that $Z$ does not uniquely determine flow dynamics
because Cases 3 and 5 that differ in $p_{\infty}$ but have same $Z$
exhibit small but noticeable differences in flow fluctuations and
mixing. In particular, an increase in supercritical $p_{\infty}$ at a
fixed $Z$ leads to a reduced normalized velocity/pressure/density/scalar
fluctuations, especially in the transition region. Therefore, the possible notion of
performing experiments at a fixed $Z$ and inferring from them information to
another state having the same $Z$ (i.e. same departure from perfect-gas behavior) but larger $p_{\infty}$, where experiments are more
challenging, may be erroneous.
Additionally, these results show that $p_{\infty}\left(\beta_{T}-1/p_{\infty}\right)$
may also not be the non-dimensional thermodynamic parameter that completely determines 
flow behavior, since despite a large change in its value from Case 3 to
Case 5 (approximately 30\% change), the results of the two cases are relatively close.

In thermodynamics, the law of corresponding states indicates that fluids at the same reduced temperature and reduced pressure have the same $Z$ and, thus, exhibit similar departure from a perfect gas behavior. However, in fluid flows, dynamic effects characterized by the flow Mach number (here $Ma_e$) are also important, in addition to the thermodynamic effects characterized by $Z$, in determining the flow behavior. Differences in $Ma_e$ arise across Cases 1--5 from the differences in the ambient speed of sound, and the effects of these differences are examined next.

\subsection{Effects of $p_{\infty}$ and $Z$ at a fixed jet-exit Mach number\label{sub:Mach0p6_results}}

Comparisons between Cases 1--4 in \S \ref{sub:Case1to4_comparison} evaluated the effects of $p_{\infty}$ and $Z$ at a fixed jet-exit (inflow) bulk velocity, $U_e$. Different ambient thermodynamic conditions in Cases 1--4 leads to different ambient speed of sound, $c_{\infty}$, and hence different jet-exit (inflow) Mach number, $Ma_e$, values as shown in table \ref{tab:Summary_of_cases}. Thus, the results discussed so far do not distinguish between the influence of Mach number and the influence of thermodynamic conditions. While a detailed assessment of the effects of Mach number is beyond the scope of this study, some conclusions may be extracted using results from Cases 2M and 4M that have the same ambient thermodynamic conditions as Cases 2 and 4, respectively, but where $U_e$ is varied to yield a $Ma_e$ of 0.6, which is the same value as that in Case 1. The change in $U_e$ from Case 2, where $Ma_e=0.58$, to Case 2M is small, but from Case 4, where $Ma_e=0.82$, to Case 4M is significant.

\begin{figure}
\begin{centering}
\includegraphics[width=6.9cm]{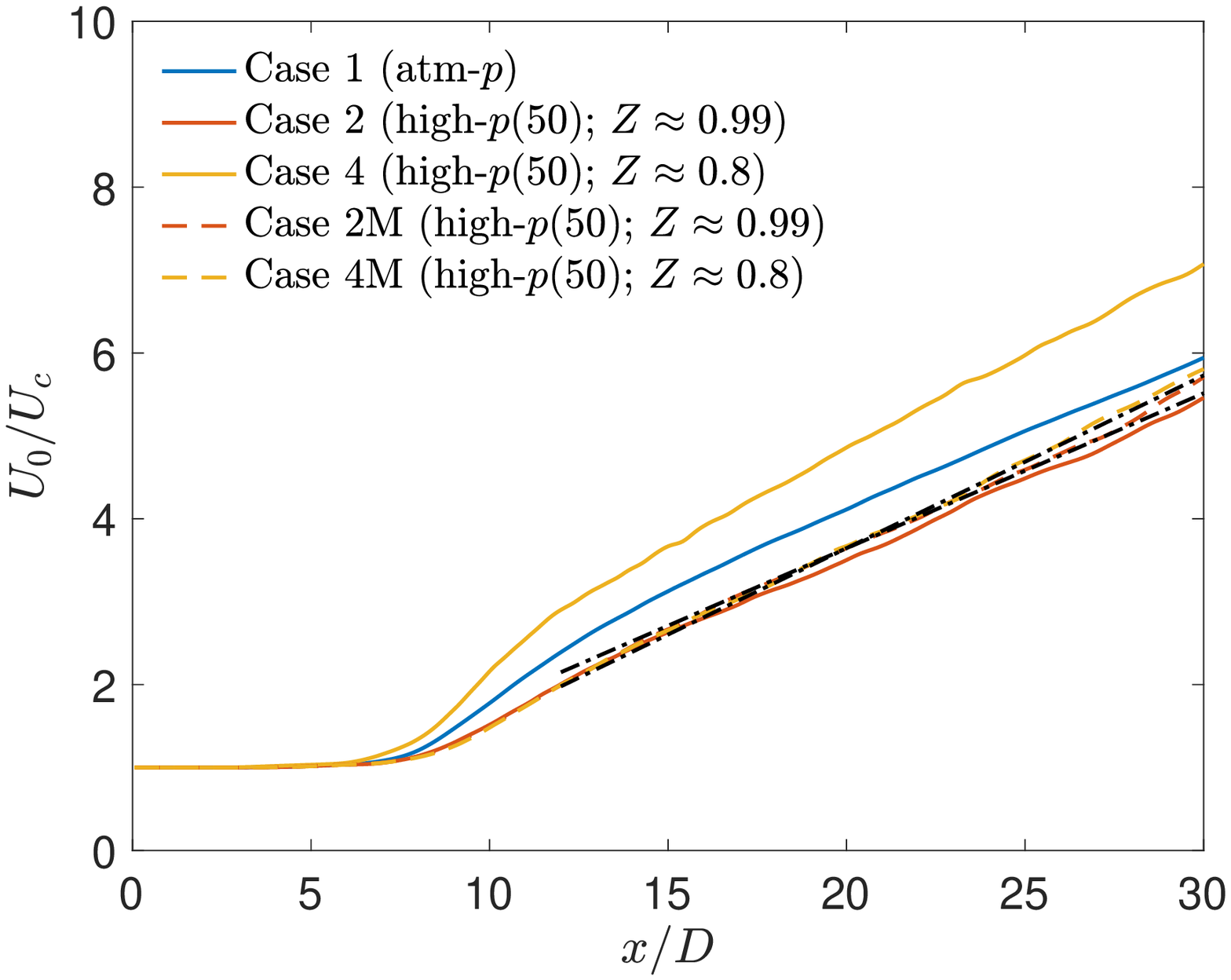}\includegraphics[width=6.9cm]{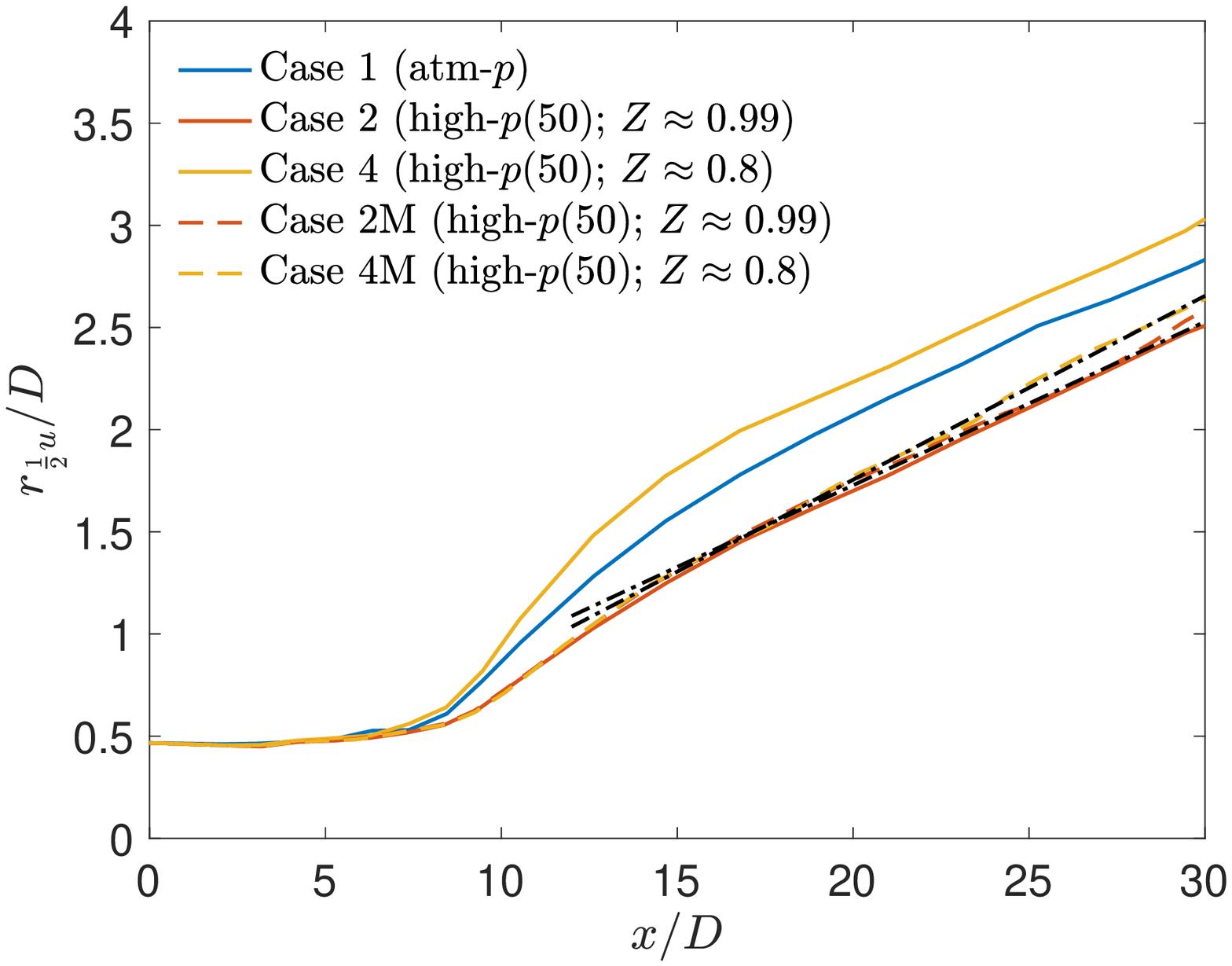}
\par\end{centering}

\begin{centering}
(a)\qquad{}\qquad{}\qquad{}\qquad{}\qquad{}\qquad{}\qquad{}\qquad{}\qquad{}\qquad{}\qquad{}\qquad{}(b)
\par\end{centering}

\caption{Fixed $U_{e}$ versus fixed $Ma_{e}$ cases: Streamwise variation
of (a) the inverse of centerline mean axial velocity ($U_{c}$) normalized
by the jet-exit centerline velocity ($U_{0}$) and (b) the velocity
half radius ($r_{\frac{1}{2}u}$) normalized by the jet diameter. The black dash-dotted lines in (a) are given by equation (\ref{eq:centerline_vel})
using $B_{u}=5.35$, $x_{0u}=0.5D$ for Case 2M, and $B_{u}=4.8$, $x_{0u}=2.5D$
for Cases 4M. The
black dash-dotted lines in (b) are given by:  
$r_{\frac{1}{2}u}/D=0.08\left(x/D+1.6\right)$ for Case 2M, and  
$r_{\frac{1}{2}u}/D=0.09\left(x/D-0.5\right)$ for Case 4M. The self-similar profiles for Cases 1, 2 and 4 are in figure \ref{fig:invUc_hfRad_Cases1to4} caption. \label{fig:invUc_hfRad_mach0p6}}
\end{figure}

The centerline mean axial-velocity, plotted in figure \ref{fig:invUc_hfRad_mach0p6}(a), shows minor change from Case 2 to 2M but notable differences between Cases 4 and 4M. The small increase in $Ma_e$ from Case 2 to Case 2M does not substantially affect the transition region behavior but slightly increases the self-similar axial-velocity decay rate, $1/B_u$ in (\ref{eq:centerline_vel}), from $1/5.4$ in Case 2 to $1/5.35$ in Case 2M, whereas the decrease in $Ma_e$ from Case 4 to Case 4M decreases the mean axial-velocity decay in the transition region as well as in the self-similar region, where $1/B_u$ reduces from $1/4.4$ in Case 4 to $1/4.8$ in Case 4M. The velocity half-radius, $r_{\frac{1}{2}u}$, showing the jet spread in figure \ref{fig:invUc_hfRad_mach0p6}(b) depicts a similar behavior, where minimal differences are observed between Cases 2 and 2M, while the jet spread in Case 4M is considerably reduced with respect to Case 4. This shows that the unique behavior of Case 4, discussed in \S \ref{sub:Case1to4_comparison}, is a combined effect of its proximity to the Widom line and the inflow Mach number. This finding provides the motivation to examine the jet flow sensitivity to inflow condition in \S \ref{sub:Inflow-effects}.

\begin{figure}
\begin{centering}
\includegraphics[width=6.9cm]{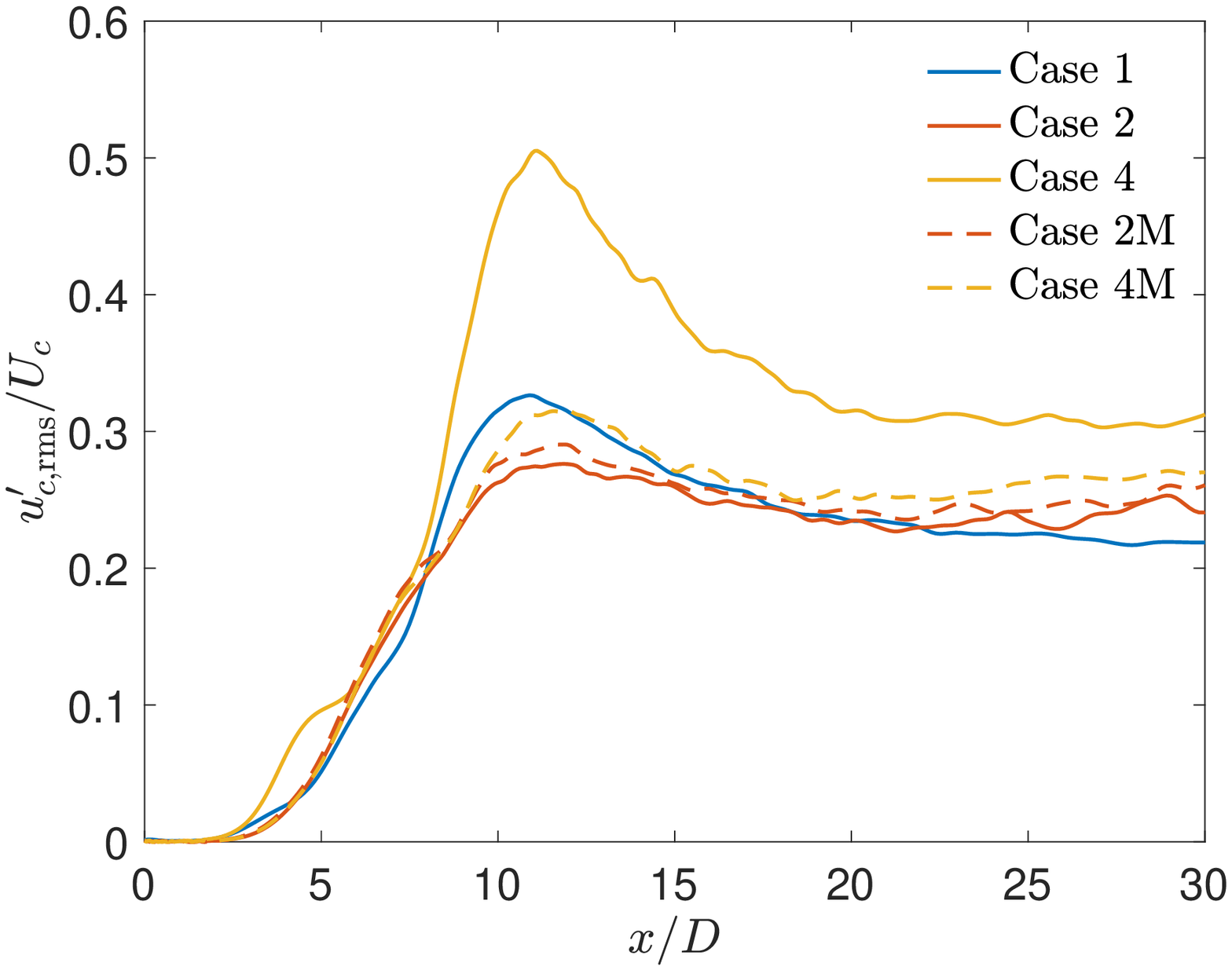}\includegraphics[width=6.9cm]{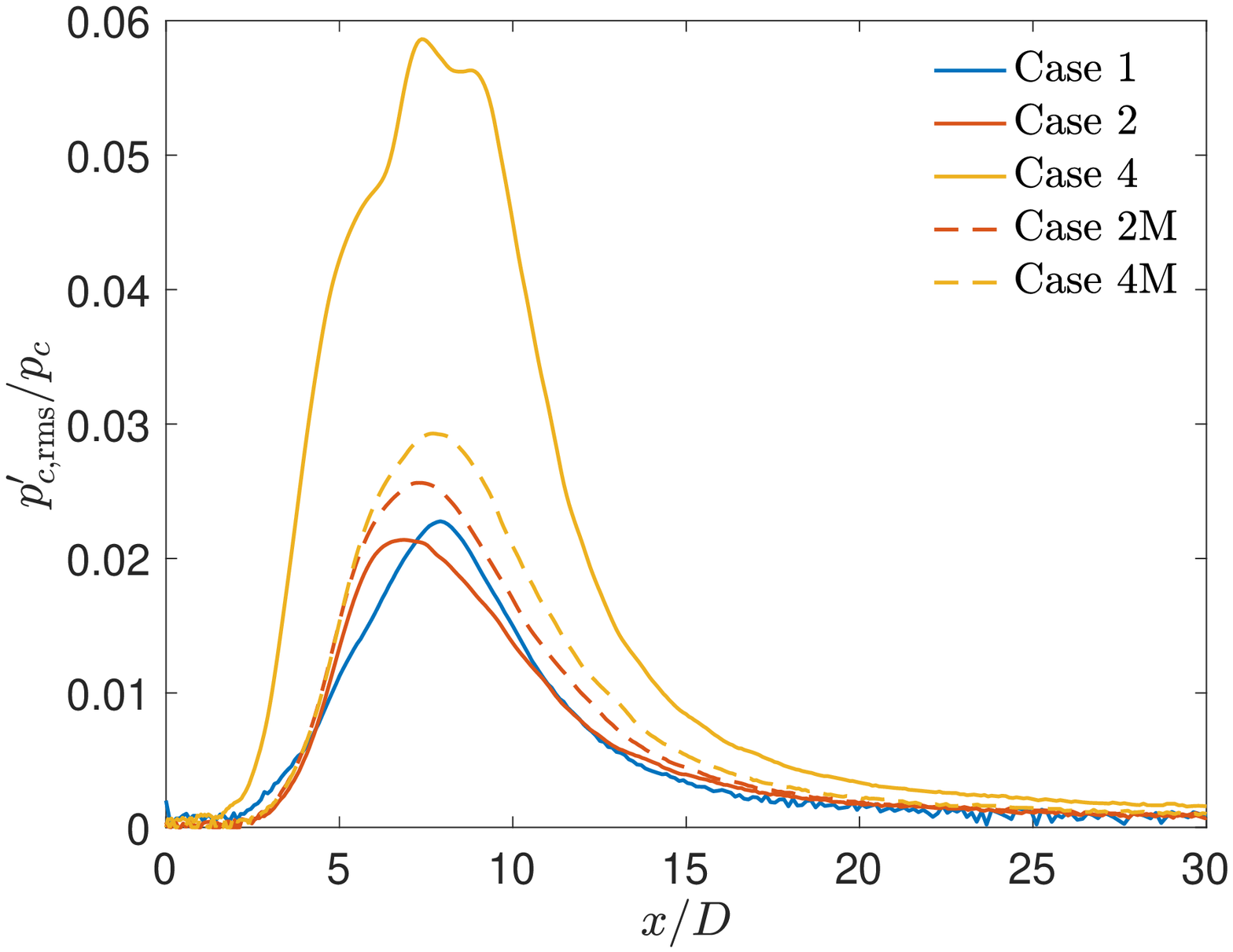}
\par\end{centering}

\begin{centering}
(a)\qquad{}\qquad{}\qquad{}\qquad{}\qquad{}\qquad{}\qquad{}\qquad{}\qquad{}\qquad{}\qquad{}\qquad{}(b)
\par\end{centering}

\caption{Fixed $U_{e}$ versus fixed $Ma_{e}$ cases: Streamwise variation
of (a) the centerline r.m.s. axial velocity fluctuations ($u_{c,\mathrm{rms}}^{'}$)
normalized by the centerline mean axial velocity ($U_{c}$) and (b)
the centerline r.m.s. pressure fluctuations ($p'_{c,\mathrm{rms}}$)
normalized by the centerline mean pressure ($p_{c}$). \label{fig:centerline_urms_prms_mach0p6}}
\end{figure}

To explore the differences observed in figure \ref{fig:invUc_hfRad_mach0p6}, the normalized velocity and pressure fluctuations among various cases are compared in figure \ref{fig:centerline_urms_prms_mach0p6}. As noted in \S \ref{subsub:velFluc_Case1to4}, the magnitude of $u'_{c,\mathrm{rms}}/U_{c}$ reflects the transfer of kinetic energy from the mean field to flucutations and, consequently, its behavior in figure \ref{fig:centerline_urms_prms_mach0p6}(a) is correlated with the mean flow behavior in figure \ref{fig:invUc_hfRad_mach0p6}(a). Larger $U_{c}$ decay rates in figure \ref{fig:invUc_hfRad_mach0p6}(a) occur in the regions of larger $u'_{c,\mathrm{rms}}/U_{c}$ in figure \ref{fig:centerline_urms_prms_mach0p6}(a). Furthermore, the increase in $Ma_e$ from Case 2 to Case 2M enhances $p'_{c,\mathrm{rms}}/p_{c}$, as seen in figure \ref{fig:centerline_urms_prms_mach0p6}(b), whereas the decrease in $Ma_e$ from Case 4 to Case 4M reduces it. At fixed $Ma_e=0.6$, $p'_{c,\mathrm{rms}}/p_{c}$ increases with decrease in $Z$ from Case 1 to Case 2M to Case 4M, indicating the role of $Z$ at a fixed $Ma_e$. Thus, figure \ref{fig:centerline_urms_prms_mach0p6}(b) in conjunction with figure \ref{fig:centerline_pfluc_pvelCorl}(a) shows that while at a fixed $U_e$, the comparative behavior of $p'_{c,\mathrm{rms}}/p_{c}$ is correlated with the value of $p_{\infty}\left(\beta_{T}-1/p_{\infty}\right)$ in table \ref{tab:Isothermal-compressibility}, this may not be the case at a fixed $Ma_e$. 

Examination of normalized fluctuating pressure-velocity correlation and third-order velocity moments (not presented here for brevity) showed that their behavior is correlated with the mean and fluctuating axial velocity behavior in figures \ref{fig:invUc_hfRad_mach0p6}(a) and \ref{fig:centerline_urms_prms_mach0p6}(a), respectively. This feature of the flow is further investigated in the next section to explain the physical mechanism by which thermodynamic and inflow conditions influence jet flow dynamics and mixing.

\noindent 

\noindent 

\subsection{Inflow effects\label{sub:Inflow-effects}}

The influence of inflow conditions on near- and far-field jet flow
statistics at atmospheric conditions has been a subject of numerous
investigations, e.g. \cite{husain1979axisymmetric}, \cite{richards1993global}, \cite{boersma1998numerical}, \cite{mi2001influence}, \cite{xu2002effect}.
Several studies have questioned the classical self-similarity hypothesis
\cite[]{townsend1980structure} that the asymptotic state of the jet
flow depends only on the rate at which momentum is added and is independent
of the inflow conditions. Those studies support the analytical result
of \cite{george1989self}, who suggested that the flow can asymptote
to different self-similar states determined by the inflow condition.
It is thus pertinent
to use the two inflow conditions described in \S \ref{sec:Inflow-condition}
to examine the uniqueness of the self-similar state at near-atmospheric
$p_{\infty}$. In contrast to past investigations in
which measurements were obtained of either the velocity or the passive
scalar field, here the inflow effects on the velocity and the scalar
field are simultaneously examined, first at perfect-gas conditions in \ref{sub:Case1n1T_comparison}, and then at high-$p$ conditions in \ref{sub:Inflow-effects-at-highP}.

\subsubsection{Inflow effects at perfect-gas condition: comparisons between Cases 1 and 1T\label{sub:Case1n1T_comparison}}

For the laminar inflow, which has a top-hat
jet-exit mean velocity profile (\ref{eq:lam_inflow_mean_vel}), the jet-exit bulk velocity is approximately $U_{e}$, and $U_{e}$ is equal to the jet-exit centerline velocity, $U_{0}$. However, for the turbulent
inflow, which has a parabolic jet-exit mean velocity profile, the jet-exit bulk velocity, $U_{e}$, is smaller than $U_{0}$.
The present study uses the same $U_{e}$ for all cases, except Cases 2M and 4M. 
As a result, $U_{0}$ is different for the laminar and turbulent inflow cases.

\begin{figure}
\begin{centering}
(a)\includegraphics[width=13cm]{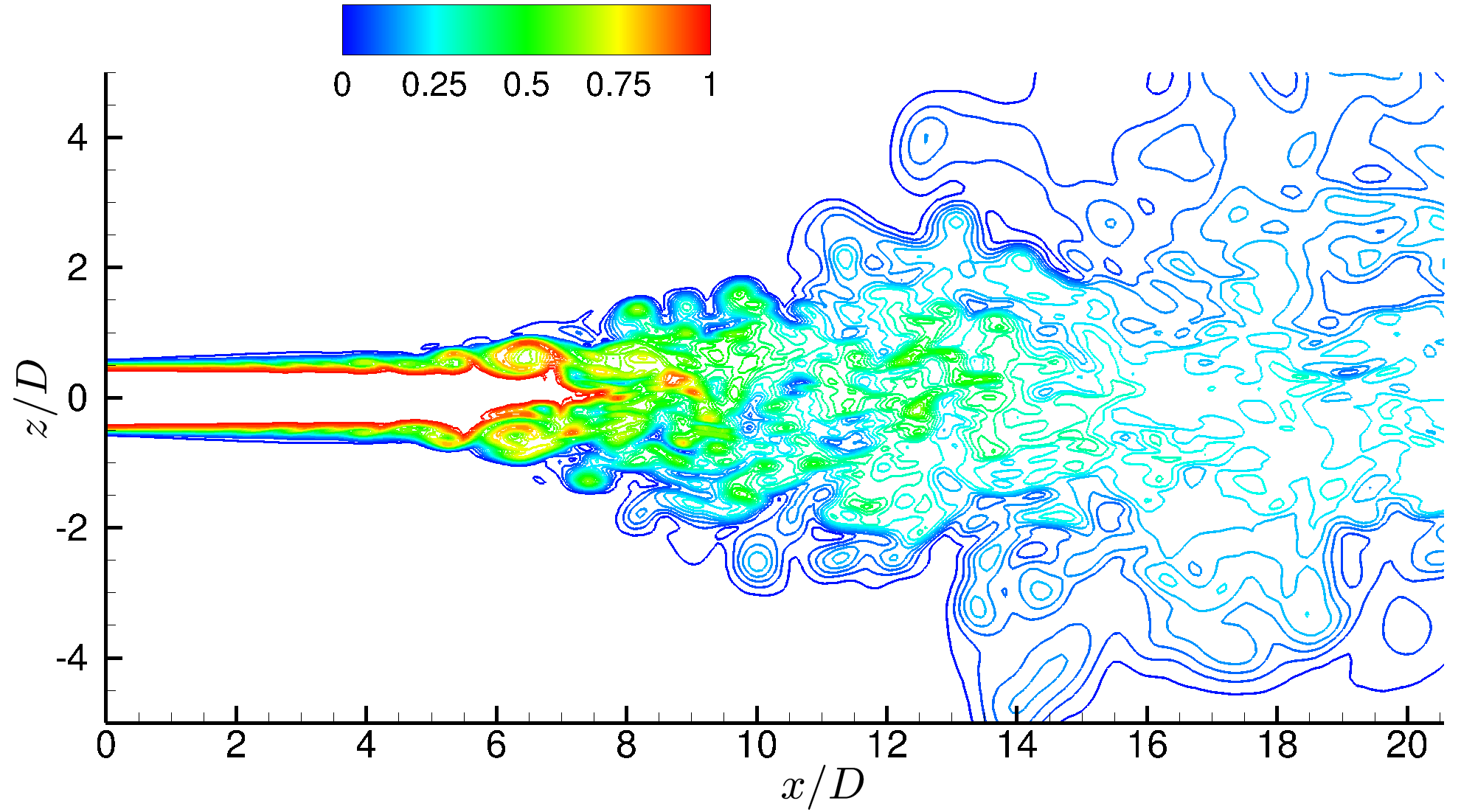}
\par\end{centering}

\begin{centering}
(b)\includegraphics[width=13cm]{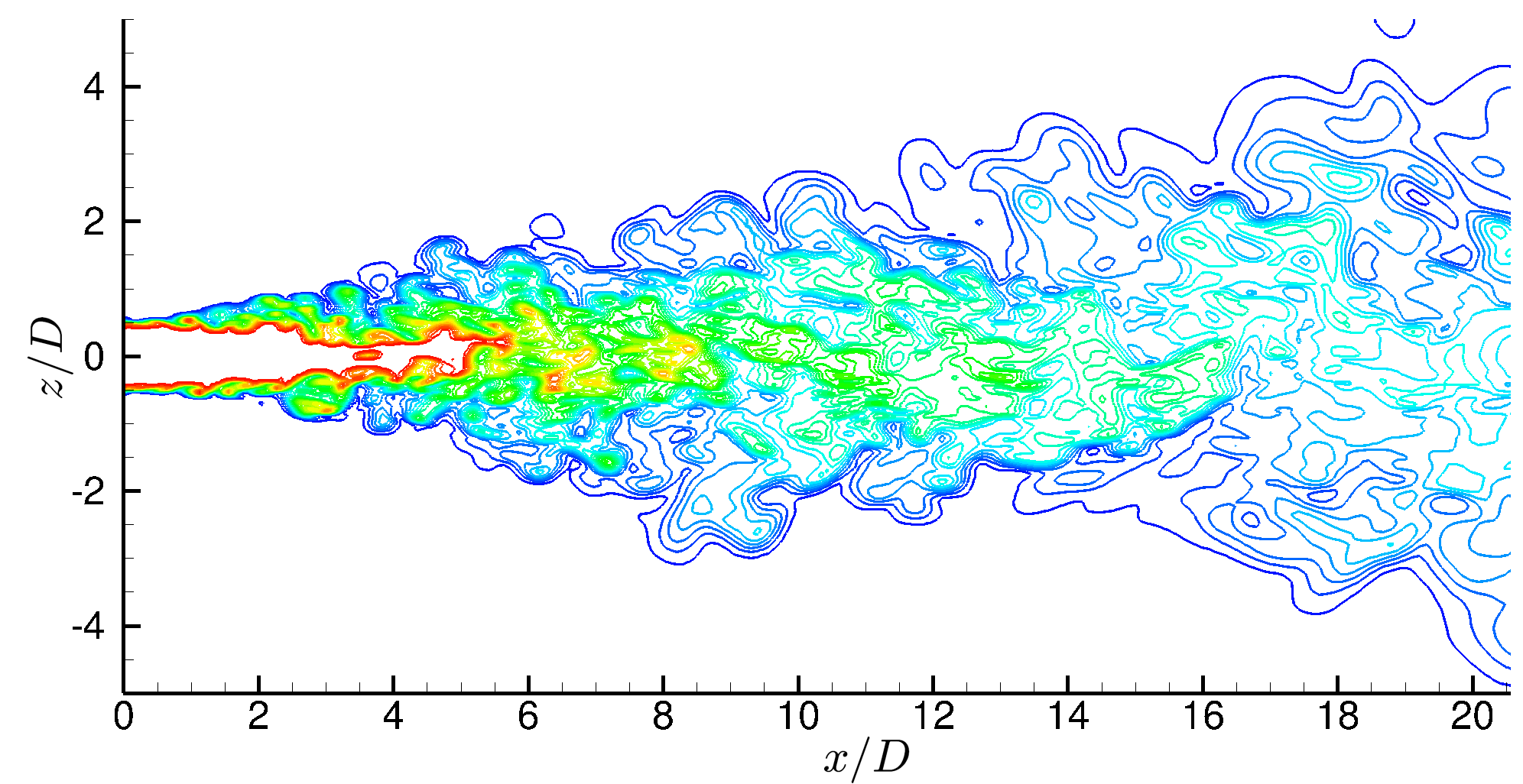}
\par\end{centering}

\caption{Near-field instantaneous scalar contours at $tU_{e}/D\approx3500$
from (a) Case 1 and (b) Case 1T showing the mixed fluid concentration.
The contour lines show 24 levels in the range $0.02\leq\xi\leq0.98$.
The legend is the same for both plots. \label{fig:scalar_contours_Case1_1T}}
\end{figure}

Figure \ref{fig:scalar_contours_Case1_1T} illustrates the near-field
scalar contours from Cases 1 and 1T at $tU_{e}/D\approx3500$. 
The rendered contour lines show the mixed fluid, defined as $0.02\leq\xi\leq0.98$.
Evidently, the near-field flow features are considerably
different for the two jets. The instabilities in the annular shear
layer that trigger vortex roll-ups appear at larger axial distance
in the jet from the laminar inflow (Case 1) than those in the jet from the pseudo-turbulent
inflow (Case 1T). The inflow disturbances in Case 1T, modeling pipe-flow
turbulence, are broadband and higher in magnitude, thus triggering
small-scale turbulence that promote axial shear-layer growth immediately
downstream of the jet exit. In contrast, the laminar inflow has small
random disturbances superimposed over the top-hat velocity profile
that trigger the natural instability frequency \cite[]{ho1981dynamics}
and dominant vortical structures/roll-up around $x/D\approx5$. The
larger axial distance required for the natural instability to take
effect in Case 1 leads to a longer potential core than in Case 1T.
However, once the instabilities take effect in Case 1, at $x/D\approx5$,
dominant vortical structures close the potential core over a short
distance, i.e. around $x/D\approx8$. In comparison, in Case 1T, the broadband
small-scale turbulence triggered immediately downstream of the jet
exit closes the potential core around $x/D\approx6$. Downstream
of the potential core collapse, an abrupt increase in the jet width
is observed in Case 1, while the jet grows gradually in Case 1T.

\noindent 

The above-discussed qualitative differences are now quantified using various velocity and scalar
statistics.

\begin{figure}
\begin{centering}
\includegraphics[width=6.9cm]{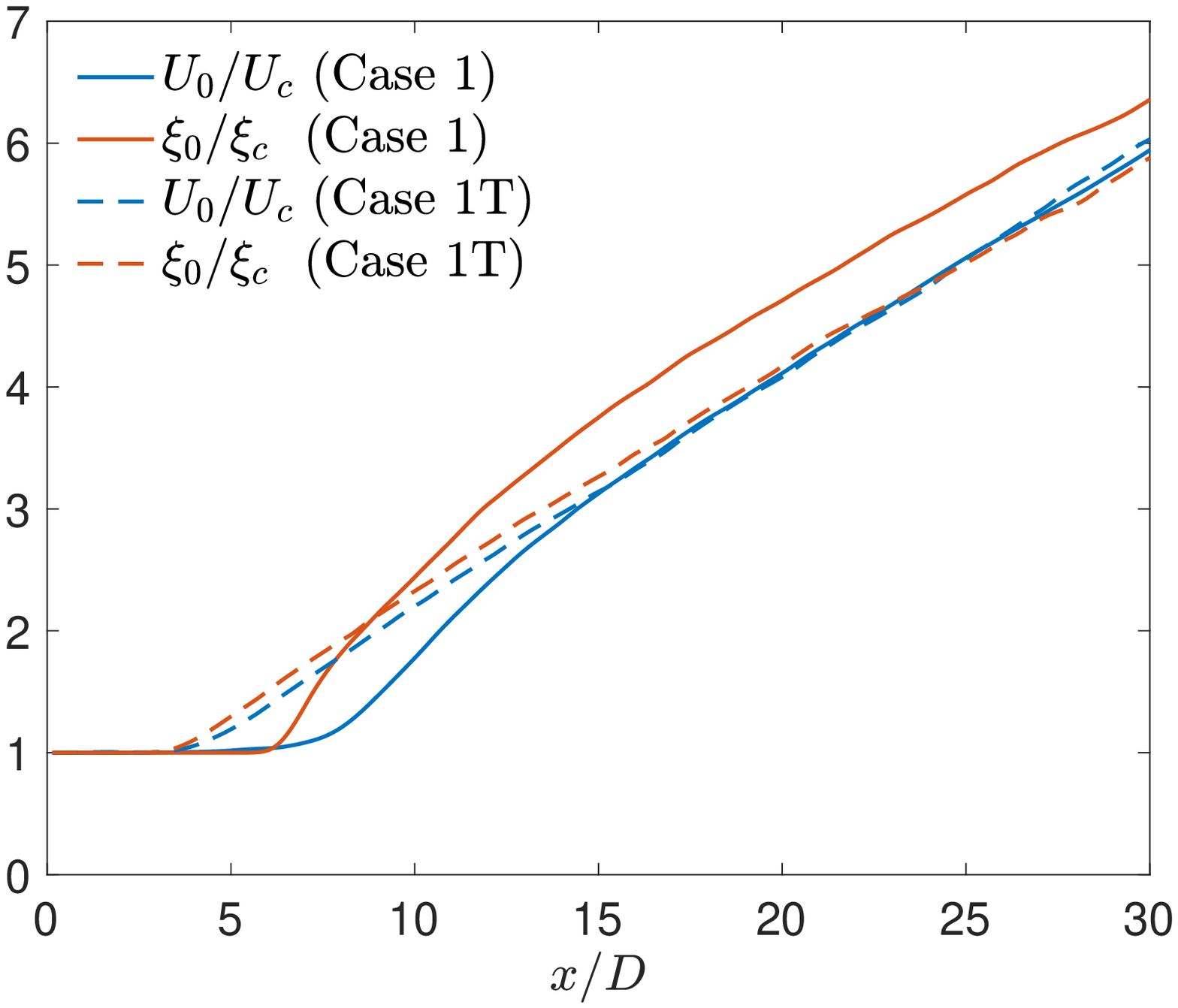}\includegraphics[width=6.9cm]{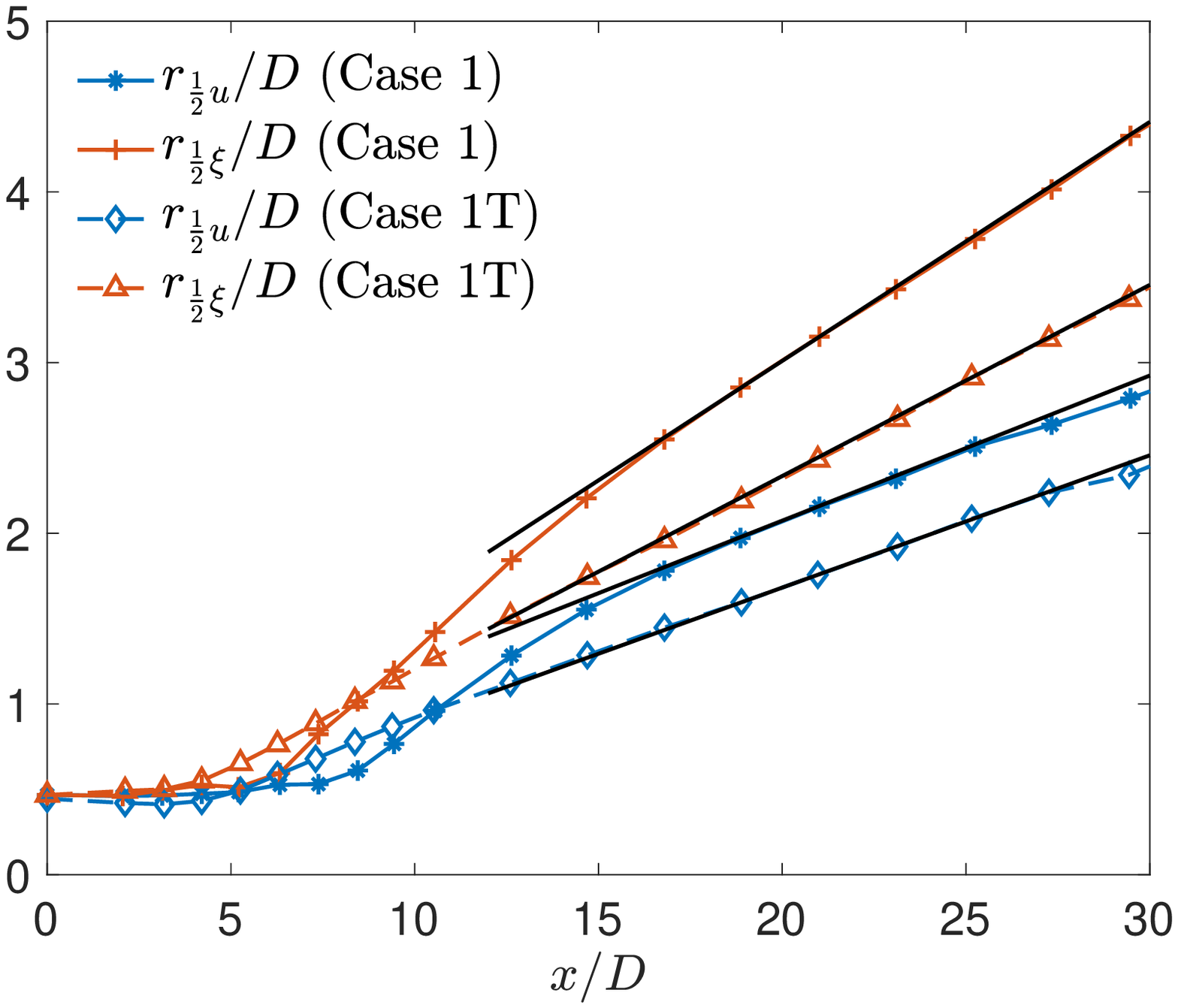}
\par\end{centering}

\begin{centering}
(a)\qquad{}\qquad{}\qquad{}\qquad{}\qquad{}\qquad{}\qquad{}\qquad{}\qquad{}\qquad{}\qquad{}\qquad{}(b)
\par\end{centering}

\caption{Case 1 and 1T comparisons: Streamwise variation of the (a) inverse
of centerline mean axial-velocity ($U_{c}$) and scalar concentration
($\xi_{c}$) normalized by the jet-exit centerline mean values and
(b) normalized velocity and scalar half radius, denoted by $r_{\frac{1}{2}u}/D$
and $r_{\frac{1}{2}\xi}/D$, respectively. The linear profile downstream of $x/D\approx15$ in (a)
can be described by equation (\ref{eq:centerline_vel}) using $B_{u}=5.5$ ($x_{0u}=-2.4D$) and $B_{\xi}=5.75$ ($x_{0\xi}=-3.5D$)
for Case 1T, whereas $B_{u}=5.5$ ($x_{0u}=-2.4D$) and $B_{\xi}=5.7$ ($x_{0\xi}=-6.8D$) for Case 1. The black solid lines in
(b) are given by: $r_{\frac{1}{2}u}/D=0.085\left(x/D+4.4\right)$ and $r_{\frac{1}{2}\xi}/D=0.14\left(x/D+1.5\right)$ for Case 1, and $r_{\frac{1}{2}u}/D=0.078\left(x/D+1.7\right)$ and $r_{\frac{1}{2}\xi}/D=0.11\left(x/D+0.85\right)$ for Case 1T.
\label{fig:invUc_hfRad_Case_1n1T}}
\end{figure}

\paragraph{Velocity and scalar statistics, and self-similarity}
A comparison of $U_{c}$ and $\xi_{c}$ axial decay between
Case 1 and Case 1T jets is
presented in figure \ref{fig:invUc_hfRad_Case_1n1T}(a). The shorter
potential core length of Case 1T leads to velocity and scalar decay
beginning upstream of that in Case 1. The difference between the axial
locations where the velocity and scalar begin to decay is noticeable
for Case 1, while it is relatively small for Case 1T
thus indicating a tighter coupling of dynamics and molecular mixing in Case 1T.
The upstream decay of the scalar, with respect to velocity, in the laminar
inflow jet (Case 1) is consistent with the observation of \cite[Figure 6]{lubbers2001simulation}
for a passive scalar diffusing at unity Schmidt number. Also noticeable
in the Case 1 results is a transition or development region,
$7\lesssim x/D\lesssim15$, where the velocity and scalar decay rates
are larger than in the asymptotic state reached further downstream.
The Case 1T results do not show a similarly distinctive
transition region, and the velocity and scalar decay rates remain approximately
the same downstream of the potential core closure. Downstream of $x/D\approx15$, 
the velocity and scalar decay rates are similar for Cases 1 and 1T. 

The velocity and scalar half-radius for Cases 1 and 1T are compared in 
figure \ref{fig:invUc_hfRad_Case_1n1T}(b).
The Case 1 jet spreads at a faster rate than Case 1T, consistent with the experimental
observations of \cite{xu2002effect} and \cite{mi2001influence}.
The decrease in velocity half-radius spreading rate from $0.085$
(Case 1) to $0.078$ (Case 1T) is consistent
with the observations of \cite{xu2002effect}, where
a decrease in spreading rate from $0.095$ for the jet issuing from
a smooth contraction nozzle to $0.086$ for the jet from a pipe nozzle was reported.
The spreading rate of $0.14$ and $0.11$ based on the scalar half-radius
for Case 1 and Case 1T, respectively, is larger than
the values of $0.11$ and $0.102$ reported by \cite{mi2001influence}
for their temperature scalar field from smooth contraction nozzle
and pipe jet, respectively, but comparable to the values of $0.13$
and $0.11$ deduced from the results of \cite{richards1993global}
for their mass-fraction scalar field from smooth contraction nozzle
and pipe jet, respectively. The profiles in figure \ref{fig:invUc_hfRad_Case_1n1T}
also show that the velocity and scalar mean fields attain self-similarity,
i.e. their centerline values decay linearly and the half-radius spreads linearly, at smaller
axial distance in Case 1T than in Case 1. In the self-similar region, the $U_{c}$ and $\xi_{c}$ decay rates
of Cases 1 and 1T are comparable, while the half-radius spreading rates differ. The non-dimensional $U_0/U_c$ have similar values for Cases 1 and 1T in the self-similar region, but since $U_0$ is different for Case 1 and Case 1T, $U_c$ differs accordingly. The results of mean velocity/scalar centerline behavior and jet spread suggest that these quantities strongly depend on the inflow condition, both in the near-field and the self-similar region, thus highlighting the importance of reporting inflow conditions in experimental studies if their data has to be used for model validation.

\begin{figure}
\begin{centering}
\includegraphics[width=6.9cm]{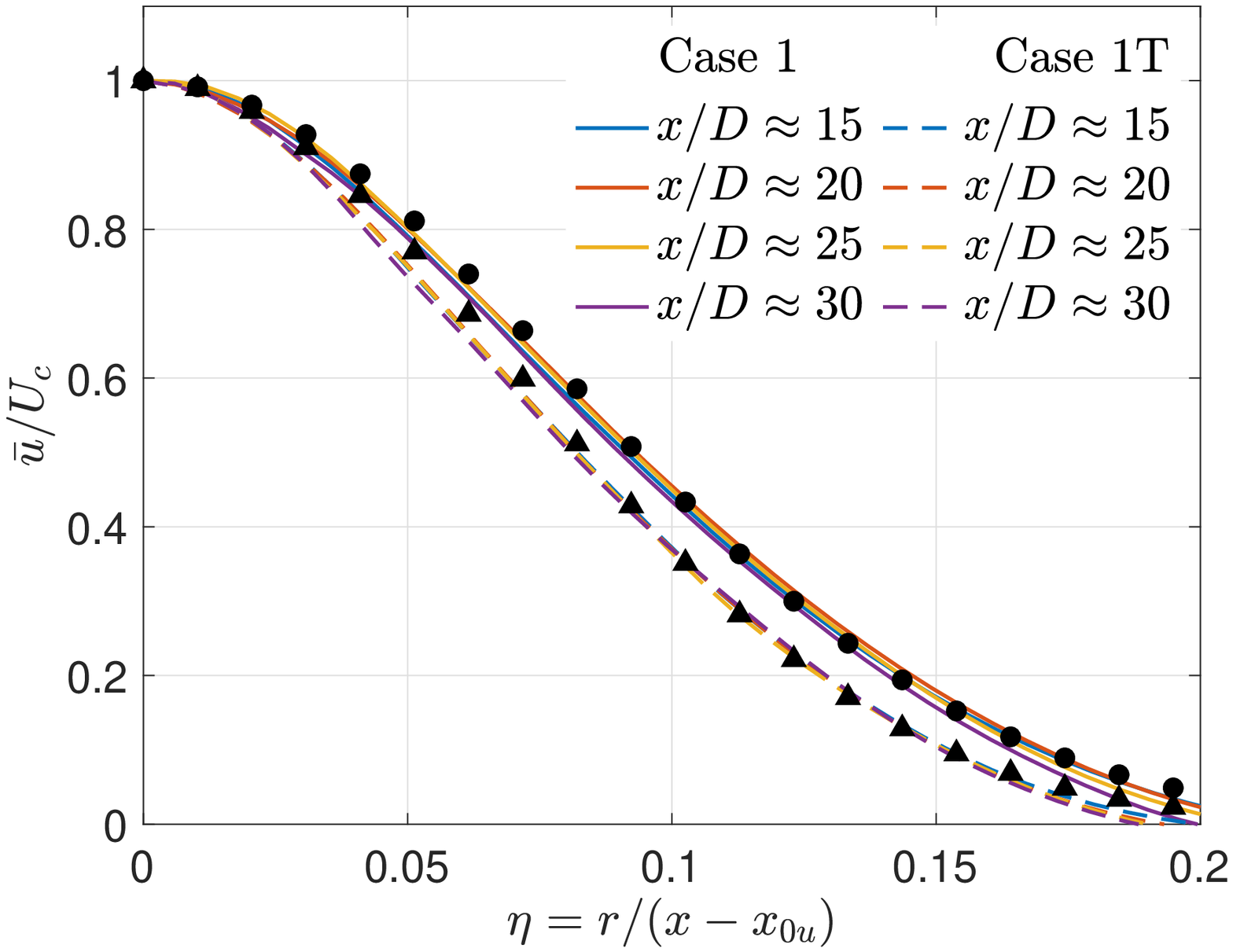}\includegraphics[width=6.9cm]{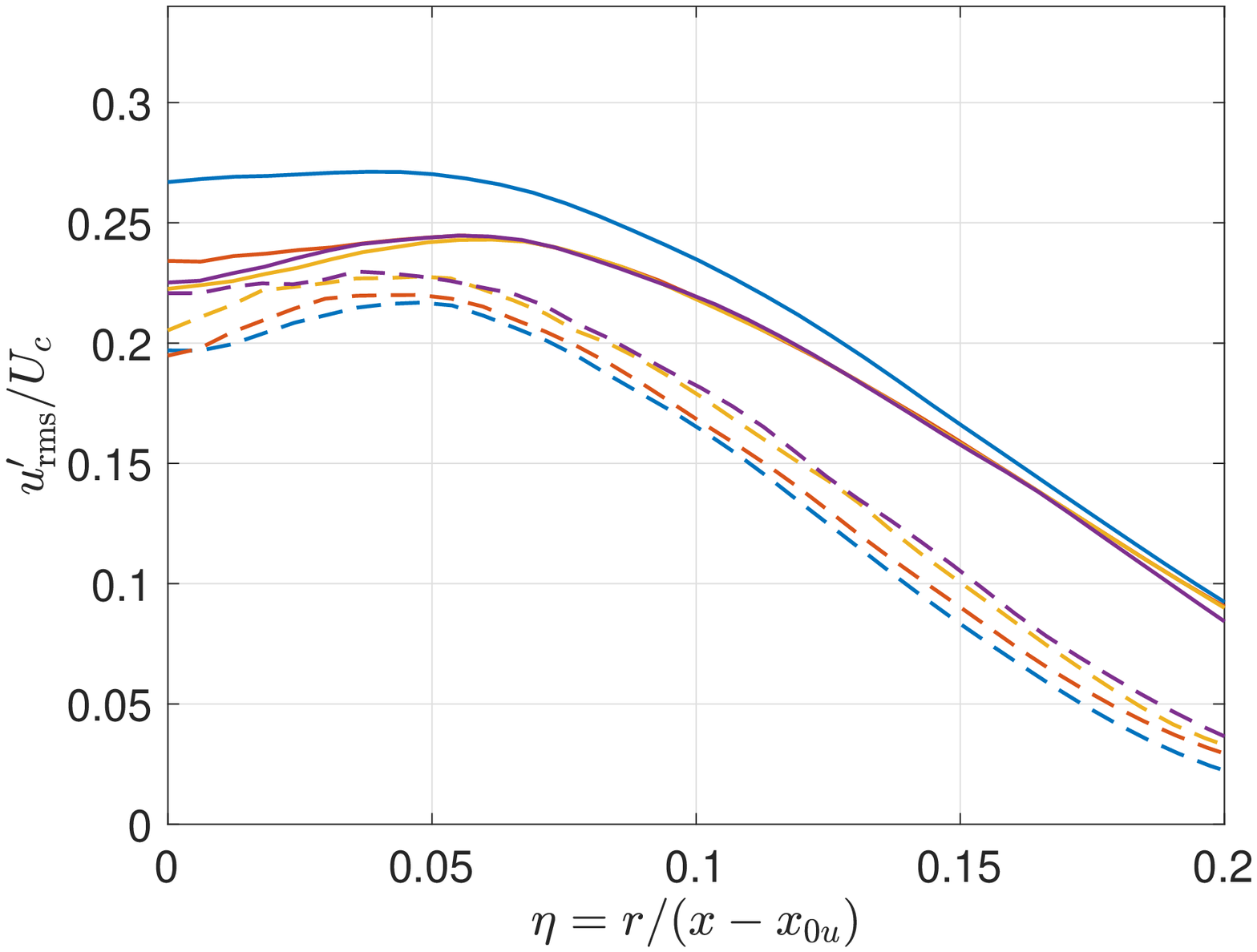}
\par\end{centering}

\begin{centering}
(a)\qquad{}\qquad{}\qquad{}\qquad{}\qquad{}\qquad{}\qquad{}\qquad{}\qquad{}\qquad{}\qquad{}\qquad{}(b)
\par\end{centering}


\caption{Case 1 and 1T comparisons: Radial profiles of (a) mean axial velocity
($\bar{u}$) and (b) r.m.s. axial velocity fluctuations ($u_{\mathrm{rms}}^{'}$)
normalized by the centerline mean velocity ($U_{c}$) 
at various
axial locations. The markers $\bullet$ and $\blacktriangle$ in (a) show
$f\left(\eta\right)$ of equation (\ref{eq:self_similar_Gauss_dist}) using $A_{u}=79.5$ and $99$, 
respectively. The legend is the same for both plots. \label{fig:Case_1_1T_velstat}}
\end{figure}

To further examine the differences between Cases 1 and 1T, and the self-similar state attained in these flows, the radial variation of axial velocity and its fluctuation is documented in figure \ref{fig:Case_1_1T_velstat}. 
Examination of figure \ref{fig:Case_1_1T_velstat}(a) shows that the
mean axial velocity attains self-similarity as near-stream as $x/D\approx15$
for both Case 1 and 1T, however, the self-similar profiles are different.
The solid markers in figure \ref{fig:Case_1_1T_velstat}(a)
show the $f\left(\eta\right)$ profiles  of (\ref{eq:self_similar_Gauss_dist}) using $A_{u}=79.5$ (circles)
and $99$ (triangles). These values are comparable to the values of
$76.5$ and $90.2$ reported by \cite{xu2002effect} for jets from
a smooth contraction and pipe nozzle, respectively.
Radial profiles of $u'_{\textrm{rms}}/U_{c}$ from Cases 1 and 1T are compared in figure
\ref{fig:Case_1_1T_velstat}(b). $u'_{\textrm{rms}}/U_{c}$ attains
self-similarity around $x/D\approx25$, a location which is further downstream than for $\overline{u}/U_{c}$, in both Cases 1 and 1T; minor
differences remain near the centerline between the profiles at $x/D\approx25$
and $30$. $u'_{\textrm{rms}}/U_{c}$ values from Case 1T are smaller than
those from Case 1 at all shown axial locations, especially away from the
centerline, consistent with the experimental observations of \cite{xu2002effect}
with laminar/turbulent inflow. Figure \ref{fig:Case_1_1T_velstat}(b)
also shows that for Case 1, $u'_{\textrm{rms}}/U_{c}$ in the near
field ($x/D\approx15$) is larger than that
in the self-similar regime, especially in the radial vicinity of the
centerline, while for Case 1T, the near field ($x/D\approx15$) values are smaller than that in the self-similar
regime. $u'_{\textrm{rms}}/U_{c}$ values in the near field are, therefore,
considerably larger with laminar inflow than with turbulent inflow. The 
behavior of other Reynolds stress components, $v'_{\textrm{rms}}/U_{c}$, 
$w'_{\textrm{rms}}/U_{c}$, and $\overline{u^{'}v^{'}}/U_{c}^{2}$, not 
shown here for brevity, is similar to that of $u'_{\textrm{rms}}/U_{c}$ in
that all of them show self-similarity around $x/D\approx25$ but the self-similar 
profiles depend on the inflow condition. Moreover, their near field values 
are much higher in Case 1 than in Case 1T, as for $u'_{\textrm{rms}}/U_{c}$. Thus, 
similar to the mean quantities in figure \ref{fig:invUc_hfRad_Case_1n1T}, the Reynolds stress 
components also show strong sensitivity to the inflow condition.  \vspace{-1.0cm}

\noindent 

\noindent 

%
%

\begin{figure}
\begin{centering}
\includegraphics[width=11cm]{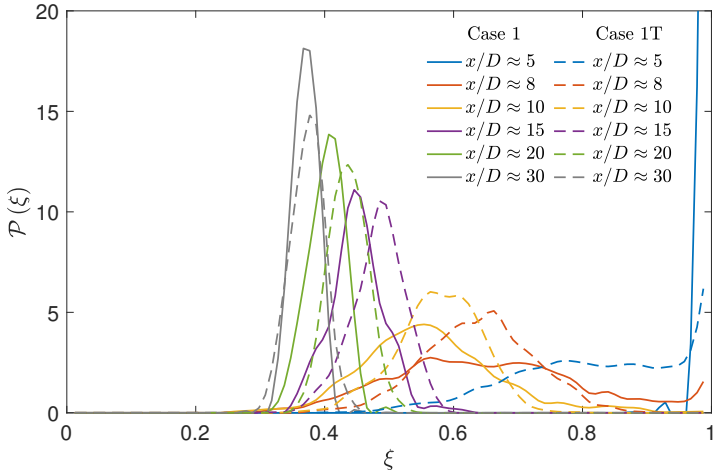}
\par\end{centering}

\caption{Case 1 and 1T comparisons: Scalar probability density function, $\mathcal{P}\left(\xi\right)$,
at various centerline axial locations. \label{fig:SclPDF_Case_1_1T}}
\end{figure}

\begin{figure}
\begin{centering}
\includegraphics[width=9cm]{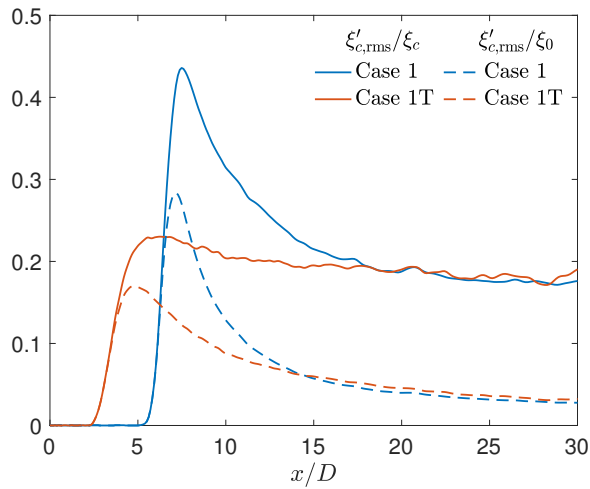}
\par\end{centering}

\caption{Case 1 and 1T comparisons: Streamwise variation of the centerline
r.m.s. scalar fluctuation ($\xi'_{c,\textrm{rms}}$) normalized by
the centerline mean scalar value ($\xi_{c}$) and the jet-exit centerline
mean scalar value ($\xi_{0}$). \label{fig:rms_SclFluc_Cases1n1T}}
\end{figure}

\paragraph{Passive scalar mixing}
To examine the differences in scalar mixing between Cases 1 and 1T, figure \ref{fig:SclPDF_Case_1_1T}
compares the scalar p.d.f., $\mathcal{P}\left(\xi\right)$,
at various locations along the jet centerline.
Significant differences are observed in the near-field p.d.f. profiles,
\textit{i.e.} for $x/D\lesssim15$. The locations $x/D\approx5$ and $8$ are approximately
the centerline location of maximum (non-dimensionalized) scalar fluctuations
for Case 1T and Case 1, respectively, as shown in figure \ref{fig:rms_SclFluc_Cases1n1T}.
Since the jet-exit centerline
mean scalar value is $\xi_{0}=1$ for all cases, normalization of $\xi'_{c,\textrm{rms}}$ with $\xi_{0}$ in figure \ref{fig:rms_SclFluc_Cases1n1T}
allows a comparison of the absolute fluctuation magnitude between
Case 1 and Case 1T. $\xi'_{c,\textrm{rms}}/\xi_{0}$
peaks when the potential core closes and, then, decreases with axial
distance for each case. In contrast, the local normalization with
$\xi_{c}$ asymptotes to a constant value at large axial distances.
$\xi'_{c,\textrm{rms}}/\xi_{c}$ exhibits a prominent hump, or a local
maximum, in the near field for Case 1, consistent with the experimental
observations in jets from a smooth contraction nozzle \cite[]{mi2001influence}.

Comparison of p.d.f. profiles at $x/D\approx5$ in figure \ref{fig:SclPDF_Case_1_1T}
between Case 1 and 1T shows pure jet fluid ($\xi=1$) for Case 1,
whereas mixed fluid with scalar concentrations ranging from $0.5$
to $1$ for Case 1T, as expected, since the potential core closes before
$x/D\approx5$ in Case 1T, but after $x/D\approx5$ in Case 1. At
$x/D\approx8$ and $10$, the p.d.f. profiles for Case 1 exhibit a
wider spread compared to that for Case 1T. Stronger large-scale vortical
structures in the near field (around $x/D\approx8$) in Case 1, as
seen in figure \ref{fig:scalar_contours_Case1_1T}(a), entrain ambient
fluid deep into the jet core resulting in larger scalar fluctuations
(see figure \ref{fig:rms_SclFluc_Cases1n1T}) and a wider distribution
of scalar concentrations at the centerline. In contrast, mixing in Case 1T
occurs through small-scale structures resulting in
weaker entrainment of ambient fluid and smaller scalar fluctuations.
P.d.f. profiles for Case 1T at $x/D\approx8$ and $10$ are, therefore,
narrower with higher peaks. Larger scalar fluctuations in the transition
region ($7\lesssim x/D\lesssim15$) of Case 1, resulting from large-scale
organized structures, cause greater mixing and, consequently, steeper
decay of the centerline mean scalar concentration $\left(\xi_{c}\right)$,
as observed in figure \ref{fig:invUc_hfRad_Case_1n1T}(a). The centerline
mean scalar concentration, indicated by the scalar value at peaks
of the p.d.f. profiles in figure \ref{fig:SclPDF_Case_1_1T}, is smaller
(or closer to the ambient scalar value of $0$) for Case 1 downstream
of $x/D\approx10$. The difference between the scalar mean values from the
two cases diminishes with axial distance. With increase in axial distance,
the jet-width (see figure \ref{fig:invUc_hfRad_Case_1n1T}(b)) increases
and the absolute centerline scalar fluctuation (see figure \ref{fig:rms_SclFluc_Cases1n1T})
diminishes. As a result, the spread of the scalar p.d.f. profile declines
and the peaks become sharper downstream of $x/D\approx10$.

\paragraph{Summary}
To conclude, both velocity and scalar statistics show sensitivity to the inflow condition. In the near field, the jet flow from laminar inflow (Case 1) is characterized by strong vortical structures leading to larger velocity/scalar fluctuations and jet spreading rate in the transition region than Case 1T. Further downstream, self-similarity is observed in velocity/scalar mean and fluctuations, but the self-similar profiles differ with the inflow, supporting the argument that they may not be universal. This indicates that a quantitative knowledge of the experimental inflow conditions is important in validating simulation results against experiments. Whether these conclusions hold at high pressures is examined next.  \vspace{-0.5cm}

\noindent 

\noindent 

\subsubsection{Inflow effects at high pressure: comparisons between Cases 1/1T, 2/2T and 4/4T\label{sub:Inflow-effects-at-highP}}
A crucial observation from \S \ref{sub:Case1to4_comparison}, where the 
influence of $p_{\infty}$ and $Z$ on jet-flow dynamics and mixing was examined,
is that $p'_{c,\textrm{rms}}/p_{c}$
decreases with increase in $p_{\infty}$ from $1$ bar (Case 1) to $50$
bar (Case 2), and increases with decrease in $Z$ from $0.99$ (Case
2) to $0.8$ (Case 4), as shown in figure \ref{fig:centerline_pfluc_pvelCorl}(a);
the velocity/scalar mean and fluctuations (figures \ref{fig:invUc_hfRad_Cases1to4}(a),
\ref{fig:rms_fluc_Cases1to4} and \ref{fig:invScl_rmsFluc_Cases1to4}),
however, follow the behavior of the normalized t.k.e. diffusive
fluxes (and not of $p'_{c,\textrm{rms}}/p_{c}$), e.g. $\frac{\overline{p'u'}/\bar{\rho}}{U_{c}^{3}}$ and $\overline{u^{\prime3}}/U_{c}^{3}$
shown in figures \ref{fig:centerline_pfluc_pvelCorl}(b) and (c), respectively.
Those observations are for laminar inflow jets, and the 
validity of those observations is here examined in pseudo-turbulent inflow jets
(inflow details in \S \ref{sec:Inflow-condition}).

To examine the effects of inflow variation at
supercritical $p_{\infty}$, Case 2 results are here compared with Case 2T,
and Case 4 with Case 4T. These results in
conjunction with those of \S \ref{sub:Case1n1T_comparison} provide an enlarged
understanding of the effect of inflow conditions at different $p_{\infty}$ and $Z$.
\vspace{-0.25cm}
\paragraph{Mean axial velocity and spreading rate}

\begin{figure}
\begin{centering}
\includegraphics[width=6.9cm]{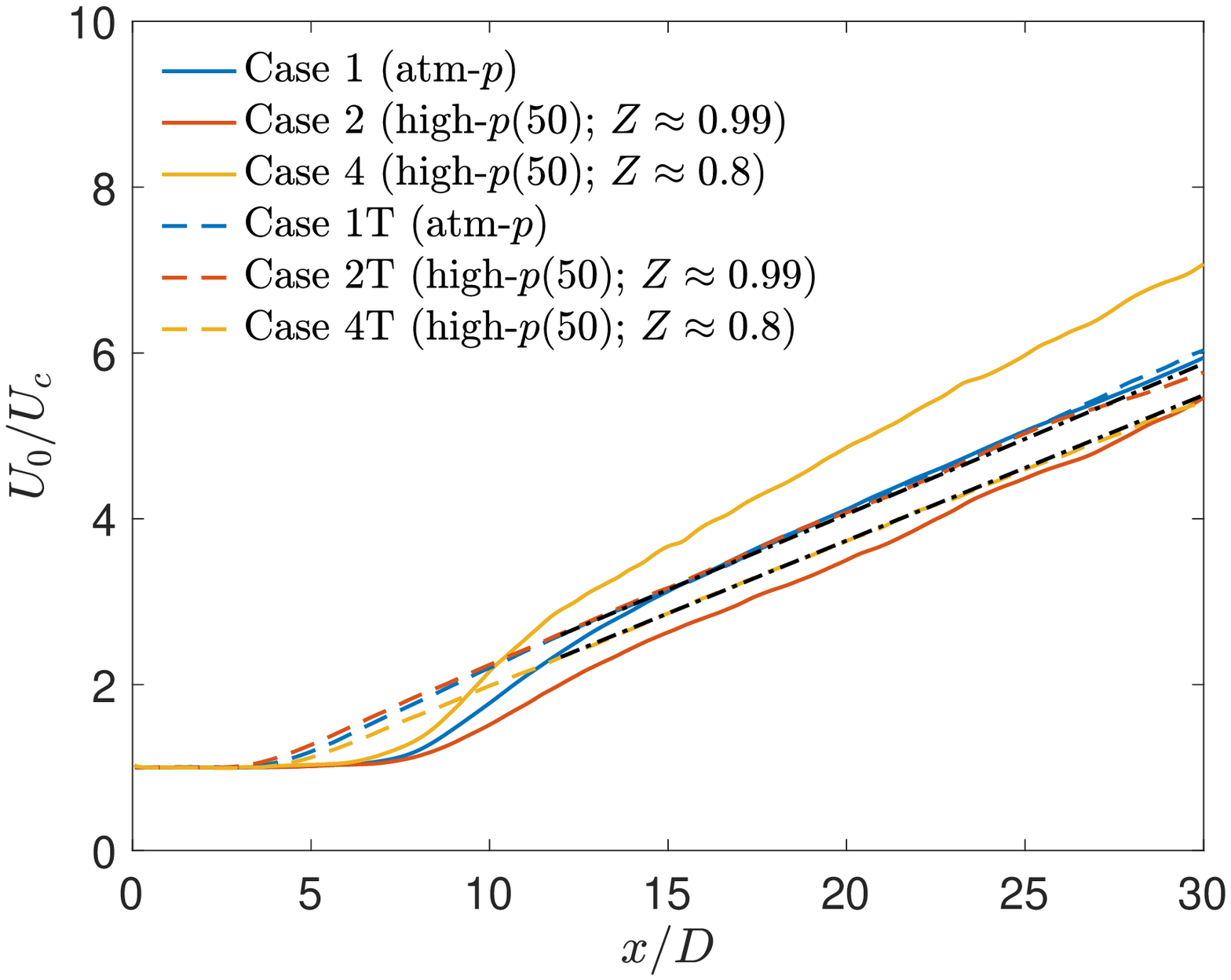}\includegraphics[width=6.9cm]{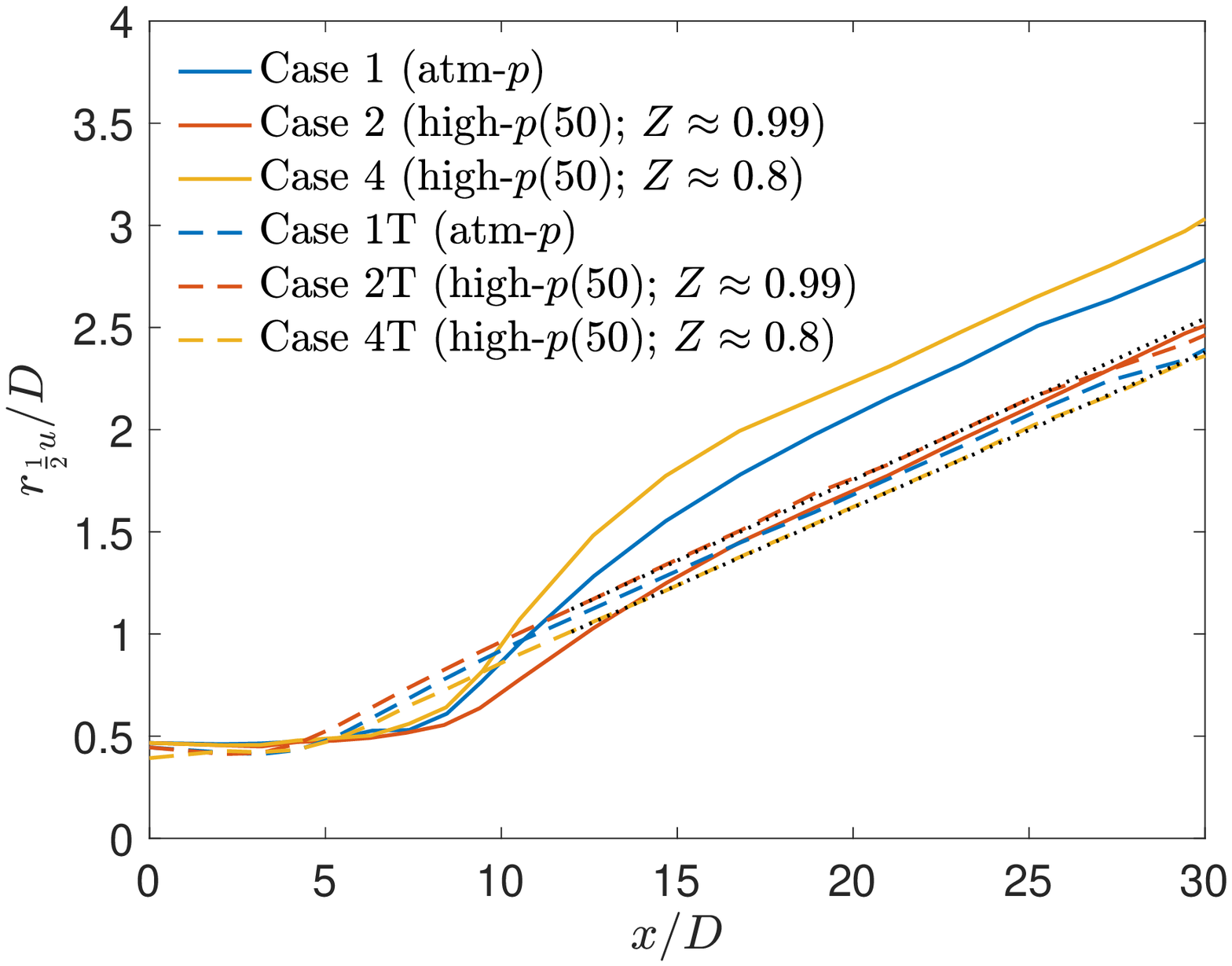}
\par\end{centering}

\begin{centering}
(a)\qquad{}\qquad{}\qquad{}\qquad{}\qquad{}\qquad{}\qquad{}\qquad{}\qquad{}\qquad{}\qquad{}\qquad{}(b)
\par\end{centering}

\caption{Comparisons of different inflow cases: Streamwise variation of the
(a) inverse of centerline mean velocity ($U_{c}$) normalized by the
jet-exit centerline velocity ($U_{0}$) and (b) velocity half-radius
($r_{\frac{1}{2}u}$). The black dash-dotted lines in (a)
are given by equation (\ref{eq:centerline_vel}) using $B_{u}=5.5$, $x_{0u}=-2.3D$
for Case 2T and $B_{u}=5.7$, $x_{0u}=-1.3D$ for Case 4T. Details of lines showing
the self-similar profile of equation (\ref{eq:centerline_vel}) for Cases 1, 2,
4 and 1T are presented in figures \ref{fig:invUc_hfRad_Cases1to4} and \ref{fig:invUc_hfRad_Case_1n1T}. The black dotted lines in 
(b) are given by $r_{\frac{1}{2}u}/D=0.079\left(x/D+2.2\right)$ for Case 2T and $r_{\frac{1}{2}u}/D=0.076\left(x/D+1.3\right)$ for Case 4T.
\label{fig:invUc_hfRad_lamTurb}}
\end{figure}

Figure \ref{fig:invUc_hfRad_lamTurb}(a) illustrates the centerline 
variation of mean axial velocity in Cases 1/1T, 2/2T and 4/4T.
In concurrence with the observation for Case 1T against Case 1, discussed
in \S \ref{sub:Case1n1T_comparison}, the pseudo-turbulent inflow
cases at supercritical pressure (Cases 2T and 4T) also exhibit a shorter
potential core than their laminar inflow counterparts (Cases 2 and
4). As a result, the axial location where the mean velocity decay
begins for Cases 1T, 2T and 4T is upstream of the corresponding location for Cases
1, 2 and 4. The laminar inflow cases 
show a distinct transition region ($7\lesssim x/D\lesssim15$) with
larger mean velocity decay rate than that further downstream in their
self-similar region. A similar change in decay rate (equal to the slope
of the plot lines) does not occur in Cases 1T, 2T and 4T, where the
slopes remain approximately the same downstream of the potential core
closure. The linear decay rate in the self-similar region is described
by $1/B_{u}$, defined by equation (\ref{eq:centerline_vel}). 
The $B_{u}$ values for various cases are included in the caption of
figure \ref{fig:invUc_hfRad_lamTurb}(a). For Cases 1 and 1T, the decay rates are the same; between
Cases 2 and 2T, the decay rate is slightly larger in the laminar inflow jet (Case 2), and between
Cases 4 and 4T, the decay rate is significantly larger in the laminar inflow jet (Case 4). 

To investigate the differences in jet spread, $r_{\frac{1}{2}u}$ from different inflow cases
are compared in figure \ref{fig:invUc_hfRad_lamTurb}(b). As expected
from smaller potential core length in the pseudo-turbulent inflow
cases, $r_{\frac{1}{2}u}$ growth in Cases 1T, 2T and 4T begins upstream of that in Cases 1, 2 
and 4. Immediately downstream of the potential core
closure, $r_{\frac{1}{2}u}$ in laminar inflow cases (Cases 1, 2 and 4)
grows at a relatively faster rate than in Cases 1T, 2T and 4T. The
linear $r_{\frac{1}{2}u}$ profiles in the self-similar region of Cases 1,
2, 4 and 1T are given in figures \ref{fig:invUc_hfRad_Cases1to4} and \ref{fig:invUc_hfRad_Case_1n1T}.
The profiles for Cases 2T and 4T 
are listed in the figure caption.
The inflow change from laminar to pseudo-turbulent
reduces the spreading rate at atmospheric as well as supercritical
conditions. The change is significant at atmospheric conditions (from
$0.085$ in Case 1 to $0.078$ in Case 1T) and relatively small for
supercritical cases (from $0.0805$ in Case 2 to $0.079$ in Case
2T and $0.077$ in Case 4 to $0.076$ in Case 4T). A noticeable feature
in the self-similar region of figure \ref{fig:invUc_hfRad_lamTurb}(b)
is the difference in $r_{\frac{1}{2}u}$ among various cases for the two inflows;
for laminar inflow, $r_{\frac{1}{2}u}$ decreases from Case 1 to Case 2 and
increases from Case 2 to Case 4, whereas the differences are comparatively
minimal between Cases 1T, 2T and 4T. In fact, $r_{\frac{1}{2}u}$ in Cases
1T and 2T are slightly larger than that in Case 4T.

The decay in mean velocity occurs in part due to the transfer of kinetic energy from the mean field to fluctuations, as discussed in \S \ref{subsub:meanVel_hfRad_Case1to4}. To determine if the differences observed here in the mean velocity are consistent with the variations in velocity fluctuations, they are examined next. \vspace{-0.75cm}

\noindent 

\noindent 
\begin{figure}
\begin{centering}
\includegraphics[width=6.9cm]{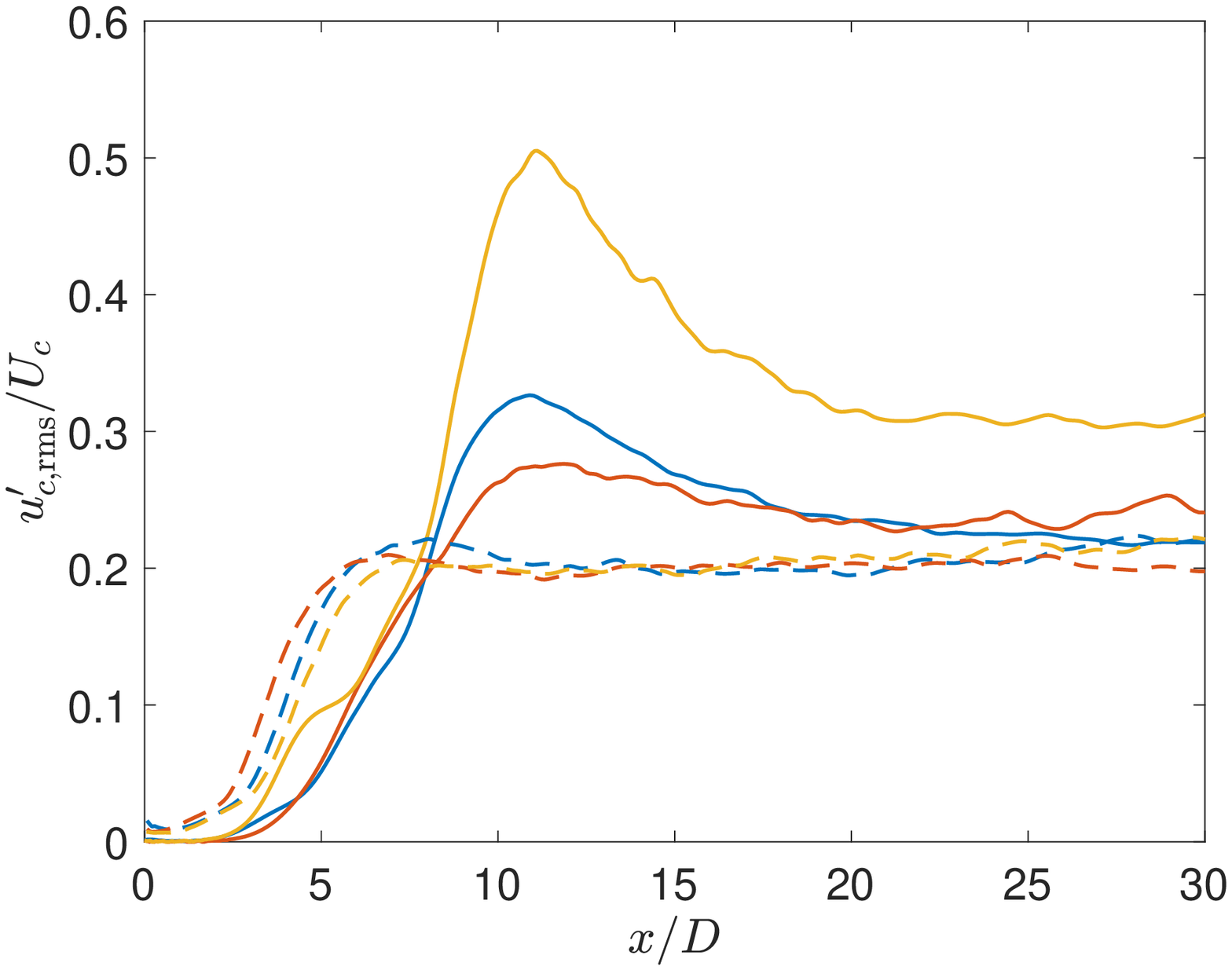}\includegraphics[width=6.9cm]{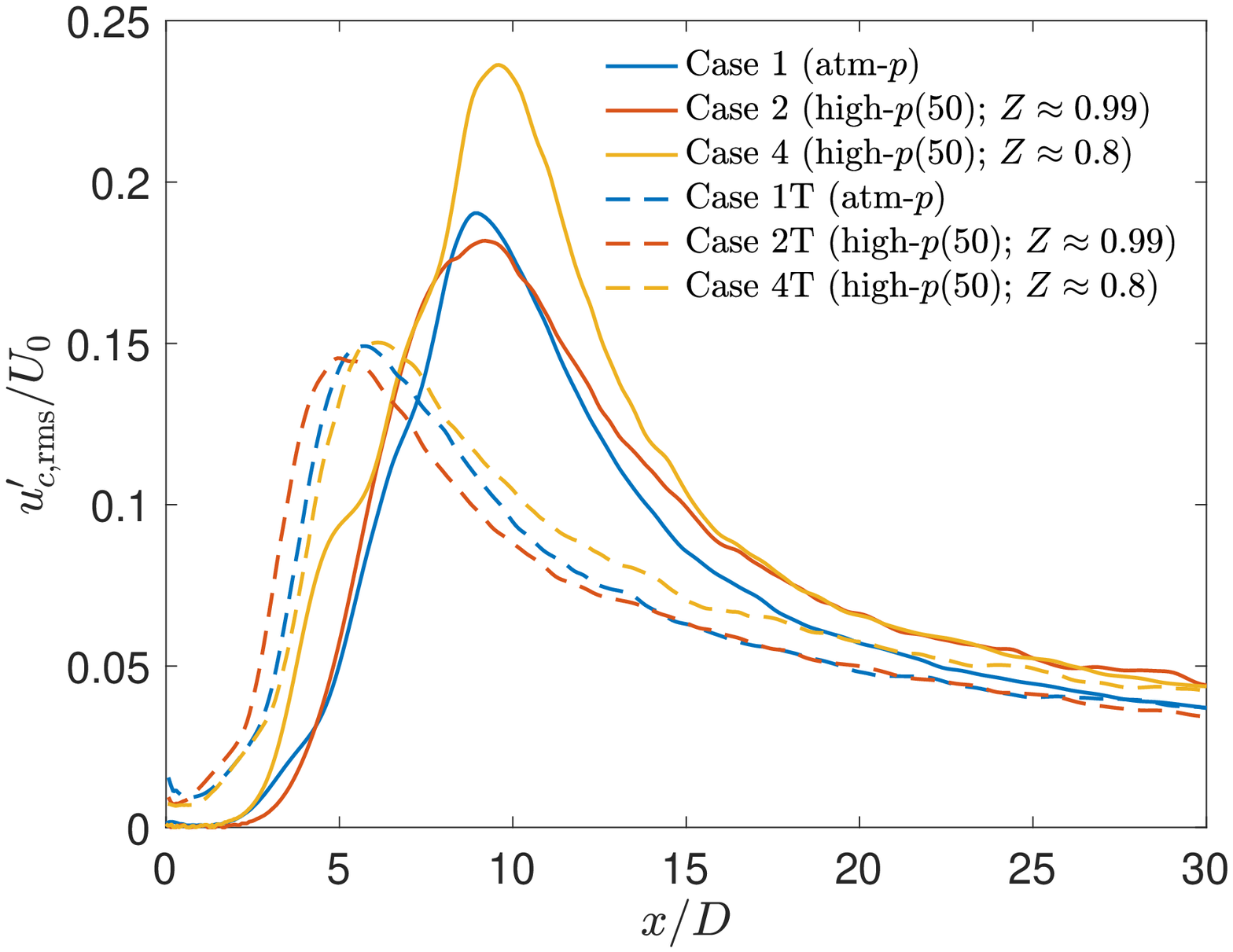}
\par\end{centering}

\begin{centering}
(a)\qquad{}\qquad{}\qquad{}\qquad{}\qquad{}\qquad{}\qquad{}\qquad{}\qquad{}\qquad{}\qquad{}\qquad{}(b)
\par\end{centering}

\caption{Comparisons of different inflow cases: Streamwise variation of the
(a) the centerline r.m.s. axial velocity fluctuations ($u_{c,\mathrm{rms}}^{'}$)
normalized by the centerline mean axial velocity ($U_{c}$), (b) $u_{c,\mathrm{rms}}^{'}$
normalized by the jet-exit centerline mean axial velocity ($U_{0}$). The legend is the same for
both plots. \label{fig:centerline_urms_lamTurb}}
\end{figure}

\paragraph{Velocity fluctuations and self-similarity}

To understand the differences observed in figure \ref{fig:invUc_hfRad_lamTurb},
the centerline variation of axial-velocity fluctuations is compared for various inflow cases 
in figure \ref{fig:centerline_urms_lamTurb}.
$u_{c,\textrm{rms}}^{\prime}/U_{c}$, presented in figure \ref{fig:centerline_urms_lamTurb}(a),
reflects the local mean energy transfer to fluctuations. As discussed
in \S \ref{sub:Case1to4_comparison}, higher $u'_{c,\textrm{rms}}/U_{c}$
implies larger mean axial-velocity decay rate or higher slope of the
line in figure \ref{fig:invUc_hfRad_lamTurb}(a). In the transition
region, the laminar inflow cases exhibit significant differences with
increase in $p_{\infty}$ (from Case 1 to Case 2) as well as with decrease
in $Z$ (from Case 2 to Case 4). In contrast, the
differences are minimal between Cases 1T, 2T and 4T. In the transition
region of these cases $\left(3\lesssim x/D\lesssim8\right)$, $u'_{c,\textrm{rms}}/U_{c}$
in Case 4T is slightly smaller than that in Cases 1T and 2T. Accordingly,
the mean axial-velocity decay rate is smaller for Case 4T in the transition
region, see figure \ref{fig:invUc_hfRad_lamTurb}(a). The difference between
the asymptotic value attained by $u'_{c,\textrm{rms}}/U_{c}$ is small
between Cases 1 and 1T, but significant at supercritical $p_{\infty}$
between Cases 2 and 2T and Cases 4 and 4T. The difference is particularly large between
Cases 4 and 4T, consistent with the large difference in their $U_c$ decay rates in 
the self-similar region of figure \ref{fig:invUc_hfRad_lamTurb}. Thus, the variation in 
axial-velocity fluctuations consistently represents the transfer of mean kinetic energy to t.k.e.

Normalizing $u'_{c,\textrm{rms}}$ with $U_{0}$, as presented
in figure \ref{fig:centerline_urms_lamTurb}(b), enables a
comparison of the absolute fluctuation magnitude for each inflow ($U_{0}$
differs for the two inflows, as discussed in \S \ref{sub:Case1n1T_comparison}).
In figure \ref{fig:centerline_urms_lamTurb}(b), $u'_{c,\textrm{rms}}/U_{0}$ decreases with axial distance, unlike
$u'_{c,\textrm{rms}}/U_{c}$ in figure \ref{fig:centerline_urms_lamTurb}(a) that asymptotes to a constant value.
The peak of $u'_{c,\textrm{rms}}/U_{0}$, attained in the transition
region, decreases with increasing $p_{\infty}$ from $1$ bar to $50$
bar and increases with decreasing $Z$ from $0.99$ to $0.8$ for
each inflow. 
The $u'_{c,\textrm{rms}}/U_{c}$ and $u'_{c,\textrm{rms}}/U_{0}$ magnitudes and their differences are larger
in the laminar inflow cases, especially in the transition region of the flow, showing that the effect 
of $p_{\infty}$ and $Z$ depends strongly on the inflow, in addition
to the ambient thermodynamic state characterized by $p_{\infty}\left(\beta_{T}-1/p_{\infty}\right)$.

\begin{figure}
\begin{centering}
\includegraphics[width=6.9cm]{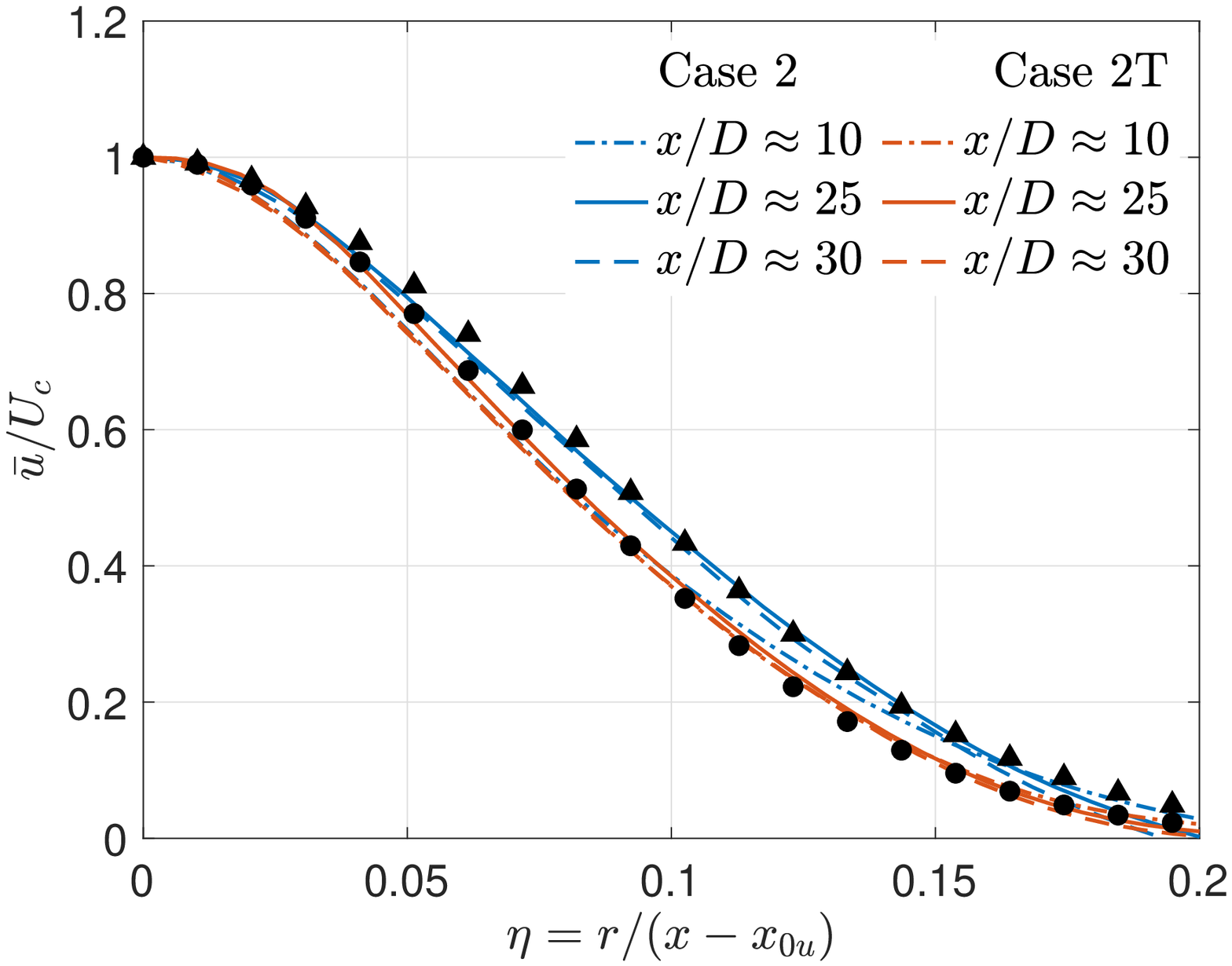}\includegraphics[width=6.9cm]{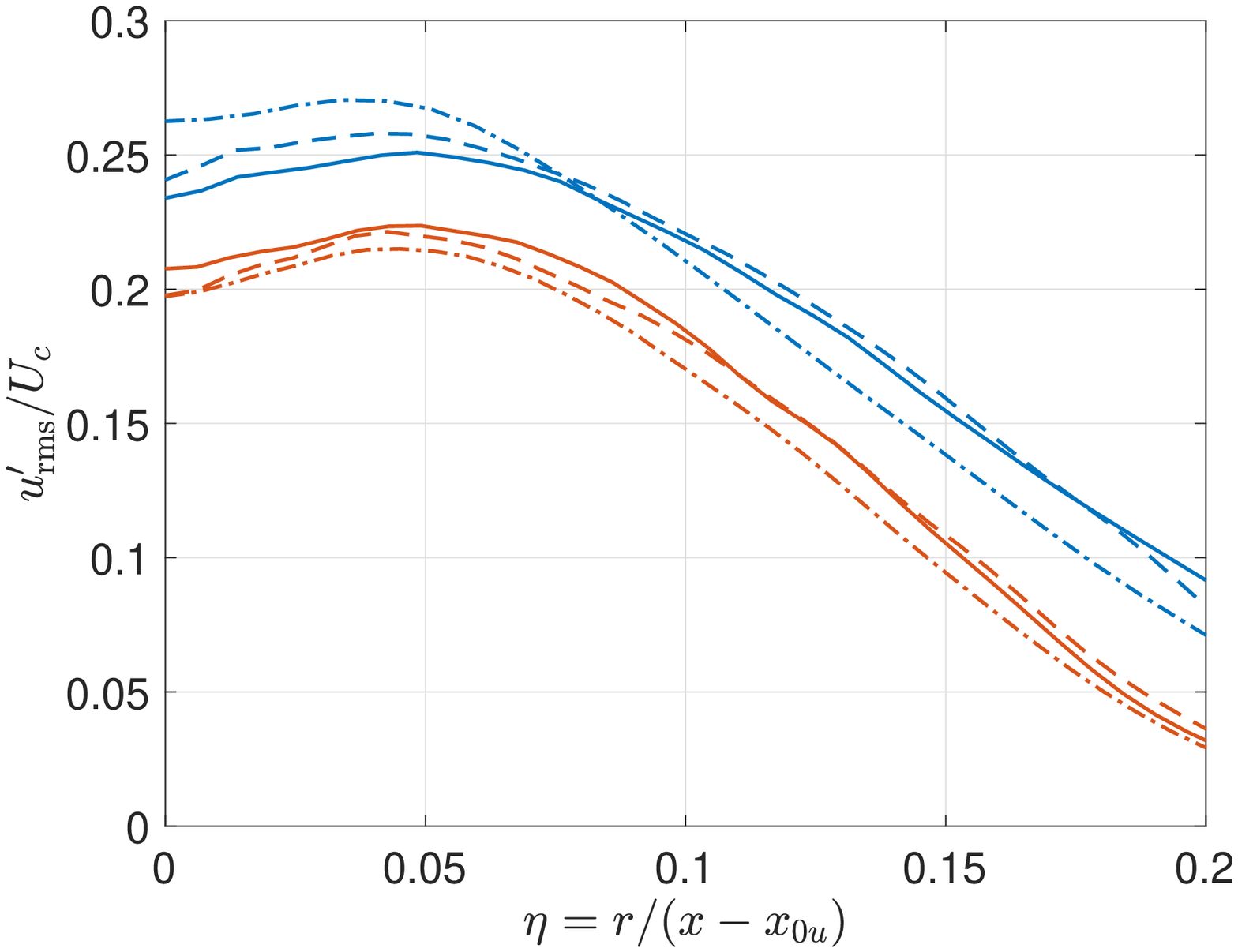}
\par\end{centering}

\begin{centering}
(a)\qquad{}\qquad{}\qquad{}\qquad{}\qquad{}\qquad{}\qquad{}\qquad{}\qquad{}\qquad{}\qquad{}\qquad{}(b)
\par\end{centering}



\caption{Case 2 and 2T comparisons:Radial profiles of (a) mean axial velocity
($\bar{u}$) normalized by the centerline mean axial velocity ($U_{c}$),
(b) r.m.s. axial-velocity fluctuations ($u'_{\mathrm{rms}}$) normalized
by the centerline mean axial velocity 
at various axial locations. The markers 
$\blacktriangle$ and $\bullet$ in (a) show
$f\left(\eta\right)$ of equation (\ref{eq:self_similar_Gauss_dist}) using $A_{u}=79.5$ and $99.2$, 
respectively. The legend is 
the same for both plots. \label{fig:Case_2_2T_rad_velstats}}
\end{figure}

\begin{figure}
\begin{centering}
\includegraphics[width=6.9cm]{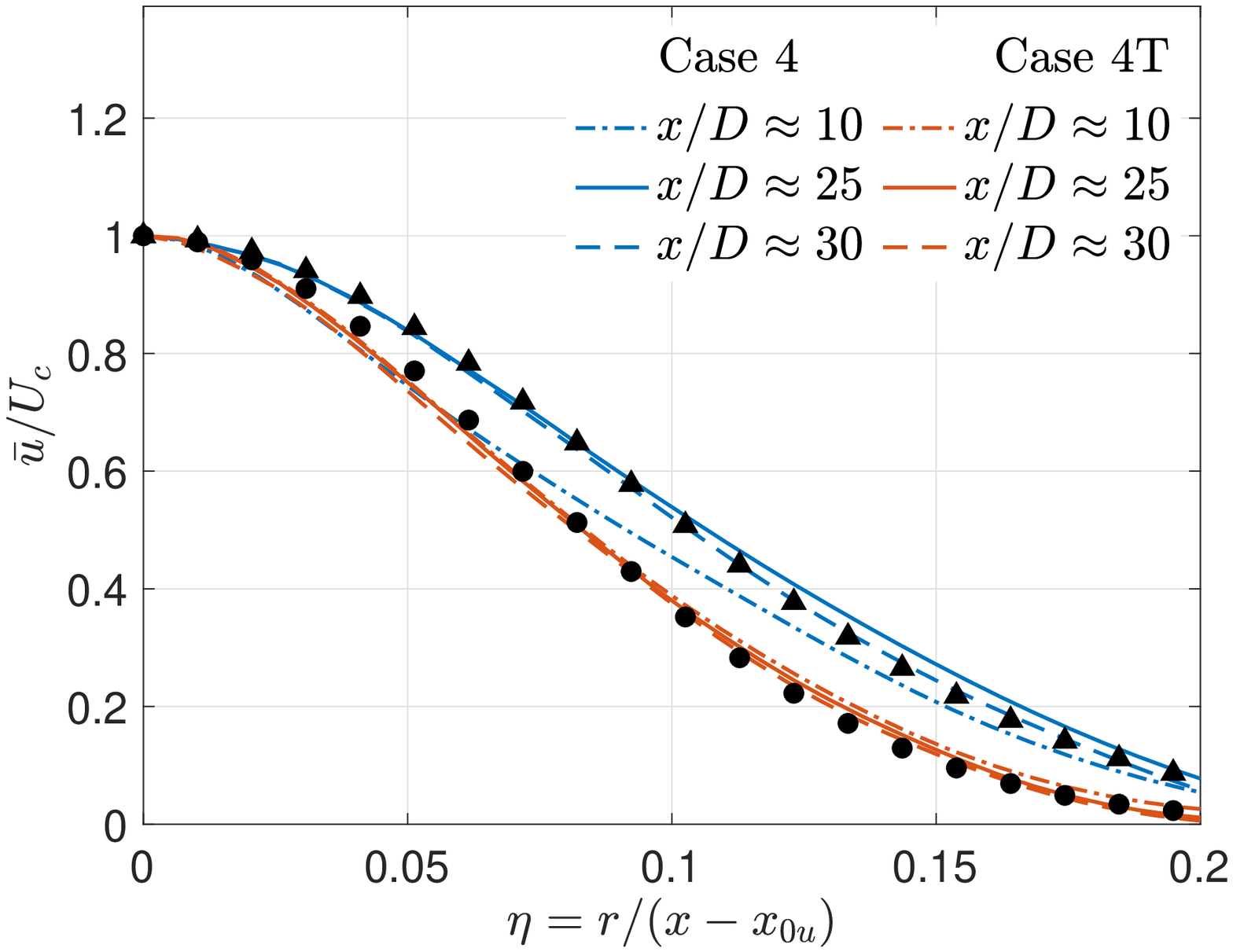}\includegraphics[width=6.9cm]{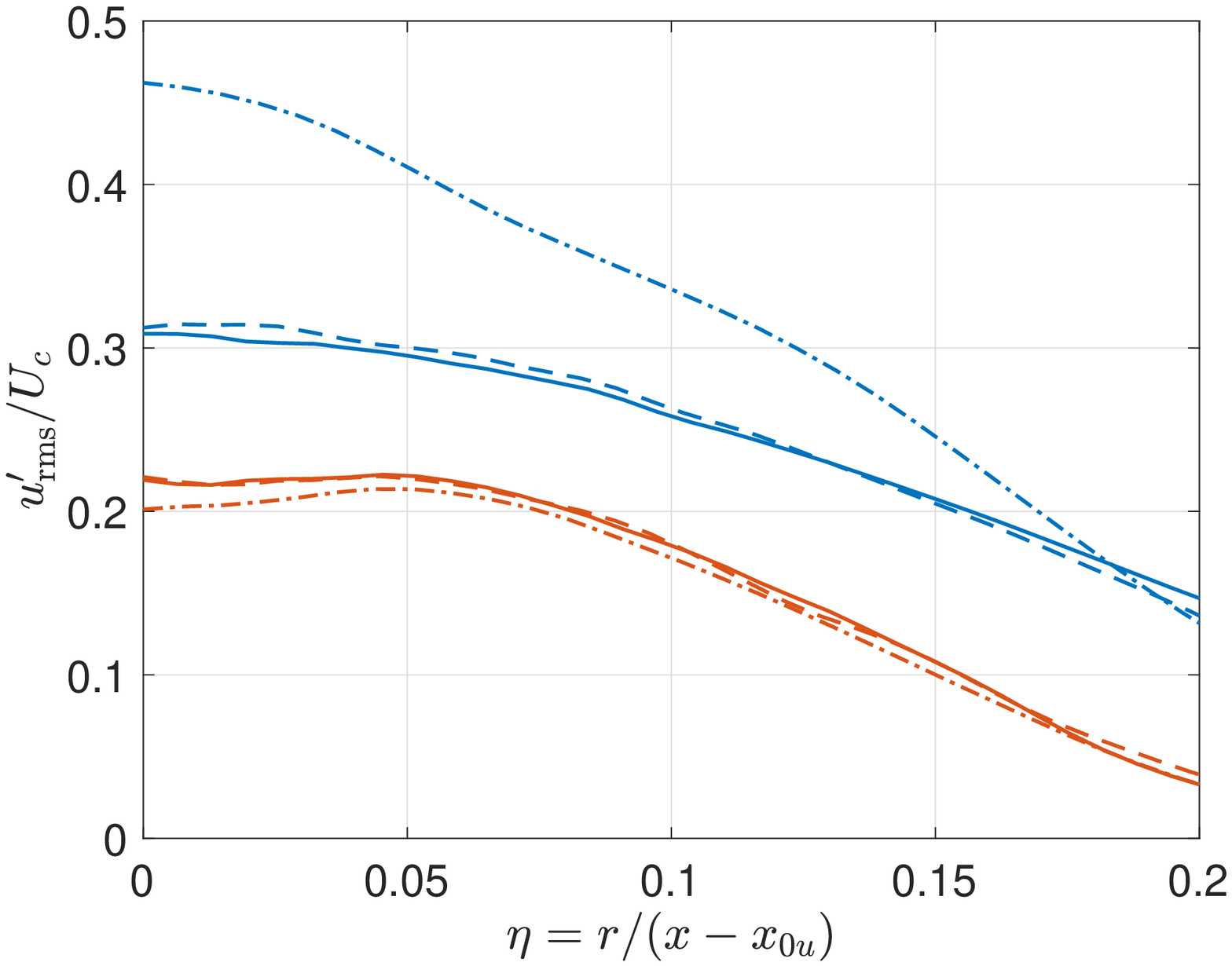}
\par\end{centering}

\begin{centering}
(a)\qquad{}\qquad{}\qquad{}\qquad{}\qquad{}\qquad{}\qquad{}\qquad{}\qquad{}\qquad{}\qquad{}\qquad{}(b)
\par\end{centering}



\caption{Case 4 and 4T comparisons:Radial profiles of (a) mean axial velocity
($\bar{u}$) normalized by the centerline mean axial velocity ($U_{c}$),
(b) r.m.s. axial-velocity fluctuations ($u'_{\mathrm{rms}}$) normalized
by the centerline mean axial-velocity 
at various axial locations. The markers 
$\blacktriangle$ and $\bullet$ in (a) show
$f\left(\eta\right)$ of equation (\ref{eq:self_similar_Gauss_dist}) using $A_{u}=64.4$ and $99.2$, 
respectively. The legend is 
the same for all plots. \label{fig:Case_4_4T_rad_velstats}}
\end{figure}

To examine self-similarity in the flow, the radial variations of mean axial velocity and r.m.s. axial-velocity flucutations 
at three axial locations are compared between Cases 2 and 2T and Cases 4 and
4T in figures \ref{fig:Case_2_2T_rad_velstats} and \ref{fig:Case_4_4T_rad_velstats},
respectively. The axial location $x/D\approx10$ lies around the jet
transition region in both inflow cases, whereas the profiles at $x/D\approx25$
and $30$ help assess self-similarity. Figure \ref{fig:Case_2_2T_rad_velstats}(a)
shows that $\bar{u}/U_{c}$ attains self-similarity upstream of $x/D\approx25$
in both cases (2 and 2T); however, the self-similar profile is different as shown by the 
least-squares fits of the Gaussian distribution, $f\left(\eta\right)$ of (\ref{eq:self_similar_Gauss_dist}),
to $x/D\approx30$ profiles, depicted as solid black markers in figure \ref{fig:Case_2_2T_rad_velstats}(a). 
$u'_{\textrm{rms}}/U_{c}$ 
profiles at $x/D\approx25$ and $30$ in figure \ref{fig:Case_2_2T_rad_velstats}(b) 
exhibit only minor differences in both Cases 2 and 2T, 
suggesting that $u'_{\textrm{rms}}/U_{c}$ 
is self-similar around $x/D\approx25$. $u'_{\textrm{rms}}/U_{c}$ is considerably 
larger in the laminar inflow case (Case 2) at all $\eta$ locations, consistent with 
the observations at atmospheric $p_{\infty}$ between Cases 1 and 1T in figure \ref{fig:Case_1_1T_velstat}(b).
Similarly, $v'_{\textrm{rms}}/U_{c}$, $w'_{\textrm{rms}}/U_{c}$, and $\overline{u'v'}/U{}_{c}^{2}$ (not shown here for brevity) are also larger in Case
2 than Case 2T and show self-similarity around $x/D\approx25$.


$\bar{u}/U_{c}$ from Cases 4 and 4T are compared in figure \ref{fig:Case_4_4T_rad_velstats}(a).
As in figures \ref{fig:Case_1_1T_velstat}(a) and \ref{fig:Case_2_2T_rad_velstats}(a) 
for Cases 1/1T and 2/2T, the self-similar profile are different for the two inflows, 
and this difference is amplified with respect to Cases 2/2T as the solid
markers in figure \ref{fig:Case_4_4T_rad_velstats}(a) that display the
Gaussian distribution, $f\left(\eta\right)$ of (\ref{eq:self_similar_Gauss_dist}), show. 
$u'_{\textrm{rms}}/U_{c}$ 
illustrated in figure \ref{fig:Case_4_4T_rad_velstats}(b) 
shows self-similarity around $x/D\approx25$ for both cases (4 and 4T) and larger
magnitude in the laminar inflow case (Case 4). $v'_{\textrm{rms}}/U_{c}$, $w'_{\textrm{rms}}/U_{c}$, 
and $\overline{u'v'}/U{}_{c}^{2}$ show similar behavior, and are not shown here for brevity.

The spatial variation of thermodynamic (pressure/density) fluctuations and their correlation with velocity fluctuations that determines the transport terms in the t.k.e. equation were used to explain the jet flow dynamics for various $p_{\infty}$ and $Z$ in \S \ref{subsub:Case1to4_prsFluc} and, therefore, these quantities are examined next to determine if they can explain the differences with inflow condition discussed above.
\vspace{-1cm}

\noindent 

\noindent 

\noindent 
\begin{figure}
\begin{centering}
\includegraphics[width=6.9cm]{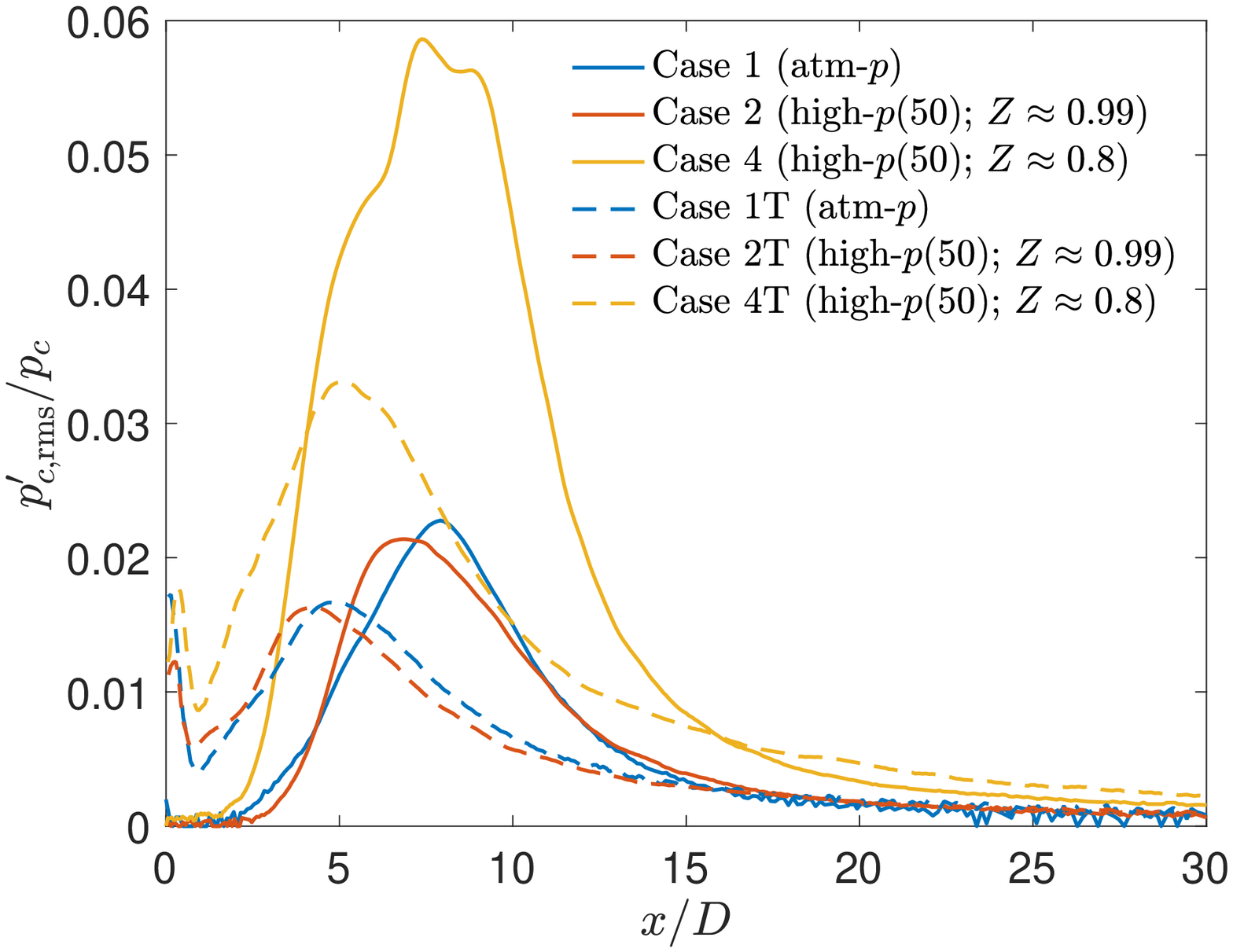}\includegraphics[width=6.9cm]{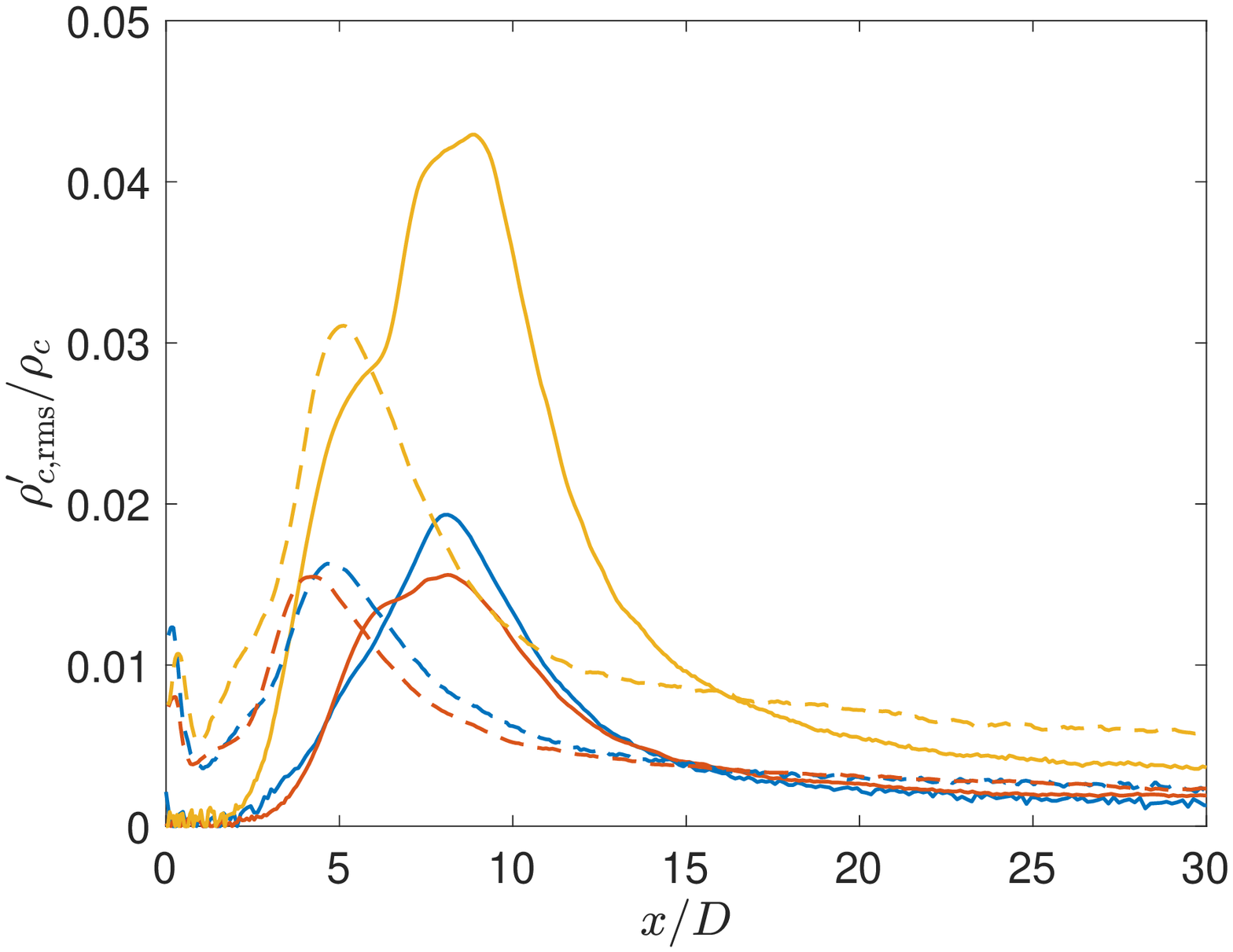}
\par\end{centering}

\begin{centering}
(a)\qquad{}\qquad{}\qquad{}\qquad{}\qquad{}\qquad{}\qquad{}\qquad{}\qquad{}\qquad{}\qquad{}\qquad{}(b)
\par\end{centering}

\caption{Comparisons of different inflow cases: Streamwise variation of the
(a) the centerline r.m.s. pressure ($p'_{c,\mathrm{rms}}$) normalized
by the centerline mean pressure ($p_{c}$) and (b) the centerline
r.m.s. density fluctuations ($\rho'_{c,\mathrm{rms}}$) normalized
by the centerline mean density ($\rho_{c}$). The legend is the same for
both plots. \label{fig:centerline_pfluc_lamTurb}}
\end{figure}

\paragraph{Pressure and density fluctuations, pressure-velocity correlation, and third-order velocity moments}

Centerline variations of $p_{c,\text{rms}}^{\prime}/p_{c}$ and $\rho_{c,\text{rms}}^{\prime}/\rho_{c}$
are presented in figures \ref{fig:centerline_pfluc_lamTurb}(a)
and (b), respectively. $p'_{c,\textrm{rms}}/p_{c}$ and $\rho'_{c,\textrm{rms}}/\rho_{c}$
are negligible at jet exit in laminar inflow cases but have significant
magnitude in pseudo-turbulent inflow cases, where it decreases with
axial distance until the potential core closes and increases in the
transition region. Variations of $p'_{c,\textrm{rms}}/p_{c}$ and $\rho'_{c,\textrm{rms}}/\rho_{c}$
with $p_{\infty}$ and $Z$ are similar for the two inflows. 
$p'_{c,\textrm{rms}}/p_{c}$ and $\rho'_{c,\textrm{rms}}/\rho_{c}$
are larger in Case 4 than in Cases 1 and 2 and, similarly, they are higher
in Case 4T than in Cases 1T and 2T. With increase in $p_{\infty}$ from 1 bar
(Cases 1 and 1T) to 50 bar (Cases 2 and 2T), the peak value of $p'_{c,\textrm{rms}}/p_{c}$
and $\rho'_{c,\textrm{rms}}/\rho_{c}$ in the transition region decreases
by a small value. The differences diminish downstream in the self-similar
region. $p'_{c,\textrm{rms}}/p_{c}$ and $\rho'_{c,\textrm{rms}}/\rho_{c}$ increase with decrease in
$Z$ from $0.99$ (Cases 2 and 2T) to $0.8$ (Cases
4 and 4T) for both inflows, especially in the transition region of
the flow. On the other hand, $u'_{c,\textrm{rms}}/U_{c}$, 
presented in figure \ref{fig:centerline_urms_lamTurb}(a),
increases with decreasing $Z$ for laminar inflow but remains
approximately the same in pseudo-turbulent inflow cases. In fact,
in the transition region, $u'_{c,\textrm{rms}}/U_{c}$ is slightly
smaller in Case 4T than Case 2T, while it is larger in Case 4 than Case
2. This anomaly with inflow change leads to contrasting mean flow
behavior, observed in figure \ref{fig:invUc_hfRad_lamTurb} in the transition region, where
the mean axial-velocity decay and jet half-radius increases from Case
2 to Case 4 but decreases from Case 2T to Case 4T. 

\begin{figure}
\begin{centering}
\includegraphics[width=6.9cm]{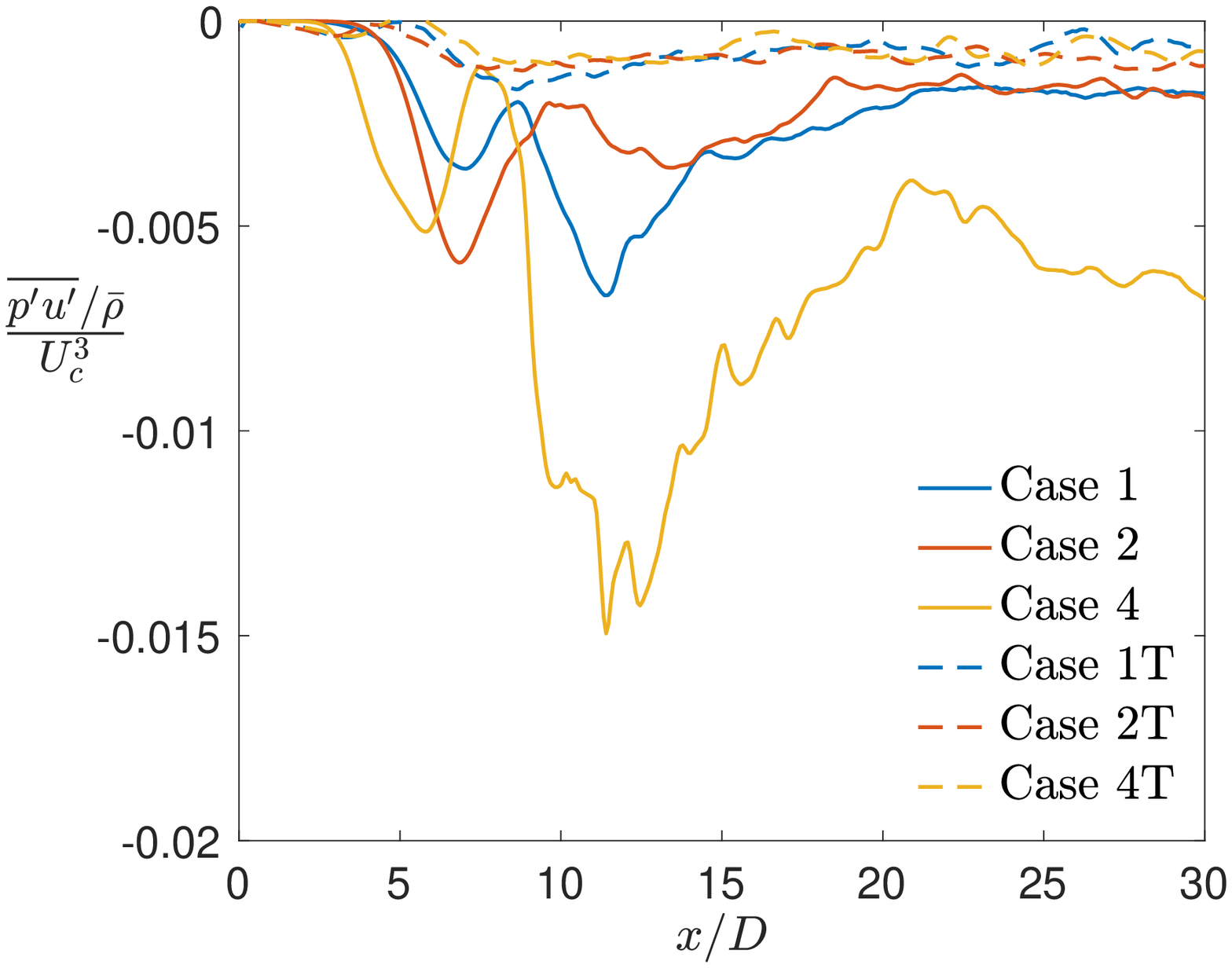}\includegraphics[width=6.9cm]{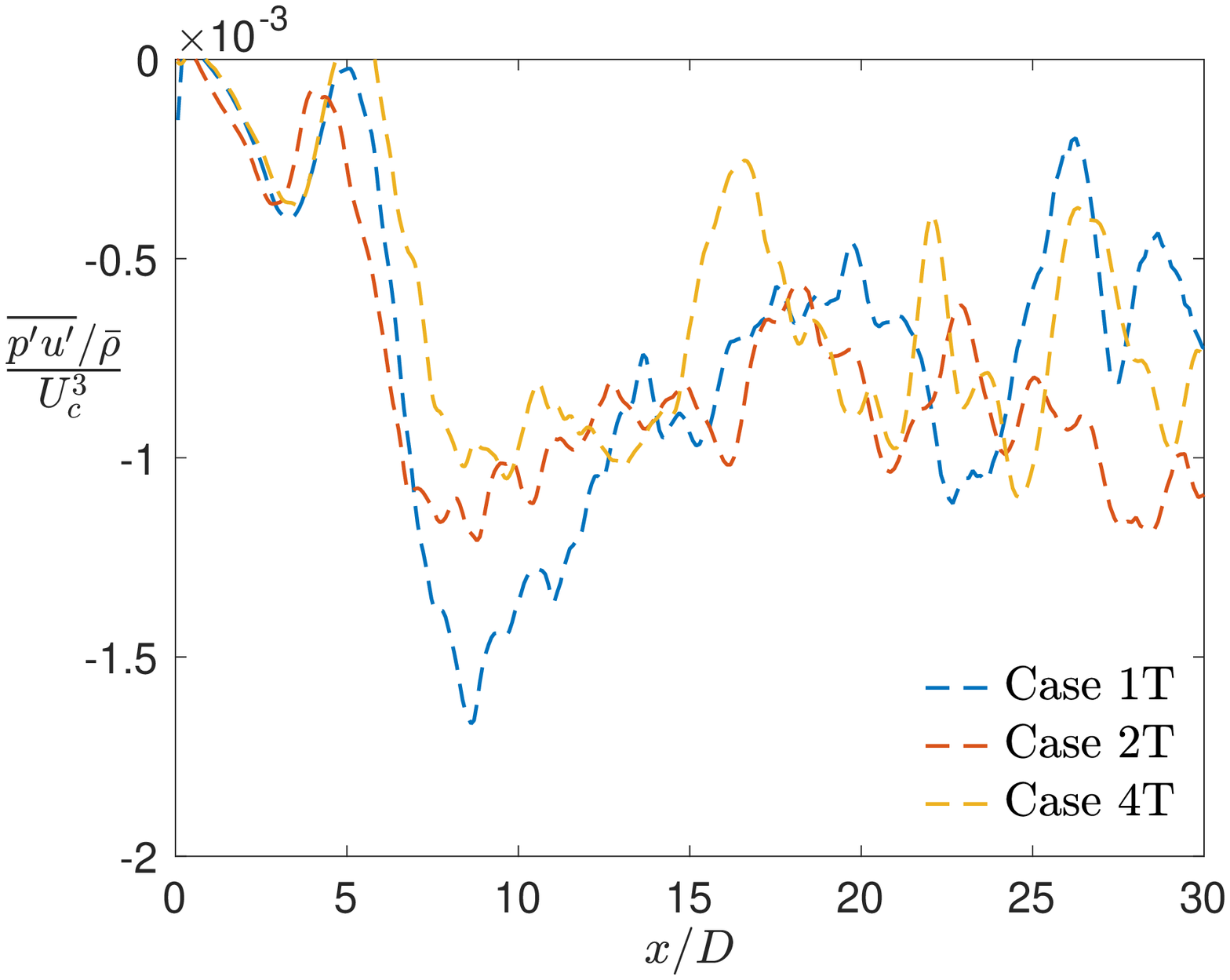}
\par\end{centering}

\begin{centering}
(a)\qquad{}\qquad{}\qquad{}\qquad{}\qquad{}\qquad{}\qquad{}\qquad{}\qquad{}\qquad{}\qquad{}\qquad{}(b)
\par\end{centering}

\caption{Comparisons of different inflow cases: Streamwise variation of the
normalized pressure-axial velocity correlation for (a) the laminar
and pseudo-turbulent inflow cases, and (b) only the pseudo-turbulent
inflow cases. Note the difference between $y$-axis scales of (a) and (b).
\label{fig:centerline_pfluc_pvelCorl_lamTurb}}
\end{figure}

\begin{figure}
\begin{centering}
\includegraphics[width=6.9cm]{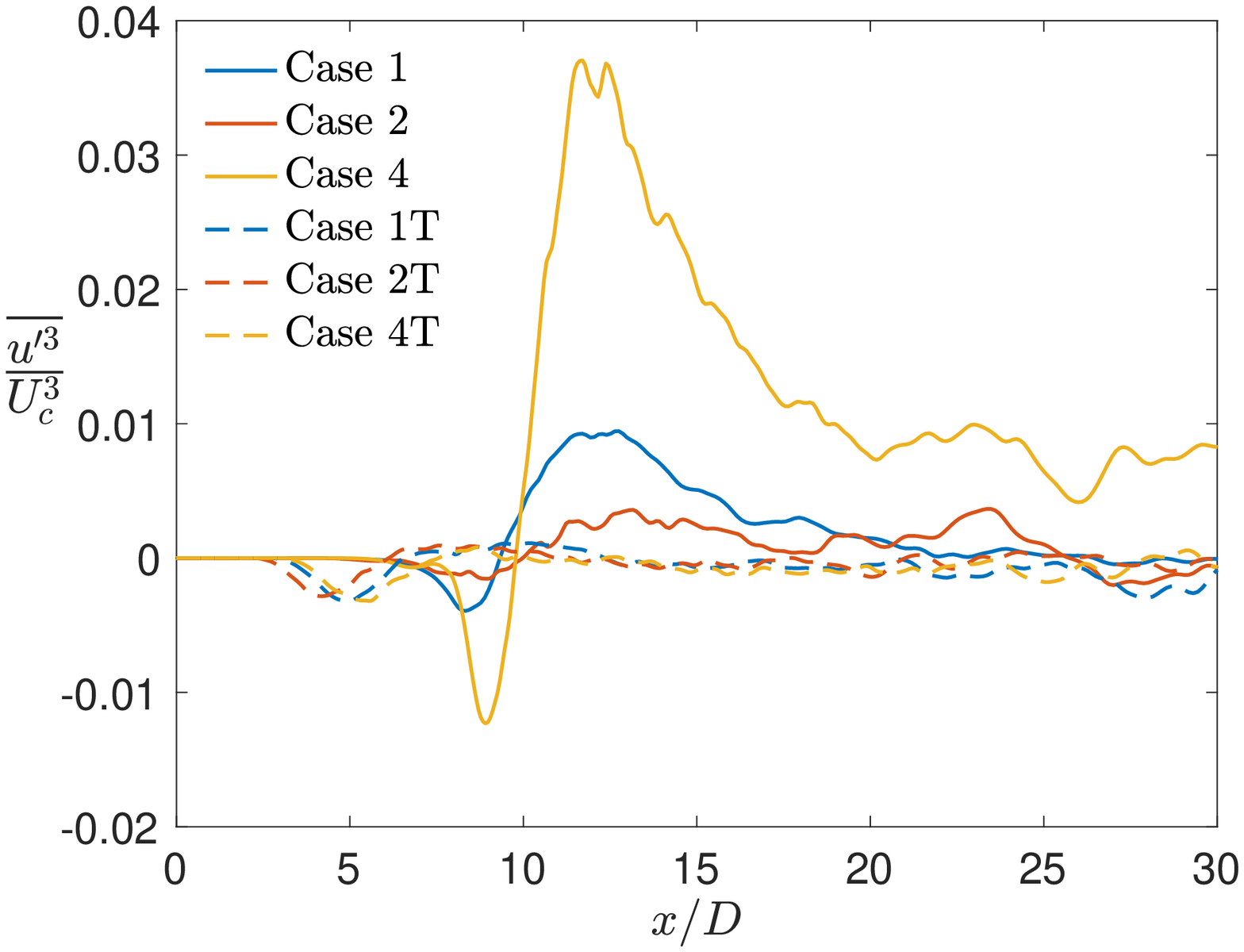}\includegraphics[width=6.9cm]{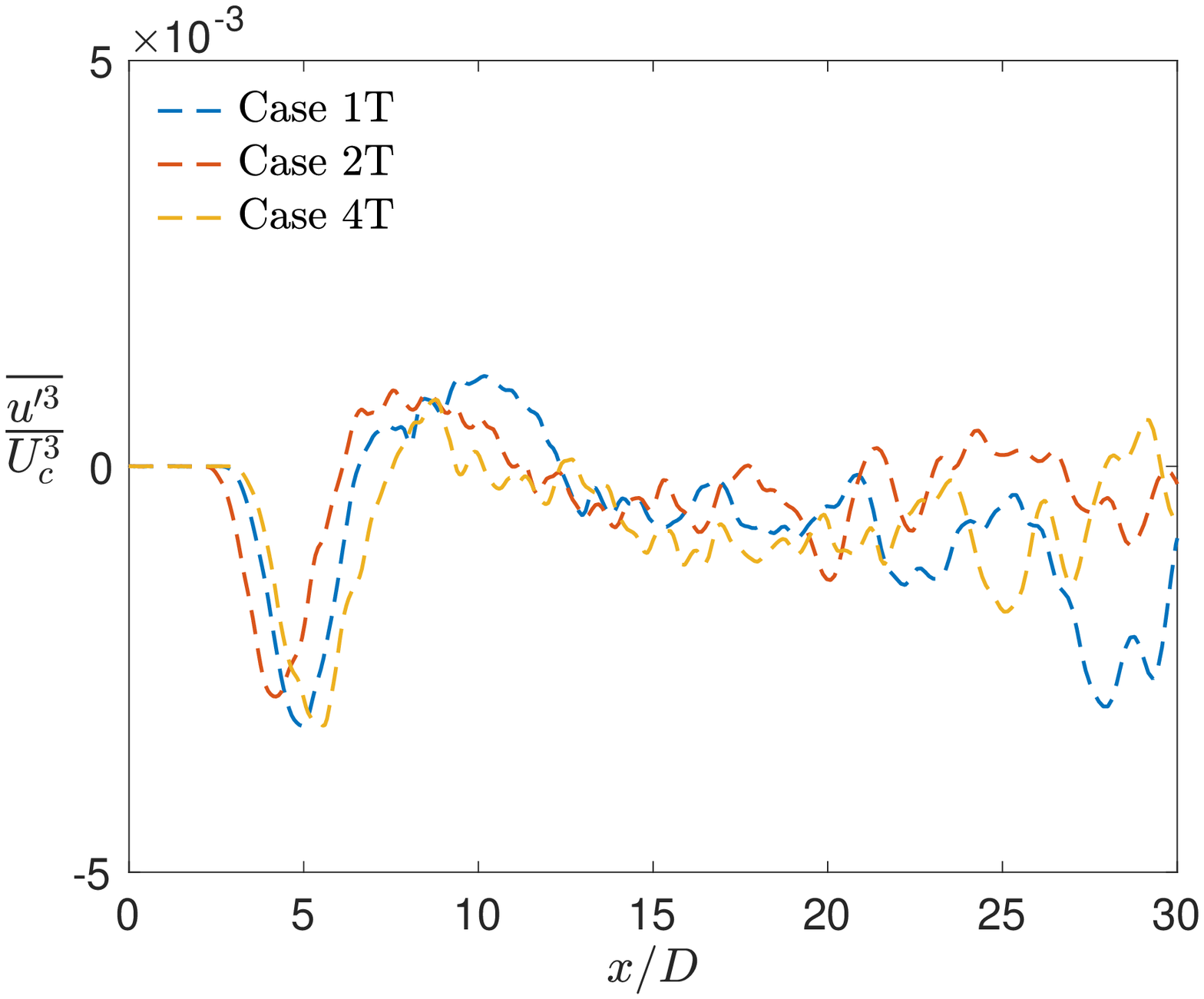}
\par\end{centering}

\begin{centering}
(a)\qquad{}\qquad{}\qquad{}\qquad{}\qquad{}\qquad{}\qquad{}\qquad{}\qquad{}\qquad{}\qquad{}\qquad{}(b)
\par\end{centering}

\caption{Comparisons of different inflow cases: Streamwise variation of $\overline{u^{\prime3}}/U_{c}^{3}$
for (a) the laminar and pseudo-turbulent inflow cases, and (b) only
the pseudo-turbulent inflow cases. Note the difference between $y$-axis scales of (a) and (b). \label{fig:centerline_ufluc3_lamTurb}}
\vspace{-0.25cm}
\end{figure}

To investigate this anomaly, the centerline variation of t.k.e.
diffusion fluxes from turbulent transport is compared in figures \ref{fig:centerline_pfluc_pvelCorl_lamTurb}
and \ref{fig:centerline_ufluc3_lamTurb}. Figure \ref{fig:centerline_pfluc_pvelCorl_lamTurb} compares $\frac{\overline{p'u'}/\bar{\rho}}{U_{c}^{3}}$,
which determines the t.k.e. diffusion due to pressure fluctuation transport. To highlight the differences among pseudo-turbulent inflow cases with suitable $y$-axis scale, figure \ref{fig:centerline_pfluc_pvelCorl_lamTurb}(b) shows only the results from Cases 1T, 2T and 4T. In the transition region, the absolute magnitude of $\frac{\overline{p'u'}/\bar{\rho}}{U_{c}^{3}}$ 
increases from Case 2 to Case 4 but decreases from Case 2T to 4T, indicating that the t.k.e. diffusion due to pressure fluctuation transport increases in the laminar inflow jet but decreases in the pseudo-turbulent inflow jet.
Further downstream, the differences are significant between Cases 2
and 4, but minimal between Cases 2T and 4T, implying that the effects of $Z$ (or the effects of ambient thermodynamic conditions closer to the Widom line) are enhanced in the laminar inflow jets. 

$\overline{u^{\prime3}}/U_{c}^{3}$, which determines t.k.e.
diffusion flux from turbulent transport of $\overline{u^{\prime2}}$, is
compared between Cases 1/1T, 2/2T and 4/4T in figure \ref{fig:centerline_ufluc3_lamTurb}(a) and
among cases 1T/2T/4T in figure \ref{fig:centerline_ufluc3_lamTurb}(b). While there
are significant differences in $\overline{u^{\prime3}}/U_{c}^{3}$ profiles
of Cases 1, 2 and 4, the differences are, again, minimal among Cases
1T, 2T and 4T. Thus, both $\frac{\overline{p'u'}/\bar{\rho}}{U_{c}^{3}}$ and $\overline{u^{\prime3}}/U_{c}^{3}$ show greater sensitivity to the ambient thermodynamic state, characterized by $p_{\infty}\left(\beta_{T}-1/p_{\infty}\right)$, in the laminar inflow jets, and their variation in figures \ref{fig:centerline_pfluc_pvelCorl_lamTurb} and \ref{fig:centerline_ufluc3_lamTurb} for various cases agrees well with the behavior of $u'_{c,\textrm{rms}}/U_{c}$
in figure \ref{fig:centerline_urms_lamTurb}(a) and the mean-flow
metrics in figure \ref{fig:invUc_hfRad_lamTurb}.\vspace{-0.75cm}

\noindent 

\noindent 

\paragraph{Passive scalar mixing}

\begin{figure}
\begin{centering}
(a) \includegraphics[width=11cm]{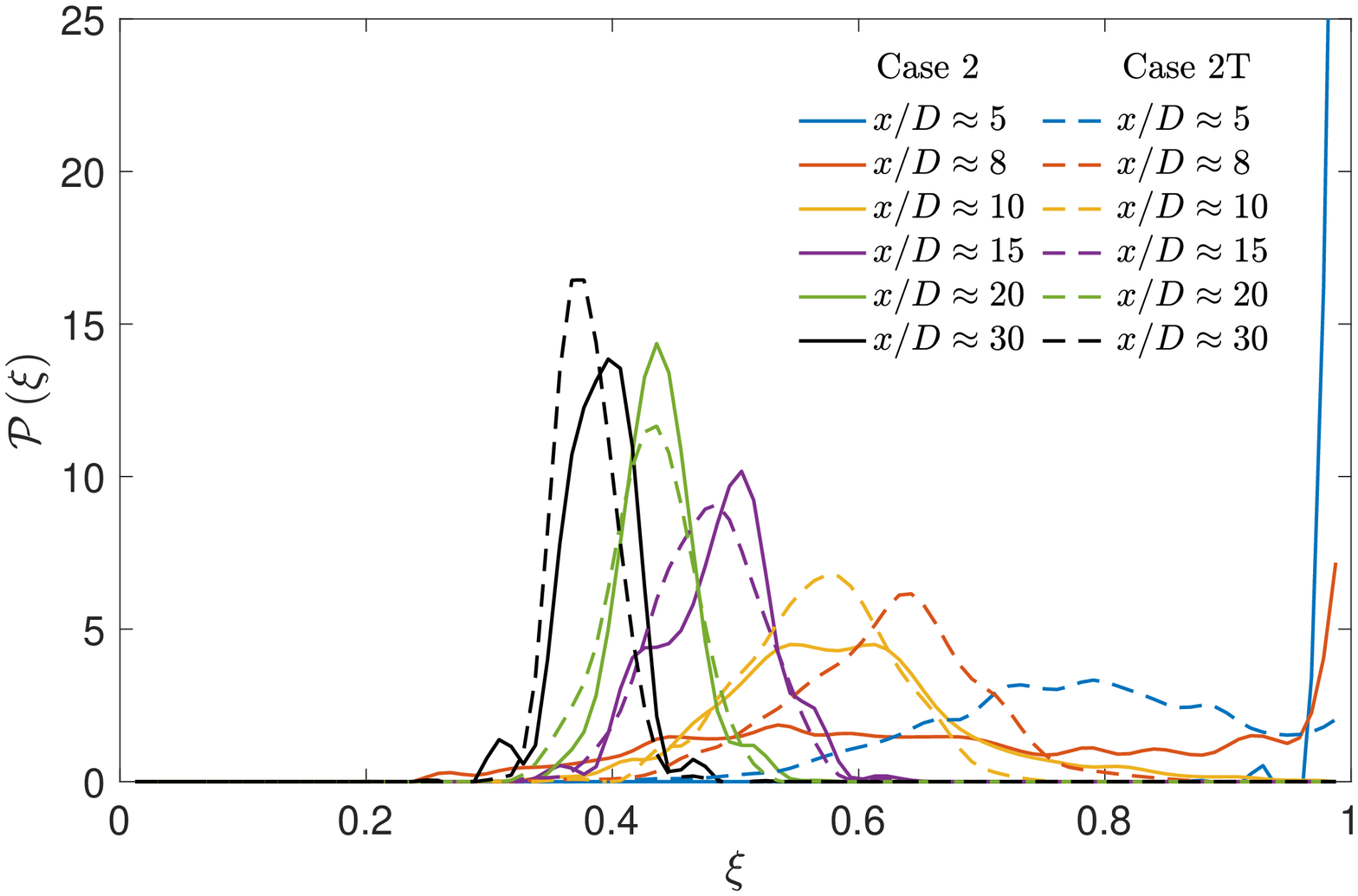}
\par\end{centering}

\begin{centering}
(b) \includegraphics[width=11cm]{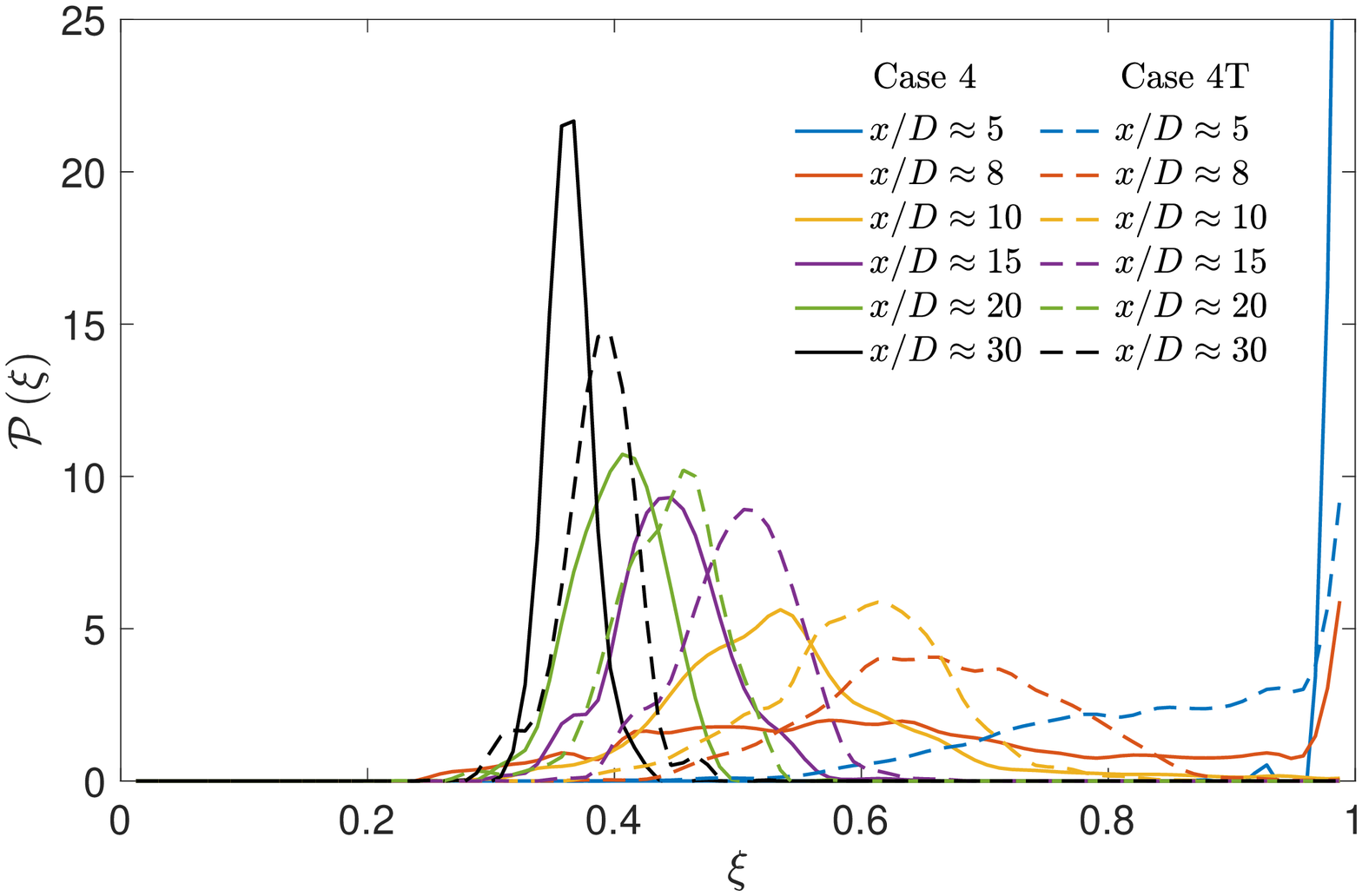}
\par\end{centering}

\caption{Scalar probability density function, $\mathcal{P}\left(\xi\right)$,
at various centerline axial locations for (a) Cases 2 and 2T, and
(b) Cases 4 and 4T. \label{fig:SclPDF_cases2T-4T}}
\end{figure}

\begin{figure}
\begin{centering}
\includegraphics[width=6.9cm]{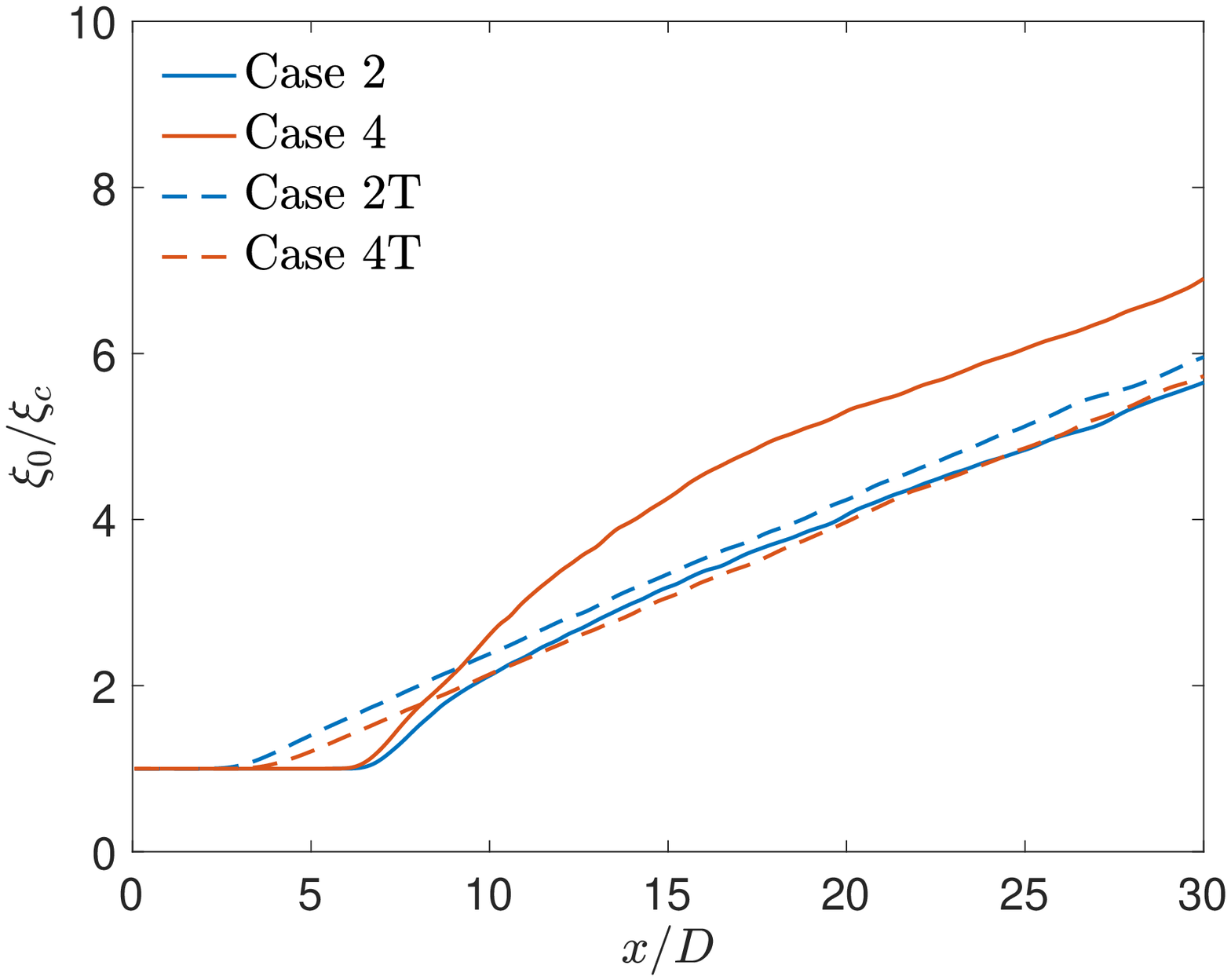}\includegraphics[width=6.9cm]{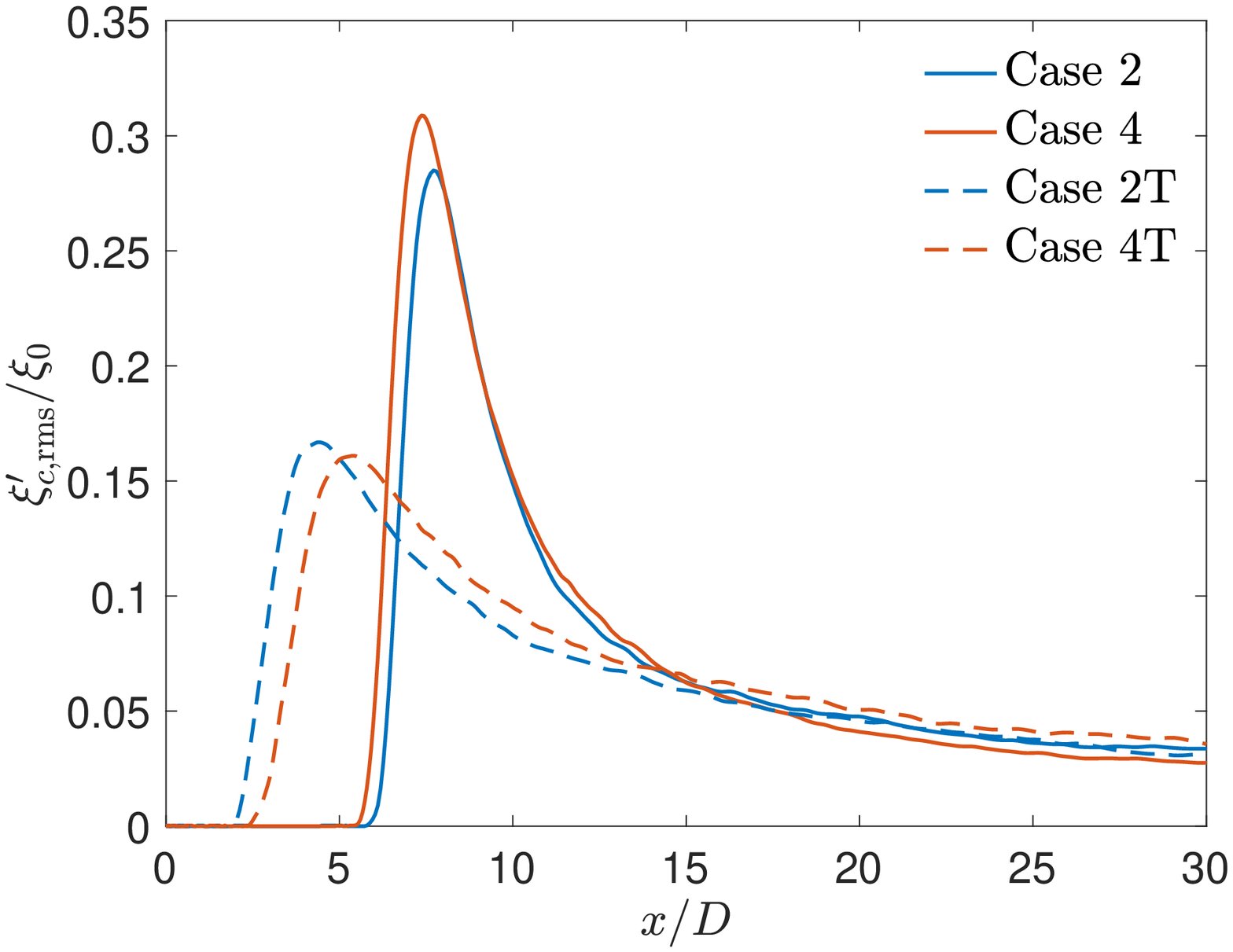}
\par\end{centering}

\begin{centering}
(a)\qquad{}\qquad{}\qquad{}\qquad{}\qquad{}\qquad{}\qquad{}\qquad{}\qquad{}\qquad{}\qquad{}\qquad{}(b)
\par\end{centering}

\caption{Streamwise variation of the (a) inverse of centerline scalar concentration
($\xi_{c}$) normalized by the jet-exit centerline value ($\xi_{0}$)
and (b) centerline r.m.s. scalar fluctuation ($\xi'_{c,\textrm{rms}}$)
normalized by the centerline jet-exit mean value ($\xi_{0}$). \label{fig:invScl_rmsFluc_Cases2n2T_4n4T}}
\end{figure}

To examine passive scalar mixing with inflow change at high pressure, 
the scalar p.d.f. is depicted in figure \ref{fig:SclPDF_cases2T-4T};
Cases 2 and 2T are compared at various $x/D$ in
figure \ref{fig:SclPDF_cases2T-4T}(a) and, similarly, Cases 4 and 4T are
compared in figure \ref{fig:SclPDF_cases2T-4T}(b).
As observed at atmospheric $p_{\infty}$ (figure \ref{fig:SclPDF_Case_1_1T}), 
the p.d.f. at $x/D\approx5$ in figure
\ref{fig:SclPDF_cases2T-4T}(a) shows pure jet fluid in the laminar
inflow case (Case 2), whereas mixed fluid in the case of pseudo-turbulent inflow (Case 2T). At $x/D\approx8$
and $10$, the p.d.f. has a wider distribution in Case 2 owing to
stronger large-scale vortical structures that yield larger normalized
scalar fluctuations, $\xi'_{c,\textrm{rms}}/\xi_{0}$, as shown in
figure \ref{fig:invScl_rmsFluc_Cases2n2T_4n4T}(b). Further downstream,
the p.d.f. profiles show minor differences, consistent with the scalar
mean and fluctuation behavior observed in figure \ref{fig:invScl_rmsFluc_Cases2n2T_4n4T}.
Thus, at supercritical $p_{\infty}$, but with ambient conditions far from the Widom line (with near-unity $Z$),
the influence of the inflow on scalar mixing is restricted to near field.
Figure \ref{fig:SclPDF_cases2T-4T}(b) shows that the situation changes when the 
ambient conditions are closer to the Widom line. Although the
differences in $\mathcal{P}\left(\xi\right)$ at $x/D\approx5$ and $8$, influenced
by the potential core length, are similar to those in figures \ref{fig:SclPDF_Case_1_1T}
and \ref{fig:SclPDF_cases2T-4T}(a), significant differences are observed
between the downstream $\mathcal{P}\left(\xi\right)$ profiles (at $x/D\approx15$, 20 and
$30$) of Case 4/4T in figure \ref{fig:SclPDF_cases2T-4T}(b),
unlike Cases 2/2T. 
For Case 4, the $\mathcal{P}\left(\xi\right)$ peaks
are further away from the jet pure fluid concentration ($\xi=1$)
and are higher than in Case 4T. The peaks in a symmetric unimodal
p.d.f. coincide with the mean value, therefore, the differences in
p.d.f. peak location and magnitude in figure \ref{fig:SclPDF_cases2T-4T}(b)
mirrors the differences observed in mean scalar values
between Case 4 and Case 4T in figure \ref{fig:invScl_rmsFluc_Cases2n2T_4n4T}(a).\vspace{-1.0cm}

\noindent 

\noindent 

\paragraph{Summary \label{sub:summary_inflowEffects}}

\begin{figure}
\begin{centering}
\includegraphics[width=6.9cm]{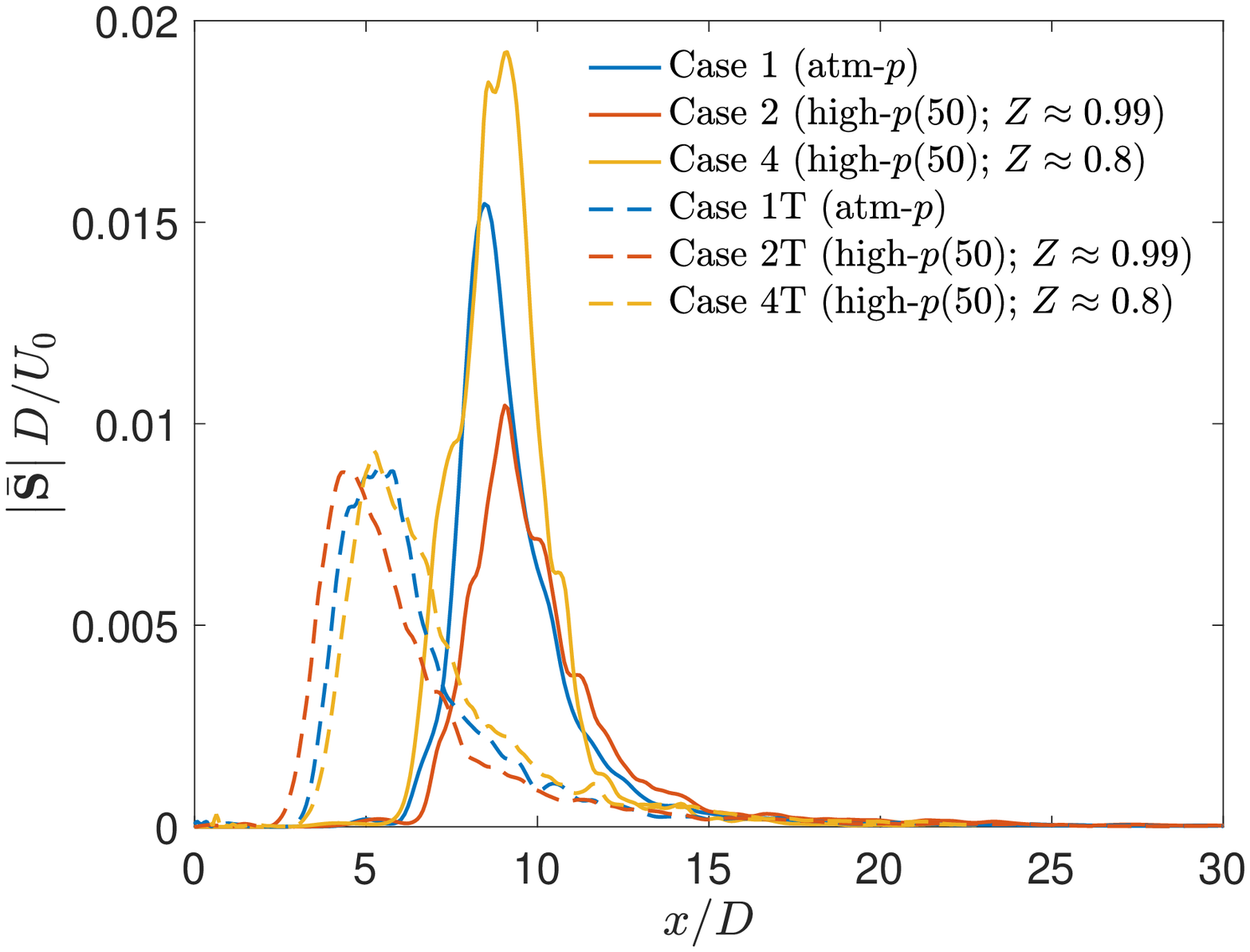}\includegraphics[width=6.9cm]{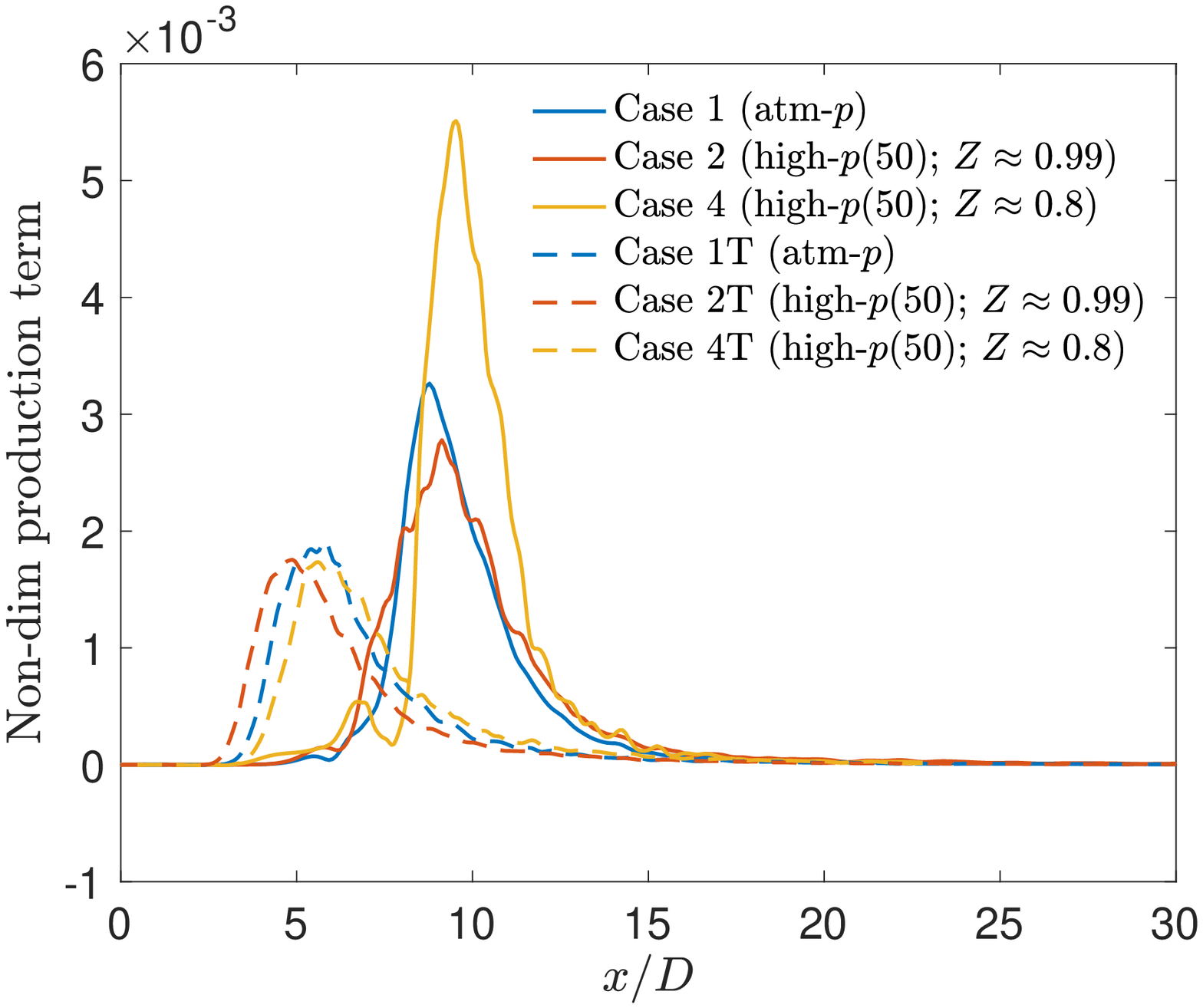}
\par\end{centering}

\begin{centering}
(a)\qquad{}\qquad{}\qquad{}\qquad{}\qquad{}\qquad{}\qquad{}\qquad{}\qquad{}\qquad{}\qquad{}\qquad{}(b)
\par\end{centering}

\caption{Centerline variation of (a) the strain rate magnitude, $\left|\bar{\mathbf{S}}\right|=\left(2\bar{S}_{ij}\bar{S}_{ij}\right)^{1/2}$,
normalized by $U_{0}/D$, where $\bar{S}_{ij}$ is the mean rate-of-strain tensor, and (b) t.k.e. production term $P=-\bar{\rho}\widetilde{u''_{i}u''_{j}}\frac{\partial\tilde{u}_{i}}{\partial x_{j}}$ normalized by $\rho_{e}$ (jet-exit density), $U_{0}$ and $D$, where tilde $\left(\tilde{\bullet}\right)$ denotes the Favre average,
$\tilde{\phi}=\overline{\rho\phi}/\bar{\rho}$, and $u''_{i}=u_{i}-\tilde{u}_{i}$. \label{fig:Str_rate_mag_highPlamTurb}}
\vspace{-0.25cm}
\end{figure}

To summarize the above results, the variation of 
$p'_{c,\textrm{rms}}/p_{c}$ and $\rho'_{c,\textrm{rms}}/\rho_{c}$ with $p_{\infty}$ and $Z$ 
is similar for both inflows (containing different perturbations but same $U_e$), as shown in figures \ref{fig:centerline_pfluc_lamTurb}(a)
and (b), and the variation can be estimated from the relative values of $p_{\infty}\left(\beta_{T}-1/p_{\infty}\right)$ for various cases.
However, 
$u'_{c,\textrm{rms}}/U_{c}$ shown in figure \ref{fig:centerline_urms_lamTurb}(a)
varies differently for the two inflows.
With decrease in $Z$, the proximity to the Widom line
increases $u'_{c,\textrm{rms}}/U_{c}$ in the laminar inflow jet, whereas
the same decrease in $Z$ has minimal influence on $u'_{c,\textrm{rms}}/U_{c}$ in
pseudo-turbulent inflow jet. This suggests that the real-gas effects, quantified by
$p_{\infty}\left(\beta_{T}-1/p_{\infty}\right)$, are strongly
felt in the laminar inflow jets that contain large-scale coherent structures which generate
regions of high mean strain rates, as shown in figure \ref{fig:Str_rate_mag_highPlamTurb}(a),
whereas the same real-gas effects minimally influence pseudo-turbulent inflow jets.
These results also showed that the 
effects of thermodynamic compressibility on jet flow dynamics cannot be directly
determined from the variation of pressure fluctuations, instead fluctuating pressure-velocity correlation and third-order velocity moments (that determine t.k.e.
diffusion fluxes from turbulent transport) are better correlated with the behavior of the velocity field. 

Studies of dynamic compressibility \cite[e.g.][]{freund2000compressibility} divide compressibility modeling efforts into \textit{explicit} and \textit{implicit} approaches. An \textit{explicit} approach targets modeling of the dilatational terms in the t.k.e. equation with the assumption that the turbulence characteristics are largely influenced by the compressible terms. In contrast, the \textit{implicit} approaches assume that compressibility influences the structure of the turbulence, which in turn changes the turbulence energetics reflected in the t.k.e. production and turbulence transport terms. In this study, the behavior of the t.k.e. production term, plotted in figure \ref{fig:Str_rate_mag_highPlamTurb}(b), and the turbulence transport terms, discussed above, follow the behavior of $u'_{c,\textrm{rms}}/U_{c}$, which suggests that \textit{implicit} modeling may be the physically correct approach for modeling of thermodynamic compressibility effects.\vspace{-0.5cm}
%
\noindent 

\section{Conclusions\label{sec:Conclusions}}

Turbulent round-jet DNS at perfect gas (subcritical) and real gas
(supercritical) conditions were performed to compare and contrast the
effects of thermodynamic and inflow condition on flow development and 
passive mixing. In all cases, the Reynolds number, $Re_{D}$, is $5000$. 
The equation of state and transport coefficient models were chosen to 
accurately represent the flow conditions. For supercritical pressures, 
the transport coefficient models were validated with the NIST database.
The thermodynamic conditions are characterized in terms of the ambient fluid
compressibility factor, $Z$, and normalized isothermal compressibility, 
$p_{\infty}\left(\beta_{T}-1/p_{\infty}\right)$.
The inflow conditions are characterized in terms of the jet-exit bulk velocity ($U_e$) or Mach 
number ($Ma_e$) and the perturbation type, namely, a laminar inflow 
(with top-hat velocity profile and small random perturbations) or a pseudo-turbulent inflow (with
statistics matching pipe-flow turbulence).

The first finding is that the mean axial velocity and Reynolds stress components 
attain self-similarity in high-$p$ fully-compressible jets, as they do in 
atmospheric-$p$ jets. However, the self-similar profiles depend on the thermodynamic and
inflow details, indicating that these details influence both the near- and far-field jet flow 
dynamics. The transition region of the flow, immediately downstream of the potential core closure, 
exhibits greater sensitivity to those details than
the self-similar region. Moreover, the laminar inflow jets show larger differences
with change in thermodynamic conditions than the pseudo-turbulent inflow jets. The laminar inflow jets exhibit
dominant coherent structures that generate a transition region with large mean strain rates,
which enhance sensitivity to real-gas effects. These coherent structures also enhance the jet spread
in laminar inflow cases compared to pseudo-turbulent inflow cases, where the jet spreads by entrainment
through small-scale turbulence.

The second important finding is that the ambient thermodynamic conditions influence
flow behavior by modifying the structure of turbulence through production and transport
of the turbulent kinetic energy (t.k.e.). In jets with a fixed $U_e$ and inflow perturbation, the normalized 
pressure/density fluctuations vary according to the value of $p_{\infty}\left(\beta_{T}-1/p_{\infty}\right)$. 
Increasing $p_{\infty}\left(\beta_{T}-1/p_{\infty}\right)$
enhances the normalized pressure/density fluctuations in both the laminar inflow and 
the pseudo-turbulent inflow jets. However, the effect on velocity/scalar fluctuations
depends on the extent of thermodynamic departure from perfect gas as well as the 
inflow perturbations. Large deviation of $Z$ from unity and of $p_{\infty}\left(\beta_{T}-1/p_{\infty}\right)$
from zero with laminar inflow 
triggers large normalized velocity/scalar fluctuations
resulting in greater mixing. However, a small departure from perfect gas 
does not change
the flow behavior significantly. Moreover, the effect of thermodynamic changes on
velocity/scalar statistics in pseudo-turbulent inflow jets is small, 
despite
significant variations in their pressure/density statistics. Velocity/scalar fluctuations 
that determine flow entrainment and mixing are not directly correlated with pressure fluctuations, 
but rather with the fluctuating pressure-velocity correlation that determines the turbulent transport 
in the t.k.e. equation due to pressure fluctuations. These
results show that the thermodynamic condition, expressed in terms of $Z$ or 
$p_{\infty}\left(  \beta_{T}-1/p_{\infty}\right)$, is insufficient to completely
determine the flow behavior that also depends on the inflow details. Thus, neither 
$Z$ nor $p_{\infty}\left(  \beta_{T}-1/p_{\infty}\right)$ are self-similarity parameters.

The third important finding is that at large departures from perfect gas, $U_e$ 
(from which the Mach number $Ma_e$ is calculated that determines the dynamic compressibility)
can strongly influence flow 
dynamics and mixing. The ambient speed of sound differs according to the thermodynamic 
condition and, hence, a fixed $U_e$ leads to different $Ma_e$ across various flow cases.
Variation of $Ma_e$ over a small subsonic range shows that an increase in $Ma_e$ with 
laminar inflow enhances velocity/scalar fluctuations leading to greater mixing. 
Variation of thermodynamic condition at a fixed $Ma_e$ 
shows that the pressure/density 
fluctuations differ according to the value of $Z$; smaller $Z$ leads to larger fluctuations. 
The velocity/scalar flucutations in high-$p$ cases also vary according to the $Z$
value. A detailed study of the effects of $Ma_e$ is beyond the scope of this study and will 
be a subject of future investigation.

The jet-flow regions and metrics sensitive to $p_{\infty}$, $Z$ 
and inflow condition, identified in this study, may guide high-$p$ experimental studies in obtaining measurements that may serve as
databases for simulation and model validation. For example, the transition region of laminar-inflow jets exhibits high sensitivity to $Z$, and thus high-resolution measurements in that region may help in evaluating high-$p$ model robustness and accuracy.
The results from this study also demonstrate that high-$p$ jet experiments performed at a fixed $Z$ 
cannot be used to infer results for jets at same $Z$ but larger $p_{\infty}$.
Moreover, the larger velocity/scalar fluctuations when the Widom line is approached in laminar inflow
jets suggest that fuel-oxidizer mixing can be enhanced by
combustor thermodynamic conditions closer to the Widom line with laminar fuel
injection. Similarly, the correlation between the
normalized pressure/density fluctuations and $p_{\infty}\left(  \beta_{T}-1/p_{\infty}\right)$ could be used to predict the thermodynamic
fluctuations in a flow at real-gas thermodynamic
conditions, without performing the flow simulations.

\section*{Acknowledgements}

This work was supported at the California Institute of Technology by the Army
Research Office under the direction of Dr. Ralph Anthenien. The computational
resources were provided by the NASA Advanced Supercomputing at Ames Research
Center under the $\mathrm{T^{3}}$ program directed by Dr. Michael Rogers.

\appendix

\section{Validation of the EOS and transport properties for high-$p$ simulations \label{sec:validation_EoS_transport}}

\begin{figure}
\begin{centering}
\includegraphics[width=6.9cm]{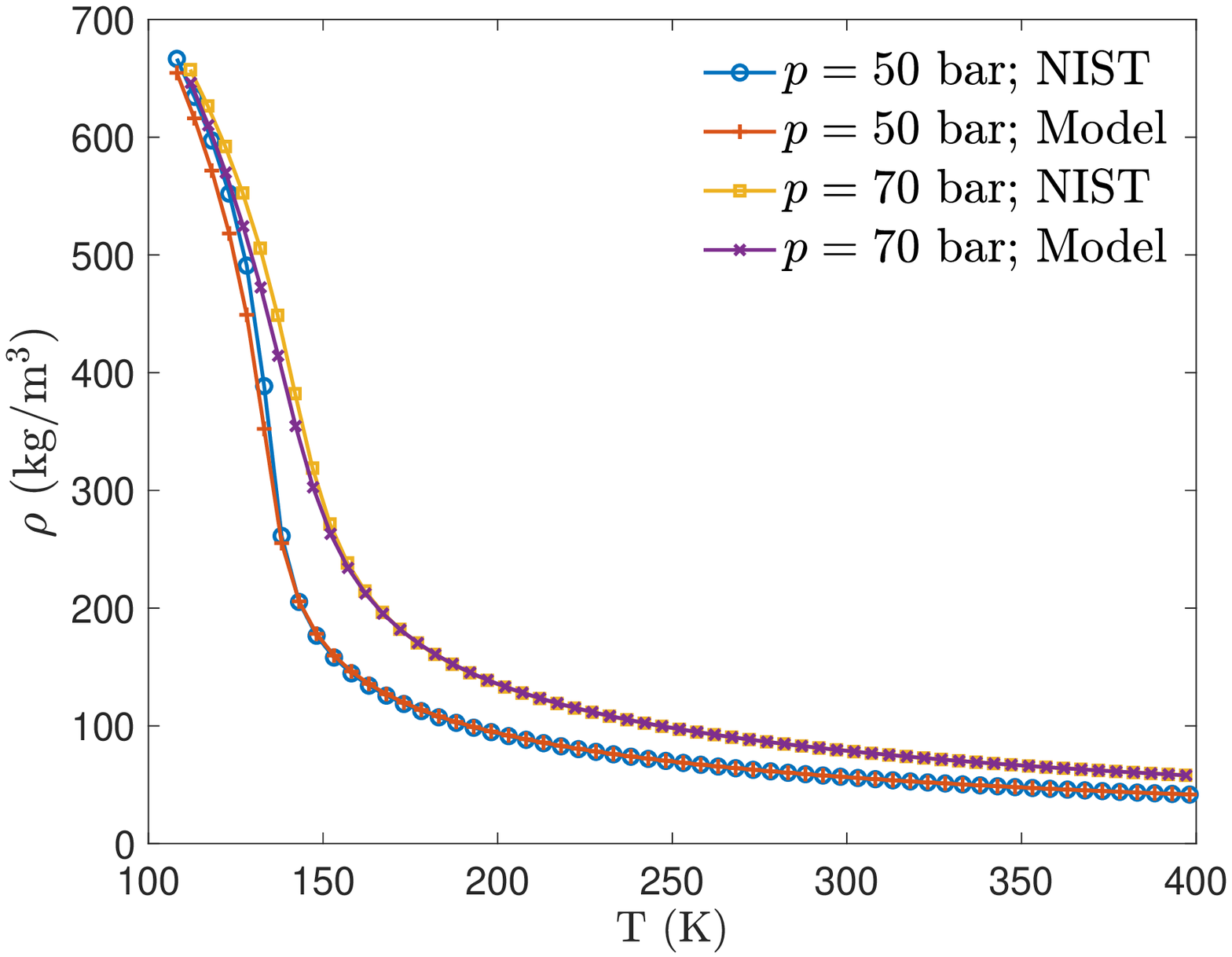}\includegraphics[width=6.9cm]{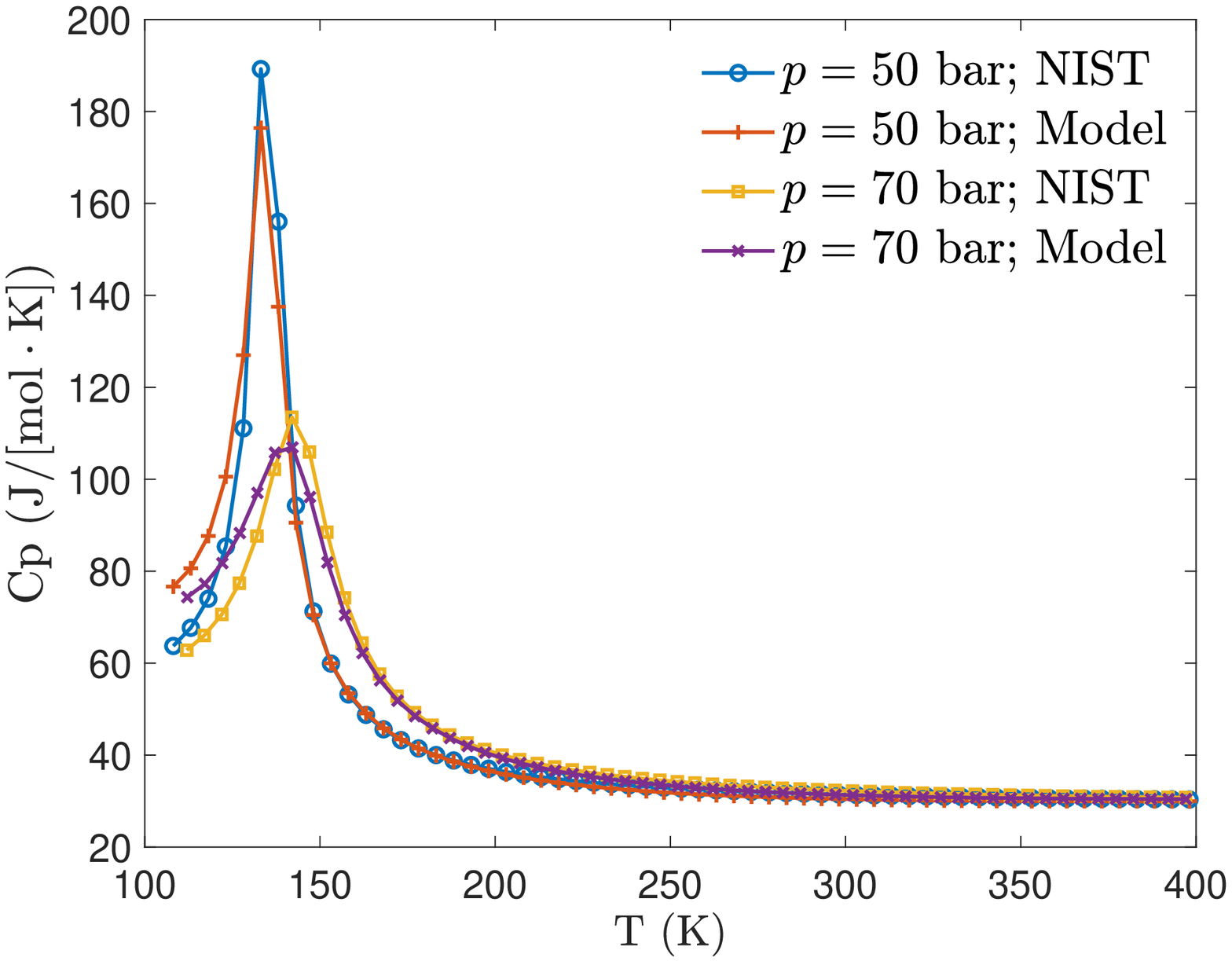}
\par\end{centering}

\begin{centering}
\qquad{}\qquad{}(a)\qquad{}\qquad{}\qquad{}\qquad{}\qquad{}\qquad{}\qquad{}\qquad{}\qquad{}\qquad{}(b)
\par\end{centering}

\begin{centering}
\includegraphics[width=6.9cm]{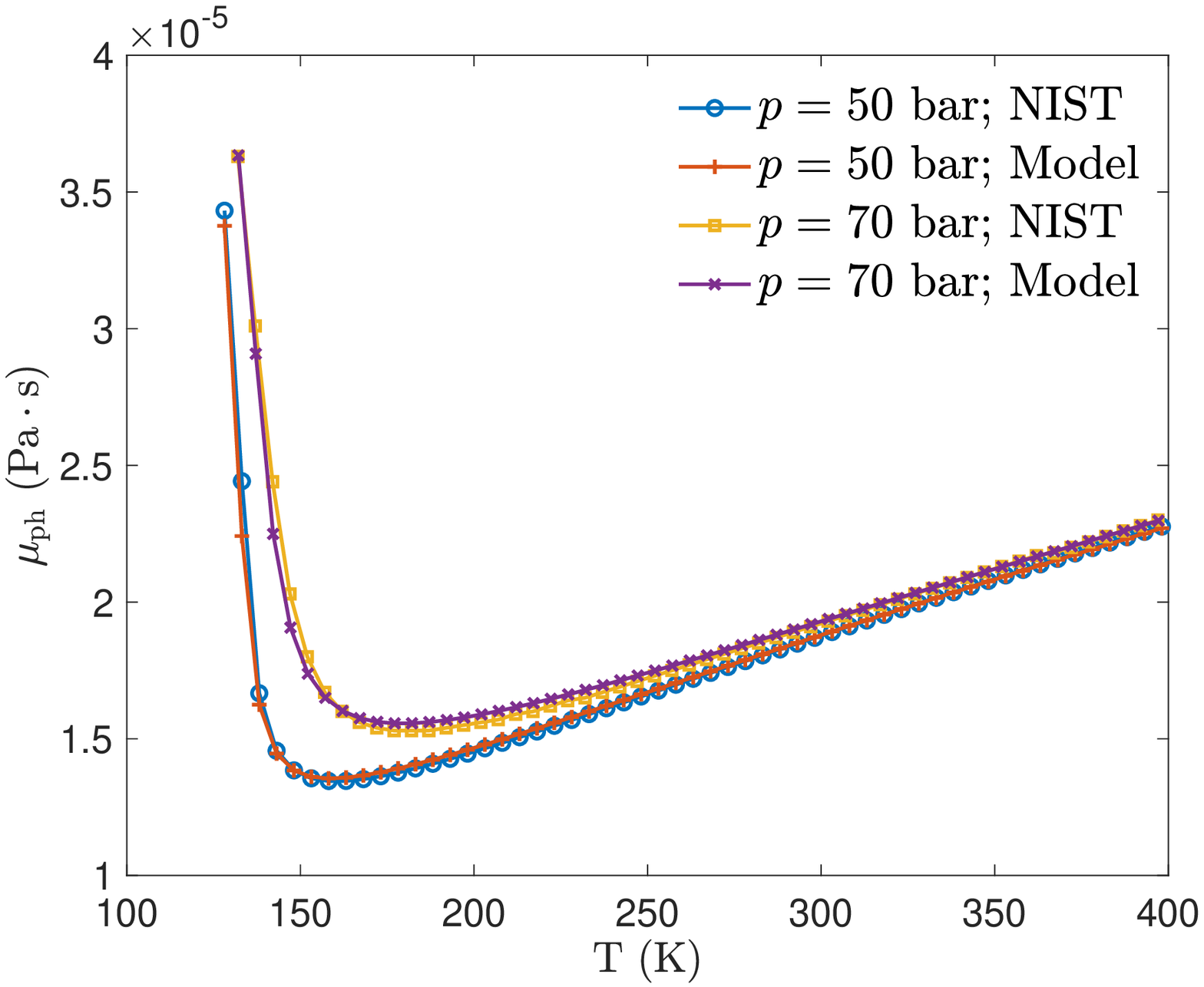}\includegraphics[width=6.9cm]{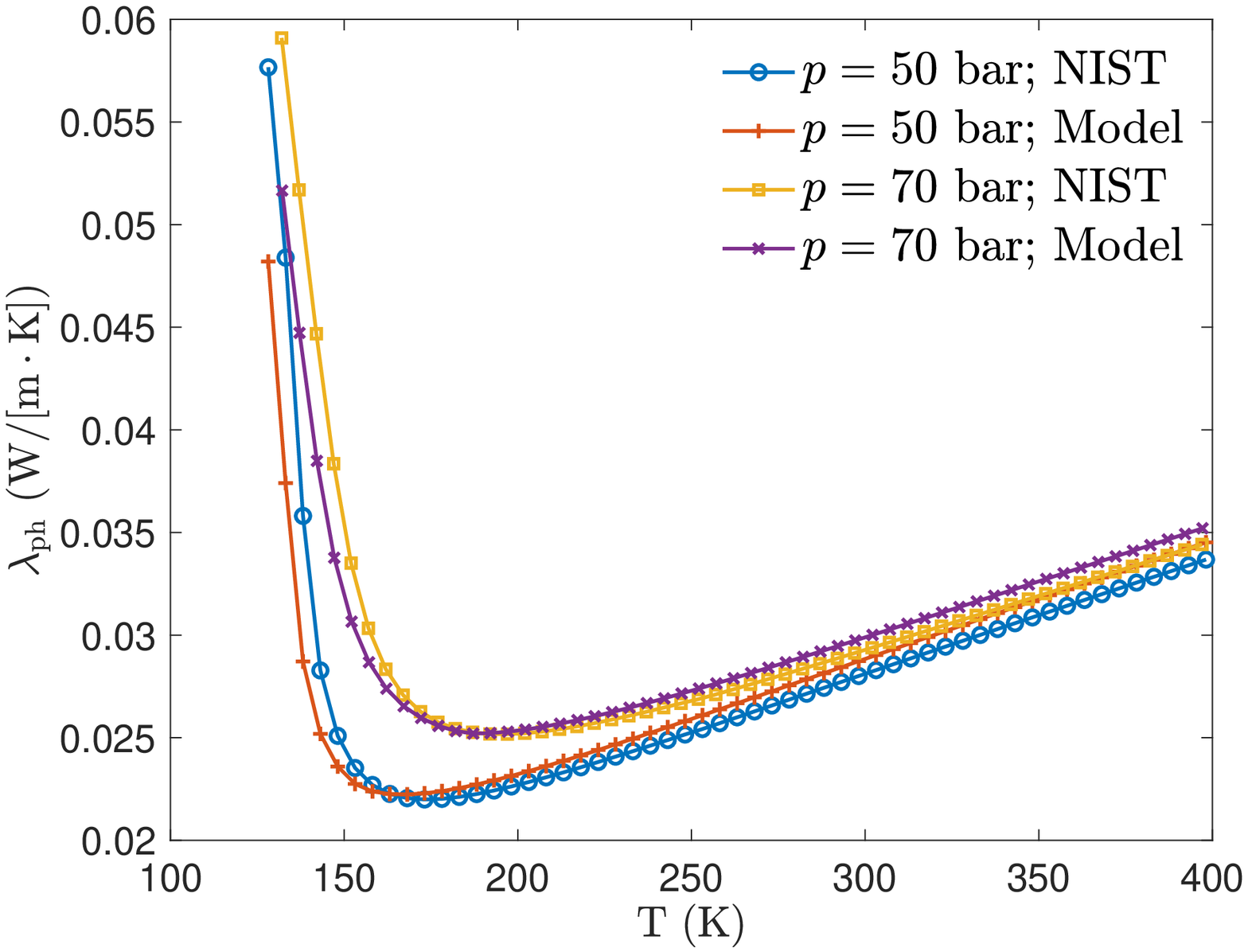}
\par\end{centering}

\begin{centering}
\qquad{}\qquad{}(c)\qquad{}\qquad{}\qquad{}\qquad{}\qquad{}\qquad{}\qquad{}\qquad{}\qquad{}\qquad{}(d)
\par\end{centering}

\caption{EOS and transport coefficients model comparison against NIST database
for pure Nitrogen at $50$ bar pressure. (a) Density, (b) Isobaric
heat capacity, (c) Viscosity, and (d) Thermal conductivity.\label{fig:NIST_vs_code_model}}
\end{figure}

To examine the robustness of the PR EOS and transport coefficient
models at supercritical conditions, figure \ref{fig:NIST_vs_code_model}
illustrates the density, isobaric heat capacity, and the transport coefficients
$\mu_{\mathrm{ph}}$ and $\lambda_{\mathrm{ph}}$ calculated from the
models described in \S \ref{sec:transport_properties}. The calculated values are compared against the National Institute of Standards and Technology
(NIST) database \cite[]{lemmon2010nist} for $\mathrm{N_{2}}$ at $p=50$ and $70$ bar in
the $T$ range 100 K - 400 K, which includes the critical temperature $T_{cr}%
=$126.2 K of N$_{2}$. As evident,
the models have good agreement with the NIST database, showing their
validity at high-$p$ conditions. The transport
coefficient models are accurate only for $T>T_{cr}$ and, thus, the comparison
of $\mu_{\mathrm{ph}}$ and $\lambda_{\mathrm{ph}}$ only spans this range.\vspace{-0.75cm}

\noindent 

\section{Grid convergence\label{sec:Grid-resolution}}
\vspace{-0.5cm}
\noindent 
\begin{figure}
\begin{centering}
\includegraphics[width=6.9cm]{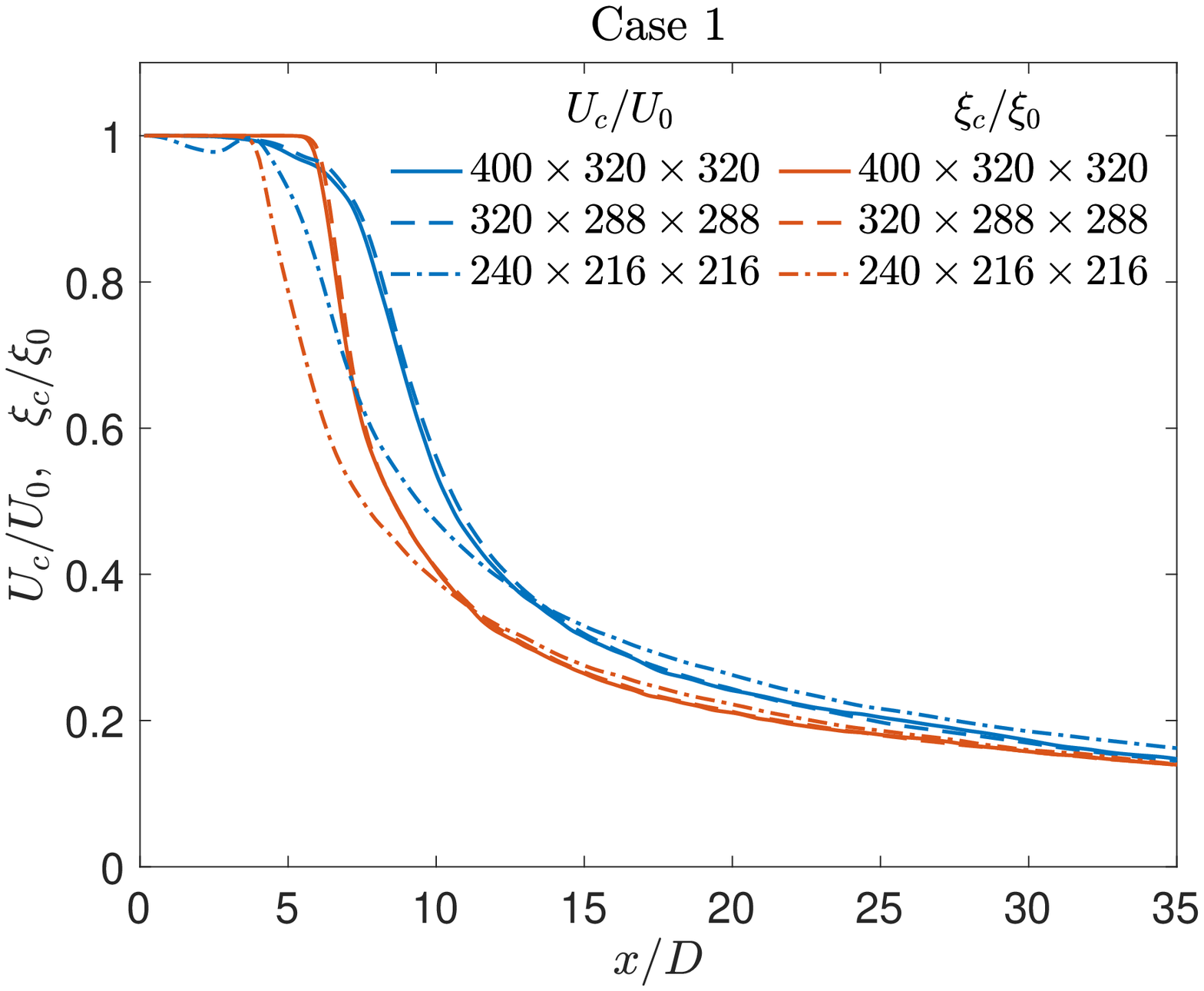}\includegraphics[width=6.9cm]{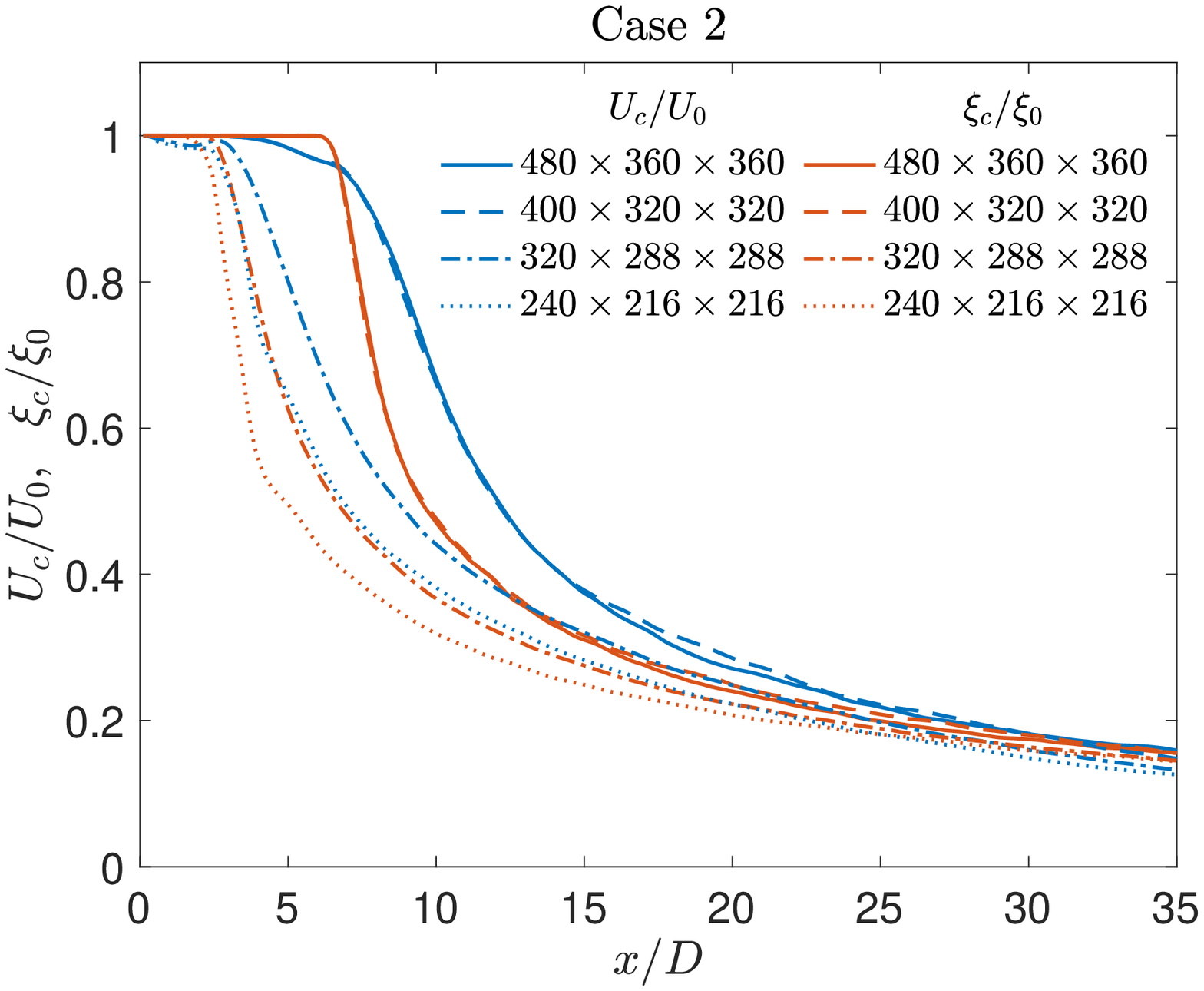}
\par\end{centering}

\begin{centering}
(a)\qquad{}\qquad{}\qquad{}\qquad{}\qquad{}\qquad{}\qquad{}\qquad{}\qquad{}\qquad{}\qquad{}\qquad{}(b)
\par\end{centering}

\begin{centering}
\includegraphics[width=6.9cm]{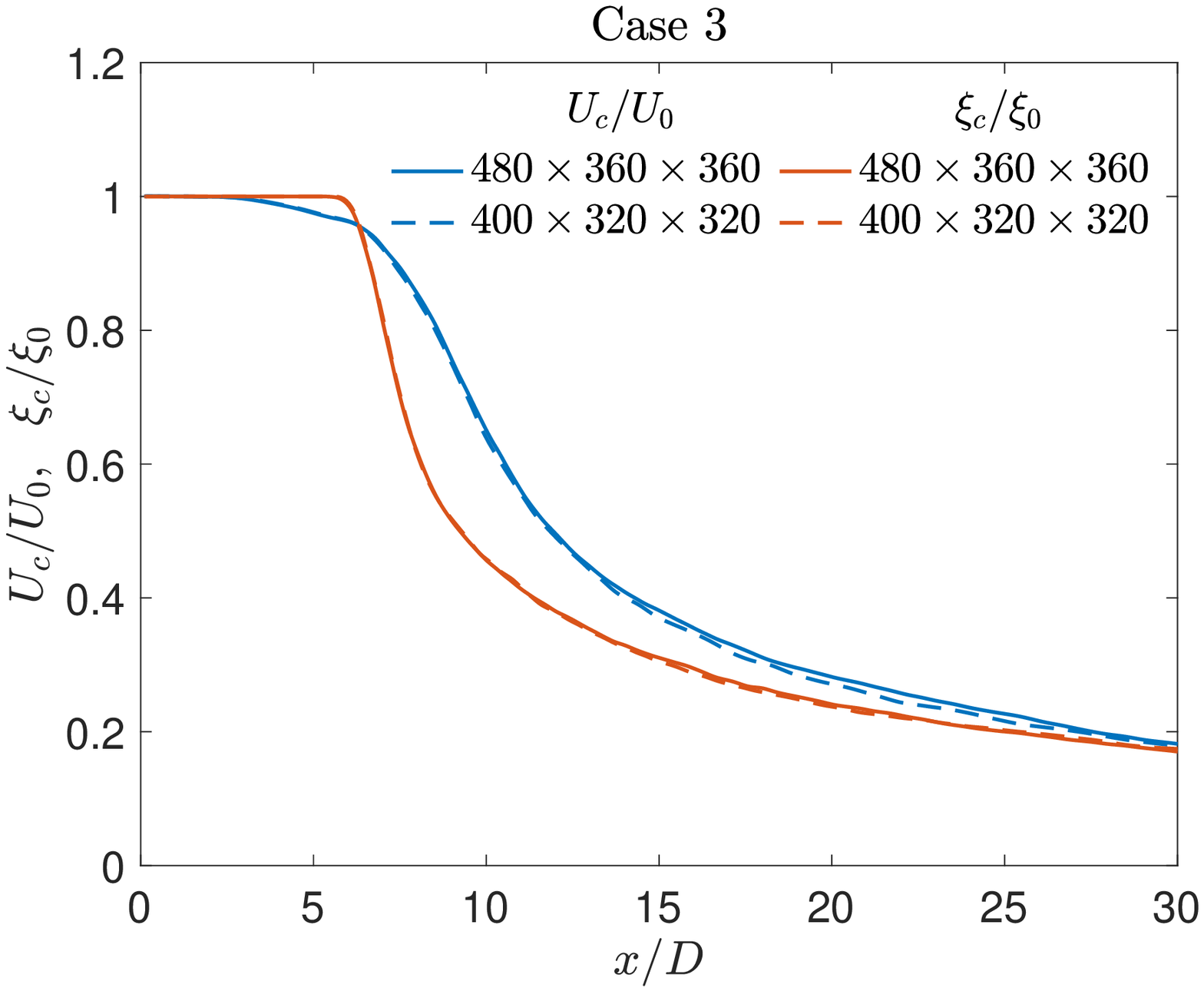}\includegraphics[width=6.9cm]{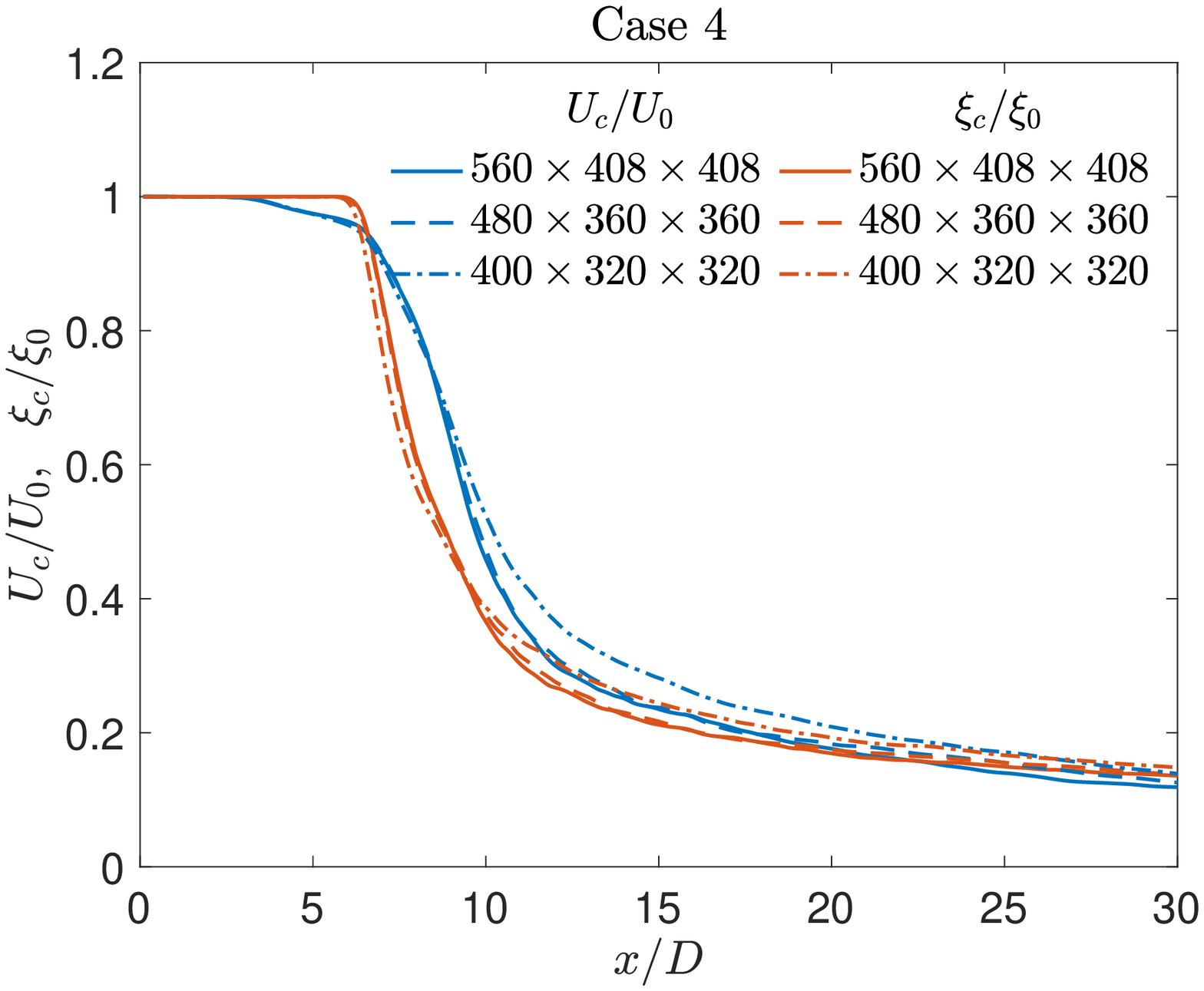}
\par\end{centering}

\begin{centering}
(c)\qquad{}\qquad{}\qquad{}\qquad{}\qquad{}\qquad{}\qquad{}\qquad{}\qquad{}\qquad{}\qquad{}\qquad{}(d)
\par\end{centering}

\begin{centering}
\includegraphics[width=6.9cm]{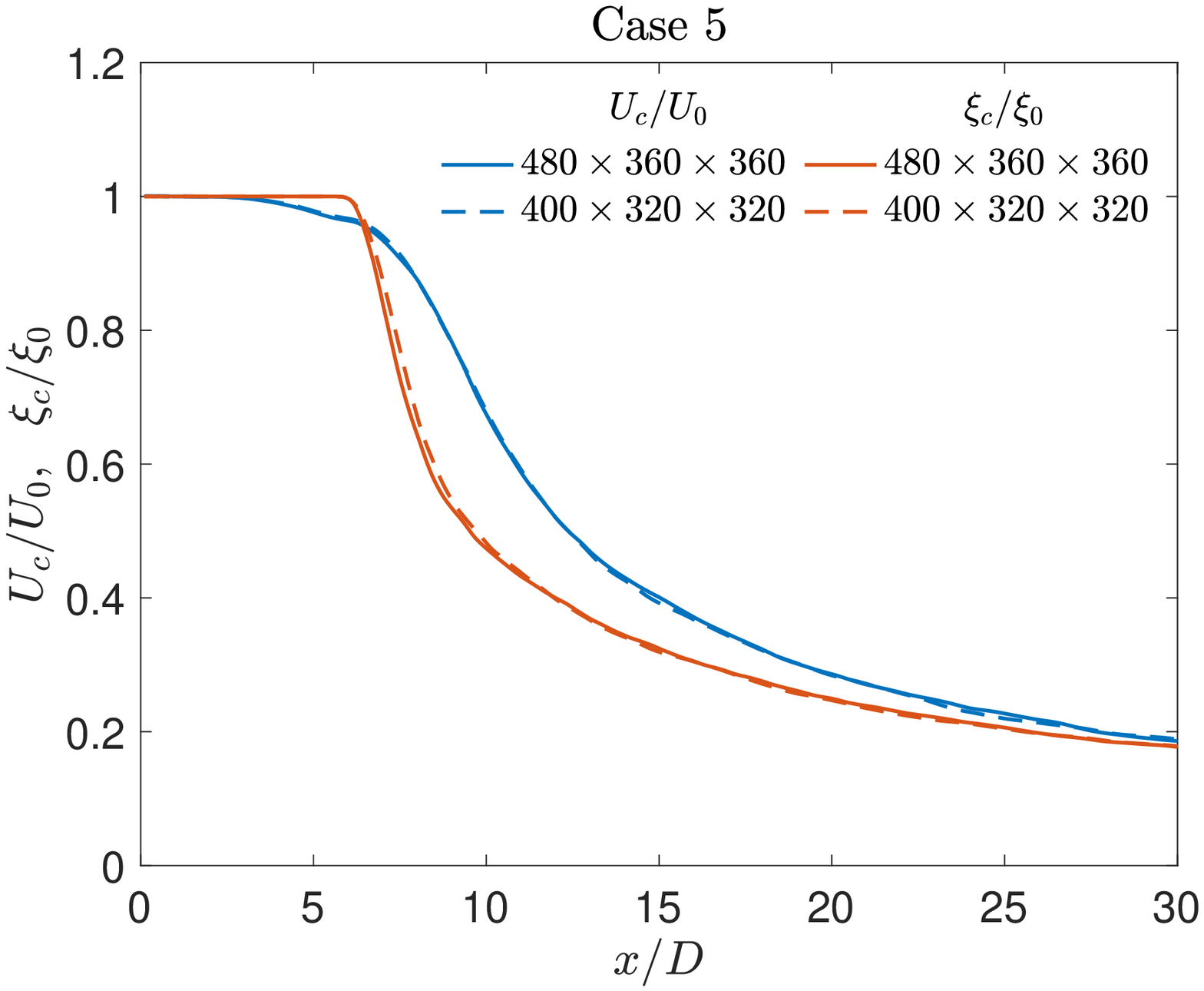}
\par\end{centering}

\begin{centering}
(e)
\par\end{centering}

\caption{Time-averaged centerline velocity ($U_{c}$) and scalar ($\xi_{c}$)
values normalized by the jet exit values $U_{0}$ and $\xi_{0}$ as
a function of axial distance for various grid resolutions. (a) Case
1, (b) Case 2, (c) Case 3, (d) Case 4, and (e) Case 5.\label{fig:grid_convergence}}
\vspace{-0.5cm}
\end{figure}

An estimate of the Kolmogorov length scale in terms of $Re_{D}$, which 
may help determine the optimal grid spacing for DNS, can be made
for incompressible (or weakly compressible) jet flows \cite[]{sharan2019numerical}. 
However, the estimates for incompressible flows may not necessarily
apply in high-pressure regimes of interest here. Therefore, to ensure
grid convergence, mean flow statistics are compared in this section
by successively refining the grid for Cases 1 to 5 (see table \ref{tab:Summary_of_cases}),
which ensures sufficient grid resolution for DNS.

Figures \ref{fig:grid_convergence}(a) to (e) compare the time-averaged
centerline velocity, $U_{c}$, and scalar concentration, $\xi_{c}$,
normalized by the jet exit values $U_{0}$ and $\xi_{0}$ as a function
of axial distance from simulations of Cases 1 to 5 with various grid
resolutions. Statistics for Case 1 at atmospheric conditions converges
at a resolution of $320\times288\times288$, whereas, for high-pressure
cases, they converge around $400\times320\times320$, except Case
4 at $Z=0.8$ having maximum deviation from perfect gas among all
cases considered that shows convergence around $480\times360\times360$.
The plots also show that in all cases the scalar concentration begins
to decay upstream of the velocity and at a faster rate than the velocity,
consistent with the observation of \cite[see Figure 6]{lubbers2001simulation}
for a passive scalar diffusing at unity Schmidt number.\vspace{-0.5cm}

\section{Validation of perfect-gas simulation: Case 1 results \label{sub:Case-1-results}}


Quantitative experimental data for supercritical jets are rare, however,
numerous measurements of high-order statistics have been made for
jets at atmospheric conditions. For comparisons,
the experimental measurements made in the self-similar region of density-matched
jets, where the jet/chamber density ratio is approximately unity, are considered here,
e.g. the velocity measurements of \cite{wygnanski1969some}, \cite{panchapakesan1993turbulence}
and \cite{hussein1994velocity}, and the passive scalar measurements of
\cite{ebrahimi1977konzentrationsfelder}, \cite{dowling1990similarity}
and \cite{mi2001influence}.

\begin{figure}
\begin{centering}
\includegraphics[width=6.9cm]{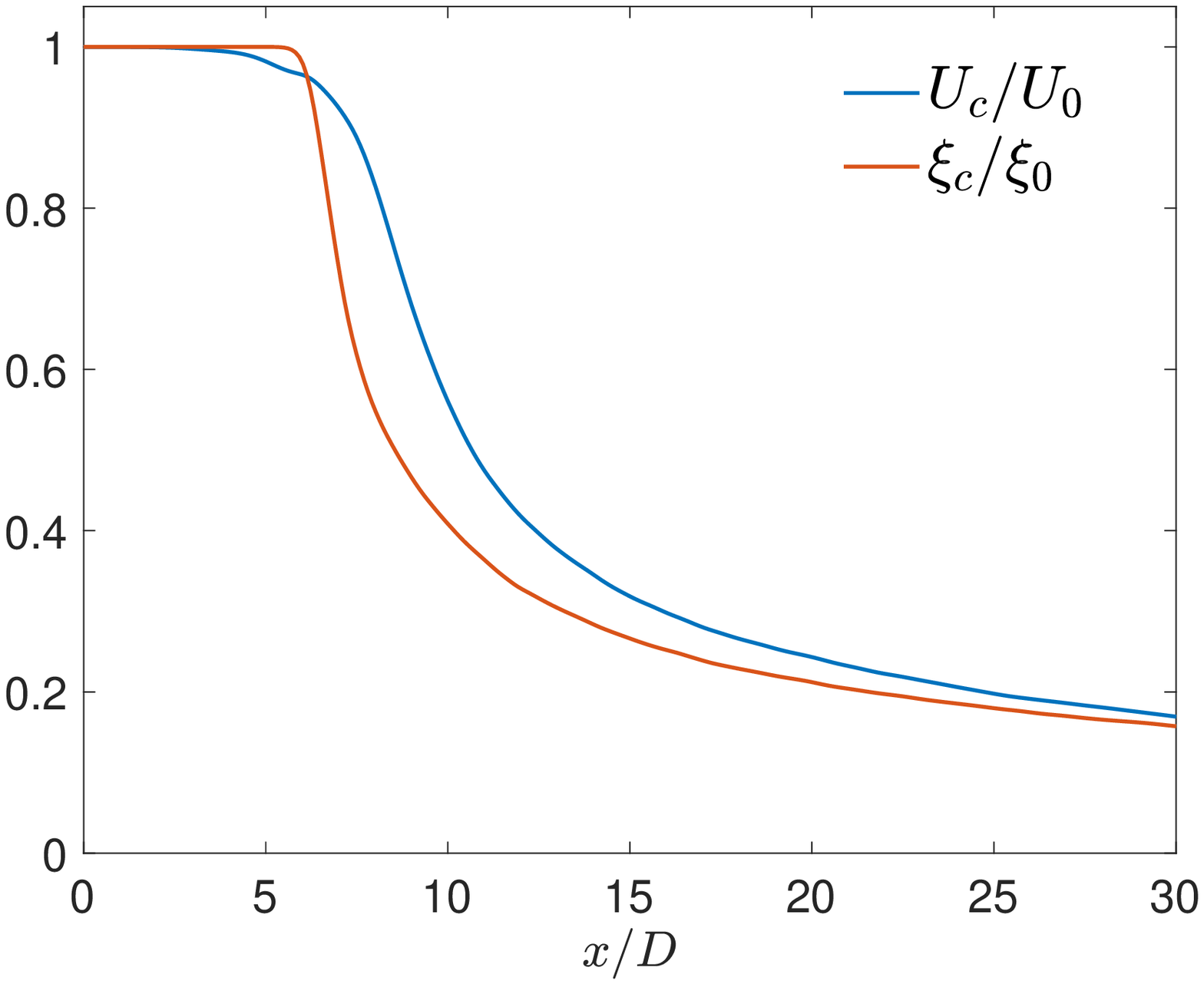}\includegraphics[width=6.9cm]{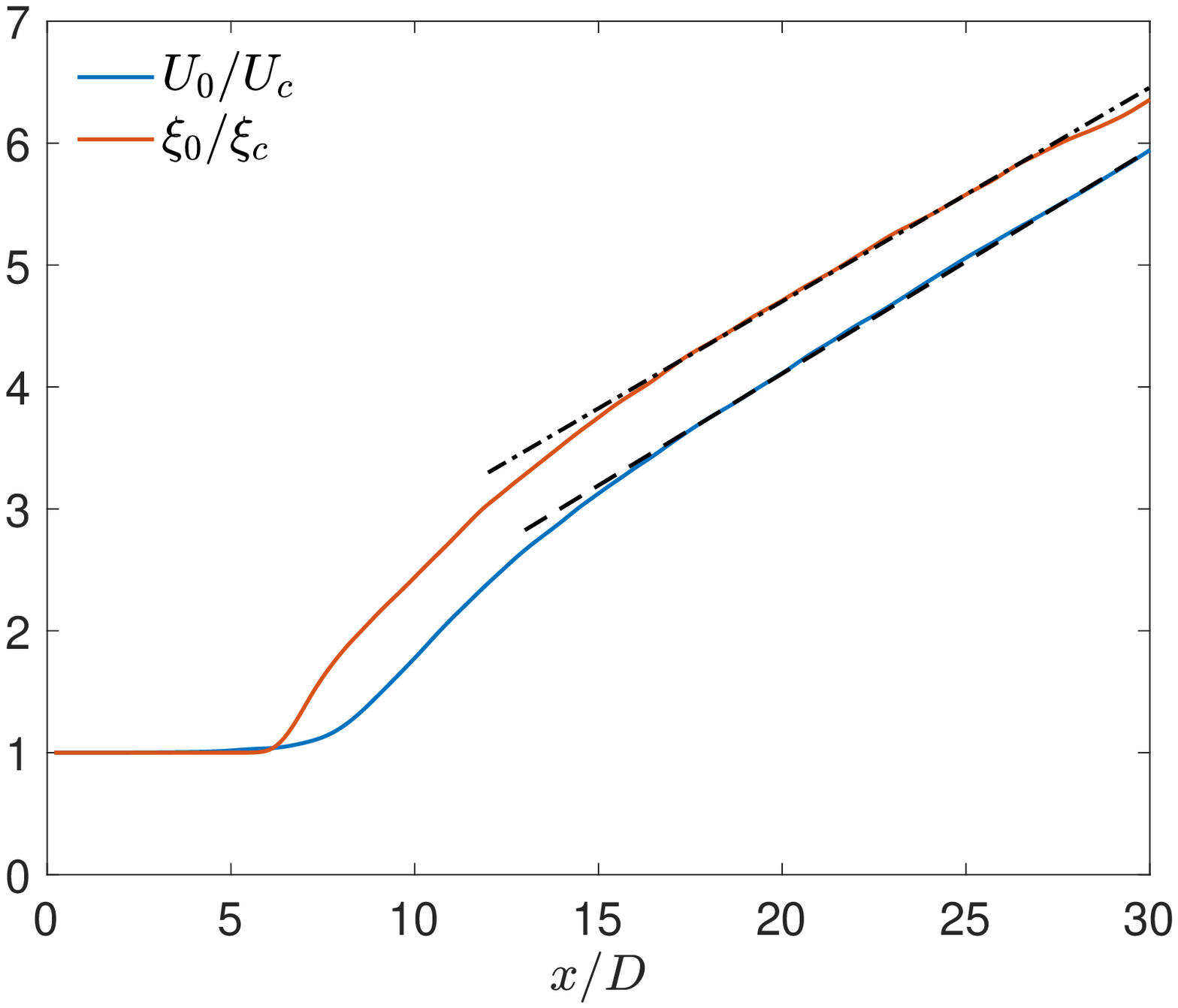}
\par\end{centering}

\begin{centering}
(a)\qquad{}\qquad{}\qquad{}\qquad{}\qquad{}\qquad{}\qquad{}\qquad{}\qquad{}\qquad{}\qquad{}\qquad{}(b)
\par\end{centering}

\caption{Case 1 results: Streamwise variation of the (a) centerline mean velocity
($U_{c}$) and scalar concentration ($\xi_{c}$) normalized by the
respective jet-exit centerline mean values, $U_{0}$ and $\xi_{0}$,
as a function of axial distance and (b) inverse of the normalized
time-averaged centerline values showing linear decay asymptotically
with axial distance. The dashed line uses $B_{u}=5.5$, $x_{0u}=-2.4D$
in (\ref{eq:centerline_vel}) and the dash-dotted line uses $B_{\xi}=5.7$,
$x_{0\xi}=-6.8D$ in (\ref{eq:centerline_scalar}). \label{fig:atmP_centerL_vel_expt_comp}}
\end{figure}

Figure \ref{fig:atmP_centerL_vel_expt_comp} shows the decay of the time-averaged
centerline velocity, $U_{c}(x)$, and centerline scalar concentration,
$\xi_{c}(x)$, normalized by the jet-exit centerline velocity, $U_{0}$ $\left(=U_{c}(0)\right)$,
and scalar concentration, $\xi_{0}$ $\left(=\xi_{c}(0)\right)$, respectively, for Case 1. In
the self-similar region, $U_{c}$ varies with the reciprocal of the
downstream distance, given by \cite[e.g.][]{hussein1994velocity}
\begin{equation}
\frac{U_{c}\left(x\right)}{U_{0}}=B_{u}\left(\frac{D}{x-x_{0u}}\right),\label{eq:centerline_vel}
\end{equation}
where $B_{u}$ is a constant and $x_{0u}$ denotes the virtual origin
derived from the centerline axial velocity. Similarly, the time-averaged
centerline scalar concentration has the form

\begin{equation}
\frac{\xi_{c}\left(x\right)}{\xi_{0}}=B_{\xi}\left(\frac{D}{x-x_{0\xi}}\right),\label{eq:centerline_scalar}
\end{equation}
where $B_{\xi}$ is a constant and $x_{0\xi}$ denotes the virtual
origin derived from the centerline scalar variation. Simultaneous
measurements of the velocity and scalar field in a single experiment
were not found in literature and, therefore, data from different studies
are used for comparing the velocity and scalar fields. The dashed
and dash-dotted lines in figure \ref{fig:atmP_centerL_vel_expt_comp}(b)
show the profiles for $B_{u}=5.5$ and $B_{\xi}=5.7$, respectively.
They are comparable to the experimentally observed values of $B_{u}=5.7$,
$6.06$ and $5.8$ by \cite{wygnanski1969some}, \cite{panchapakesan1993turbulence}
and \cite{hussein1994velocity}, respectively, and of $B_{\xi}=5.78$
and $5.11$ by \cite{ebrahimi1977konzentrationsfelder} and \cite{dowling1990similarity},
respectively. \vspace{-0.5cm}
\noindent 

\noindent 
\begin{figure}
\begin{centering}
\includegraphics[width=11cm]{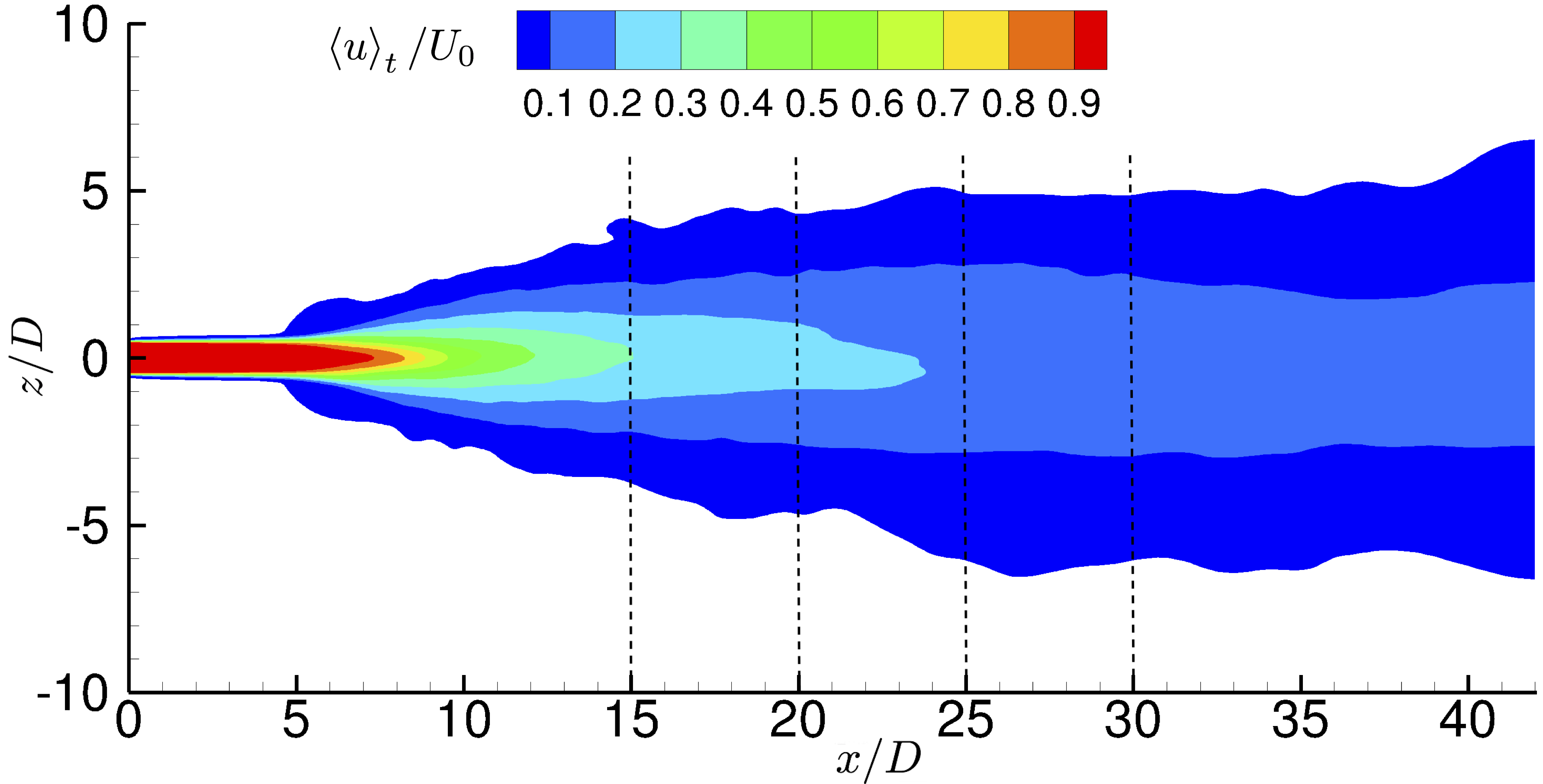}
\par\end{centering}

\caption{Time-averaged axial velocity contours for Case 1. The dashed lines
show axial locations where statistics are azimuthally averaged. \label{fig:axial_loc_for_avg}}
\end{figure}

\noindent 
\begin{figure}
\begin{centering}
\includegraphics[width=6.9cm]{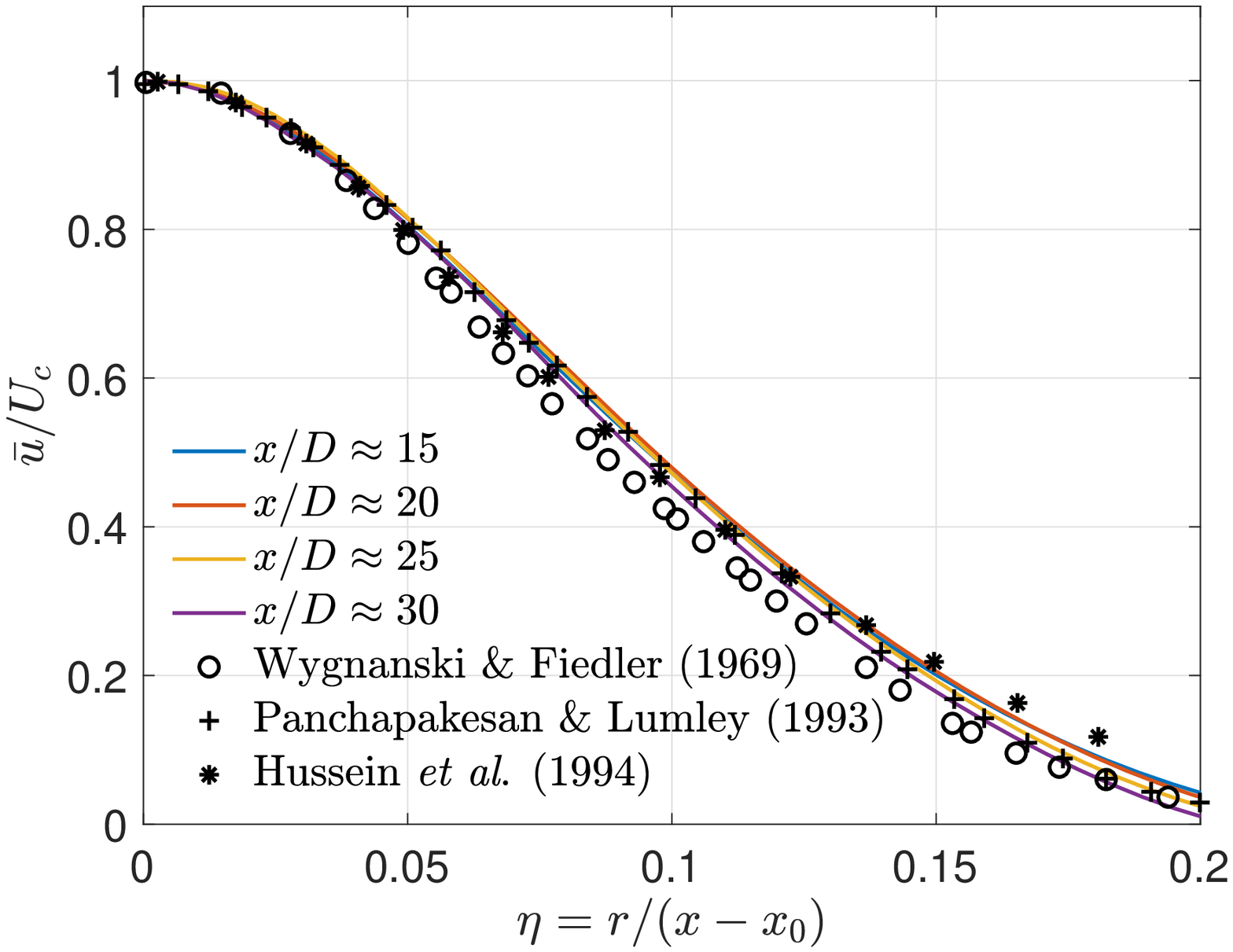}\includegraphics[width=6.9cm]{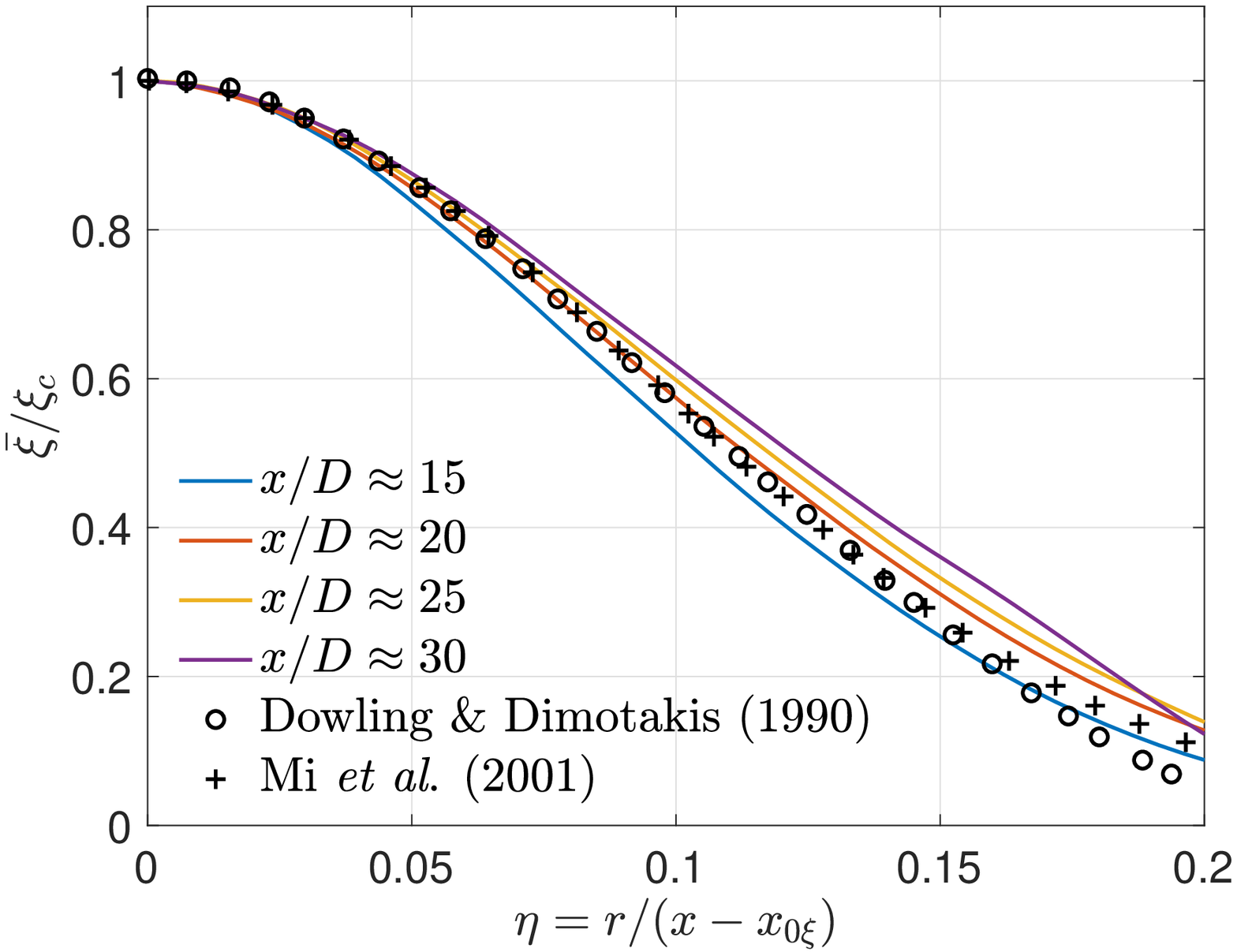}
\par\end{centering}

\begin{centering}
\qquad{}(a)\qquad{}\qquad{}\qquad{}\qquad{}\qquad{}\qquad{}\qquad{}\qquad{}\qquad{}\qquad{}\qquad{}(b)
\par\end{centering}

\caption{Case 1 results: (a) Mean axial velocity ($\bar{u}$) normalized by the
centerline mean velocity ($U_{c}$) and (b) mean scalar concentration ($\bar{\xi}$) normalized
by the centerline mean scalar value at various axial locations plotted
as a function of similarity coordinates. The velocity profiles are compared against the self-similar
profiles from the experiments of \cite{wygnanski1969some}, \cite{panchapakesan1993turbulence}
and \cite{hussein1994velocity}, whereas the scalar profiles are compared against the experimental profiles of \cite{dowling1990similarity} and
\cite{mi2001influence}. \label{fig:atmP_mean_axial_vel_w_expt}}
\end{figure}

Contours of $\left\langle u\right\rangle _{t}/U_{0}$ in the $y/D=0$ plane are
depicted in figure \ref{fig:axial_loc_for_avg}, with dashed lines showing the
axial locations where $\bar{u}/U_{c}$ and $\bar{\xi}/\xi_{c}$ are plotted in figure \ref{fig:atmP_mean_axial_vel_w_expt}(a) and (b), respectively. 
The azimuthal averages for calculations of $\bar{u}$ and $\bar{\xi}$ using (\ref{eq:average_defn}) 
are performed by interpolation of the time-averaged
Cartesian-grid solution to a polar grid at respective axial locations. 
The variation of $\bar{u}/U_{c}$ with the similarity
coordinate, $\eta=r/\left(  x-x_{0u}\right)$, is compared
with the self-similar profile from experiments in figure
\ref{fig:atmP_mean_axial_vel_w_expt}(a). The simulation
results at various axial locations agree well with each other
and with the experimental profiles, indicating that the mean axial
velocity becomes self-similar around $x/D\approx15$ for this case. $\bar{\xi}/\xi_{c}$
at various $x/D$ locations is compared against the self-similar profiles of 
\cite{dowling1990similarity} and \cite{mi2001influence} in figure \ref{fig:atmP_mean_axial_vel_w_expt}(b).
As evident from the figure, there are minor differences between the
profiles at various axial locations, suggesting that the mean scalar
concentration is not fully self-similar, but close to self-similarity
around $x/D\approx30$.

\begin{figure}
\begin{centering}
\includegraphics[width=6.9cm]{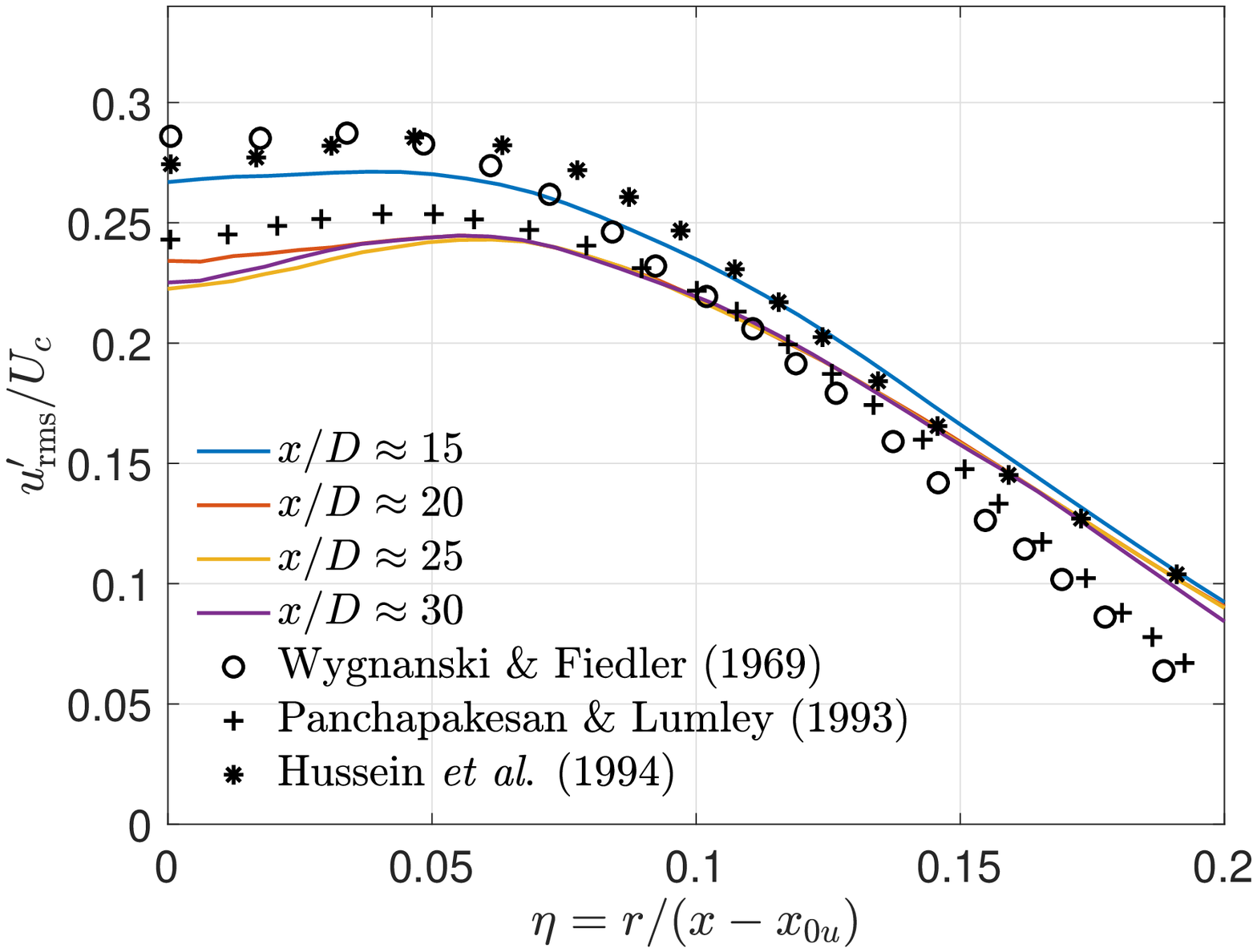}\includegraphics[width=6.9cm]{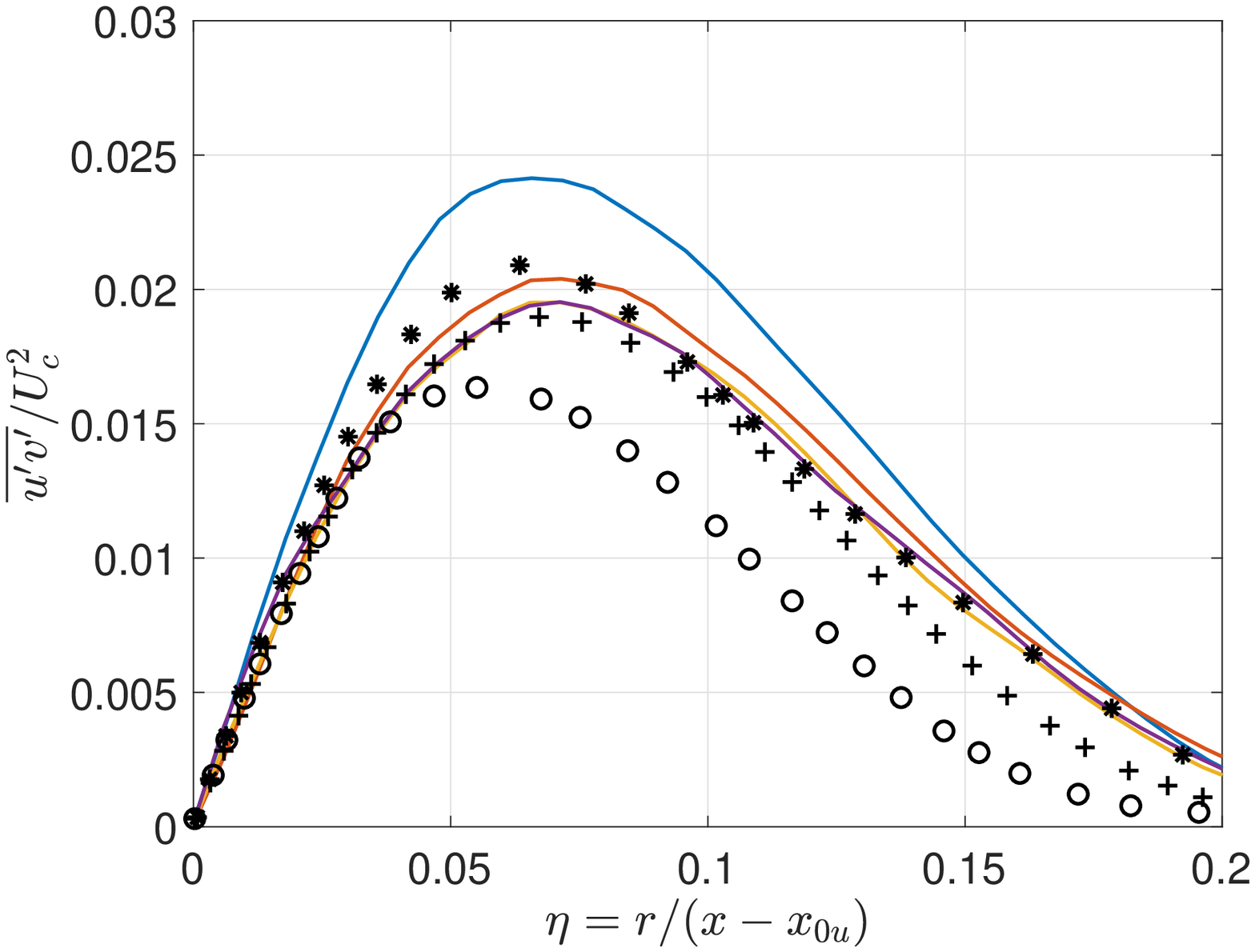}
\par\end{centering}

\begin{centering}
(a)\qquad{}\qquad{}\qquad{}\qquad{}\qquad{}\qquad{}\qquad{}\qquad{}\qquad{}\qquad{}\qquad{}\qquad{}(b)
\par\end{centering}



\caption{Case 1 results: (a) r.m.s. axial velocity fluctuations ($u_{\mathrm{rms}}^{'}$),
and (b) Reynolds stress ($\overline{u^{'}v^{'}}$) normalized by the
centerline mean velocity at various axial locations plotted as a function
of similarity coordinates compared against the self-similar profiles
from the experiments of \cite{wygnanski1969some}, \cite{panchapakesan1993turbulence}
and \cite{hussein1994velocity}. The legend is the same for both plots.
\label{fig:atmP_Rstress_w_expt}}
\vspace{-0.25cm}
\end{figure}

To further examine the velocity field, figure \ref{fig:atmP_Rstress_w_expt}(a) and (b) show the radial variation of normalized
root-mean-square (r.m.s.) axial-velocity fluctuation and Reynolds stress 
compared against the self-similar profile from experiments.
The simulation profiles 
at $x/D\approx25$ and $x/D\approx30$, in figure \ref{fig:atmP_Rstress_w_expt},
agree well with each other, indicating that these quantities attain
self-similarity downstream of $x/D\approx20$. The simulation self-similar profiles also lie
within the experimentally observed self-similar profiles of these
quantities.\vspace{-0.25cm}

\noindent 

The favorable comparisons between simulation results and experimental
measurements indicate that the governing equations with perturbed
laminar inflow and the numerical method accurately simulate the jet
exiting a smooth contracting nozzle at atmospheric-$p$. More comparisons
are not included here for brevity, but can be found in \cite{sharan2021direct}.


%
%
%
%
\vspace{-0.5cm}
\bibliographystyle{jfm}
\bibliography{main}

\end{document}